\documentclass[a4paper,11pt]{article}
\usepackage{jcappub} 
\usepackage{footnote}
\makesavenoteenv{table}
\usepackage{lineno}
\usepackage{subcaption}
\usepackage{multirow}
\usepackage{soul}
\usepackage{orcidlink}
\usepackage{hanging}
\usepackage{aas_macros}

\title{\boldmath The Linear Point Standard Ruler with DESI DR1 and DR2 Data}

\newcommand{\desilike}{{\tt desilike}}

\newcommand{\abacussummit}{{\tt AbacusSummit}}
\newcommand{\abacus}{{\tt Abacus}}
\newcommand{\abacust}{{\tt Abacus}-{\tt 2}}
\newcommand{\ezmocks}{{\tt EZmocks}}
\newcommand{\bgs}{{\tt BGS}}
\newcommand{\lrgelg}{{\tt LRG3}+{\tt ELG1}}
\newcommand{\elgo}{{\tt ELG1}}
\newcommand{\elgt}{{\tt ELG2}}
\newcommand{\elgs}{{\tt ELGs}}
\newcommand{\elg}{{\tt ELG}}
\newcommand{\lrg}{{\tt LRG}}
\newcommand{\lrgo}{{\tt LRG1}}
\newcommand{\lrgt}{{\tt LRG2}}
\newcommand{\lrgth}{{\tt LRG3}}
\newcommand{\lrgs}{{\tt LRGs}}

\newcommand{\qsos}{{\tt QSOs}}
\newcommand{\complete}{{\tt complete}}
\newcommand{\altmtl}{{\tt altmtl}}
\newcommand{\ffa}{{\tt ffa}}

\newcommand{\qisolp}{$\alpha_\mathrm{iso,LP}$}
\newcommand{\qisobao}{$\alpha_\mathrm{iso,BAO}$}

\newcommand{\sigmaper}{\Sigma_{\perp}}
\newcommand{\sigmapar}{\Sigma_{\parallel}}

\newcommand{\slp}{$s_\mathrm{LP}$}

\newcommand{\lcdm}{$\Lambda$CDM}
\newcommand \hMpc {$h^{-1}\mathrm{Mpc}$}

 \emailAdd{navya.uberoi@yale.edu}
 \emailAdd{farnik.nikakhtar@yale.edu}

\author[a]{{N.~Uberoi}\orcidlink{0000-0002-7517-9629},}
\author[a]{{F.~Nikakhtar}\orcidlink{0000-0002-3641-4366},}
\author[a]{{N.~Padmanabhan}\orcidlink{0000-0002-2885-8602},}
\author[b]{{R.~K.~Sheth}\orcidlink{0000-0002-2330-0917},}
\author[c]{{J.~Aguilar},}
\author[d]{{S.~Ahlen}\orcidlink{0000-0001-6098-7247},}
\author[e,f]{{D.~Bianchi}\orcidlink{0000-0001-9712-0006},}
\author[g]{{D.~Brooks},}
\author[h,i]{{F.~J.~Castander}\orcidlink{0000-0001-7316-4573},}
\author[c]{{T.~Claybaugh},}
\author[c]{{A.~Cuceu}\orcidlink{0000-0002-2169-0595},}
\author[j]{{A.~de la Macorra}\orcidlink{0000-0002-1769-1640},}
\author[k]{{Arjun~Dey}\orcidlink{0000-0002-4928-4003},}
\author[l,m]{{Biprateep~Dey}\orcidlink{0000-0002-5665-7912},}
\author[g]{{P.~Doel},}
\author[n,o]{{J.~E.~Forero-Romero}\orcidlink{0000-0002-2890-3725},}
\author[h,p,i]{{E.~Gaztañaga}\orcidlink{0000-0001-9632-0815},}
\author[c,q]{{S.~Gontcho A Gontcho}\orcidlink{0000-0003-3142-233X},}
\author[r]{{G.~Gutierrez},}
\author[s,t,u]{{K.~Honscheid}\orcidlink{0000-0002-6550-2023},}
\author[v]{{C.~Howlett}\orcidlink{0000-0002-1081-9410},}
\author[w]{{M.~Ishak}\orcidlink{0000-0002-6024-466X},}
\author[k]{{R.~Joyce}\orcidlink{0000-0003-0201-5241},}
\author[x]{{D.~Kirkby}\orcidlink{0000-0002-8828-5463},}
\author[c]{{T.~Kisner}\orcidlink{0000-0003-3510-7134},}
\author[g]{{O.~Lahav},}
\author[u]{{C.~Lamman}\orcidlink{0000-0002-6731-9329},}
\author[c]{{M.~Landriau}\orcidlink{0000-0003-1838-8528},}
\author[y]{{L.~Le~Guillou}\orcidlink{0000-0001-7178-8868},}
\author[z,aa]{{M.~Manera}\orcidlink{0000-0003-4962-8934},}
\author[s,ab,u]{{P.~Martini}\orcidlink{0000-0002-4279-4182},}
\author[k]{{A.~Meisner}\orcidlink{0000-0002-1125-7384},}
\author[ac,aa]{{R.~Miquel},}
\author[p]{{S.~Nadathur}\orcidlink{0000-0001-9070-3102},}
\author[ad,ae,af]{{W.~J.~Percival}\orcidlink{0000-0002-0644-5727},}
\author[c,ag,ah]{{C.~Poppett},}
\author[ai]{{F.~Prada}\orcidlink{0000-0001-7145-8674},}
\author[aj]{{I.~P\'erez-R\`afols}\orcidlink{0000-0001-6979-0125},}
\author[ak]{{G.~Rossi},}
\author[al,am,an]{{L.~Samushia}\orcidlink{0000-0002-1609-5687},}
\author[ao]{{E.~Sanchez}\orcidlink{0000-0002-9646-8198},}
\author[c]{{D.~Schlegel},}
\author[ap,aq]{{M.~Schubnell},}
\author[c]{{J.~Silber}\orcidlink{0000-0002-3461-0320},}
\author[k]{{D.~Sprayberry},}
\author[aq]{{G.~Tarl\'{e}}\orcidlink{0000-0003-1704-0781},}
\author[k]{{B.~A.~Weaver},}
\author[ar]{{H.~Zou}\orcidlink{0000-0002-6684-3997},}

\affiliation[a]{Physics Department, Yale University, P.O. Box 208120, New Haven, CT 06511, USA}
\affiliation[b]{Center for Particle Cosmology, University of Pennsylvania, PA 19104, USA}
\affiliation[c]{Lawrence Berkeley National Laboratory, 1 Cyclotron Road, Berkeley, CA 94720, USA}
\affiliation[d]{Department of Physics, Boston University, 590 Commonwealth Avenue, Boston, MA 02215 USA}
\affiliation[e]{Dipartimento di Fisica ``Aldo Pontremoli'', Universit\`a degli Studi di Milano, Via Celoria 16, I-20133 Milano, Italy}
\affiliation[f]{INAF-Osservatorio Astronomico di Brera, Via Brera 28, 20122 Milano, Italy}
\affiliation[g]{Department of Physics \& Astronomy, University College London, Gower Street, London, WC1E 6BT, UK}
\affiliation[h]{Institut d'Estudis Espacials de Catalunya (IEEC), c/ Esteve Terradas 1, Edifici RDIT, Campus PMT-UPC, 08860 Castelldefels, Spain}
\affiliation[i]{Institute of Space Sciences, ICE-CSIC, Campus UAB, Carrer de Can Magrans s/n, 08913 Bellaterra, Barcelona, Spain}
\affiliation[j]{Instituto de F\'{\i}sica, Universidad Nacional Aut\'{o}noma de M\'{e}xico,  Circuito de la Investigaci\'{o}n Cient\'{\i}fica, Ciudad Universitaria, Cd. de M\'{e}xico  C.~P.~04510,  M\'{e}xico}
\affiliation[k]{NSF NOIRLab, 950 N. Cherry Ave., Tucson, AZ 85719, USA}
\affiliation[l]{Department of Astronomy \& Astrophysics, University of Toronto, Toronto, ON M5S 3H4, Canada}
\affiliation[m]{Department of Physics \& Astronomy and Pittsburgh Particle Physics, Astrophysics, and Cosmology Center (PITT PACC), University of Pittsburgh, 3941 O'Hara Street, Pittsburgh, PA 15260, USA}
\affiliation[n]{Departamento de F\'isica, Universidad de los Andes, Cra. 1 No. 18A-10, Edificio Ip, CP 111711, Bogot\'a, Colombia}
\affiliation[o]{Observatorio Astron\'omico, Universidad de los Andes, Cra. 1 No. 18A-10, Edificio H, CP 111711 Bogot\'a, Colombia}
\affiliation[p]{Institute of Cosmology and Gravitation, University of Portsmouth, Dennis Sciama Building, Portsmouth, PO1 3FX, UK}
\affiliation[q]{University of Virginia, Department of Astronomy, Charlottesville, VA 22904, USA}
\affiliation[r]{Fermi National Accelerator Laboratory, PO Box 500, Batavia, IL 60510, USA}
\affiliation[s]{Center for Cosmology and AstroParticle Physics, The Ohio State University, 191 West Woodruff Avenue, Columbus, OH 43210, USA}
\affiliation[t]{Department of Physics, The Ohio State University, 191 West Woodruff Avenue, Columbus, OH 43210, USA}
\affiliation[u]{The Ohio State University, Columbus, 43210 OH, USA}
\affiliation[v]{School of Mathematics and Physics, University of Queensland, Brisbane, QLD 4072, Australia}
\affiliation[w]{Department of Physics, The University of Texas at Dallas, 800 W. Campbell Rd., Richardson, TX 75080, USA}
\affiliation[x]{Department of Physics and Astronomy, University of California, Irvine, 92697, USA}
\affiliation[y]{Sorbonne Universit\'{e}, CNRS/IN2P3, Laboratoire de Physique Nucl\'{e}aire et de Hautes Energies (LPNHE), FR-75005 Paris, France}
\affiliation[z]{Departament de F\'{i}sica, Serra H\'{u}nter, Universitat Aut\`{o}noma de Barcelona, 08193 Bellaterra (Barcelona), Spain}
\affiliation[aa]{Institut de F\'{i}sica d’Altes Energies (IFAE), The Barcelona Institute of Science and Technology, Edifici Cn, Campus UAB, 08193, Bellaterra (Barcelona), Spain}
\affiliation[ab]{Department of Astronomy, The Ohio State University, 4055 McPherson Laboratory, 140 W 18th Avenue, Columbus, OH 43210, USA}
\affiliation[ac]{Instituci\'{o} Catalana de Recerca i Estudis Avan\c{c}ats, Passeig de Llu\'{\i}s Companys, 23, 08010 Barcelona, Spain}
\affiliation[ad]{Department of Physics and Astronomy, University of Waterloo, 200 University Ave W, Waterloo, ON N2L 3G1, Canada}
\affiliation[ae]{Perimeter Institute for Theoretical Physics, 31 Caroline St. North, Waterloo, ON N2L 2Y5, Canada}
\affiliation[af]{Waterloo Centre for Astrophysics, University of Waterloo, 200 University Ave W, Waterloo, ON N2L 3G1, Canada}
\affiliation[ag]{Space Sciences Laboratory, University of California, Berkeley, 7 Gauss Way, Berkeley, CA  94720, USA}
\affiliation[ah]{University of California, Berkeley, 110 Sproul Hall \#5800 Berkeley, CA 94720, USA}
\affiliation[ai]{Instituto de Astrof\'{i}sica de Andaluc\'{i}a (CSIC), Glorieta de la Astronom\'{i}a, s/n, E-18008 Granada, Spain}
\affiliation[aj]{Departament de F\'isica, EEBE, Universitat Polit\`ecnica de Catalunya, c/Eduard Maristany 10, 08930 Barcelona, Spain}
\affiliation[ak]{Department of Physics and Astronomy, Sejong University, 209 Neungdong-ro, Gwangjin-gu, Seoul 05006, Republic of Korea}
\affiliation[al]{Abastumani Astrophysical Observatory, Tbilisi, GE-0179, Georgia}
\affiliation[am]{Department of Physics, Kansas State University, 116 Cardwell Hall, Manhattan, KS 66506, USA}
\affiliation[an]{Faculty of Natural Sciences and Medicine, Ilia State University, 0194 Tbilisi, Georgia}
\affiliation[ao]{CIEMAT, Avenida Complutense 40, E-28040 Madrid, Spain}
\affiliation[ap]{Department of Physics, University of Michigan, 450 Church Street, Ann Arbor, MI 48109, USA}
\affiliation[aq]{University of Michigan, 500 S. State Street, Ann Arbor, MI 48109, USA}
\affiliation[ar]{National Astronomical Observatories, Chinese Academy of Sciences, A20 Datun Road, Chaoyang District, Beijing, 100101, P.~R.~China}

\abstract{The linear point \slp, a purely geometric feature in the monopole of the two-point correlation function, has been proposed as an alternative standard ruler that, unlike the peak in the correlation function, is more robust to late-time non-linear effects at the percent level.  
In light of improved simulations and high-quality data, we revisit this claimed robustness and test whether it persists at the sub-percent precision demanded by current surveys, using the linear point as an alternative to the template-based fitting approaches employed in standard BAO analyses.
We present the first linear point measurements on galaxy samples from the first and second data releases (DR1 and DR2) of the Dark Energy Spectroscopic Instrument (DESI). We convert the linear point to a dimensionless parameter \qisolp, defined as the ratio of the fiducial linear point to the observed value, analogous to the isotropic BAO scaling parameter \qisobao\ obtained from template-based BAO fits. Using the second generation of \abacussummit\ mock catalogs for DESI DR1 (\abacust), we find that post-reconstruction linear point measurements are more precise than pre-reconstruction, with 15-60\% smaller uncertainties. 
We compare the mean \qisolp\ across the 25 \abacust\ mocks for each tracer to the mean \qisobao\ from the DESI template-based fits, and find a systematic offset which we attribute to the smearing of the correlation function, and thereby the linear point, in the non-linear regime. This shift is small relative to the method's original percent-level robustness, but becomes significant at the sub-percent level.
We propose a sample-dependent correction to \qisolp, as a function of the isotropic damping parameter $\Sigma_\mathrm{iso}$, that mitigates this offset; the resulting $\alpha_\mathrm{iso,LP}^\mathrm{corrected}$ values show excellent agreement with \qisobao\ in both mocks and data, particularly post-reconstruction. This agreement comes at the price of the model-independence that motivated the linear point in the first place, since both reconstruction and the proposed correction are cosmology-dependent.
We conclude that, at present precision, the linear point is best regarded as a complementary geometric cross-check to BAO analyses rather than a strict alternative to template-based fitting approaches  -- one that grows more valuable as DESI hints at time-evolving dark energy, where independent geometric probes of the expansion history help test its robustness.
}

\begin{document}
\maketitle
\flushbottom
\section{Introduction}
\label{sec:intro}
 In the early universe, photons and baryons were coupled in a hot, dense plasma. In regions with slight overdensities of baryons, gravity and radiation pressure would counteract each other, resulting in pressure waves propagating through this plasma. These baryon acoustic oscillations (BAO) were halted at redshift $z\approx 1100$ when the universe cooled enough for the photons to decouple.
These oscillations imprint a characteristic peak in the matter (and galaxy) two-point correlation function\footnote{In the power spectrum (the Fourier transform of the correlation function), these manifest as a series of oscillations.} at a scale corresponding to the sound horizon at the drag epoch \cite{peebles, bao_intro_bassett}.
The location of the peak in the correlation function serves as a standard ruler and is one of the most important probes used to study the expansion history of the universe \cite{bao_intro_eisenstein, bao_intro_weinberg}. 
The BAO imprint was first observed in the SDSS \cite{sdss} and 2dFGRS \cite{2dFGRS} surveys, with progressively higher significance measurements  made in subsequent spectroscopic surveys like BOSS \cite{boss}, eBOSS \cite{eBOSS}, 6dFGRS \cite{6dFGRS}, WiggleZ \cite{wigglez}, DES \cite{des_y6}, and more recently the Dark Energy Spectroscopic Instrument (DESI, \cite{desi2016_i, desi2024_iii, desi_dr2_validation}). These surveys have used galaxy clustering to derive cosmological parameters from the BAO imprint in two-point statistics \cite{sdss_DA_H0, desi2024_vi}. Future large-scale structure surveys like Euclid \cite{euclid_bao}, Roman \cite{roman}, and Spec-S5 \cite{specs5} will seek to further constrain the expansion history of the universe. 

However, like most standard rulers, the BAO signal is an imperfect one. It is sensitive to non-linear effects due to late-time structure formation, and as such the peak in the correlation function is smeared \cite{eisenstein2007, seo2008, crocce_scoccimarro}. Simple peak finding algorithms are therefore insufficient to constrain the BAO scale, as the position of the peak shifts by more than 2\%. Surveys have therefore used 
physically-motivated templates, initialized with a fiducial cosmology, that fit the correlation function to extract the BAO scale \cite{eisenstein2007_recon, sdss_anderson2012, sdss_vargas-magana2014, sdss_beutler2016, sdss_ross2016, sdss_vargas-magana2018, sdss_gil-marin2020, chen2024_bao}. Density field reconstruction \cite{eisenstein2007_recon, padmanabhan2008, padmanabhan2009} has been extensively implemented in galaxy surveys \cite{sdss_recon, desi2024_iii, optimal_recon} and has been shown to be incredibly effective in removing the shifts in the BAO peak caused by non-linearities and galaxy biasing, reducing the systematic errors in BAO measurements \cite{seo2008, seo2010, schmittfull2015, ding2018}. 
These methods require assuming a fiducial cosmology to convert galaxy redshifts to distances as well as a fiducial value for the growth rate of cosmic structure and the linear bias of the galaxy sample \cite{eisenstein2007_recon, Zeldovich}. 
While it is in principle a cosmology-dependent pipeline, significant work has been done to show that both template-based fits and standard reconstruction are robust and that the effects of assuming a fiducial cosmology are minimal \cite{Sherwin_2019, carter2020, bernal2020, desi_recon_fiducial_cosmo}. 
However, it is desirable to examine the effects of alternative BAO fitting schemes, especially in light of the most recent findings by DESI hinting at evolving dark energy \cite{DESI_DR1, desi2024_vii, desi_dr2_bao, desi_dr2_extendedDE}. 
This includes considering alternative standard rulers like the linear point (the focus of this work) and the zero-crossing scale \cite{zero_crossing}, and alternative reconstruction algorithms like Laguerre reconstruction \cite{nikakhtar2021, nikakhtar2021_halo}, optimal transport reconstruction \cite{ot_levy2021, ot_nikakhtar2022, ot_von_Hausegger2022, ot_nikakhtar2023, ot_nikakhtar2024}, iterative algorithms \cite{schmittful2017_iterative, hada_einsenstein2018_iterative, seo2022_iterative, chen2024_iterative}, and hybrid methods involving machine learning \cite{chen2023_cnn, shallue2023_cnn, parker2025_ml}.

The linear point $s_\mathrm{LP}$ \cite{anselmi2016_lp, anselmi2017_sdss} is defined as the mean of the locations of the peak and the preceding dip in the monopole of the two-point correlation function. It is a purely geometric scale whose value can be inferred using a simple polynomial fit to a narrow region of the correlation function \cite{anselmi2018_validation}. Its utility as a standard ruler is motivated by its weak sensitivity to smearing effects like late-time non-linearities, redshift-space distortions, and scale-dependent bias, which was argued to potentially mitigate the need for density-field reconstruction altogether. 
In its original formulation, a single cosmology- and redshift-independent adjustment of 0.5\% to the measured linear point was proposed to absorb any residual sensitivity to these effects.
To further reduce the dependence on the fiducial cosmology assumed to calculate the correlation function, the linear point is often redefined as $y_\mathrm{LP}\equiv s_\mathrm{LP}/D_V$, where $D_V$ is the isotropic volume distance for the fiducial cosmology. 
This is analogous to how the standard BAO measurement is often denoted as the ratio of the BAO sound horizon scale $r_d$ and the angular diameter distance $D_A$.

The linear point pipeline was validated in \cite{anselmi2018_validation}, and the first measurements of the linear point on data were made on the LOWZ and CMASS samples from the twelfth data release of the BOSS experiment \cite{boss_dr12} in \cite{anselmi2017_sdss}. Cosmological inference using the linear point was first described in \cite{anselmi2019_distance_inference}, and preliminary measurements of the Hubble constant $H_0$ and the matter density $\Omega_m$ using BOSS DR12 linear point measurements were presented in \cite{he2023}.
With the improved simulations and high-precision data now available, it is critical to revisit the linear point and test its robustness as a standard ruler in the era of precision cosmology. This is the central aim of the present work. 

In what follows, we distinguish between two distinct ways of using the linear point. 
The first is close to the original \cite{anselmi2016_lp, anselmi2017_sdss, anselmi2018_validation, anselmi2019_distance_inference, anselmi2023_cosmological_forecasts, odwyer2020}: to provide an estimate of the distance scale in as cosmology-independent a way as possible. This means that no cosmology-dependent template shapes are used when estimating the peak and dip scales in the two-point correlation function measured in the {\em pre}-reconstructed field, and the only correction for non-linear effects is to multiply the raw $s_{\rm LP}$ estimate by a cosmology- and redshift-independent factor of 1.005 as done in \cite{anselmi2017_sdss, anselmi2018_validation, anselmi2019_distance_inference}. We show that this mode leaves a residual offset relative to template-based BAO measurements, large enough to matter at the precision dictated by DESI.
The second is to use the linear point as an alternative to template-based estimates of the distance scale for providing constraints that are complementary to standard BAO analyses. In this mode we relax the strict cosmology-independence ideal: the 1.005 factor is allowed to become cosmology- and redshift-dependent, and estimating the linear point in the {\em post}-reconstructed field becomes useful. We show that this does recover agreement with BAO, but at the cost of the model-independence that motivated the first. To compare the linear point against BAO, we introduce a dimensionless parameter \qisolp, analogous to the isotropic BAO dilation parameter \qisobao\ and defined as the ratio of the fiducial (theoretical) linear point to its measured value. This ratio is our primary basis for comparison in this work.

In this work, we present linear point measurements made on the bright galaxy survey (\bgs), luminous red galaxy (\lrg), and emission line galaxy (\elg) samples from the first and second data releases (DR1 and DR2) of the DESI survey. In Section \ref{sec:data_methods}, we describe the data and mocks used in this work, along with a brief discussion of the methods used in the primary DESI analysis pipeline. In Section \ref{sec:linear_point}, we introduce the linear point, discuss its robustness to non-linear effects and smearing, detail the pipeline, and present our treatment of the fiducial linear point and the damping correction.
In Section \ref{sec:mocks}, we use mock catalogs generated with the DESI fiducial cosmology to investigate the impact of reconstruction on the linear point and compare the measurements to isotropic BAO results obtained on the same correlation functions using the fixed-template pipeline.
We then present the linear point measurements for DESI DR1 and DR2 samples in Section \ref{sec:results}, and summarize our findings and discuss future prospects in Section \ref{sec:conclusion}.

\section{Data and Methods}
\label{sec:data_methods}
The Dark Energy Spectroscopic Instrument (DESI) survey is a Stage IV dark energy experiment designed to measure the expansion history and growth of structure at redshifts $z<3$. The instrument is a multi-fiber spectrograph installed on the Mayall 4m telescope located at the Kitt Peak National Observatory in Arizona \cite{desi2022_instrument}. 
DESI is conducting an eight-year survey \cite{SurveyOps.Schlafly.2023} covering about 17,000 deg$^2$  of the sky, and can obtain simultaneous spectra of almost 5000 objects over a $\sim$3$^\circ$ field \cite{desi2016_ii, Corrector.Miller.2023, FiberSystem.Poppett.2024}.
The full survey is expected to measure spectra of 63 million galaxies and quasars \cite{desi2023_spectra_processing}, compared to the initial forecasts of 39 million.
The first data release (DR1, \cite{DESI_DR1}) includes observations from 5.7 million galaxies and quasars made in the first year of survey operations starting May 14, 2021, after a successful validation phase \cite{desi2024_validation}, continuing until June 14, 2022 and covering an effective volume of roughly 18 Gpc$^3$. Now, with redshifts of more than 14 million galaxies and quasars, the second data release sample (DR2) is by far the largest spectroscopic galaxy sample to date, covering a cumulative effective volume of over 42 Gpc$^3$. DESI adjusts its observing schedule based on conditions, running a ``bright-time" program focused on the Bright Galaxy Survey \cite{Hahn_2023_bgs} and a ``dark-time" program that targets luminous red galaxies (\lrgs\ \cite{Zhou_2023_lrg}), emission-line galaxies (\elgs\ \cite{Raichoor_2023_elg}), and quasars (\qsos\ \cite{Chaussidon_2023_qso}). A detailed overview of the spectroscopy, target selection, and creation of large-scale structure catalogs is provided in \cite{desi2023_spectra_processing, desi2023_target_selection, Anand_2024,Ross_2025_LSS}.

In this section, we describe the DESI DR1 survey targets and mock catalogs used in this work, along with a brief discussion on the methods used to measure the two-point correlation function, perform reconstruction, compute covariance matrices, and perform BAO template fits.

\subsection{Survey targets and mock catalogs}
\label{sec:targets_mocks}

We use identical samples of the bright galaxy survey (\bgs),  luminous red galaxies (\lrgs), and emission-line galaxies (\elgs) from the first and second data releases (DR1 and DR2) used in BAO analysis \cite{desi2024_iii, desi_dr2_bao}. We list the redshift cuts, effective redshift, and linear bias for each target in Table~\ref{tab:targets}. 
Our analysis also makes use of mock catalogs simulating large-scale structure in DR1 with DESI survey geometry. Below is a brief description of these mocks; we refer the reader to \cite{desi2024_iii} and references within for more details:
\begin{itemize}
    \item The \textbf{\abacus\ mocks} \cite{abacus} derive from the \abacussummit\ N-body simulation suite \cite{abacussummit} and produce highly accurate non-linear structure in the DESI footprint. In this analysis, we use the 25 base simulation boxes for each tracer with a combined volume of 200 $h^{-3}$Gpc$^{3}$, generated using the Planck 2018 $\Lambda$CDM cosmology \cite{planck2018}. The \abacus\ mocks used in this work were produced in two generations for DR1 analyses -- the first generation used a very early version of the DESI early data release (DESI-EDR \cite{edr}) to find the best fit halo occupation distribution model, whereas the second generation (\abacust\ hereafter) used the final DESI-EDR after correcting for all the systematics and including a detailed model for DESI focal plane effects.
    These mocks are further produced in three variations of fiber assignment completeness: \complete\ (no fiber assignment), \altmtl\ (full fiber assignment pipeline \cite{altmtl}), and \ffa\ (fast fiber assignment, a sampling-based process that is quicker to implement than the full pipeline \cite{ffa}). 
    \item The \textbf{Effective Zel'dovich mocks (\ezmocks)} \cite{EZmock} use the Zel'dovich approximation \cite{Zeldovich} to produce 1000 computationally cheap simulation boxes. 
    While they may not have accurate non-linear physics, they are effective in calculating covariances between the 25 \abacust\ mocks. 
    The latter is calculated using \ezmocks\ with a box side of 2 $h^{-1}$Gpc to match the size of \abacust\ boxes, whereas realizations with a box side of 6 $h^{-1}$Gpc are used to validate covariance matrices for the full survey volume.
    Because of the large effective volume of these mocks, only the \ffa\ pipeline has been applied to them. These mocks have been shown to have good agreement with large-scale clustering \cite{sdss_ezmock}. 
\end{itemize}
In this work, we use DR1 \abacust\ mocks with full fiber assignment (\altmtl) for validation\footnote{DR2 mocks were not available at the time this analysis was performed; however, a recent work \cite{desi_dr2_validation} suggests strong consistency between BAO measurements made on DR1 and DR2 mocks, with smaller uncertainties in the latter. We therefore expect our validation tests to be consistent with DR2 mocks as well.}, and use \ezmocks\ for numerically computing covariance matrices when needed. We further discuss covariance matrices in the context of linear point analyses in Section \ref{sec:methods}.

\begin{table}
    \centering
    
    \begin{tabular}{|l|c|c|c|}
    \hline
    Tracer & Redshift range & Effective redshift $z_\mathrm{eff}$ & Linear bias $b$ \\
    \hline
    \bgs & 0.1--0.4 & 0.30 & 1.5 \\ 
    \lrgo & 0.4--0.6 & 0.51 & 2.0 \\
    \lrgt & 0.6--0.8 & 0.71 & 2.0 \\
    \lrgth & 0.8--1.1 & 0.92 & 2.0 \\
    \lrgelg & 0.8-1.1 & 0.93 & 1.6 \\
    \elgo & 0.8--1.1 & 0.95 & 1.2 \\
    \elgt & 1.1--1.6 & 1.32 & 1.2 \\
    \hline
    \end{tabular}
    
    \caption{Fiducial redshift range, effective redshift, and linear bias for each DESI target sample used in this analysis. The choice of these parameters is motivated in \cite{desi2024_ii, optimal_recon}.}
    
    \label{tab:targets}
\end{table}

\subsection{Methods}
\label{sec:methods}
In this section, we describe the methods used in the DESI DR1 and DR2 BAO analyses to measure two-point clustering statistics in configuration-space, perform reconstruction, and use the template-based pipeline to derive BAO measurements from the correlation function. 

\subsubsection{Two-point correlation function measurements}

Two-point clustering measurements on DESI DR1 and DR2 samples are discussed in \cite{desi2024_ii, desi_dr2_bao}, with the DR2 measurements validated using mock catalogs in \cite{desi_dr2_validation}.
In this work, we use the configuration-space measurements, namely the $l=0$ monopole component of two-point correlation function. The correlation function was calculated using the Landy-Szalay estimator \cite{LandySzalay} and its modified version post-reconstruction \cite{modifiedLandySzalay}.
Galaxies were weighted using `FKP' weights, inspired by \cite{FKP} and described in detail in \cite{desi2024_iii, DESI_DR1} in the context of DESI DR1 measurements and validated in \cite{desi_dr2_validation} for DR2 BAO measurements. Clustering measurements were combined over the North and South Galactic Caps. 
The two-point correlation function multipoles were computed using {\tt pycorr}\footnote{\url{https://github.com/cosmodesi/pycorr}}, a Python wrapper for the pair-counting code {\tt corrfunc} \cite{corrfunc1, corrfunc2}. 
These measurements can be computed for a variety of bin widths; in this analysis, we sample the correlation function with a 4 \hMpc bin width and justify our choice in Appendix \ref{sec:optimization}.
The linear theory correlation function for the DESI fiducial cosmology is predicted using {\tt CLASS}\footnote{\url{https://github.com/lesgourg/class_public}} \cite{class}. 

\subsubsection{Reconstruction}
DESI BAO analyses use a modification of the standard density-field reconstruction algorithm originally proposed in \cite{eisenstein2007_recon} and improved in \cite{padmanabhan2009}. 
An overview describing additional reconstruction algorithms in the context of DESI BAO results is presented in \cite{chen2024_bao_algos}. 
We use the iterative Fast Fourier Transform (iFFT) algorithm first presented in \cite{burden2015_ifft} and extensively described in \cite{optimal_recon} as the optimal reconstruction algorithm for DESI BAO analyses.
In line with the BAO results presented in \cite{desi2024_iii, desi_dr2_bao}, we use the ${\tt RecSym}$ convention \cite{white2015_recsym}, which shifts tracers and randoms in LSS catalogs by the same amount to preserve redshift-space distortions in post-reconstruction clustering. The smoothing scale used in the reconstruction pipeline is prescribed in \cite{optimal_recon}.
Reconstruction is numerically implemented using {\tt pyrecon}\footnote{\url{https://github.com/cosmodesi/pyrecon}}, a Python package developed by the DESI collaboration that offers a range of reconstruction algorithms and conventions.

\subsubsection{BAO pipeline}

The DESI BAO fitting pipeline is designed to extract the BAO feature in two-point clustering measurements by combining a physically motivated theory model from quasi-linear theory and a parameterized model to marginalize over non-linearities \cite{desi2024_iii}. The observed power spectrum in Fourier space is modeled as a function of the smooth (no-wiggles) component and the BAO (wiggles) component damped by non-linear evolution. This template is Hankel-transformed to configuration space to yield the template for the correlation function multipoles. This template inherits model parameters from the power spectrum template, namely BAO dilation parameters, damping effects, linear galaxy bias, and growth of structure. The Python package \desilike \footnote{\url{https://github.com/cosmodesi/desilike}} provides a framework for writing DESI BAO theory templates and likelihoods, which are implemented using JAX \cite{jax}. Posterior estimates are made by analytically marginalizing over broadband parameters using MCMC sampling with the EMCEE package \cite{emcee}. All model parameters in the pipeline are listed in \cite{desi2024_iii} and are initialized with flat priors, with the exception of the transverse and line-of-sight damping parameters, $\Sigma_\perp$ and $\Sigma_\parallel$ respectively, which are sampled from Gaussian priors. The means and standard deviations of these Gaussian priors are derived using a combination of theoretical calculations and many realizations of measurements of the cross-correlation between pre- and post-reconstruction density fields in \abacust\ mocks; these values are listed in \cite{desi2024_iii} and restated in Table \ref{tab:damping_parameters}.  
By attempting to undo the effects of non-linearities in structure formation, reconstruction aims to reduce the damping parameters; this is evident in Table \ref{tab:damping_parameters}.
However, the effects of residual damping in the post-reconstruction field will affect comparisons between linear point and standard BAO measurements, which we discuss later in this work.

The DESI BAO measurements are made in terms of Alcock-Paczynski-like parameters \cite{alcock_paczynski, padmanabhan2008}, namely isotropic and anisotropic BAO dilation parameter ($\alpha_\mathrm{iso}$ and $\alpha_\mathrm{AP}$ respectively). These are often redefined in terms of the apparent size of the BAO standard ruler perpendicular ($\alpha_\perp$) and parallel ($\alpha_\parallel$) to the line of sight:
\begin{equation}
        \alpha_\mathrm{iso} = \left(\alpha_\perp^2 \alpha_\parallel\right)^\frac{1}{3} , \,\,\,\,\,\,\; \alpha_\mathrm{AP}=\alpha_\parallel/\alpha_\mathrm{\
        \perp}
        \label{eq:qiso_bao}
    \end{equation}
These are further defined in terms of the angular diameter distance $D_A(z)$ and the Hubble parameter $H(z)$:
\begin{equation}
    \alpha_\parallel=\frac{H^\mathrm{fid}(z)r_d^\mathrm{fid}}{H(z)r_d}, \,\,\,\,\,\,\; \;\alpha_\perp=\frac{D_A(z)r_d^\mathrm{fid}}{D_A^\mathrm{fid}(z)r_d}
    \label{eq:qpar_bao}
\end{equation}
Here, the superscript `$\mathrm{fid}$' refers to quantities measured in the fiducial cosmology, and $r_d$ is the comoving scale of the BAO feature. 

\subsection{Covariance matrices}
In this work, we use both analytic and numeric covariance matrices for the two-point correlation function. 

Analytic covariance matrices for DESI are calculated using the {\tt RascalC} code in configuration space \cite{analytic_cov_oconnell2016, analytic_cov_oconnell2019, analytic_cov_philcox2019, rascalc2020} and using {\tt TheCov} in Fourier space \cite{Alves2024prep, Wadekar:2019rdu, Kobayashi:2023vpu}. 
Since this work is performed exclusively in configuration space, we use the former. These covariance matrices are extensively validated in \cite{rascalc_validation}. At the time of writing this paper, these covariance matrices are only available for correlation functions sampled with a bin width 4 \hMpc. A full discussion of analytical covariance matrices in the context of DR1 data is presented in \cite{rascalc2024}.
Numerical covariances are calculated using the 1000 \ezmocks\ and can be computed at any integer bin width $\Delta s\geq1$ \hMpc.
The two flavors of covariance matrices are compared in \cite{analytic_numeric_cov_comparison} and are found to have a good level of agreement in configuration space. DESI Y1 BAO analyses are therefore performed using the analytic covariance matrices, which additionally allow tuning the covariances to match observed clustering in data and avoid discrepancies between clustering in mocks and data.

All previous linear point analyses have used the Gauss-Poisson approximation for covariance matrices \cite{eppur2008, gaussPoisson2016, paranjape2022}, so using the numerical ones represents a change in methodology. We find no significant difference between linear points measured using numerical and analytical covariances with 4 \hMpc\ bin widths, which is expected given the excellent agreement found between the two flavors of covariances in \cite{analytic_numeric_cov_comparison}. 
We also refer the reader to \cite{eigencov2024} for a discussion of the dependence on bin size in the context of the linear point.

\begin{table*}
    \renewcommand{\arraystretch}{1.2}
    \centering
    \begin{tabular}{| c | c | c| c | c| c | c | c|}
        \hline
        Parameter & Recon & \bgs\ & \lrgs & \elgs \\
        \hline
        $\sigmaper$ [\hMpc]  & Pre  & $6.5\pm1.0$& $4.5\pm 1.0$ & $4.5\pm 1.0$ \\
        $\sigmapar$ [\hMpc] & Pre & $10.0\pm2.0$& $9.0\pm 2.0$ & $8.5\pm 2.0$ \\
        $\Sigma_\mathrm{iso}$ [\hMpc] & Pre & $7.5\pm0.9$& $5.7\pm 0.9$ & $5.6\pm 0.9$ \\
        \hline
        $\sigmaper$ [\hMpc] & Post & $3.0\pm1.0$ & $3.0\pm 1.0$ & $3.0\pm 1.0$ \\
        $\sigmapar$ [\hMpc] & Post & $8.0\pm2.0$&$6.0\pm 2.0$ & $6.0\pm 2.0$ \\
        $\Sigma_\mathrm{iso}$ [\hMpc] & Post & $4.2\pm0.9$&$3.8\pm 0.9$ & $3.8\pm 0.9$ \\
        \hline
    \end{tabular}
    \caption{Mean values and standard deviations of the Gaussian priors for the non-linear BAO damping parameters across and along the line of sight ($\sigmaper$ and $\sigmapar$, respectively) used in the DESI BAO fitting pipeline, and the corresponding isotropic damping scale  $\Sigma_\mathrm{iso}\approx\left(\Sigma_\perp^2\Sigma_\parallel\right)^\frac{1}{3}$.
    }
    \label{tab:damping_parameters}
\end{table*}

\section{The Linear Point}
\label{sec:linear_point}
\begin{figure}
    \centering
    \includegraphics[width=0.75\linewidth]{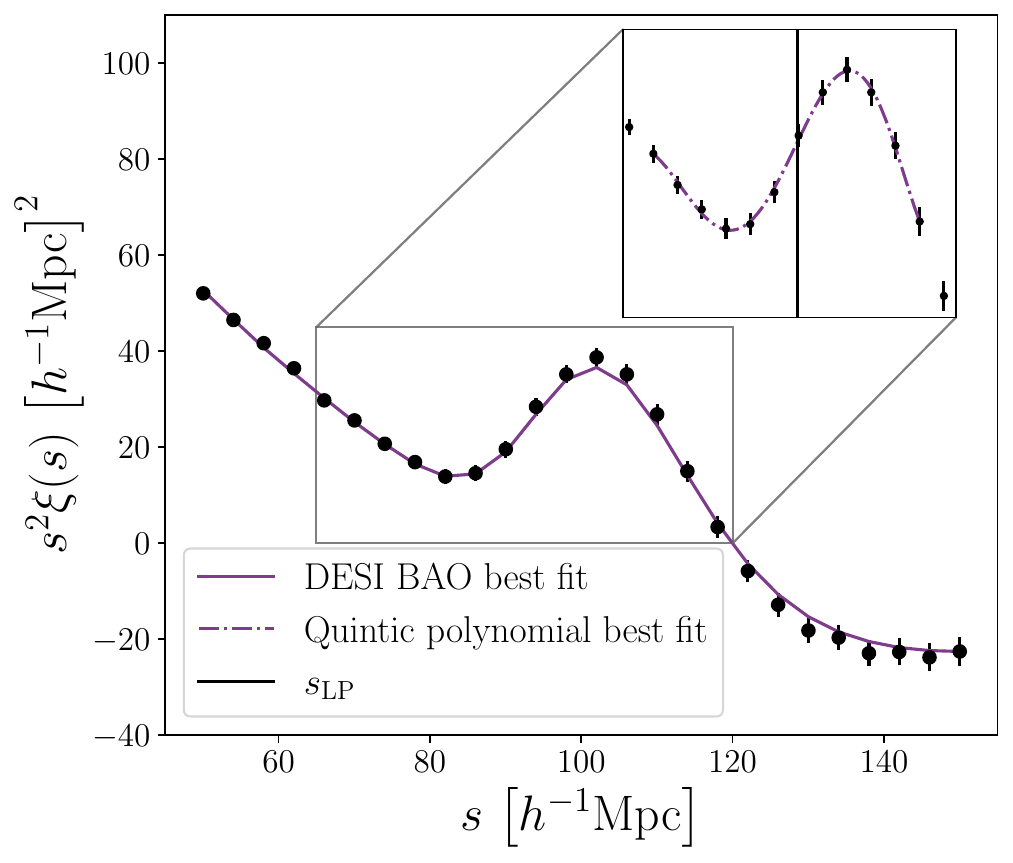}
    \caption{Average two-point correlation function $\xi(s)$ of the 25 \lrgo\ \abacust\ mocks, shown over the typical BAO fitting range 50–150 \hMpc. Points show the mock-averaged measurement with error bars giving the uncertainty on the mean; the solid line is the best-fit DESI BAO template. The inset zooms into 70–115 \hMpc, the narrower range over which $\xi(s)$ is well approximated by an odd-degree polynomial: the dot-dashed line is a fifth-degree polynomial fit, and the vertical line marks the resulting linear point \slp. Following convention, the $y$-axis shows $s^2\xi(s)$ rather than $\xi(s)$ to enhance the visibility of the BAO feature; however, all analyses in this work are performed on $\xi(s)$ itself unless otherwise noted (see Appendix \ref{sec:alteratives}).}
    \label{fig:inset}
\end{figure}

The linear point is a geometric feature of the monopole of the two-point correlation function, defined as the mean of the distance scales corresponding to the peak in the correlation function and the preceding dip:
\begin{equation}
    s_\mathrm{LP} \equiv \frac{s_\mathrm{peak} + s_\mathrm{dip}}{2}
\label{eq:lp_definition}
\end{equation}
Since this is a purely geometric scale, its value can be inferred without assuming an underlying model for the correlation function. The utility of the linear point lies in its robustness to smearing in the correlation function monopole due to non-linear structure formation and redshift space distortions, compared to the location of the BAO peak. The linear point was proposed as a purely geometric standard ruler to provide an alternative to cosmology-informed template fits which are typically used to extract the BAO location in the correlation function, thus making the calculation as cosmology- and model-agnostic as possible. This implies that we do not need to assume a template or model to fit the correlation function; rather, a model-independent fit over the region encompassing the peak and the dip (such that Eq.~\ref{eq:lp_definition} can be computed) is sufficient. Figure \ref{fig:inset} illustrates this polynomial fit and compares it to the much wider range used in the template-based fitting method.

Linear point measurements are often quoted in the form of an angle-like quantity \cite{anselmi2017_sdss, anselmi2023_cosmological_forecasts}:
\begin{equation}
    y_\mathrm{LP} \equiv \frac{s_\mathrm{LP}}{D_V},
    \label{eq:yLP}
\end{equation}
where $D_V$ is the isotropic volume distance defined as follows:
\begin{equation}
    D_V(z)\equiv\left((1+z)^2D_A(z)^2\frac{cz}{H(z)}\right)^\frac{1}{3}
    \label{eq:D_V}
\end{equation} 
This is analogous to the angular size of the BAO standard ruler, defined as the ratio of the sound horizon at the drag epoch $r_d$ and the angular diameter distance $D_A$. The angular form of the linear point measurements $y_\mathrm{LP}$ further reduces the dependence on fiducial cosmology, since the fiducial and true cosmologies are related as follows: \cite{anselmi2017_sdss, anselmi2018_validation, anselmi2019_distance_inference, anselmi2023_cosmological_forecasts, odwyer2020}
\begin{equation}
    \frac{s_\mathrm{LP}^\mathrm{fid}}{D_V^\mathrm{fid}}\approx \frac{s_\mathrm{LP}^\mathrm{true}}{D_V^\mathrm{true}}
    \label{eq:ylp_equality}
\end{equation}
It is therefore natural to define a dimensionless parameter for the linear point, analogous to the BAO dilation parameters stated in Eqs. \ref{eq:qiso_bao} and \ref{eq:qpar_bao}, as the ratio of the quantities stated above:
\begin{equation}
    \alpha_\mathrm{iso}=\frac{y_\mathrm{LP}^\mathrm{fid}}{y_\mathrm{LP}^\mathrm{true}}=\frac{s_\mathrm{LP}^\mathrm{fid}}{D_V^\mathrm{fid}}\cdot\frac{D_V^\mathrm{true}}{s_\mathrm{LP}^\mathrm{true}}
    \label{eq:qiso_lp_longform}
\end{equation}
The subscript `iso' follows the convention used in BAO analyses to indicate the isotropic size of the standard ruler; since the linear point has no analog in the quadrupole, only the isotropic dilation is defined, and we drop the anisotropic decomposition used in BAO analyses. With this definition and the relation stated in Eq. \ref{eq:ylp_equality}, we naturally expect $\alpha_\mathrm{iso}\approx1$ if the fiducial cosmology used to calculate the correlation function is the same as (or indistinguishable from) the true underlying cosmology. However, the observed angular size of the linear point is related to measured linear point $s_\mathrm{LP}^\mathrm{obs}$ (simplified to $s_\mathrm{LP}$ hereon) as follows:
\begin{equation}
    y_\mathrm{LP}^\mathrm{obs}= \frac{s_\mathrm{LP}}{D_V^\mathrm{fid}}
    \label{eq:yLP_obs}
\end{equation}
To calculate the isotropic size of the \textit{observed} linear point standard ruler relative to the fiducial cosmology, we insert the above equation into Eq. \ref{eq:qiso_lp_longform} to obtain:
\begin{equation}   \alpha_\mathrm{iso}=\frac{s_\mathrm{LP}^\mathrm{fid}}{s_\mathrm{LP}},
    \label{eq:qiso_lp}
\end{equation}
where the fiducial volume distances cancel.
From here on, we refer to the quantity defined in Eq.~\ref{eq:qiso_bao} as $\alpha_\mathrm{iso,BAO}$ and the quantity defined in Eq.~\ref{eq:qiso_lp} as $\alpha_\mathrm{iso,LP}$. The calculation of the fiducial linear point is detailed later in Section \ref{sec:s_fid}.

\subsection{Robustness of the linear point}
\label{sec:robustness}
The motivation behind using the linear point as an alternative standard ruler is its robustness to non-linear effects. To a good approximation, the non-linear correlation function, $\xi_\mathrm{NL}$, can be written as a Gaussian convolution of the linear theory correlation function $\xi_\mathrm{lin.th.}$ with an isotropic smoothing kernel $\Sigma_\mathrm{iso}$ \cite{crocce_scoccimarro, nikakhtar2021_halo}\footnote{The convolution produces a modified Bessel function of the first kind, $I_0$. In this equation, $i_0$ is the modified \textit{spherical} Bessel function, and is related to $I_0$ in the same way that the spherical Bessel function $j_0$ is related to $J_0$. See \cite{arfken} for more details.}:
\begin{equation}
    \xi_\mathrm{NL}(s)=\int_0^\infty\frac{dr\, r^2}{\Sigma_\mathrm{iso}^3} \frac{e^{-\frac{(r^2+s^2)}{2\Sigma_\mathrm{iso}^2}}}{\sqrt{2\pi}}\,2i_0\left(\frac{rs}{\Sigma_\mathrm{iso}^2}\right) \xi_\mathrm{lin.th.}(r),
    \label{eq:convolution}
\end{equation}
This smoothing kernel can be approximated by an isotropic damping parameter, related to the transverse and line-of-sight damping parameters ($\Sigma_\perp$ and $\Sigma_\parallel$) used in template-based BAO fitting as follows:
\begin{equation}
    \Sigma_\mathrm{iso} \approx \left(\Sigma_\mathrm{\perp}^2\Sigma_\mathrm{\parallel}\right)^\frac{1}{3}.
    \label{eq:sigma_iso}
\end{equation}
This smoothing makes the linear point in $\xi_\mathrm{NL}$ differ slightly from that in $\xi_\mathrm{lin.th.}$; we denote the latter, unsmeared value by $s_\mathrm{LP}^\mathrm{lin.th.}$, a fixed property of the cosmology (defined formally in Section~\ref{sec:s_fid}). 
To illustrate, for the Planck 2018 cosmology, the linear point exhibits a 0.5\% deviation from its linear theory value when $\Sigma_\mathrm{iso}\sim3.5$ \hMpc, with larger $\Sigma_\mathrm{iso}$ producing larger differences.
For this reason, \cite{anselmi2016_lp} propose multiplying the value measured in the evolved field by a factor of 1.005 and argue that this obviates the need for reconstruction altogether, in the sense that any remaining difference from the linear theory value will always be sub-percent for sufficiently small $\Sigma_\mathrm{iso}$.
This makes the linear point a useful alternative standard ruler \cite{anselmi2016_lp, anselmi2017_sdss, anselmi2018_validation}.
Although sub-percent, this shift is large enough that any cosmological parameters inferred from the uncorrected linear point would inherit its disagreement with BAO, without reconstruction and correction to remove it. We show in later sections that modeling the shift is necessary for the linear point to agree with isotropic BAO results.

The DESI BAO fitting pipeline uses Gaussian priors on the transverse and line-of-sight damping parameters, $\Sigma_\mathrm{\perp}$ and $\Sigma_\mathrm{\parallel}$, to initialize the correlation function template. Using the means and standard deviations of these priors, we can estimate the corresponding isotropic damping parameter $\Sigma_\mathrm{iso}$ from Eq. \ref{eq:sigma_iso}; we list these values for pre- and post-reconstruction templates in Table \ref{tab:damping_parameters}. The pre-reconstruction damping in DESI \bgs\ evaluates to $\Sigma_\mathrm{iso}\sim7.5$ \hMpc\ and that for \lrgs\ and \elgs\ is $\Sigma_\mathrm{iso}\sim5.7$ \hMpc, which will reduce the linear point from its fiducial value by up to 1\% \cite{nikakhtar2021}.   
Multiplying by 1.005 will not entirely undo this shift, potentially compromising the constraining value of linear point measurements in the pre-reconstructed field.
In fact, the (non-zero) post-reconstruction value of $\Sigma_\mathrm{iso}$ suggests that the linear point in the reconstructed correlation functions would still deviate at the 0.5\% level from $s_\mathrm{LP}^\mathrm{lin.th.}$. We explore this further in the following sections.

\subsection{Estimating the linear point}
\label{sec:calc_lp}
We take advantage of the purely geometric nature of the linear point scale and use a linear combination of polynomials as our fitting function for the correlation function over a narrow range that includes the peak and the dip. Previous works \cite{anselmi2018_validation} have shown that an odd-degree polynomial interpolation is a valid approximation for the correlation function in the region roughly between 60 \hMpc\ and 120 \hMpc, and \cite{paranjape2022} provide a Bayesian framework for determining the appropriate order polynomial.

Using the DESI DR1 and DR2 correlation functions and associated covariance matrices, we can use $\chi^2$ likelihood minimization to fit a centered and scaled polynomial to the data points:
\begin{equation}
    \xi_0(s)=\sum_{i=0}^n a_i \left(\frac{s-s_0}{\sigma}\right)^i,
\label{eq:poly_interp}
\end{equation}
where $a_i$ are the fitting coefficients of an $n$th degree polynomial, $s_0=93$ \hMpc\ is chosen to be roughly in the middle of the fitting range for centering, and $\sigma=15$ \hMpc\ is an arbitrary scaling parameter. Centering and scaling the polynomial fit ensures that the fitting coefficients $a_i$ are unaffected by any floating point errors when such coefficients are too small.

We find the locations of the peak ($s_\mathrm{peak}$) and the dip ($s_\mathrm{dip}$) in the correlation function by analytically computing the roots of the polynomial interpolation and calculating the linear point, $s_\mathrm{LP}$, using Eq.~\ref{eq:lp_definition}. 
We use a Monte-Carlo like approach to calculate the error on a single measurement of the linear point. Using the covariance matrix of the best-fit coefficients, we generate 1000 Gaussian random samples of the coefficients centered around the best-fit values. We compute the linear point for each realization and report their standard deviation as the error on the original measurement. This approach was shown to be more robust than standard error propagation in \cite{he2023}.

The linear point pipeline has a set of adjustable parameters including the order of the polynomial fitting function, $n$; the range of the correlation function over which to fit the polynomial; and the spacing of the correlation function measurements, $\Delta s$. These parameters were optimized and the pipeline validated for a BOSS-like survey in \cite{anselmi2018_validation} using QPM mocks \cite{qpm}. 
We make adjustments as needed to optimize the pipeline for DR1 and DR2 DESI correlation functions and covariance matrices.
We find that a quintic polynomial ($n=5$) fit to the correlation function calculated with bin width $\Delta s=4$ \hMpc\ in the range 70-115 \hMpc\ is the most appropriate choice. We justify these choices in Appendix \ref{sec:optimization}.

It is possible in some cases that the chosen polynomial interpolation does not yield an identifiable peak and/or dip in the range of the correlation function that we consider. It is also possible that multiple peaks and/or dips are identified, some of which could have no physical meaning and would not yield a reliable linear point estimate.
To get a reliable linear point measurement, we identify the local minimum (maximum) closest to $s_0$ as the dip (peak); spatially, the next (previous) root is therefore identified as the peak (dip). This ensures that we choose the peak that is most likely to capture the BAO signal. We reject any analytical roots that are found outside the fitting range.
This can result in failure to identify the BAO feature in the monopole of the correlation function which may have otherwise been measured with the template-based BAO fitting approach. We explore this further in Appendix \ref{sec:bgs_bad}.

\subsubsection{The linear point in linear theory}
\label{sec:s_fid}
Knowing the fiducial linear point is essential for constraining cosmological parameters using the linear point estimated from observed correlation functions. The dimensionless parameter defined in Eq. \ref{eq:qiso_lp} can be directly compared to the isotropic BAO measurements, and any deviation of this ratio from unity can be attributed to a true cosmology that differs from the assumed fiducial cosmology.

We distinguish two related quantities. 
The \textit{linear theory} linear point, $s_\mathrm{LP}^\mathrm{lin.th.}$, is the linear point of the unsmeared linear theory correlation function. Because linear evolution rescales the amplitude of $\xi_\mathrm{lin.th.}$ without altering its shape, the peak and dip locations -- and hence $s_\mathrm{LP}^\mathrm{lin.th.}$ -- are essentially independent of redshift.
We generate the linear theory power spectrum at $z=0$ for the fiducial Planck 2018 $\Lambda$CDM cosmology using {\tt CLASS} and compute the correlation function $\xi_\mathrm{lin.th.}$ by Fourier transformation. The \textit{fiducial} linear point, $s_\mathrm{LP}^\mathrm{fid}$, is by contrast the linear point expected in the non-linear regime in the fiducial cosmology -- that is, after $\xi_\mathrm{lin.th.}$ is smeared by the isotropic damping $\Sigma_\mathrm{iso}$ appropriate to a given sample and redshift (Eq. \ref{eq:convolution}). The two coincide only in the absence of smearing: $s_\mathrm{LP}^\mathrm{fid}=s_\mathrm{LP}^\mathrm{lin.th.}$ when $\Sigma_\mathrm{iso}\approx0$.

We noted in Section \ref{sec:robustness} that the linear point deviates under non-linear damping -- for the DESI \bgs\ sample, this results in a 1\% (0.6\%) shift pre- (post-)reconstruction, whereas for \lrgs\ and \elgs\ this shift is 0.9\% (0.5\%) pre- (post-)reconstruction. This means that the observed linear point \slp\ would result in inflated values of \qisolp\ if Eq. \ref{eq:qiso_lp} is evaluated with $s_\mathrm{LP}^\mathrm{fid}=s_\mathrm{LP}^\mathrm{lin.th.}$, i.e. neglecting the smearing.  Given our two distinct views of how the linear point can be used, we have two distinct ways of accounting for smearing. 
\begin{itemize}
    \item Only work with the measured \slp\ or the ratio \qisolp$\equiv s_\mathrm{LP}^\mathrm{lin.th.}$/\slp, (perhaps with \slp\ multiplied by 1.005), comparing against the fixed linear theory value in the fiducial cosmology, as originally proposed. This preserves the model independence of the pipeline, but at a cost: the resulting measurements would be biased relative to template-based BAO measurements, which would propagate into any cosmological interpretation. One would then have to accept this bias as the price of model independence, one that becomes difficult to justify in the era of precision cosmology.
    \item Modify the linear point \slp\ measured on the observed pre- (post-)reconstruction correlation function by including a sample-dependent multiplicative factor; for example, for the DESI \lrg\ and \elg\ samples, this factor would be 1.009 (1.005), and for the \bgs\ sample, this would be 1.01 (1.006).  This enables a more direct comparison with standard BAO analyses when the measurements are converted to \qisolp. Equivalently, we can use the smearing-aware fiducial linear point $s_\mathrm{LP}^\mathrm{fid}$ in Eq. \ref{eq:qiso_lp}, so that it reflects the effects of damping while keeping \slp\ unchanged. When the pre- and post-reconstruction damping parameters $\Sigma_\mathrm{iso}$ from Table \ref{tab:damping_parameters} are used, these two approaches are mathematically equivalent. The latter, however, lets us leave the measurements untouched and instead adjust the reference value against which they are compared. Additionally, as we will see later in this section, it allows us to propagate the uncertainty in $\Sigma_\mathrm{iso}$ into the errors on the modified linear point ratios.
\end{itemize}
In this work, we correct the \qisolp\ measurements using the latter approach -- re-defining the fiducial linear point by a redshift- and bias-dependent factor in the smeared (non-linear) regime. 

There is however a third possibility that involves an alternative reconstruction algorithm in which we use the shape of the observed non-linear correlation function to derive the linear correlation function. This is done by undoing the Gaussian smearing from Eq. \ref{eq:convolution}, subject to a (de)convolution kernel which we derive from the pre-reconstruction prior on $\Sigma_\mathrm{iso}$. When the linear point is measured on the resulting linear correlation function, no correction is required to either \slp\ or $s_\mathrm{LP}^\mathrm{fid}$ when calculating \qisolp. 
We discuss the third approach, known as Laguerre reconstruction, in Appendix \ref{sec:laguerre}; we focus on it among alternative reconstruction methods because it operates directly on the correlation function and is designed to complement the linear point estimator.

We can quantify the effect of damping on the fiducial linear point by performing a Gaussian convolution of the linear theory correlation function using the relevant isotropic kernel $\Sigma_\mathrm{iso}$. 
We generate 1000 realizations of $\Sigma_\mathrm{iso}$ sampled from the Gaussian priors mentioned in Table \ref{tab:damping_parameters}; for each, we generate a non-linear correlation function $\xi_\mathrm{NL}(s)$ using Eq. \ref{eq:convolution}.
We calculate the linear point for each realization and compute the mean and standard deviation of the resulting sample.
We take the mean of the sample to be smearing-corrected fiducial linear point, and its standard deviation as its uncertainty; we list the pre- and post-reconstruction values of $s_\mathrm{LP}^\mathrm{fid}$ in Table \ref{tab:slp_fid}. We define $\alpha_\mathrm{iso,LP}^\mathrm{corrected}$ as the ratio of the smeared fiducial linear point to the observed linear point.

\begin{table}[]
\renewcommand{\arraystretch}{1.2}
    \centering
    \begin{tabular}{c|c|c|}
     \cline{2-3}
     & $\Sigma_\mathrm{iso}$ [\hMpc] & $s_\mathrm{LP}^\mathrm{fid}$ [\hMpc] \\
      \hline
       \multicolumn{1}{|l|}{Linear theory } & 0 & 93.01 \\
       \hline
       \multicolumn{1}{|l|}{\bgs, Pre-recon.}  & $7.5\pm0.9$ & $92.04\pm0.03$ \\
       \multicolumn{1}{|l|}{\lrgs, Pre-recon.} & $5.7\pm0.9$ & $92.13\pm0.44$ \\
       \multicolumn{1}{|l|}{\elgs, Pre-recon.} & $5.6\pm0.9$ & $92.11\pm0.12$ \\
       \hline
       \multicolumn{1}{|l|}{\bgs, Post-recon.}  & $4.2\pm0.9$ & $92.34\pm0.19$ \\
       \multicolumn{1}{|l|}{\lrgs, Post-recon.} & $3.8\pm0.9$ & $92.46\pm0.17$ \\
       \multicolumn{1}{|l|}{\elgs, Post-recon.} & $3.8\pm0.9$ & $92.41\pm0.19$ \\
       \hline
    \end{tabular}
    \caption{Smearing-corrected fiducial linear point $s_\mathrm{LP}^\mathrm{fid}$ for each DESI tracer, computed at the corresponding isotropic smearing scale $\Sigma_\mathrm{iso}$ in the pre- and post-reconstruction regimes. The unsmeared linear theory value is listed for reference. 
    }
    \label{tab:slp_fid}
\end{table}

\section{Linear Point Measurements in \abacust\ DR1 Mocks}
\label{sec:mocks}
In this section, we run the linear point pipeline on \abacust\ mock catalogs for DESI DR1 samples, and compare the linear point standard ruler to the BAO measurements derived using the standard template-based pipeline. 
We validate the robustness of the linear point to non-linear effects and explore whether standard reconstruction is necessary to correct for some of the resulting smearing.

\begin{table}[]
    \centering
    \begin{tabular}{|l|c|c|c|c|c|}
    \hline
         Tracer & Redshift & Recon & $\langle s_\mathrm{LP} \rangle$ [\hMpc] & $\langle\alpha_\mathrm{iso,LP}\rangle$ & $\langle\alpha_\mathrm{iso,LP}^\mathrm{corrected}\rangle$ \\
         \hline
        \bgs & 0.1-0.4 & Pre & $90.62 \pm 1.08$ & $1.026 \pm 0.012$ & $1.016\pm0.012$ \\
        \lrgo & 0.4-0.6 & Pre & $91.52 \pm 0.61$ & $1.016 \pm 0.007$ & $1.006\pm0.008$\\
        \lrgt & 0.6-0.8 & Pre & $91.52 \pm 0.39$ & $1.009 \pm 0.004$ & $1.000\pm0.006$\\
        \lrgth & 0.8-1.1 & Pre & $91.52 \pm 0.46$ & $1.016 \pm 0.005$ & $1.007\pm0.007$ \\
        \elgo & 0.8-1.1 & Pre & $91.46 \pm 1.03$ & $1.017 \pm 0.011$ & $1.007\pm0.012$\\
        \elgt & 1.1-1.6 & Pre & $91.46 \pm 0.36$ & $1.011 \pm 0.004$ & $1.001\pm0.004$\\
        \hline
        \bgs & 0.1-0.4 & Post & $92.64 \pm 0.47$ & $1.004 \pm 0.005$ & $0.997\pm0.005$\\
        \lrgo & 0.4-0.6 & Post & $92.60 \pm 0.34$ & $1.008 \pm 0.004$ & $1.002\pm0.004$\\
        \lrgt & 0.6-0.8 & Post & $92.60 \pm 0.22$ & $1.003 \pm 0.002$ & $0.997\pm0.003$\\
        \lrgth & 0.8-1.1 & Post & $92.60 \pm 0.23$ & $1.004 \pm 0.003$ & $0.998\pm0.003$\\
        \elgo & 0.8-1.1 & Post & $92.61 \pm 0.69$ & $1.004 \pm 0.007$ & $0.998\pm0.008$\\
        \elgt & 1.1-1.6 & Post & $92.61 \pm 0.31$ & $1.001 \pm 0.003$ & $0.994\pm0.004$\\
        \hline
    \end{tabular}
    \caption{Mean linear point measurements over the 25 \abacust\ mocks for each DR1 tracer, pre- and post-reconstruction. For each tracer and redshift bin we list the mean measured linear point \slp, the mean uncorrected ratio \qisolp\ (i.e., without correcting for the smearing of the fiducial linear point $s_\mathrm{LP}^\mathrm{fid}$ due to non-linear evolution), and the mean corrected ratio $\alpha_\mathrm{iso,LP}^\mathrm{corrected}$ (using the modified $s_\mathrm{LP}^\mathrm{fid}$ from Table \ref{tab:slp_fid}). Quoted uncertainties are the errors on the mean over the 25 mocks.
    }
    \label{tab:LPs_abacus}
\end{table}

\begin{figure}
    \centering
    \begin{tabular}{cc}
        \includegraphics[width=0.45\linewidth]{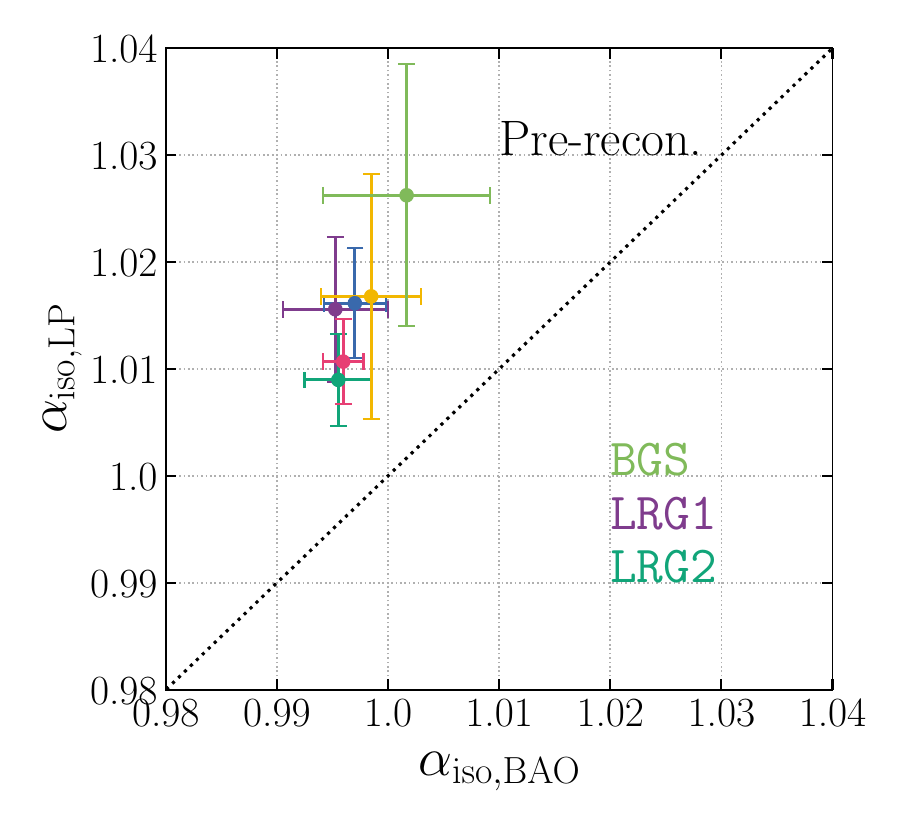} &  \includegraphics[width=0.45\linewidth]{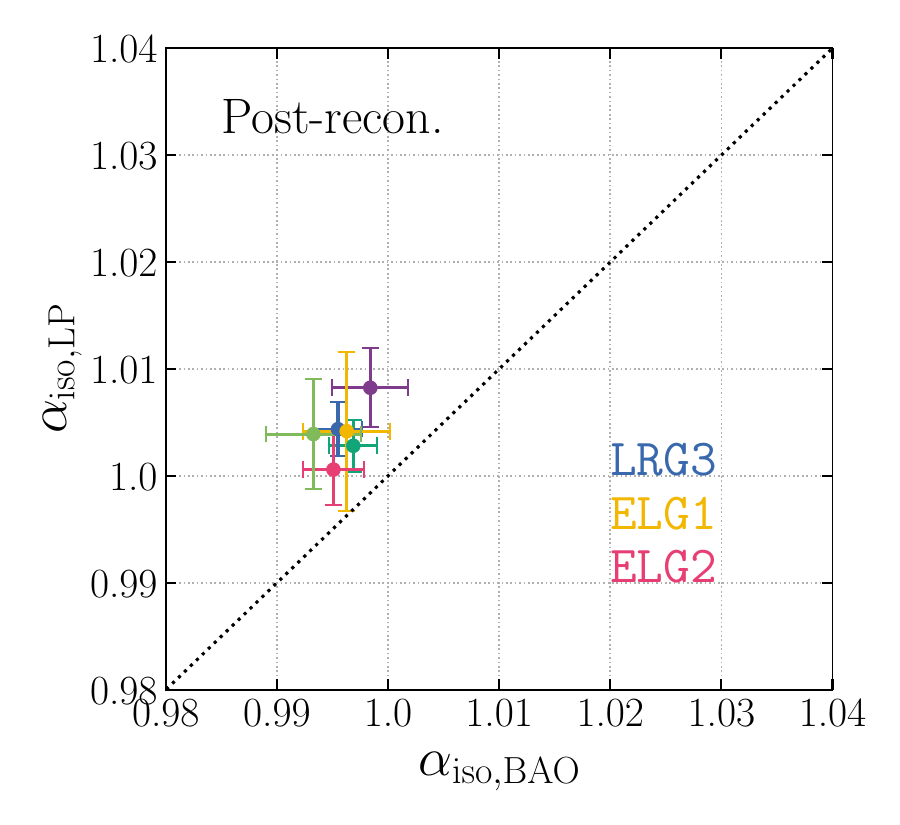}
    \end{tabular}
    
    \caption{
    Comparison of the mean isotropic BAO dilation parameter \qisobao\ with the mean linear point ratio \qisolp\ (the ratio of the fiducial to the measured linear point, defined in Eq. \ref{eq:qiso_lp}), for \bgs, \lrg, and \elg\ \abacust\ mock catalogs, pre-reconstruction \emph{(left)} and post-reconstruction \emph{(right)}. Points show the mean over the 25 mocks and error bars show the uncertainty on the mean; each tracer is annotated in text of the same color as its point. The dotted line marks $x=y$, along which the two measures would ideally agree. A systematic offset from this line is present for all tracers, larger pre-reconstruction than post but nonzero in both cases. Both panels share the same $x$ and $y$-axis ranges to make the reduction in error-bar size and offset from pre- to post-reconstruction directly visible.
    }
    \label{fig:qiso_bao_lp_mocks_uncorrected}
\end{figure}

For each tracer, we measure the linear point on each of the 25 \abacust\ mock correlation functions and convert these measurements to \qisolp\ using Eq. \ref{eq:qiso_lp}, first with $s_\mathrm{LP}^\mathrm{fid}=s_\mathrm{LP}^\mathrm{lin.th.}=93.01$ \hMpc. We present the mean linear point measurements and mean \qisolp\ values for each set of 25 \abacust\ mocks in Table \ref{tab:LPs_abacus}. The error bars in each column represent the error on the mean, which is the standard deviation of the sample divided by the square root of the number of samples.
In Figure \ref{fig:qiso_bao_lp_mocks_uncorrected}, we plot the mean values of \qisolp\ for each tracer against the mean isotropic BAO measurements, \qisobao, made using the BAO template pipeline. The left panel shows these measurements pre-reconstruction (i.e., \qisolp\ is the linear point ratio as originally intended, without multiplying by 1.005), and the right panel depicts post-reconstruction measurements (as a first step for estimating consistency with standard BAO analyses in reconstructed fields). The error bars in both panels show the errors on the mean.

\begin{figure}
    \centering
    \begin{tabular}{cc}
        \includegraphics[width=0.45\linewidth]{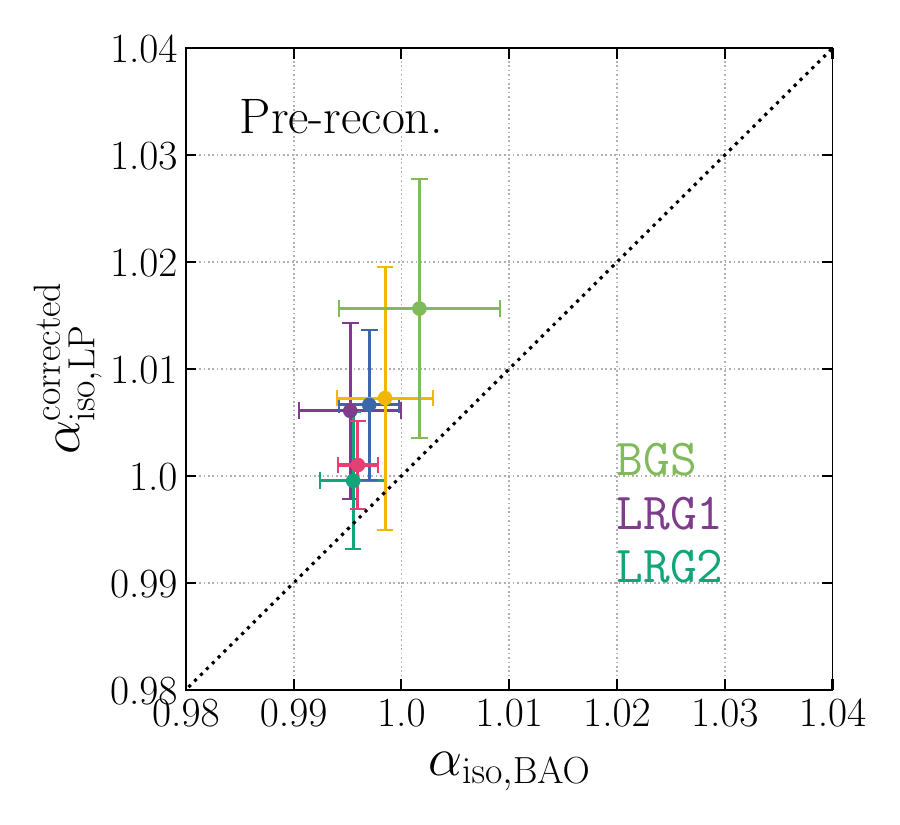} &  \includegraphics[width=0.45\linewidth]{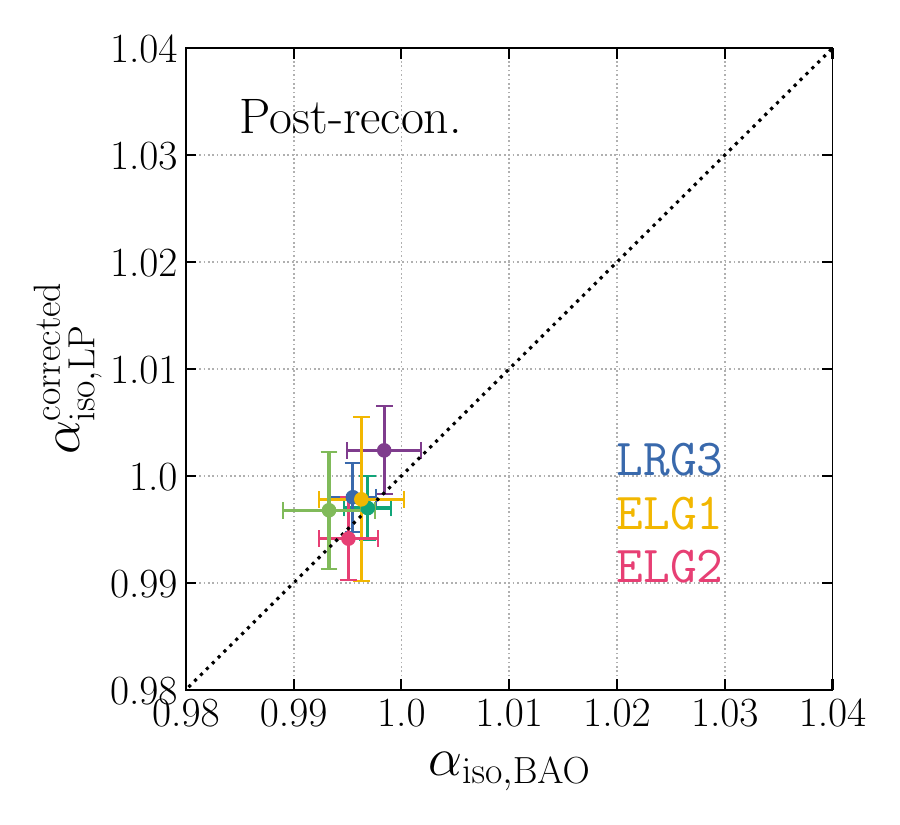}
    \end{tabular}
    
    \caption{
    Same as Figure \ref{fig:qiso_bao_lp_mocks_uncorrected}, but using the corrected linear point ratios $\alpha_\mathrm{iso,LP}^\mathrm{corrected}$ obtained by accounting for the smearing of the fiducial linear point due to late-time non-linear evolution (Table \ref{tab:slp_fid}, motivated in Section \ref{sec:robustness}). The correction shifts the $y$-axis values downward and slightly inflates the $y$ error bars, bringing $\alpha_\mathrm{iso,LP}^\mathrm{corrected}$ into closer agreement with \qisobao\ along the $x=y$ line, particularly post-reconstruction.}
    \label{fig:qiso_bao_lp_mocks_corrected}
\end{figure}

In both pre- and post-reconstruction regimes, the linear point measurements are systematically biased compared to BAO measurements made using the template-based pipeline, despite being otherwise well correlated. As pointed out in Section \ref{sec:robustness}, this is likely a result of non-linear damping effects that inflate the value of \qisolp\ by roughly 1\% (0.5\%) due to damping in the  pre- (post-)reconstruction regime. 
Naively, we expect the pre-reconstruction \qisolp\ to always be larger than unity and to be larger at lower redshift.  The left-side panel in Figure \ref{fig:qiso_bao_lp_mocks_uncorrected} shows that \qisolp$\ge 1$, but any trend with redshift is difficult to discern given the error bars.
Since the pre-reconstruction \qisolp\ measurements are all greater than 1.005, the 0.5\% correction to the linear point proposed in \cite{anselmi2016_lp} will neither fully correct this offset nor bring the measurements into agreement with the post-reconstruction values. 

The right-hand panel shows that \qisolp\ is substantially closer to unity in the post-reconstructed field.  Here, some of the discrepancy from unity, and with respect to \qisobao, arises from the fact that $\Sigma_\mathrm{iso}\ne 0$ in the post-reconstruction field.  To address this, we correct each \qisolp\ measurement using the sample-dependent value of $s_\mathrm{LP}^\mathrm{fid}$ which reflects the bias in the linear point that arises from smearing in the fiducial cosmology.  This bias is larger in the pre-reconstructed fields, but recall that it is non-zero post-reconstruction as well.

Figure~\ref{fig:qiso_bao_lp_mocks_corrected} shows the result of using these modified (corrected) \slp\ values.   
We see that the offset from the \qisobao\ values is reduced, with linear point measurements much closer to unity, indicating better agreement with the fiducial cosmology.  
However, we note an increase in the scatter in the measurements up to 29\% (48\%) in the pre- (post-)reconstruction regime when compared to the uncorrected \qisolp\ measurements. These arise from propagating the uncertainties on the appropriate value of $\Sigma_{\rm iso}$ into $s_\mathrm{LP}^\mathrm{fid}$, as listed in Table \ref{tab:slp_fid}. More precise measurements yield a larger relative increase in the size of the error bars in Figure \ref{fig:qiso_bao_lp_mocks_corrected}; as we will see in later sections, this increase is therefore relatively modest at the precision of DR1 and DR2 data themselves. Regardless, this is in principle a drawback of the correction regime introduced in Section \ref{sec:linear_point}.

The above correction, when applied to the pre-reconstruction linear point measurements, should in principle completely undo the effects of non-linearities and yield measurements consistent with linear theory, given the correct prior on the isotropic damping scale $\Sigma_\mathrm{iso}$. It is evident from the difference in the left and right-side panels in Figure \ref{fig:qiso_bao_lp_mocks_corrected} that this is not the case. The scatter in pre-reconstruction \qisolp\ measurements on mocks is considerably larger than post-reconstruction, with mean values that remain more than 0.5\% offset from unity for most tracers. The larger scatter is likely a consequence of poorer polynomial fits pre-reconstruction: the BAO signal-to-noise is substantially lower, degrading the quality of the fit to the correlation function across the 25 mocks. This trend is absent post-reconstruction, where the improved signal-to-noise yields tighter fits and correspondingly smaller scatter (see \cite{eigencov2024} for why this is expected).
This points to a limitation of the linear point pipeline in the strongly-damped regime.
Reconstruction is therefore a crucial step in achieving higher precision, lower scatter, and better agreement with isotropic BAO measurements. However, this reliance on reconstruction, combined with the need for a sample-dependent correction, compromises the strictly model-independent nature of the linear point pipeline once we seek sub-percent precision and agreement with BAO results derived from template-based fits.

\section{DESI Linear Point Measurements and Comparisons}
\label{sec:results}
In this section, we present the linear point measurements for the \bgs, \lrg, and \elg\ targets in the first and second DESI data releases (DR1 and DR2) using two-point correlation functions and covariance matrices from \cite{desi2024_iii, desi_dr2_bao}. 
We compare the linear point measurements to BAO measurements made using the DESI template-based BAO fitting pipeline \cite{desi2024_iii, desi2024_v, desi_dr2_bao}.

\subsection{DR1 results}

\begin{figure}
    \centering
    \begin{tabular}{ccc}
       *\hspace{-.85cm}\includegraphics[width=0.33\textwidth]{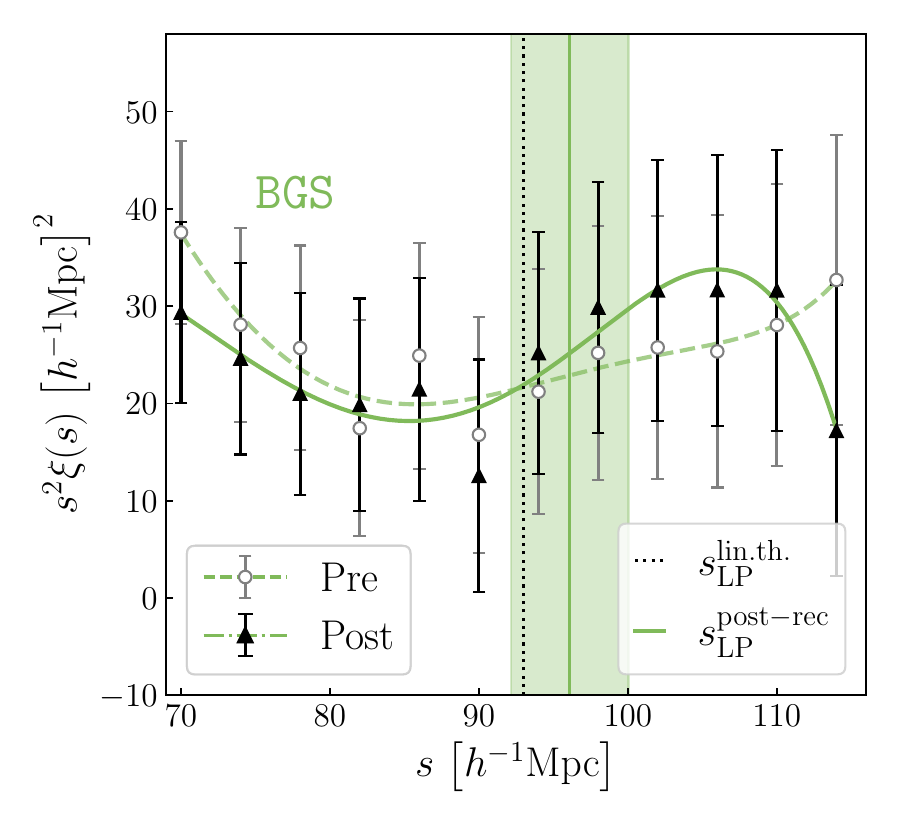} &
       \includegraphics[width=0.33\textwidth]{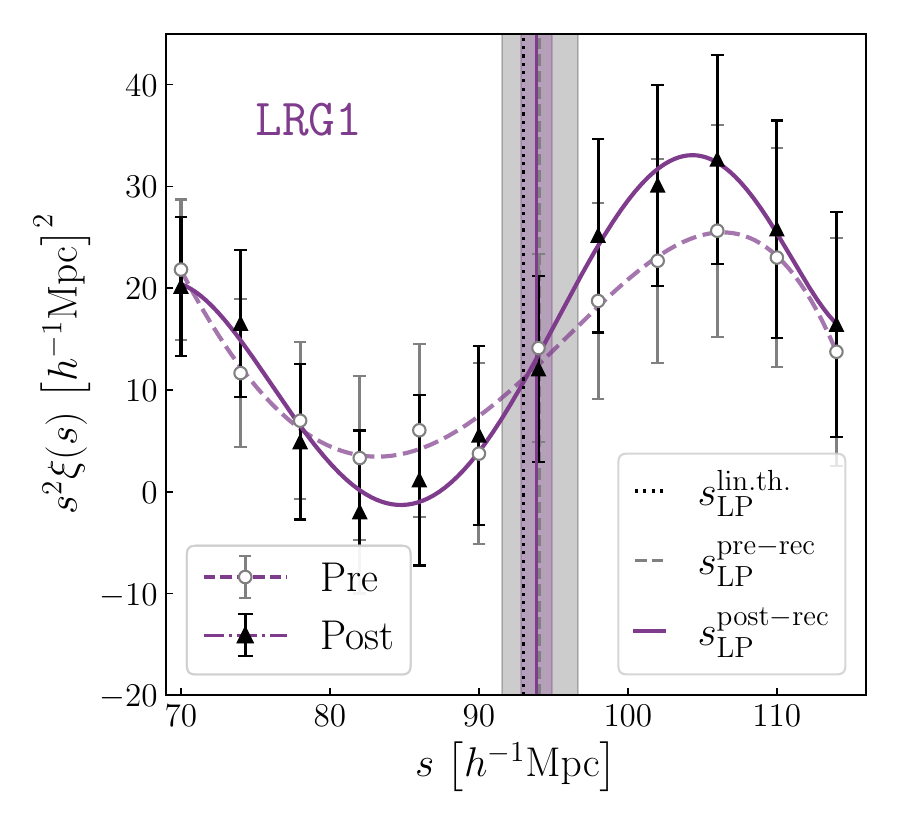}  & \includegraphics[width=0.33\textwidth]{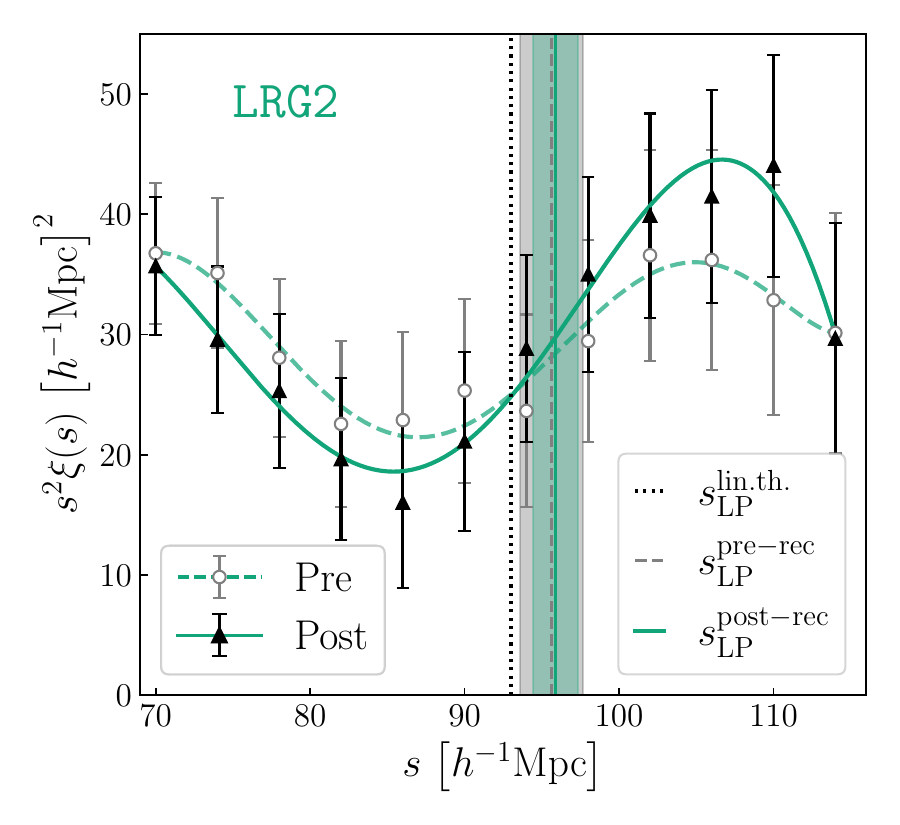} \\
    \end{tabular}
    \begin{tabular}{cc}
    *\hspace{-.7cm}\includegraphics[width=0.33\textwidth]{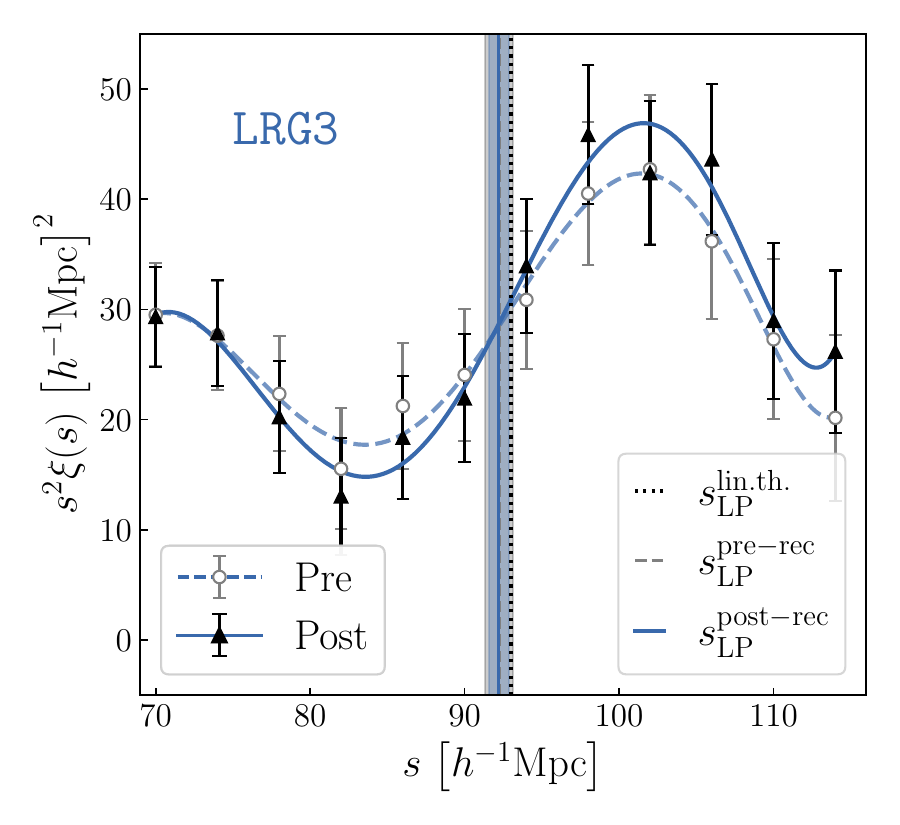} &
    \includegraphics[width=0.33\textwidth]{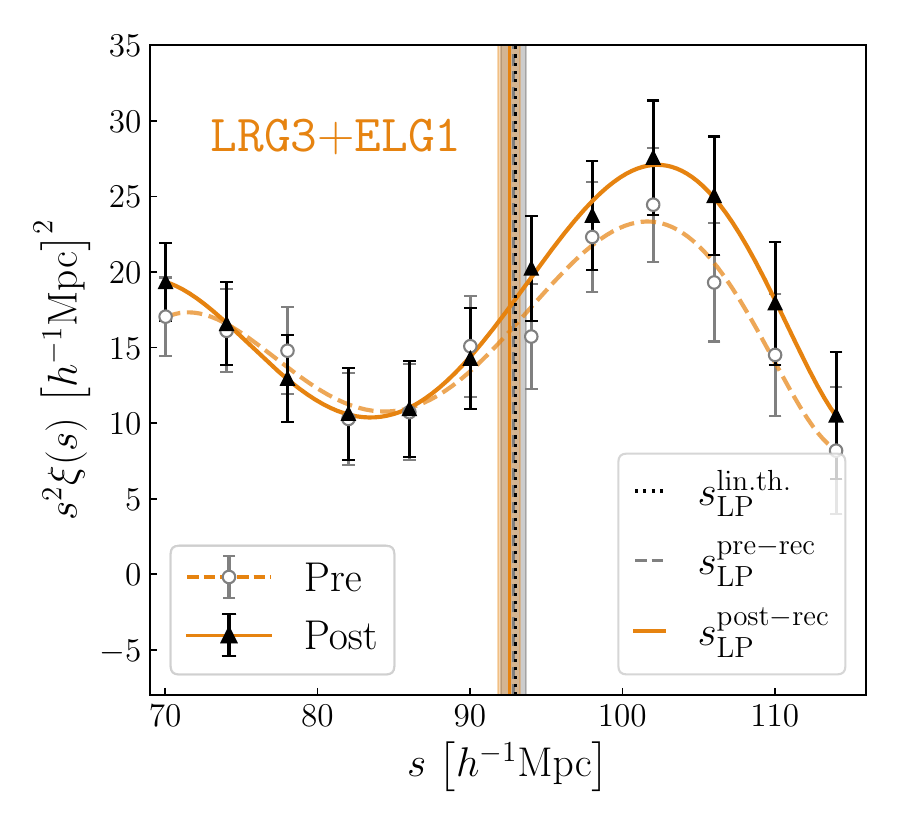} \\
    *\hspace{-.7cm}\includegraphics[width=0.33\textwidth]{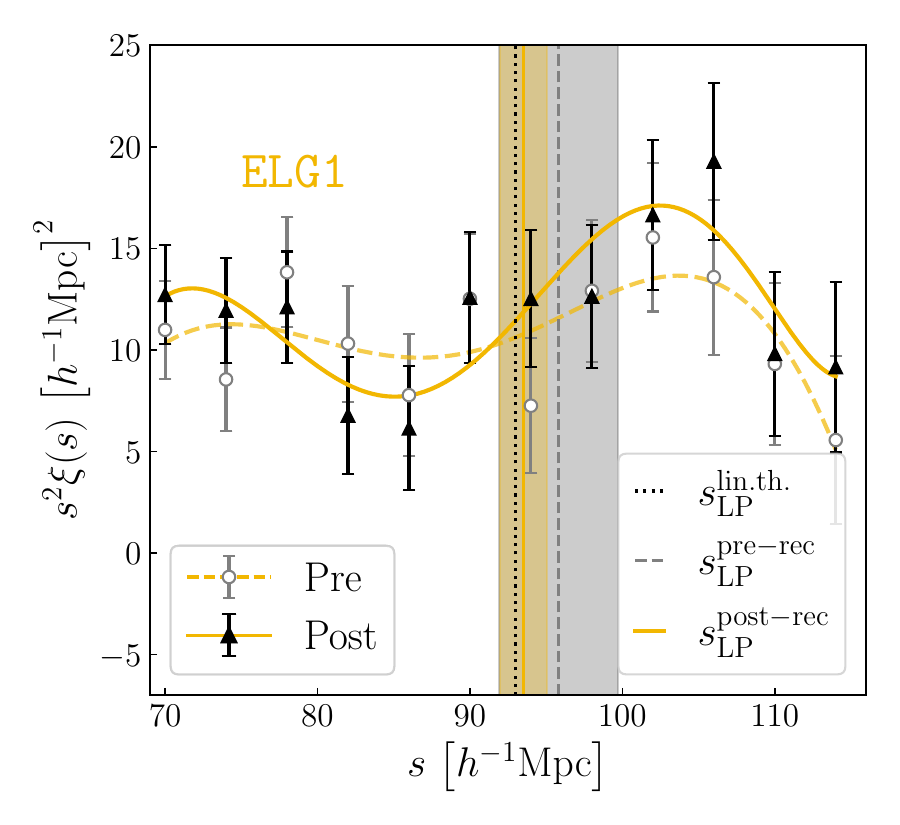} &
    \includegraphics[width=0.33\textwidth]{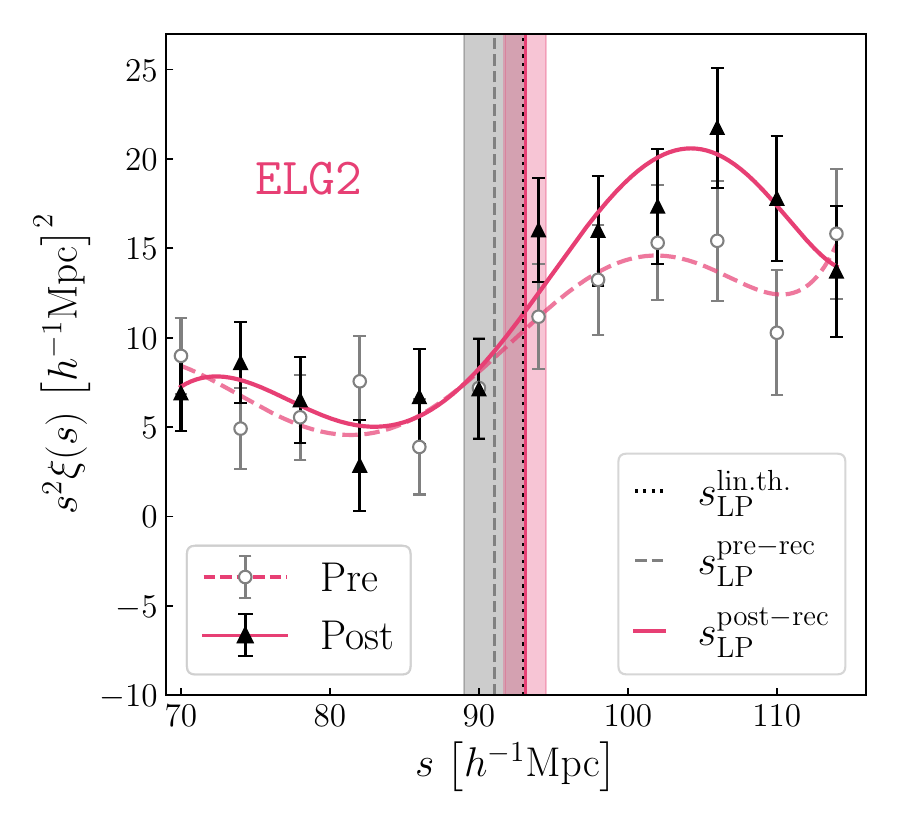}
    \end{tabular}
    \caption{Two-point correlation functions for DESI DR1 tracers, one per panel, before (open circles) and after reconstruction (solid triangles). Overplotted curves show the best-fit polynomial interpolations over the linear point fitting range (70 \hMpc$<s<$ 115 \hMpc) before (dashed) and after (dotted-dashed) reconstruction; all curves are multiplied by $s^2$ following convention. Vertical lines mark the measured linear point \slp, pre- (gray) and post-reconstruction (colored), with shaded bands giving the $1\sigma$ uncertainty. The dotted vertical line marks the linear point of the unsmeared linear theory correlation function, computed with {\tt CLASS} for the DESI fiducial cosmology (identical in all panels). The linear point could not be measured on the pre-reconstruction \bgs\ correlation function, so the first panel omits the gray line and its shaded band. Measured linear points for all tracers are listed in Table \ref{tab:Y1_results}.
    }
    \label{fig:Y1_LP}
\end{figure}

\begin{table}
    \centering
    \resizebox{\columnwidth}{!}{%
    \begin{tabular}{|l|c|c|c|c|c|c|}
    \hline
        \multicolumn{1}{|c|}{Tracer} & Redshift & Recon & $s_\mathrm{LP}$ [\hMpc] & $\alpha_\mathrm{iso, LP}$ & $y^\mathrm{obs}_\mathrm{LP}$ & $\chi^2$/dof\\
        \hline
        \lrgo & 0.4-0.6 & Pre & $94.09 \pm 2.14$ & $0.989 \pm 0.023$ & $0.0740 \pm 0.0017$ & 3.33/6\\
        \lrgt & 0.6-0.8 & Pre & $95.60 \pm 1.70$ & $0.973 \pm 0.017$ & $0.0584 \pm 0.0010$ & 4.25/6\\
        \lrgth & 0.8-1.1 & Pre & $92.24 \pm 0.75$ & $1.008 \pm 0.008$ & $0.0471 \pm 0.0004$& 2.89/6\\
        \lrgelg & 0.8-1.1 & Pre & $92.83 \pm 0.82$ & $1.002 \pm 0.009$ & $0.0465 \pm 0.0004$ & 8.53/6 \\ 
        \elgo & 0.8-1.1 & Pre & $96.00 \pm 3.61$ & $0.969 \pm 0.036$ & $0.0480 \pm 0.0018$& 12.62/6\\
        \elgt & 1.1-1.6 & Pre & $90.27 \pm 1.69$ & $1.022 \pm 0.019$ & $0.0372 \pm 0.0007$& 6.66/6\\
        \hline
        \bgs & 0.1-0.4 & Post & $96.09 \pm 3.95$ & $0.968 \pm 0.040$ & $0.1397 \pm 0.0057$ & 4.93/6 \\
        \lrgo & 0.4-0.6 & Post & $93.87\pm 0.87$ & $0.991 \pm 0.009$ & $0.0738 \pm 0.0007$& 3.48/6\\
        \lrgt & 0.6-0.8 & Post & $95.86 \pm 1.24$ & $0.970 \pm 0.013$ & $0.0585 \pm 0.0008$& 4.82/6\\
        \lrgth & 0.8-1.1 & Post & $92.21 \pm 0.54$ & $1.009 \pm 0.006$ &$0.0471 \pm 0.0003$ & 6.50/6\\
        \lrgelg & 0.8-1.1 & Post & $92.57 \pm 0.73$ & $1.005 \pm 0.008$ & $0.0463 \pm 0.0004$ & 0.96/6 \\
        \elgo & 0.8-1.1 & Post & $93.52 \pm 1.25$ & $0.995 \pm 0.013$ & $0.0468 \pm 0.0006$& 9.58/6\\
        \elgt & 1.1-1.6 & Post & $93.11 \pm 1.08$ & $0.999 \pm 0.012$ &$0.0384 \pm 0.0004$ & 7.48/6\\
        \hline
    \end{tabular}%
    }
    \caption{Linear point measurements on the \bgs, \lrg, and \elg\ correlation functions from the first DESI data release (DR1). For each tracer and redshift bin we list the measured linear point \slp, the corresponding isotropic ratio \qisolp\ (without correcting for the smearing of the fiducial linear point due to non-linear evolution), the angular size of the linear point standard ruler $y^\mathrm{obs}_\mathrm{LP}$ (Eq. \ref{eq:yLP_obs}), and the reduced chi-squared $\chi^2$/dof of the polynomial fit. The linear point could not be measured on the pre-reconstruction \bgs\ correlation function; that entry is omitted and discussed in Appendix \ref{sec:bgs_bad}.
    } 
    \label{tab:Y1_results}
\end{table}
We present the \slp\ estimates for correlation function measurements for DR1 samples in Table \ref{tab:Y1_results}, along with the \qisolp\ and $y_\mathrm{LP}$ measurements and the $\chi^2$ values corresponding to each fit\footnote{The $\chi^2$ value for the \lrgelg\ tracer post-reconstruction is observed to be smaller than that of other tracers. We estimate that this occurs in about 1\% of samples; we therefore consider this a reasonable value.}. 
We plot the pre- and post-reconstruction correlation functions, zoomed in on the region used in the linear point pipeline with quintic polynomial fits, in Figure \ref{fig:Y1_LP}, along with the measured linear points (gray band around dashed vertical line, and colored band around solid vertical line, respectively). The vertical dotted line shows $s_\mathrm{LP}^{\rm lin.th.}$. 

\begin{figure}[]
    \centering
    \begin{tabular}{cc}
       \includegraphics[width=0.45\linewidth]{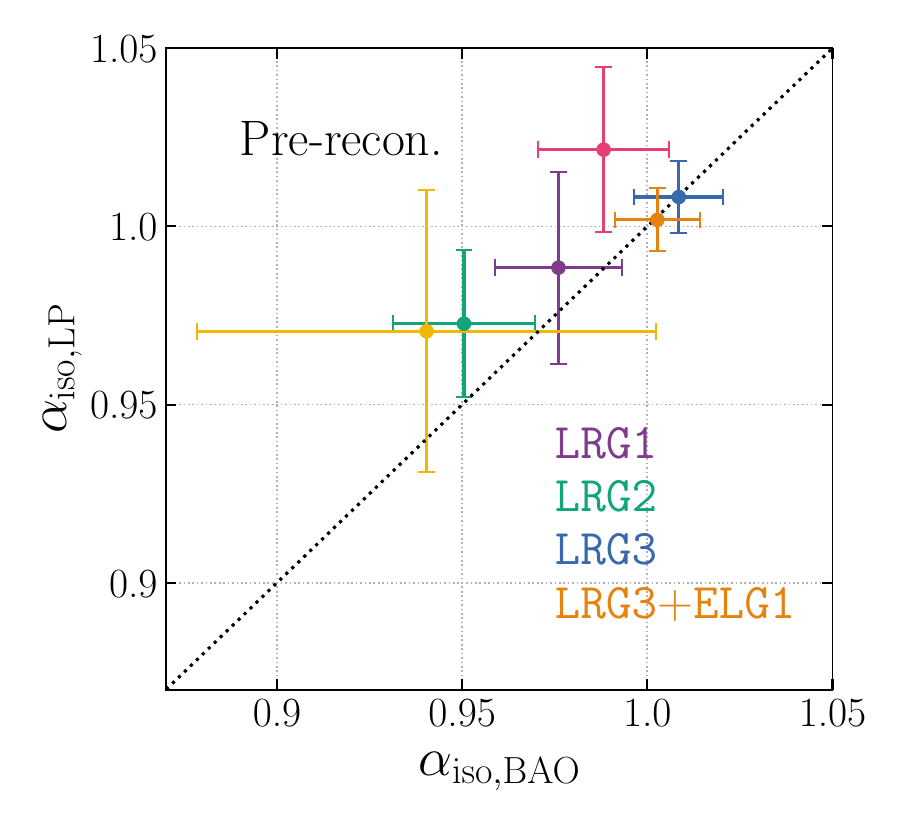}  &  \includegraphics[width=0.45\linewidth]{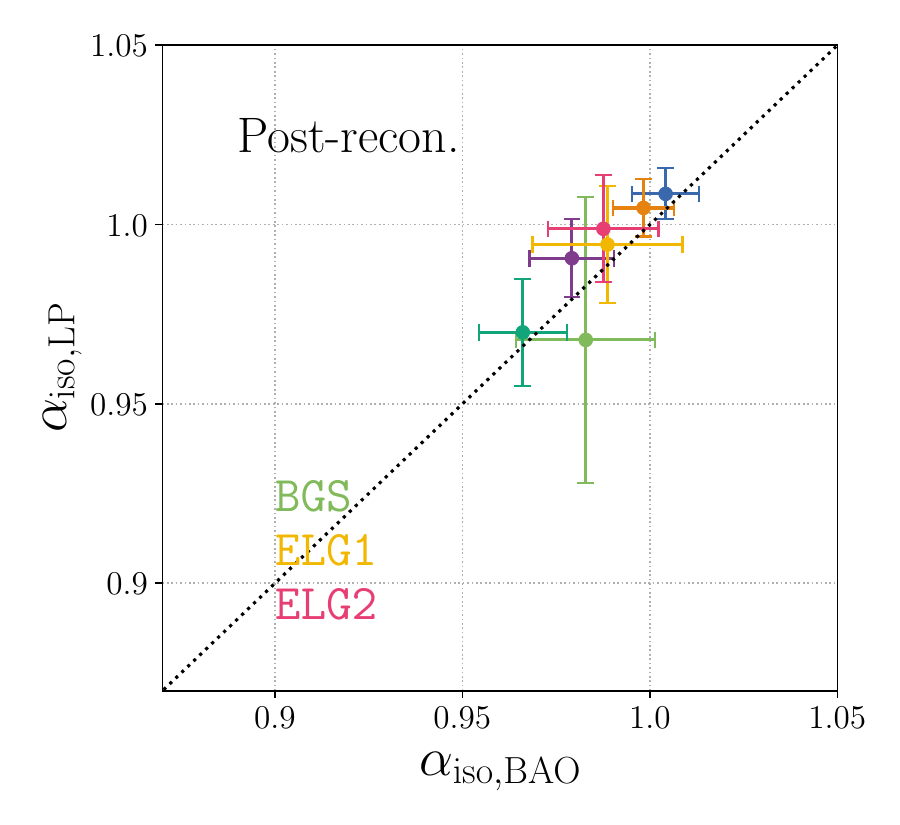}
    \end{tabular}
    \caption{Linear point measurements on the DESI DR1 tracers, converted to the uncorrected ratio \qisolp\ and plotted against the isotropic BAO dilation parameter \qisobao\ from \cite{desi2024_iii}, pre-reconstruction (left) and post-reconstruction (right). Each tracer is annotated in text of the same color as its point. Error bars on \qisolp\ are computed using the sampling approach of Section \ref{sec:calc_lp}, while those on \qisobao\ are outputs of the standard DESI fitting pipeline \cite{desi2024_iii, desi2024_v}. The dotted line marks $x=y$, along which the two measures agree exactly. The two sets of measurements agree well, especially post-reconstruction; note that, unlike in the mocks, the \qisolp\ uncertainties on data are smaller than those on \qisobao\ (see Appendix \ref{sec:errors}). This is the data analog of Figure \ref{fig:qiso_bao_lp_mocks_uncorrected}; the corrected ratio is shown in Figure \ref{fig:qiso_bao_lp_Y1_corrected}. Uncorrected \qisolp\ values for all tracers are listed in Table \ref{tab:Y1_results}.
}
    \label{fig:qiso_bao_lp_Y1_uncorrected}
\end{figure}

\begin{figure}[]
    \centering
    \begin{tabular}{cc}
       \includegraphics[width=0.45\linewidth]{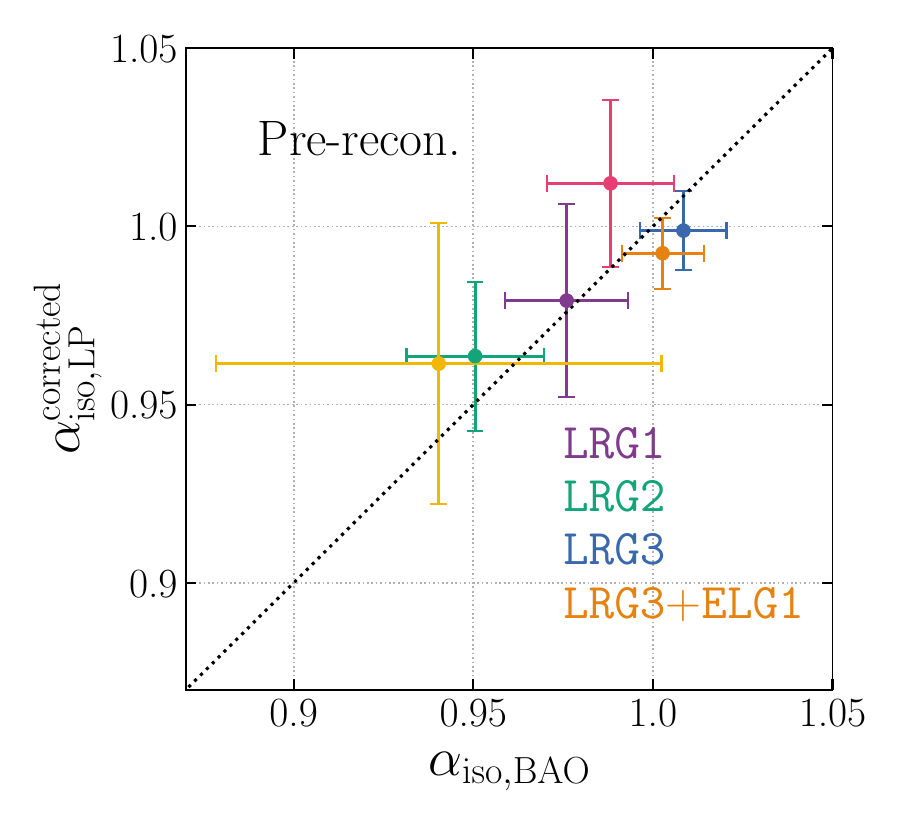}  &  \includegraphics[width=0.45\linewidth]{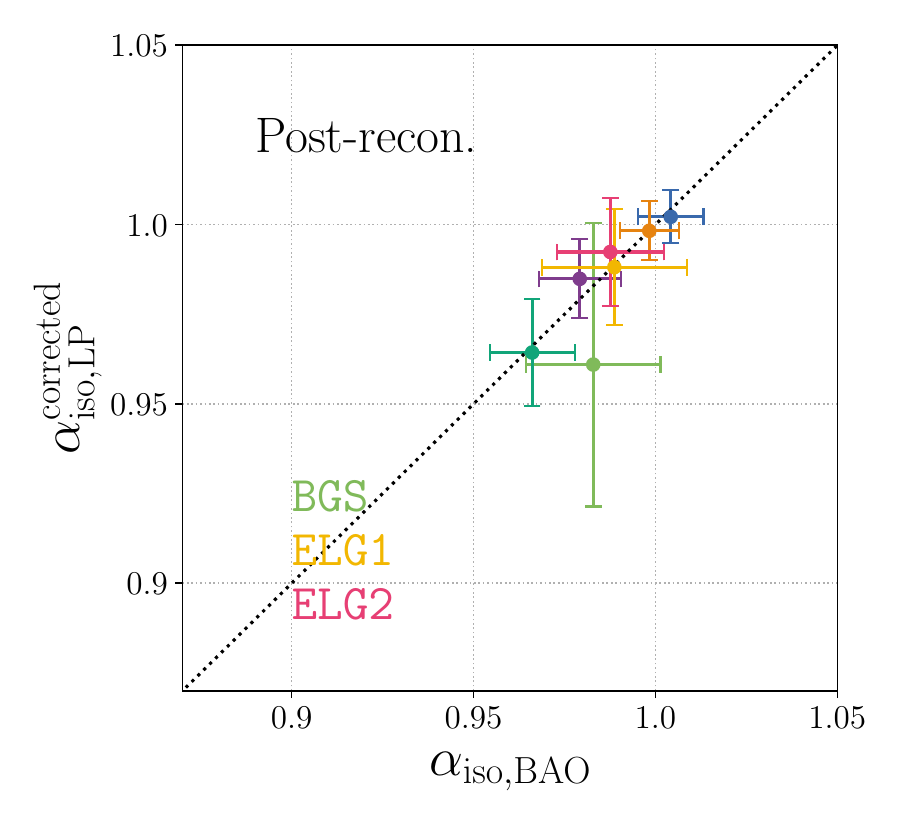}
    \end{tabular}
    \caption{Same as Figure \ref{fig:qiso_bao_lp_Y1_uncorrected}, but using the corrected ratio $\alpha_\mathrm{iso,LP}^\mathrm{corrected}$, which accounts for the smearing of the fiducial linear point due to late-time non-linear evolution (Table \ref{tab:slp_fid}; motivated in Section \ref{sec:robustness}). The correction shifts the linear point ratio on the $y$-axis downward by the same amount as in the mocks, but the shift is less visually apparent here because the single-realization data uncertainties are larger than the mock error-on-the-mean.}
    \label{fig:qiso_bao_lp_Y1_corrected}
\end{figure}

We note that we were not able to measure a linear point on the pre-reconstruction \bgs\ correlation function. We present its polynomial fit in Figure \ref{fig:Y1_LP}, where we see that no peak was identified within the fitting range. This highlights a major drawback of the linear point pipeline -- when the correlation function is poorly constrained due to low signal-to-noise ratio and/or large errors, a polynomial is no longer an adequate fitting function and the linear point feature cannot be identified. In these cases, reconstruction becomes essential for improving the signal-to-noise ratio and measuring the linear point. We explore this in more detail in Appendix \ref{sec:bgs_bad}, and discuss alternative fitting functions in Appendix \ref{sec:alteratives}. 

The naive expectation is that the linear point measured in pre-reconstructed fields (gray bands) should be slightly smaller at lower $z$. This was difficult to see in the mocks, and is not obviously the case in DR1; the higher redshift \elgt\ sample especially, appears to be shifted to smaller values than the others. (Note that we have {\em not} multiplied any of our \slp\ values by 1.005 as suggested by \cite{anselmi2016_lp, anselmi2017_sdss}.)  The \elgo\ sample has relatively noisier clustering measurements, and \elgt\ has low completeness and irregular footprint for DESI DR1 \cite{desi2024_ii, desi2024_iii}. This increased noise is expected to distort the polynomial interpolation especially of the pre-reconstruction correlation functions (see $\chi^2$ for pre-recon \elgo\ in Table \ref{tab:Y1_results}), ultimately leading to larger estimates of uncertainties on the linear point (this is especially evident for \elgo\ in Fig \ref{fig:Y1_LP}).

The linear point \qisolp\ measurements are compared against the DR1 isotropic BAO dilation parameter measurements \qisobao\ \cite{desi2024_iii} in Figure \ref{fig:qiso_bao_lp_Y1_uncorrected}. The error bars on \qisolp\ are generated using the sampling approach discussed in Section \ref{sec:calc_lp}, while the errors on \qisobao\ are an output of the standard fitting pipeline \cite{desi2024_iii, desi2024_v}.  Figure~\ref{fig:qiso_bao_lp_Y1_uncorrected} is the data analog of Figure~\ref{fig:qiso_bao_lp_mocks_uncorrected}; the corrected \qisolp\ values, analogous to Figure~\ref{fig:qiso_bao_lp_mocks_corrected}, are plotted against the same \qisobao\ measurements in Figure~\ref{fig:qiso_bao_lp_Y1_corrected}.
Similar to what we saw in the mocks, we see excellent agreement between the two sets of measurements, especially post-reconstruction, which also improves the precision of \qisolp\ measurements.  
The uncertainties on the \slp\ measurements on data, however, are smaller than those measured on the isotropic BAO scale, in contrast to mocks, where the \qisolp\ errors (scatter in measurements) were larger than \qisobao\ scatter.  We study the discrepancy between mock errors and data errors in Appendix \ref{sec:errors}.
Nevertheless, it is striking that the pre-reconstruction \qisolp\ values are slightly smaller than unity; in the mocks they were slightly larger. Likewise, in the corrected, post-reconstruction values, the estimated distance scale seems, if anything, to be slightly larger than the fiducial cosmology value, in agreement with the \qisobao\ values.  With this in mind, we now consider estimates in DESI DR2.

\subsection{DR2 results}

\begin{figure}
    \centering
    \begin{tabular}{ccc}
       *\hspace{-.85cm}\includegraphics[width=0.33\textwidth]{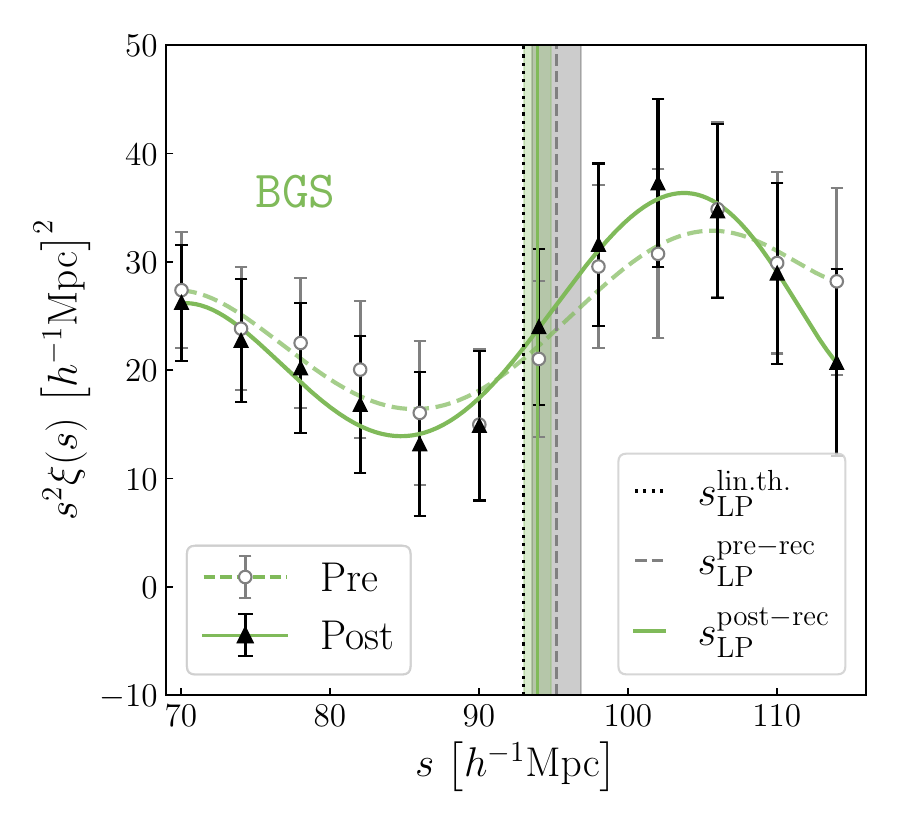} &
       \includegraphics[width=0.33\textwidth]{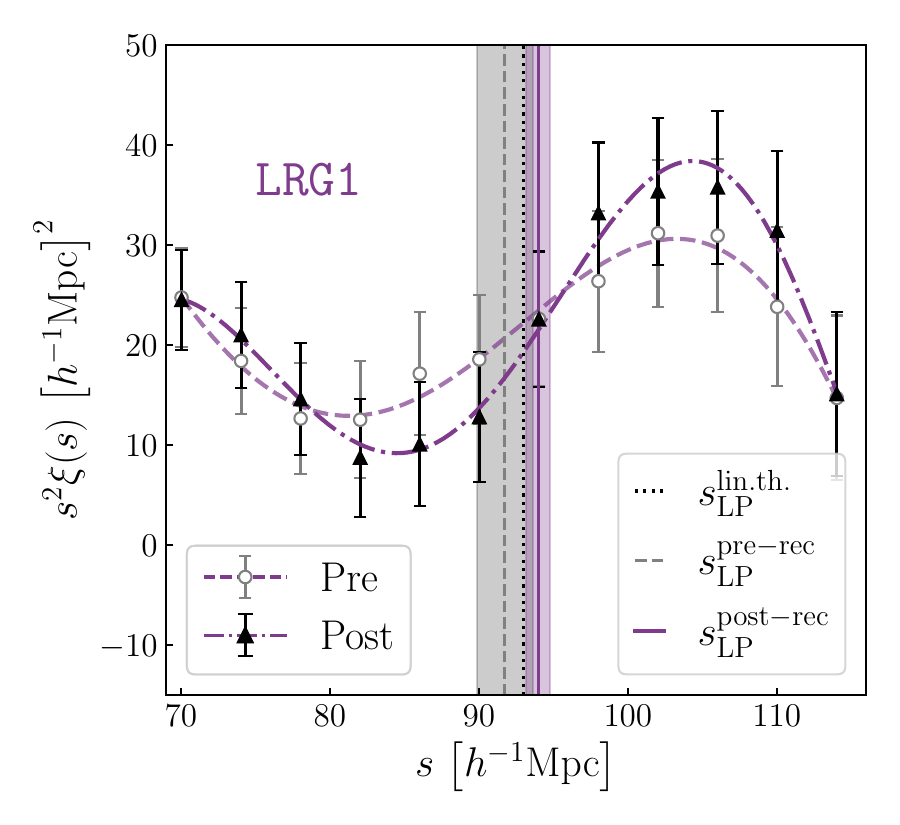}  & \includegraphics[width=0.33\textwidth]{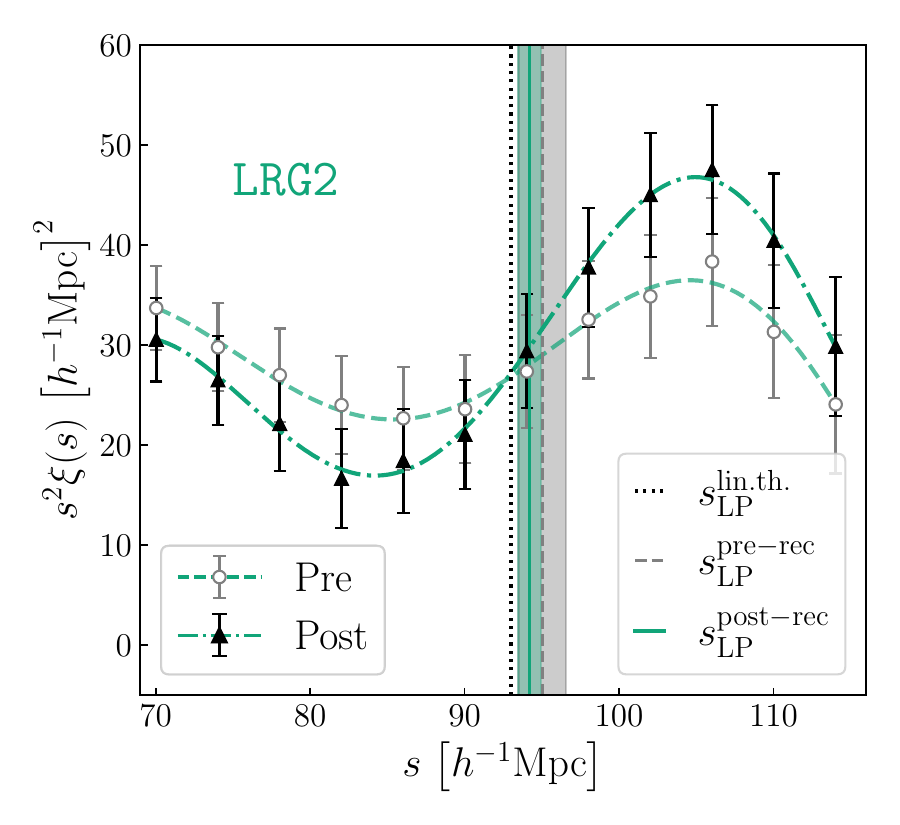} \\
    \end{tabular}
    \begin{tabular}{cc}
    *\hspace{-.7cm}\includegraphics[width=0.33\textwidth]{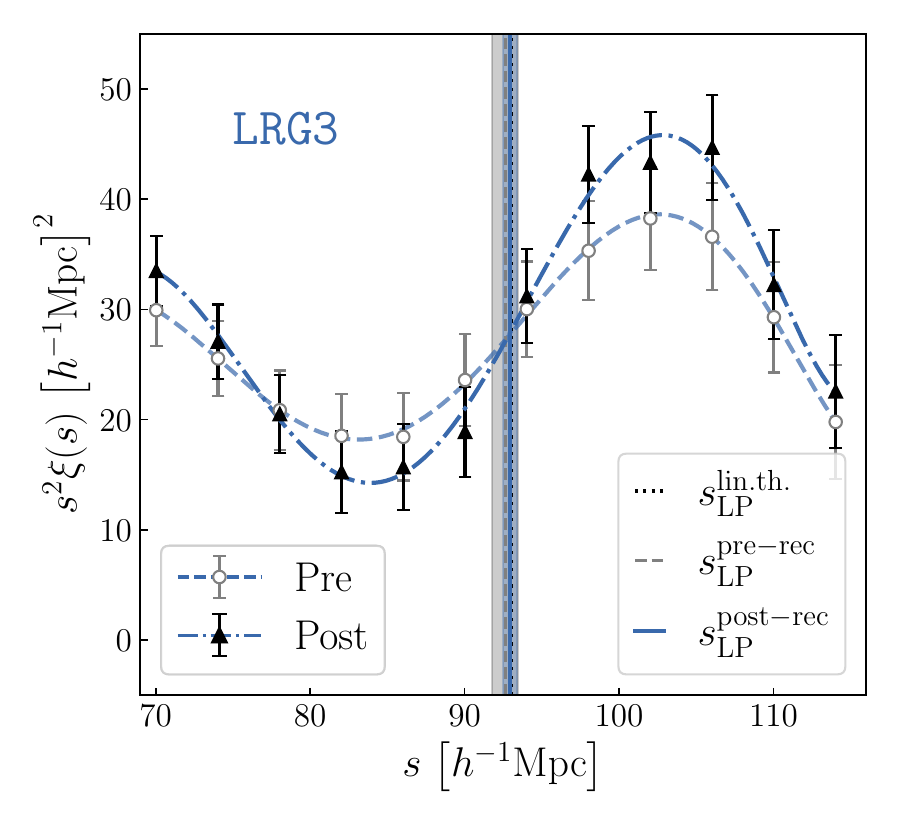} &
    \includegraphics[width=0.33\textwidth]{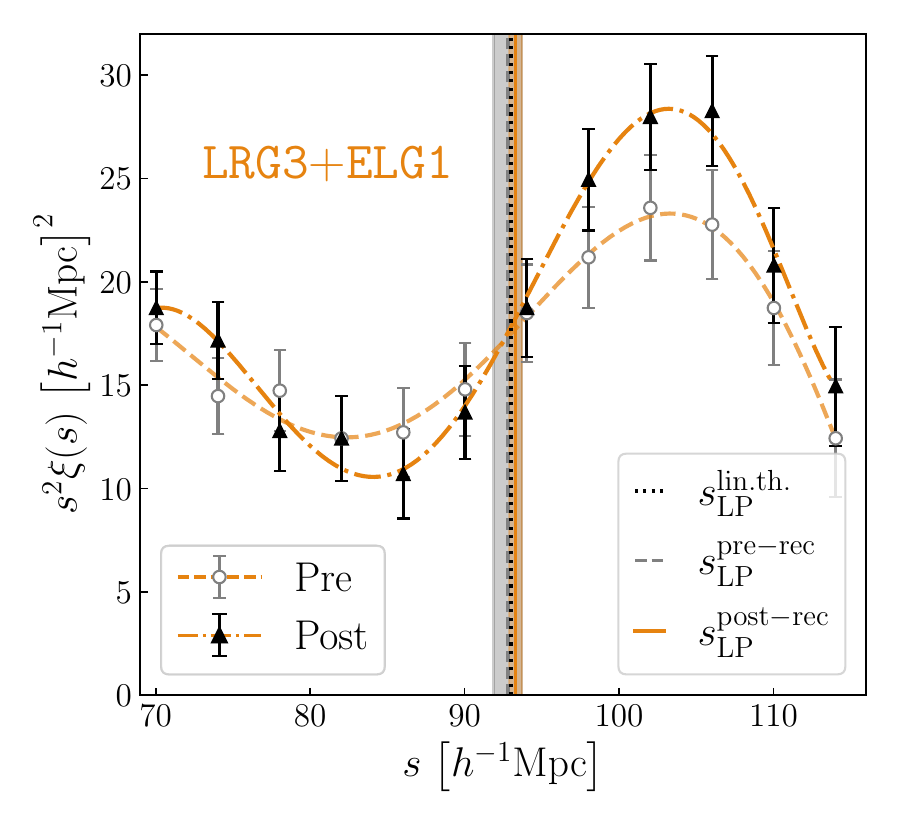} \\
    *\hspace{-.7cm}\includegraphics[width=0.33\textwidth]{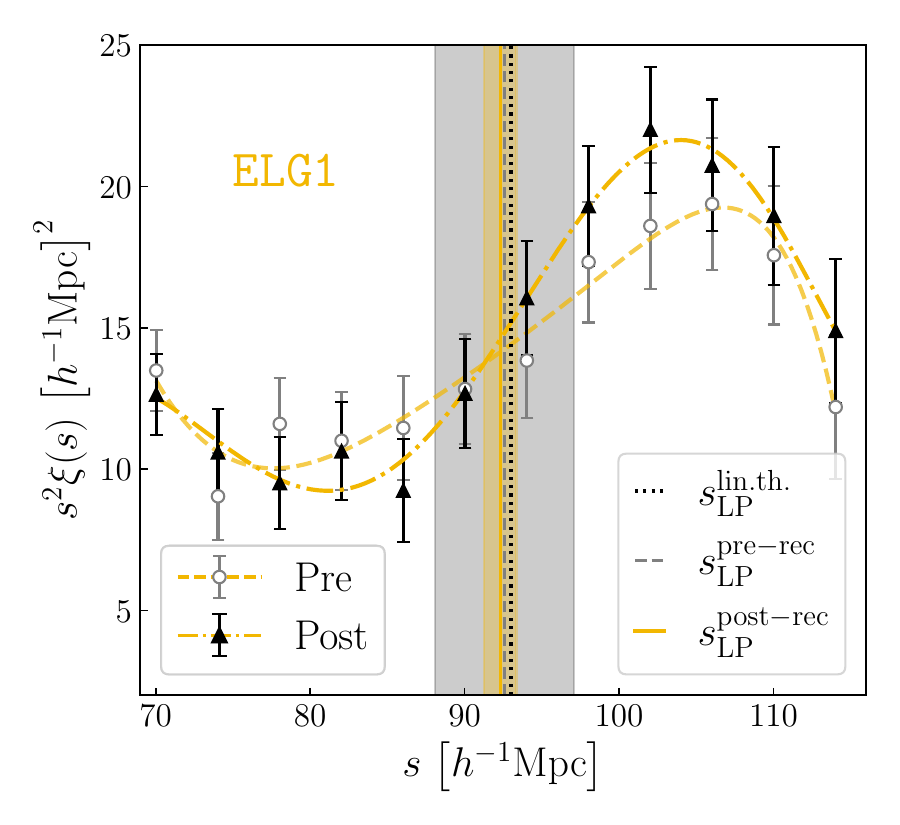} &
    \includegraphics[width=0.33\textwidth]{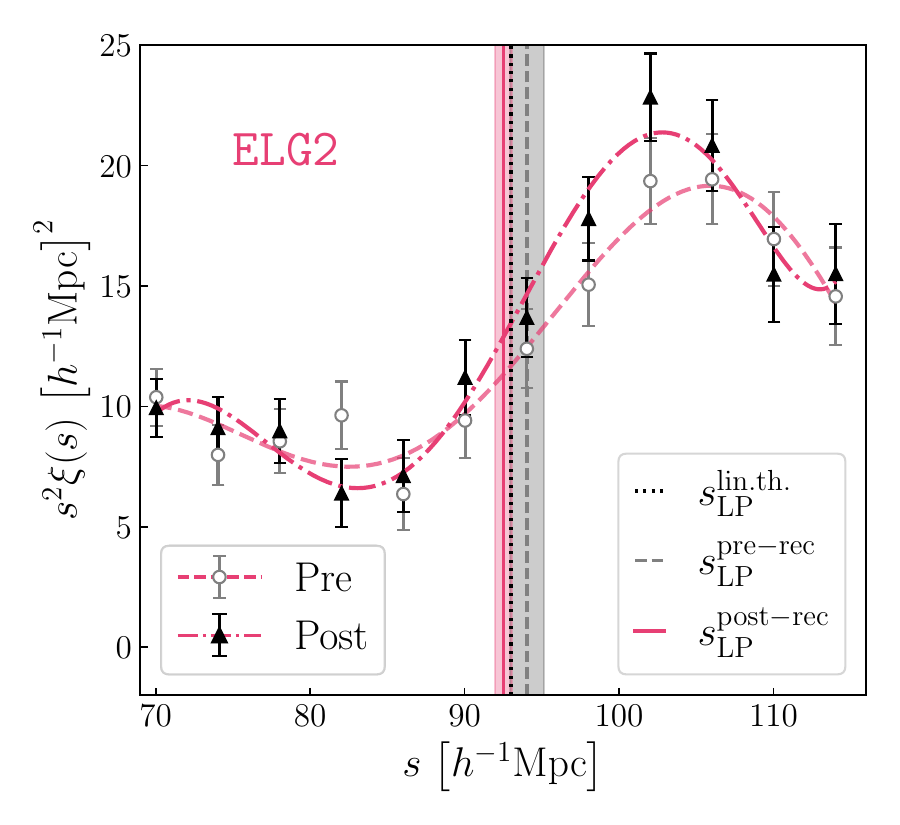}
    \end{tabular}
    \caption{Two-point correlation functions for the DESI DR2 tracers, one per panel, with the best-fit quintic polynomial interpolation shown as dashed (pre-reconstruction) and dot-dashed (post-reconstruction) lines. The black dotted vertical line marks the linear point in linear theory, $s_\mathrm{LP}^\mathrm{lin.th.}$ , for the fiducial Planck 2018 \lcdm\ cosmology. The dashed gray and solid colored vertical lines mark the measured linear points pre- and post-reconstruction respectively, with the shaded bands giving the $1\sigma$ uncertainties.}
    \label{fig:Y3_LP}
\end{figure}

The linear point measurements on the DR2 \bgs, \lrg, and \elg\ samples are listed in Table \ref{tab:Y3_results}, along with the corresponding ratio \qisolp\ and angular size $y_\mathrm{LP}^\mathrm{obs}$ values and best-fit reduced $\chi^2$ of the polynomial fit\footnote{The pre-reconstruction \lrgth\ $\chi^2$ is unusually small relative to the other tracers. We estimate that this occurs in roughly 0.1\% of samples. This may reflect an overestimated covariance matrix, though we do not have a way to assess its statistical significance directly. The post-reconstruction $\chi^2$ for the same tracer is, however, reasonable. Conversely, the post-reconstruction \lrgelg\ $\chi^2$ is large, a value we estimate to occur about 4\% of the time.}. The corresponding correlation functions are shown in Figure \ref{fig:Y3_LP}, each with its best-fit quintic polynomial interpolation and the measured linear point marked as a vertical line. 

\begin{table}
    \centering
    \resizebox{\columnwidth}{!}{%
    \begin{tabular}{|l|c|c|c|c|c|c|}
    \hline
            \multicolumn{1}{|c|}{Tracer} & Redshift & Recon & $s_\mathrm{LP}$ [\hMpc] & $\alpha_\mathrm{iso, LP}$ & $y_\mathrm{LP}$ & $\chi^2$/dof\\
            \hline
            \bgs & 0.1-0.4 & Pre & $95.21 \pm 1.53$ & $0.977\pm  0.016$ & $0.1193\pm   0.0019$ & 13.80/6\\
            \lrgo & 0.4-0.6 &  Pre & $91.74 \pm 1.69$ & $1.014\pm  0.019$ & $0.0721\pm 0.0013$& 4.78/6\\
            \lrgt & 0.6-0.8 & Pre & $95.02 \pm 1.44$ & $0.979\pm  0.015$ & $0.0580\pm0.0009$ & 4.27/6\\
            \lrgth & 0.8-1.1 & Pre & $92.62 \pm 0.80$ & $1.004\pm  0.009$ & $0.0473\pm0.0004$ & 0.45/6\\
            \lrgth+\elgo & 0.8-1.1 & Pre & $92.81 \pm 0.89$ & $1.002\pm  0.010$ & $0.0464\pm0.0004$ & 11.40/6\\
            \elgo & 0.8-1.1 & Pre & $92.59 \pm 3.06$ & $1.005\pm  0.033$ & $0.0463\pm0.0015$ & 9.80/6\\
            \elgt & 1.1-1.6 & Pre & $94.03 \pm 1.09$ & $0.989\pm  0.012$ & $0.0388\pm0.0005$ & 13.46/6\\
            \hline
            \bgs &  0.1-0.4 & Post & $93.91 \pm 0.86$ & $0.990\pm  0.009$ & $ 0.1176\pm0.0011$ & 8.44/6\\
            \lrgo & 0.4-0.6 &  Post & $93.98 \pm 0.78$ & $0.990\pm  0.008$ & $0.0739\pm0.0006$& 6.89/6\\
            \lrgt & 0.6-0.8 & Post & $94.19 \pm 0.72$ & $0.987\pm  0.008$ & $0.0575\pm0.0004$ & 3.68/6\\
            \lrgth & 0.8-1.1 & Post & $92.94 \pm 0.46$ & $1.001\pm  0.005$ & $0.0475\pm0.0002$ & 7.90/6\\
            \lrgth+\elgo & 0.8-1.1 & Post & $93.31 \pm 0.42$ & $0.997\pm  0.004$ & $0.0467\pm 0.0002$ & 13.73/6\\
            \elgo & 0.8-1.1 & Post & $92.31 \pm 0.99$ & $1.008\pm  0.011$ & $0.0462\pm0.0005$ & 4.64/6\\
            \elgt & 1.1-1.6 & Post & $92.52 \pm 0.55$ & $1.005\pm  0.006$ & $0.0382\pm0.0002$ & 8.74/6\\
            \hline
    \end{tabular}
    }
    \caption{Linear point measurements on the pre- and post-reconstruction correlation functions of the DESI second data release (DR2) tracers \cite{desi_dr2_bao}. For each tracer and redshift bin we list the measured linear point \slp, the corresponding isotropic ratio \qisolp\ (uncorrected), the angular size of the linear point standard ruler $y_\mathrm{LP}^\mathrm{obs}$, and the reduced chi-squared $\chi^2$/dof of the polynomial fit.
    }
\label{tab:Y3_results}
\end{table}

\begin{figure}
    \centering
    \begin{tabular}{cc}
       \includegraphics[width=0.45\linewidth]{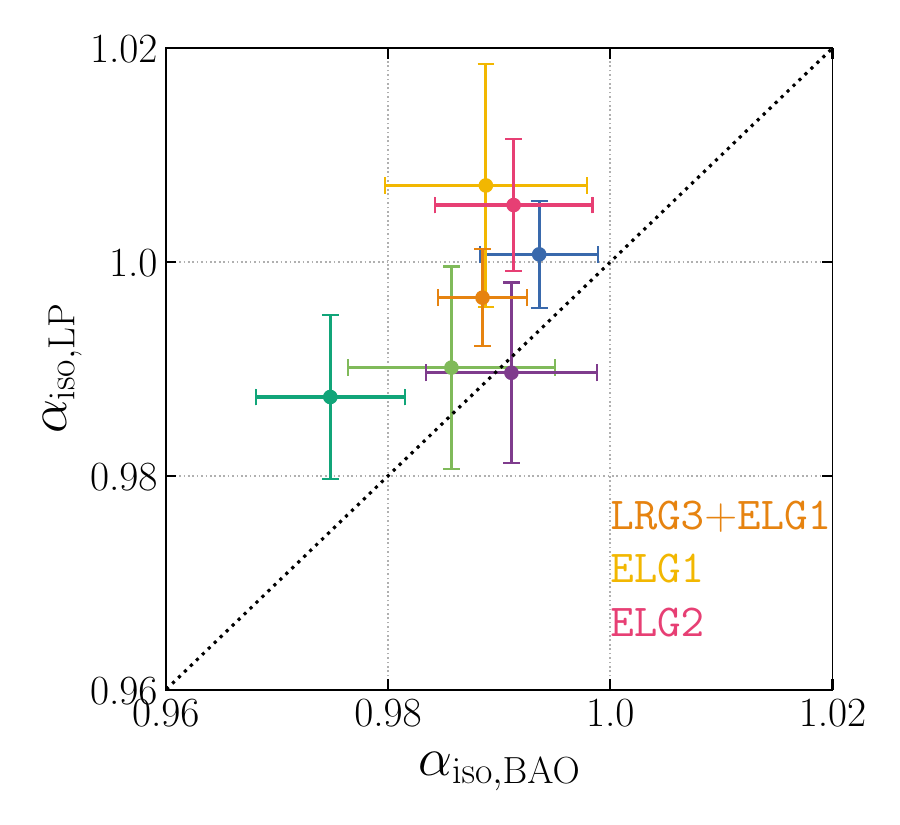}  & \includegraphics[width=0.45\linewidth]{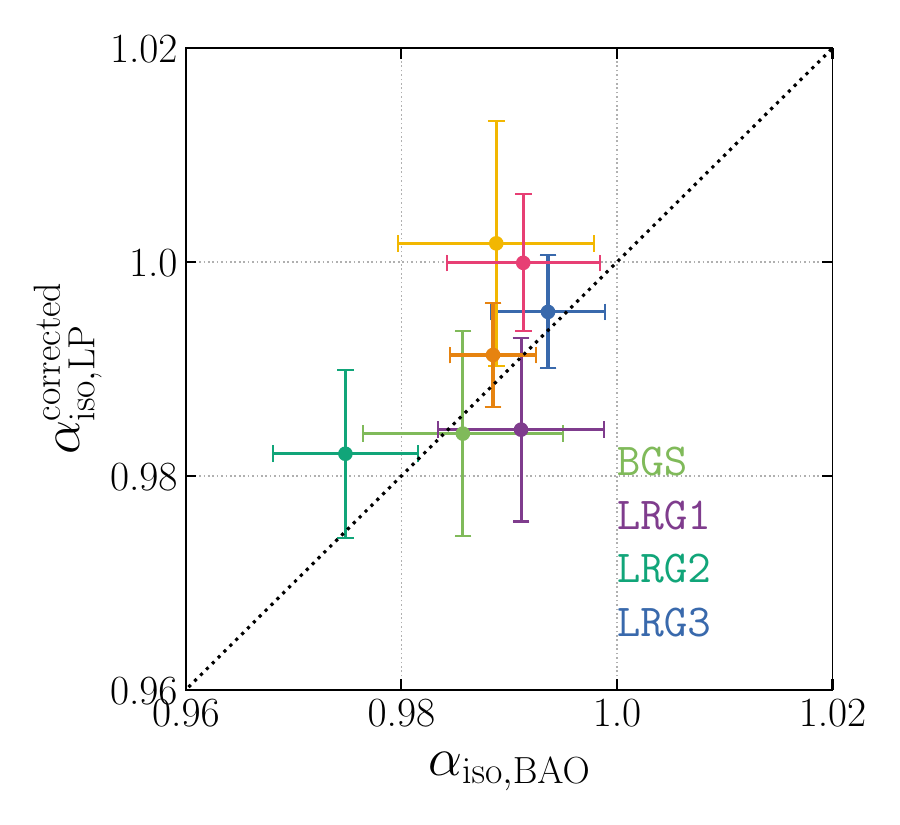}
    \end{tabular}  
    \caption{Post-reconstruction linear point measurements on DESI DR2 tracers, converted to the ratio \qisolp\ and compared against the isotropic BAO measurements \qisobao\ from \cite{desi_dr2_bao}, before (left) and after (right) applying the correction for the smearing of the fiducial linear point (Table \ref{tab:slp_fid}). Each tracer is annotated in the text of the same color as its point, and the dotted line marks $x=y$ along which the two measurements agree exactly. Error bars on \qisolp\ are computed using the sampling approach of Section \ref{sec:calc_lp}, while those on \qisobao\ are outputs of the standard DESI fitting pipeline and are quoted in \cite{desi_dr2_bao}. Note that the axis range here is $\sim3\times$ smaller than in the DR1 comparison (Figures \ref{fig:qiso_bao_lp_Y1_uncorrected} and \ref{fig:qiso_bao_lp_Y1_corrected}), reflecting the improved precision of the DR2 measurements.}   
    \label{fig:qiso_bao_lp_Y3}
\end{figure}

We compare \qisolp against the DR2 isotropic BAO measurements \qisobao\ from \cite{desi_dr2_bao} in Figure \ref{fig:qiso_bao_lp_Y3}. As the published DR2 BAO measurements are post-reconstruction, we restrict this comparison to the post-reconstruction case. The left panel uses the unmodified fiducial value $s_\mathrm{LP}^\mathrm{fid}=s_\mathrm{LP}^\mathrm{lin.th.}$ to calculate \qisolp, whereas the right panel applies the smearing correction to $s_\mathrm{LP}^\mathrm{fid}$ from Table \ref{tab:damping_parameters}. As with DR1, the two measurements agree well, and the correction removes the bias visible in the uncorrected (left panel) \qisolp\ values.  Improved DR2 correlation function measurements yield smaller uncertainties on both the linear point and BAO measurements than in DR1, consistent with the tighter axis range in Figure \ref{fig:qiso_bao_lp_Y3} relative to its DR1 counterpart.

\begin{figure}
    \centering
    \begin{tabular}{cc}
    \hspace{-0.5cm}\includegraphics[width=0.46\textwidth]{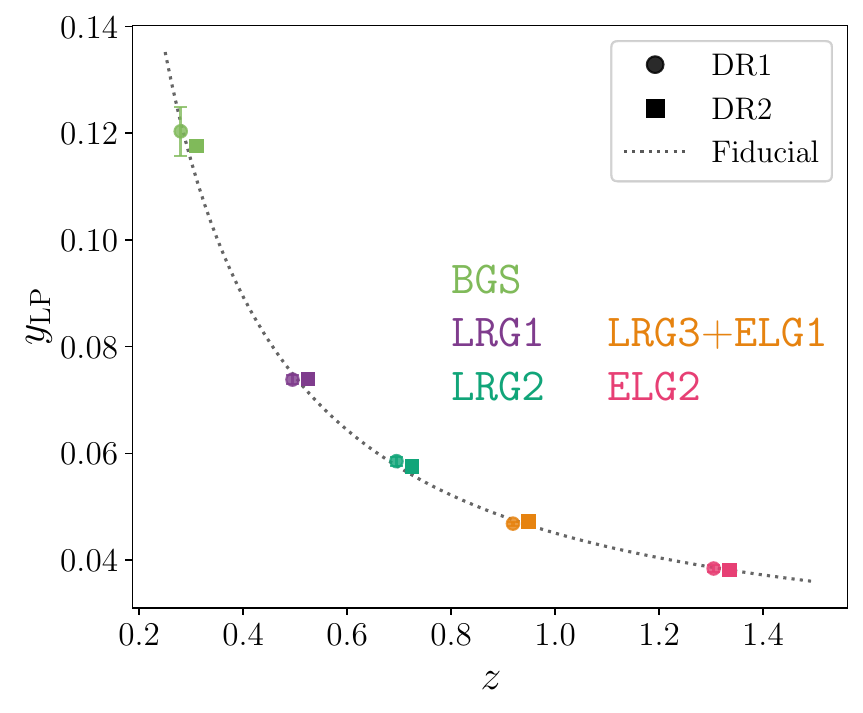} &
    \includegraphics[width=0.46\textwidth]{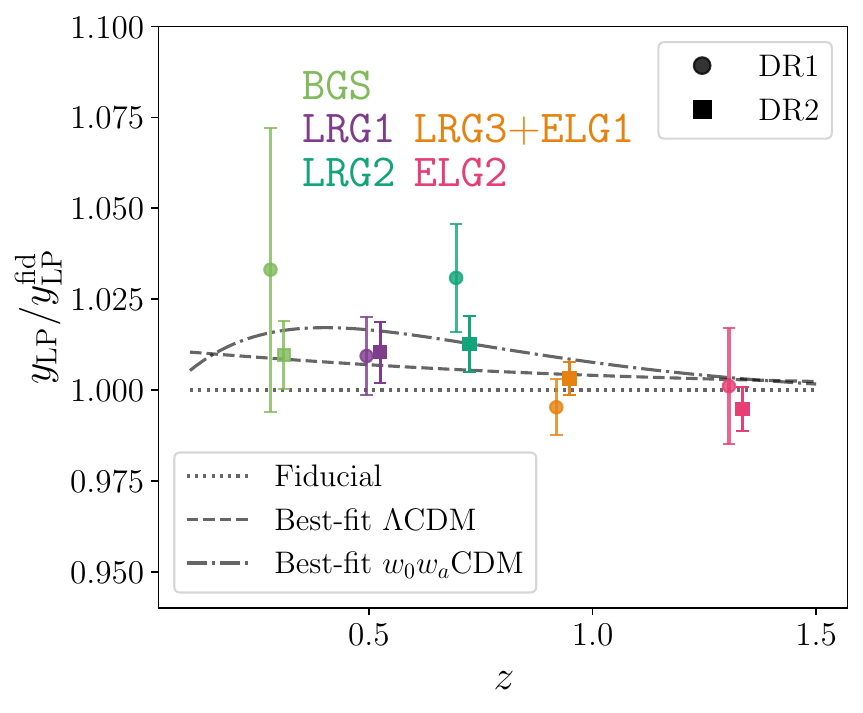}
    \end{tabular}  
    \caption{The angular form of the linear point standard ruler, $y_\mathrm{LP}\equiv s_\mathrm{LP}/D_V$ , as a function of redshift $z$. The left panel shows the post-reconstruction $y_\mathrm{LP}$ measurements for the DR1 (circles) and DR2 (squares) tracers considered in this work; the dotted line is the fiducial prediction $y_\mathrm{LP}^\mathrm{fid}$ for the \abacussummit\ base cosmology (i.e. Planck 2018 \lcdm). The right panel shows the same measurements normalized by $y_\mathrm{LP}^\mathrm{fid}$, so that the fiducial prediction becomes unity by construction; to this panel we add $y_\mathrm{LP}$ computed for the DESI DR2 best-fit \lcdm\ (dashed) and $w_0w_a$CDM (dot-dashed) cosmologies \cite{desi_dr2_bao}, likewise normalized. This panel is analogous to the top-left panel of Figure 13 in \cite{desi_dr2_bao}.} 
    \label{fig:ylp_z}
\end{figure}

The observable angular size of the linear point standard ruler, $y_\mathrm{LP}^\mathrm{obs}$ (hereafter simply $y_\mathrm{LP}$), is of particular interest, since it is analogous to the observable angular size of the BAO standard ruler. 
In the left panel of Figure \ref{fig:ylp_z}, we plot the post-reconstruction linear point standard ruler $y_\mathrm{LP}\equiv s_\mathrm{LP}/D_V$ as a function of redshift for the DR1 (circles) and DR2 (squares) tracers considered in this work. We calculate the isotropic volume distance $D_V$ for the fiducial cosmology by first obtaining the angular diameter distance $D_A$ using {\tt CLASS} and applying Eq. \ref{eq:D_V}, and plot the fiducial prediction $y_\mathrm{LP}^\mathrm{fid}\equiv s_\mathrm{LP}^\mathrm{fid}/D_V$ as a dotted curve. 
The right panel presents the same measurements normalized by $y_\mathrm{LP}^\mathrm{fid}$: a Hubble diagram analogous to the top left panel of Figure 13 in \cite{desi_dr2_bao}. 
To this panel we add two theoretical predictions: $y_\mathrm{LP}$ as a function of redshift computed for the DESI DR2 best-fit \lcdm\ cosmology (dashed) and best-fit $w_0w_a$CDM cosmology with time-varying dark energy (dotted-dashed) \cite{desi_dr2_bao}. Each measurement is obtained by computing the linear point as described in Section~\ref{sec:s_fid} and $D_V$ from {\tt CLASS} as above, and is likewise normalized by $y_\mathrm{LP}^\mathrm{fid}$ (with the fiducial fixed to Planck 2018 \lcdm).
As expected, the DR2 measurements are more precise than those from DR1. The current measurements cannot distinguish \lcdm\ from $w_0w_a$CDM on the basis of this figure alone; we do not pursue a full cosmological analysis of the linear point here, since the reconstruction and correction required for agreement with BAO leave little to be gained from an independent inference (see Section \ref{sec:conclusion}). We also note that none of the measurements in either panel have been multiplied by 1.005 or otherwise modified to account for smearing, which is expected to impact the consistency we expect to see with the DESI BAO results. Additionally, with the linear point we can only measure the isotropic dilation parameter of the standard ruler $\alpha_\mathrm{iso}$, whereas the template-based BAO pipeline is able to extract more information by measuring the anisotropic parameter $\alpha_\mathrm{AP}$ as well (see remaining panels in Figure 13 in \cite{desi_dr2_bao}). This further limits the scope of cosmological inference using the linear point. Nonetheless, Figure \ref{fig:ylp_z} qualitatively corroborates the results of \cite{desi_dr2_bao}, illustrating how the linear point could serve as a complementary, geometric cross-check of the BAO distance scale in future analyses.

\section{Conclusion and Discussion}
\label{sec:conclusion}

The linear point was proposed as an alternative to the BAO standard ruler on account of its increased robustness to non-linear effects in the two-point correlation function \cite{anselmi2016_lp, anselmi2017_sdss}. Defined as the average of the scales corresponding to the BAO peak and the preceding dip in the correlation function, it is a purely geometric standard ruler, since its value depends only on the shape of the correlation function and avoids the need for cosmology-dependent template fitting. In this work, we measured the linear point using the correlation function measurements from the first and second data releases (DR1 and DR2) of DESI, and used it to test whether this model-independence survives at the precision dictated by current surveys.

We validated the model-independent linear point pipeline -- a simple polynomial fit to a narrow section of the correlation function -- using \abacust\ DR1 mock catalogs, in both the evolved and reconstructed fields where available.

The linear point measurements in the pre-reconstructed field, converted to \qisolp\ without any modifications for damping, show shifts of up to 1\% when compared to the analogous template-based BAO measurements.  
Even after applying the redshift- and cosmology-independent multiplicative correction of 1.005 prescribed in \cite{anselmi2016_lp, anselmi2017_sdss}, a residual offset remains large enough to systematically bias constraints on the cosmological distance scale. We therefore determined that in its strictly model-independent mode, the linear point does not recover the template-based BAO scale at the accuracy and precision DESI demands.

We consequently considered two techniques that relax the strictly model-independent nature of the linear point pipeline: (a) applying physically motivated sample-dependent shifts to the pre-reconstructed measurements, and (b) performing linear point analyses in the post-reconstructed field.
For the former, we presented a theoretically-motivated method to correct for smearing due to non-linear damping: convolving the linear theory correlation function with a Gaussian kernel sampled from a prior on the isotropic damping scale to obtain a smeared fiducial non-linear correlation function, computing the corresponding fiducial linear point, and using the ratio of this value and the measurement as the corrected \qisolp.
As for the latter, we found that standard reconstruction removes much of the non-linear smearing, yielding linear point measurements with increased precision and accuracy. 
A non-zero damping remained even post-reconstruction, however, shifting the measurements in mocks by up to 0.5\% relative to the fiducial cosmology expectation.

We presented the linear point measurements made on DESI DR1 correlation functions in Table \ref{tab:Y1_results} and Figure \ref{fig:Y1_LP}, and those on DR2 tracers in Table \ref{tab:Y3_results} and Figure \ref{fig:Y3_LP}. 
When converted to \qisolp\ and further corrected for the smearing using the physically-motivated correction above ($\alpha_\mathrm{iso,LP}^\mathrm{corrected}$), we found excellent agreement between the linear point and isotropic template-based BAO in DR1 and DR2 samples \cite{desi2024_iii, desi_dr2_bao}. 
This agreement, however, comes at a price. Both the damping correction and reconstruction are cosmology-dependent, so the agreement is achieved only by giving up the model-independence that motivated the linear point. Propagating the uncertainty on the damping scale into \qisolp\ inflates its errors, though by only up to 4\% in data. The correction thus degrades the model-independence and, to a smaller degree, the precision of the measurement.

The offset in the uncorrected linear point measurements compared to the template-based BAO measurements implies that the strictly model-independent linear point will systematically bias inferred cosmological parameters if non-linear smearing is unaccounted for. Yet, accounting for damping is precisely what reintroduces the cosmological dependence the linear point was proposed to mitigate in the first place. 
We also found that the linear point pipeline is more sensitive to the signal-to-noise ratio of the BAO peak than template-based fitting approaches; no linear point was measured in the DR1 \bgs\ correlation function pre-reconstruction, where a polynomial failed to identify a peak and a dip. The linear point therefore retains its appeal as a geometric probe, but at the sub-percent precision of DESI, it is not a strictly model-independent alternative to template-based BAO; it is most useful as a complementary geometric cross-check of the BAO distance scale, used in tandem with reconstruction.
For this reason, we did not carry the corrected linear point measurements through to cosmological parameter inference in this work. Once the measurements have been reconstructed and corrected to agree with the template-based BAO results, they retain little constraining power independent of BAO, and a separate inference would largely reproduce the existing BAO constraints through a noisier pipeline. A further limitation is that the linear point yields only the isotropic dilation parameter $\alpha_\mathrm{iso}$, whereas the template-based BAO pipeline additionally recovers the anisotropic parameter and thus more distance information. 

Several directions remain open. The Gaussian smearing correction we proposed did not fully undo the non-linear offset in the pre-reconstruction regime, which leaves room for improvement. More detailed geometric modeling of the correlation function shape under non-linear evolution may capture the shift more accurately and, in turn, tighten the resulting constraints. It would be valuable to test whether alternative reconstruction schemes with smaller residual damping -- for instance, the recent methods based on optimal transport \cite{ot_levy2021, ot_nikakhtar2022, ot_von_Hausegger2022, ot_nikakhtar2023, ot_nikakhtar2024} -- reduce the need for an explicit correction, and whether the Laguerre reconstruction discussed in Appendix \ref{sec:laguerre} recovers the linear theory linear point without any correction at all. Finally, while \cite{he2023} has shown that $H_0$ and $\Omega_m$ can be extracted from linear point measurements, our results suggest that such an inference on DESI data would be informative chiefly as a consistency test against the BAO-derived parameters, rather than as an independent constraint. The hints at time-evolving dark energy \cite{desi_dr2_bao, desi_dr2_extendedDE} motivate such tests, and Figure \ref{fig:ylp_z} offers a first look at the Hubble diagram which qualitatively tracks both the best-fit \lcdm\ and $w_0w_a$CDM predictions, consistent with a similar figure in \cite{desi_dr2_bao}. As BAO measurements push into territory with significant implications for fundamental physics, geometric probes of the expansion history that rely on different modeling assumptions offer a useful, if not fully independent, check against systematics.

\section{Data Availability}
The data used in this analysis is part of the DESI Data Release 1, available at \url{https://data.desi.lbl.gov/doc/releases/dr1/}, and Data Release 2, which will be made public (details in \url{https://data.desi.lbl.gov/doc/releases/}). 

\acknowledgments
We thank Xinyi Chen and Uendert Andrade for helpful comments and suggestions as internal reviewers. 
NU and NP are supported in part by DoE DE-SC0017660. FN gratefully acknowledges support from the Yale Center for Astronomy and Astrophysics Prize Postdoctoral Fellowship.
RKS is grateful to the ICTP for its hospitality during the summer of 2024. 

This material is based upon work supported by the U.S. Department of Energy (DOE), Office of Science, Office of High-Energy Physics, under Contract No. DE–AC02–05CH11231, and by the National Energy Research Scientific Computing Center, a DOE Office of Science User Facility under the same contract. Additional support for DESI was provided by the U.S. National Science Foundation (NSF), Division of Astronomical Sciences under Contract No. AST-0950945 to the NSF’s National Optical-Infrared Astronomy Research Laboratory; the Science and Technology Facilities Council of the United Kingdom; the Gordon and Betty Moore Foundation; the Heising-Simons Foundation; the French Alternative Energies and Atomic Energy Commission (CEA); the National Council of Humanities, Science and Technology of Mexico (CONAHCYT); the Ministry of Science, Innovation and Universities of Spain (MICIU/AEI/10.13039/501100011033), and by the DESI Member Institutions: \url{https://www.desi.lbl.gov/collaborating-institutions}. Any opinions, findings, and conclusions or recommendations expressed in this material are those of the author(s) and do not necessarily reflect the views of the U. S. National Science Foundation, the U. S. Department of Energy, or any of the listed funding agencies.

The authors are honored to be permitted to conduct scientific research on I'oligam Du'ag (Kitt Peak), a mountain with particular significance to the Tohono O’odham Nation.

\appendix

\section{Optimizing Free Parameters in the Linear Point Pipeline}
\label{sec:optimization}

The linear point pipeline consists of a model-independent fit to a narrow region of the two-point correlation function. In \cite{anselmi2018_validation}, a detailed validation process suggested that a quintic polynomial ($n=5$) fit to the correlation function calculated with a bin width $\Delta s=3$ \hMpc\ in the range 60-120 \hMpc\ was the most optimal choice of free parameters in the pipeline.
Strictly speaking, their validation tests are applicable to a BOSS-like survey. However, the Bayesian analysis in \cite{paranjape2022} suggests that the same choices should remain appropriate at least for DESI \bgs, \lrg, and potentially \elg\ samples. 
In this section, we make adjustments to the recommended optimal parameters in \cite{anselmi2018_validation} for DESI correlation functions and covariance matrices to optimize the pipeline. 

Since we scale and center the correlation function measurements using $s_0$ and $\sigma$ in Eq. \ref{eq:poly_interp}, our pipeline is largely unaffected by the range of the correlation function, as long as a polynomial interpolation remains a valid approximation for the general shape. 
We choose to fit over the range 70-115 \hMpc\ in our pipeline, since it is more symmetric with respect to the linear point in linear theory for the fiducial \lcdm\ cosmology and retains the polynomial nature of the correlation function. 

\begin{figure}
    \centering
    \begin{tabular}{cc}
      \hspace{-0.35cm}\includegraphics[width=0.5\columnwidth]{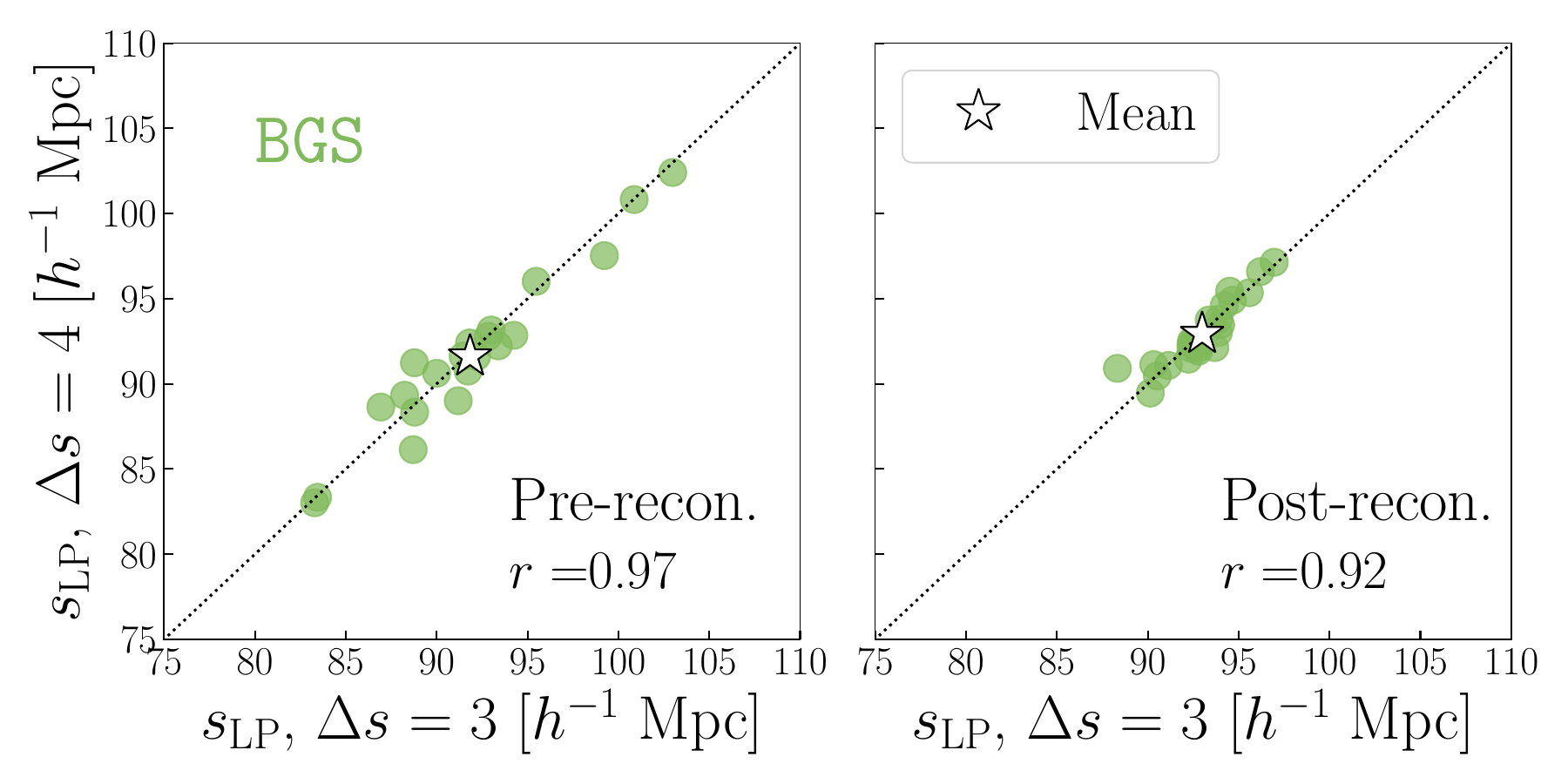} &
      \hspace{-0.25cm}\includegraphics[width=0.5\columnwidth]{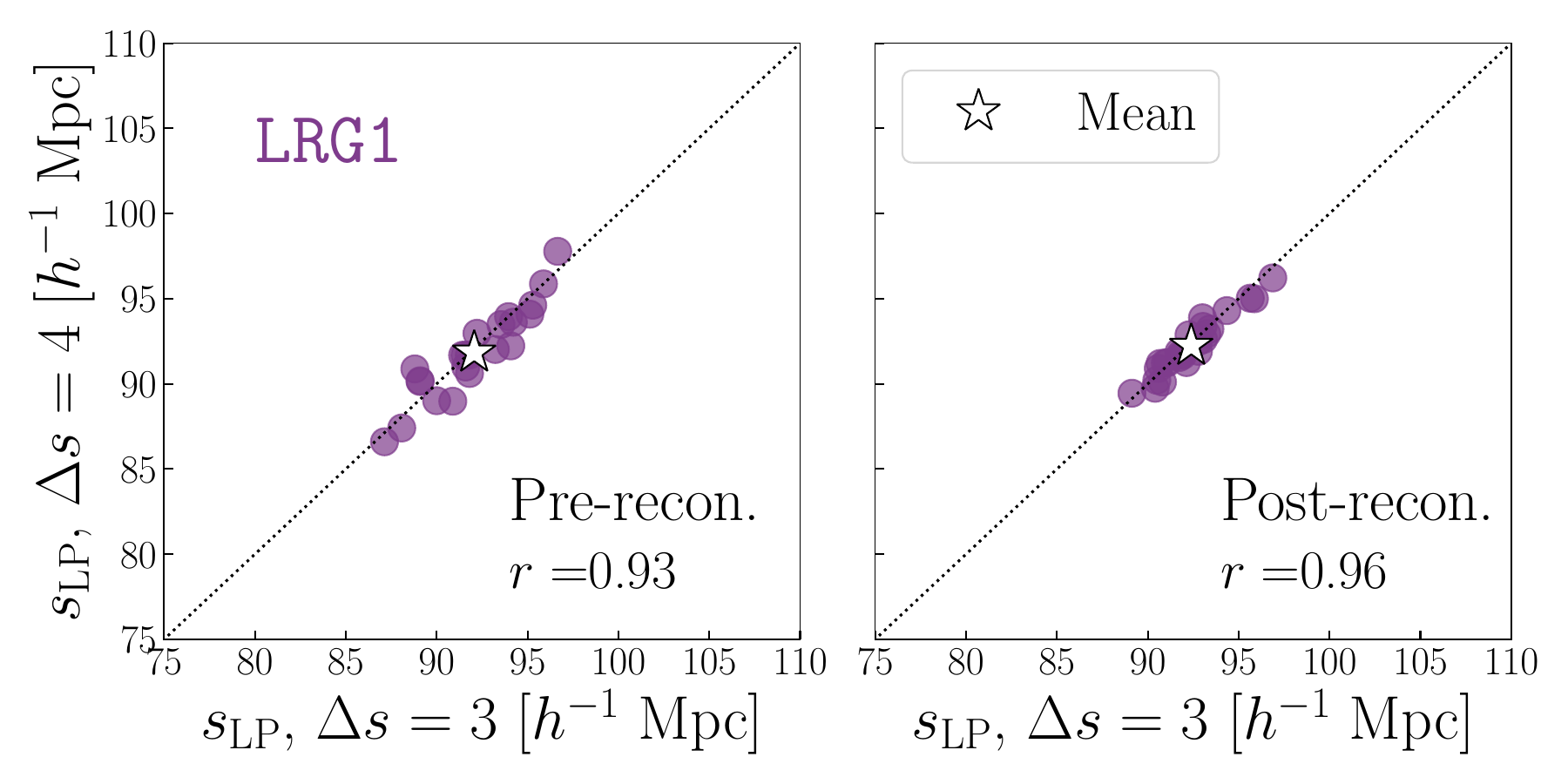} \\
      \hspace{-0.35cm}\includegraphics[width=0.5\columnwidth]{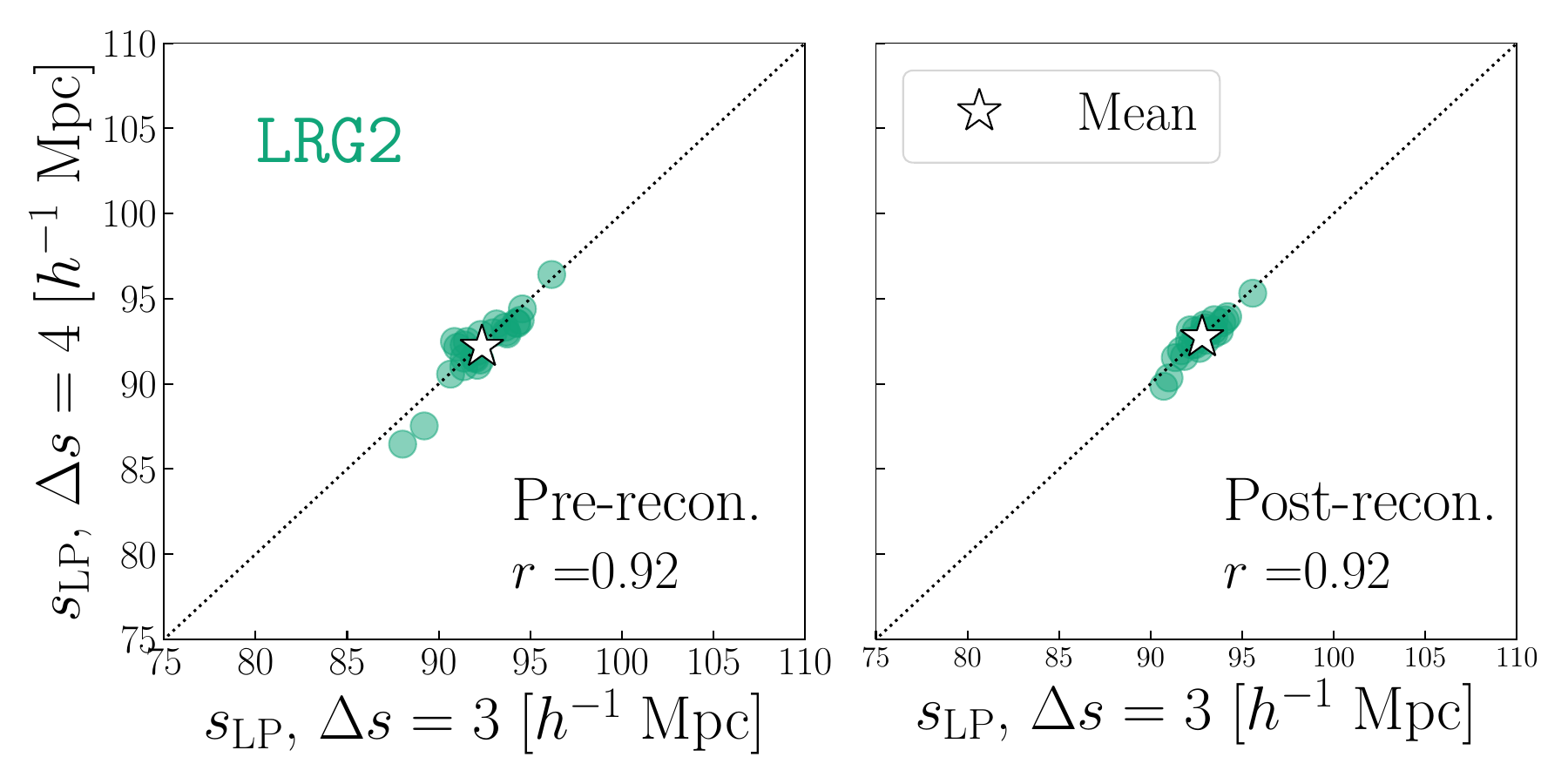} &
      \hspace{-0.25cm}\includegraphics[width=0.5\columnwidth]{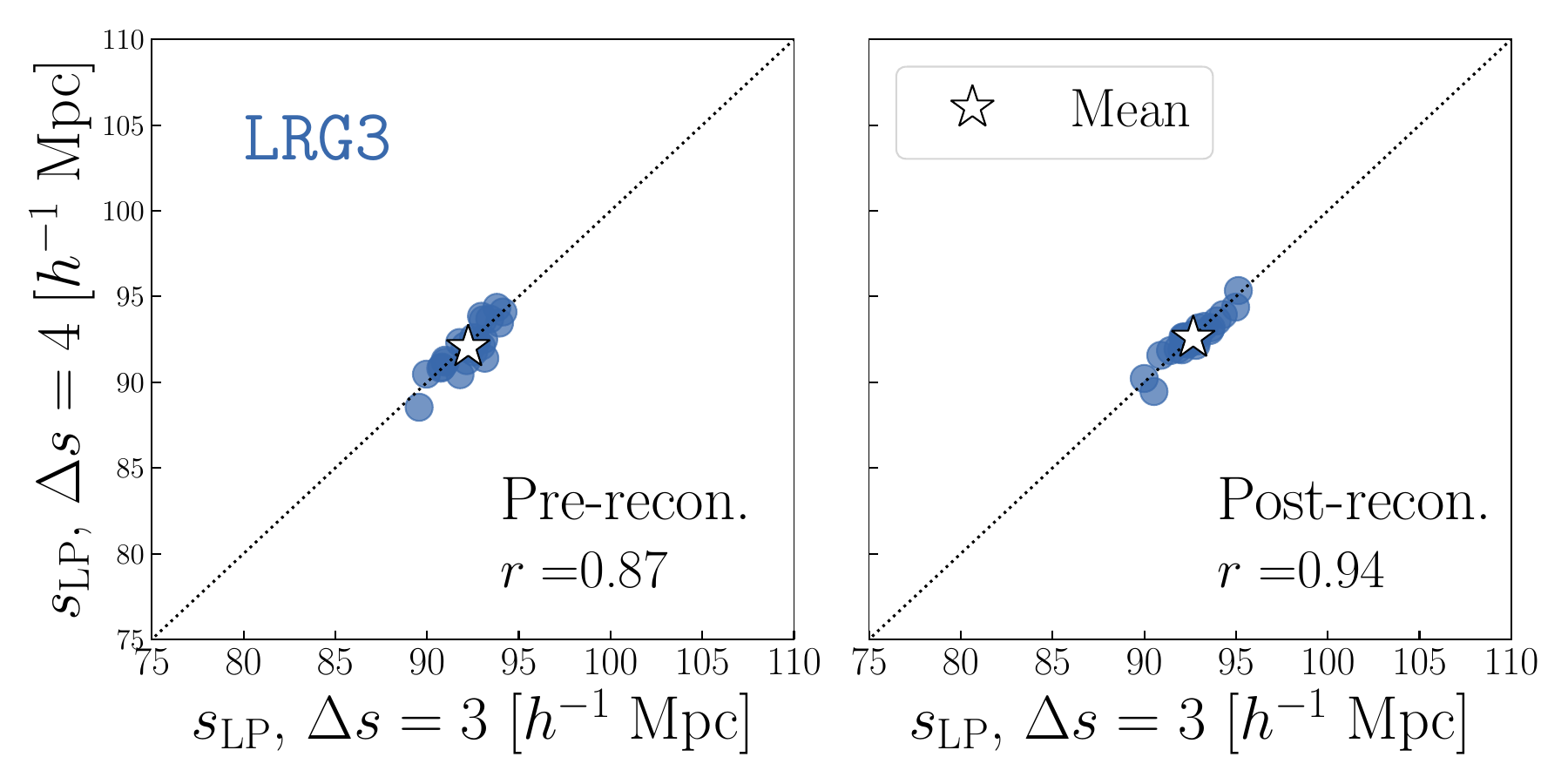} \\
      \hspace{-0.35cm}\includegraphics[width=0.5\columnwidth]{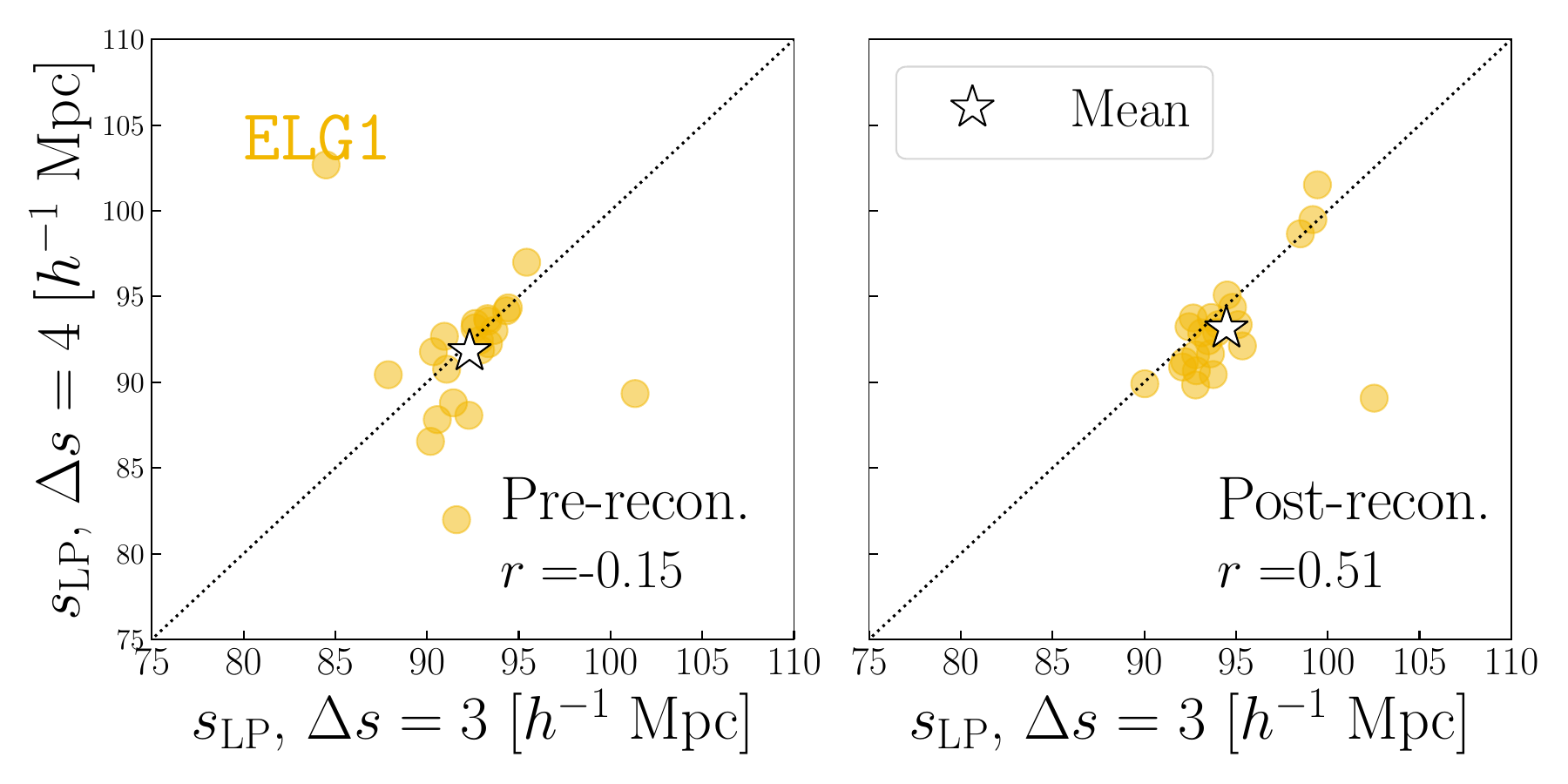} &
      \hspace{-0.25cm}\includegraphics[width=0.5\columnwidth]{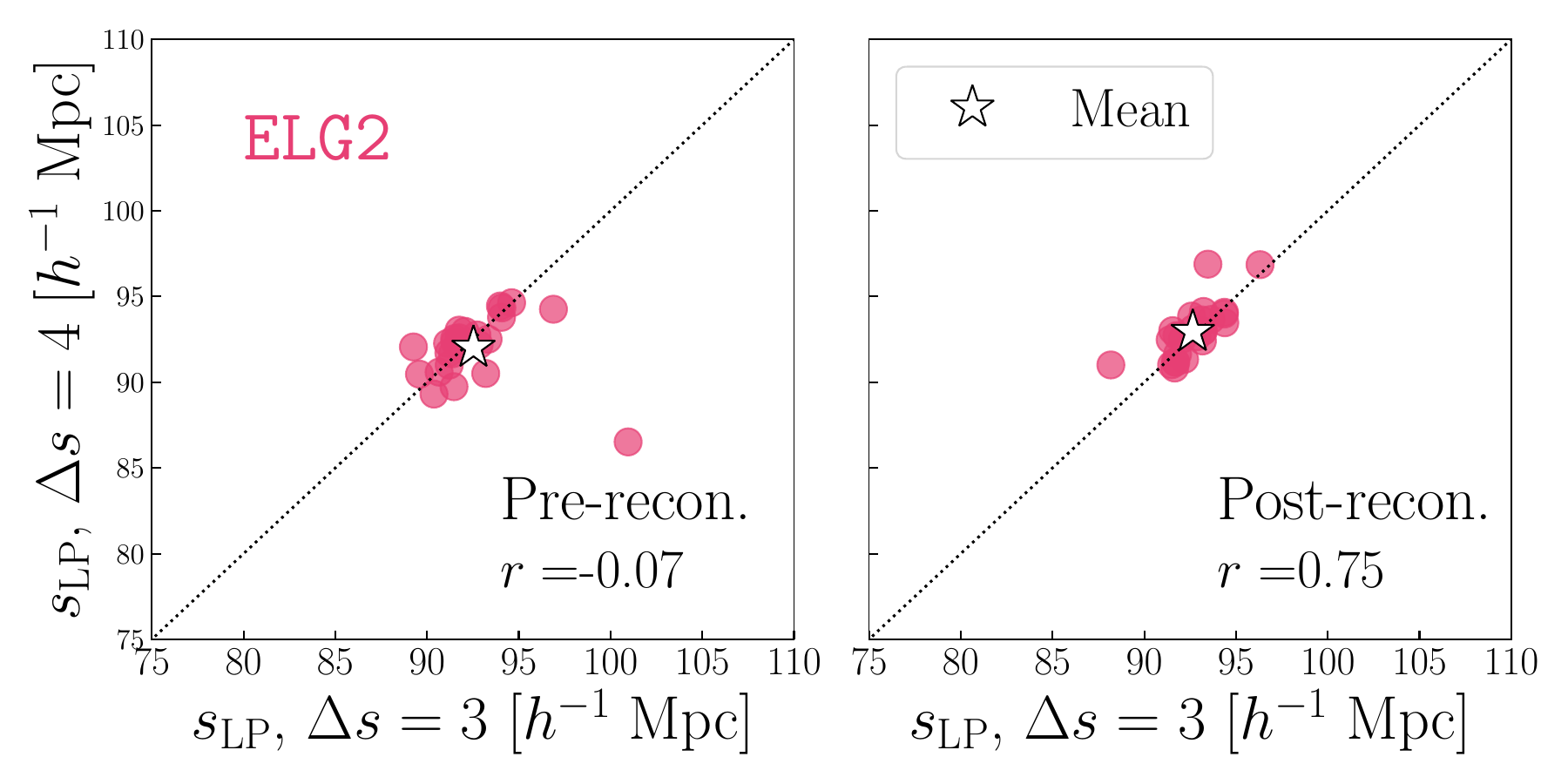}
    \end{tabular}
    \caption{Linear point measurements on correlation functions measured on \abacust\ mocks with bin widths $\Delta s = 3$ \hMpc\ and 4 \hMpc. We use the numerical covariance matrix calculated using the 1000 \ezmocks\ for the 3 \hMpc\ binned correlation functions and the semi-analytical covariance matrix calculated using the {\tt RascalC} code for the 4 \hMpc\ binned correlation functions. In both cases, we use a fifth-degree polynomial interpolation to calculate the linear point. The left panel for each tracer depicts linear points calculated pre-reconstruction and the right panel, post-reconstruction. We additionally plot the average of the 25 mocks in each panel (white star marker), and annotate the Pearson correlation coefficient $r$ between the two sets of measurements.}
    \label{fig:nbins3_vs_nbins4}
\end{figure}

\begin{figure}
    \centering
    \begin{tabular}{cc}
      \hspace{-0.25cm}\includegraphics[width=0.5\columnwidth]{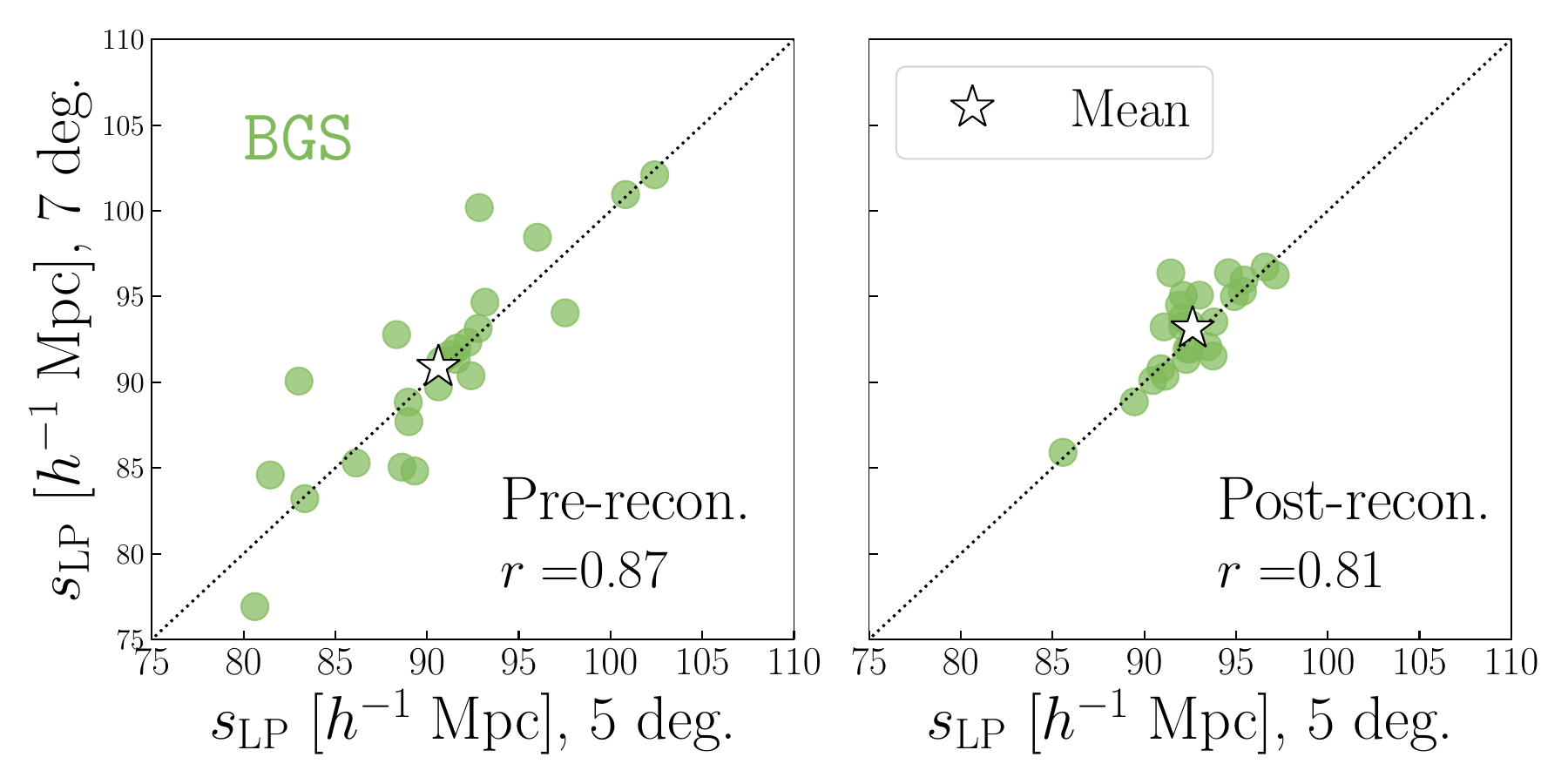} &
      \hspace{-0.25cm}\includegraphics[width=0.5\columnwidth]{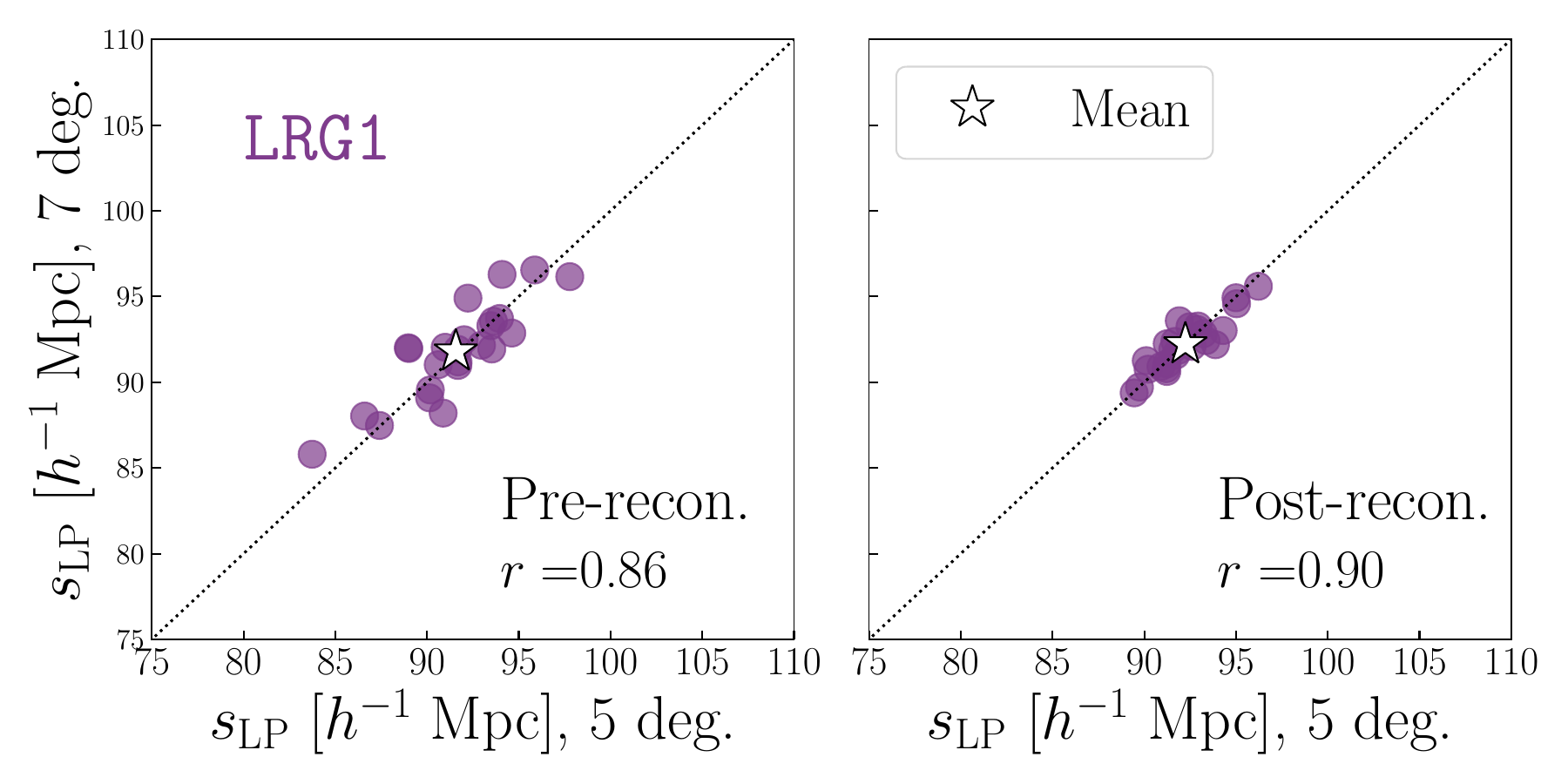} \\
      \hspace{-0.25cm}\includegraphics[width=0.5\columnwidth]{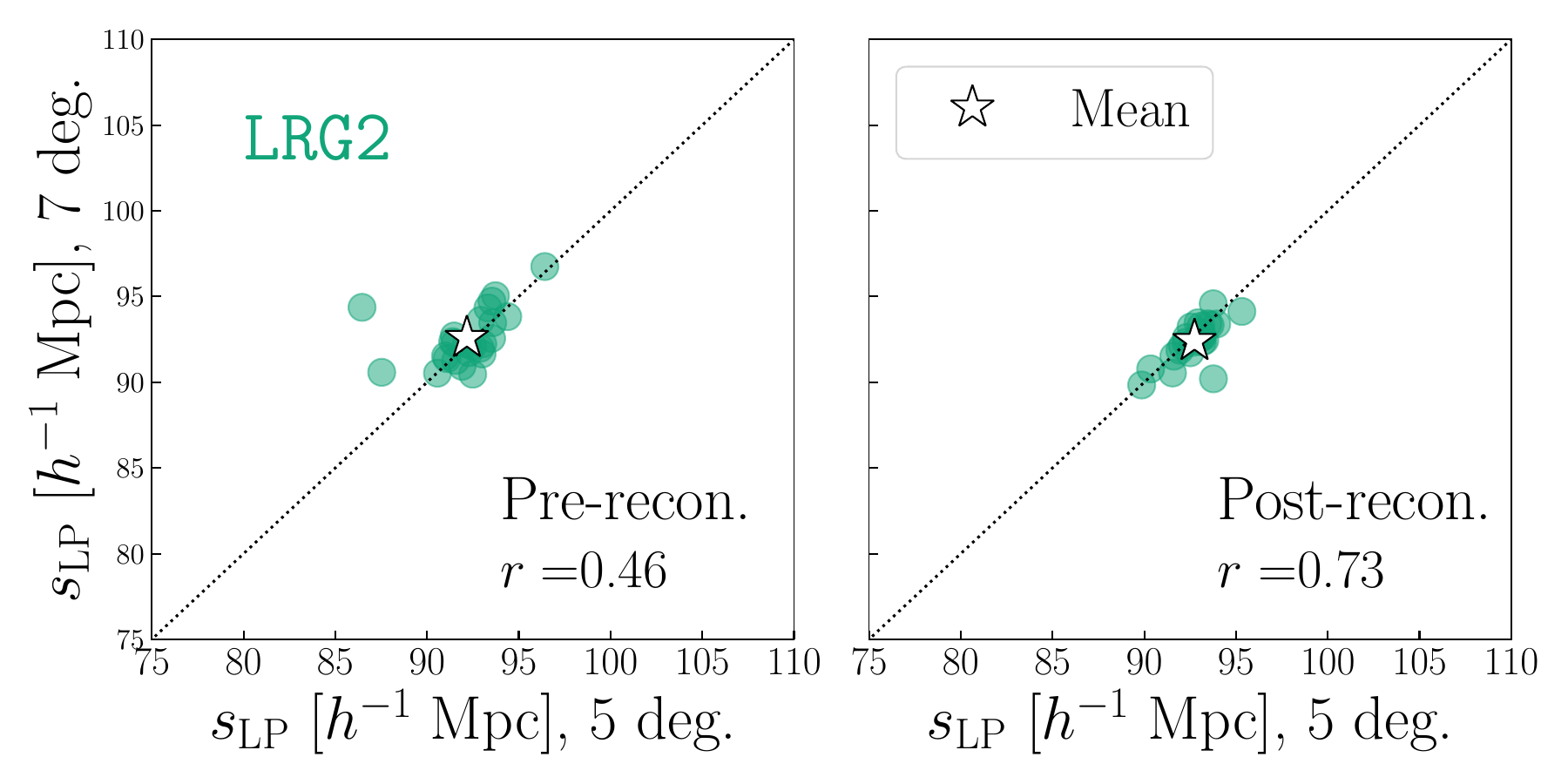} &
      \includegraphics[width=0.5\columnwidth]{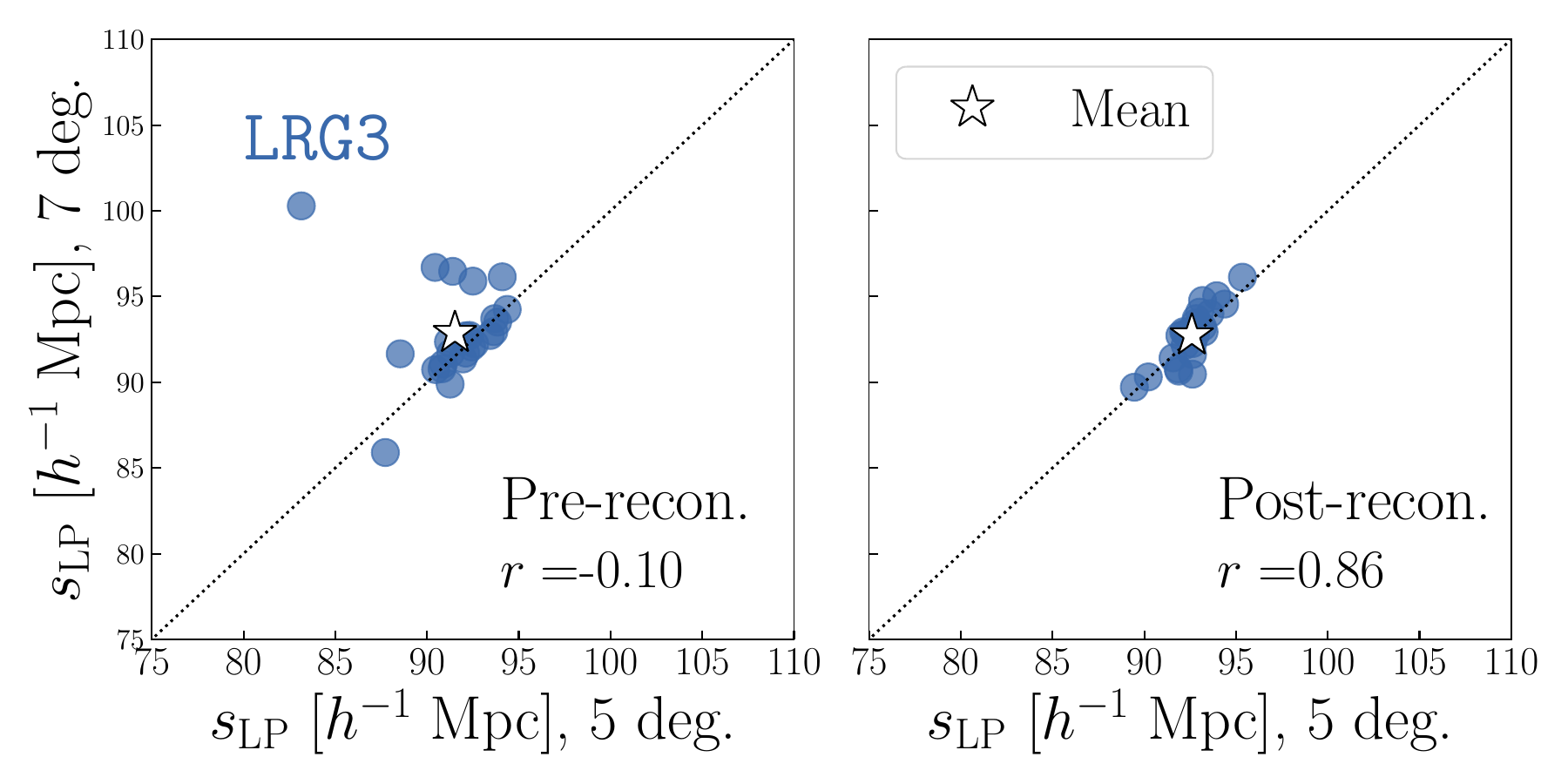} \\
      \hspace{-0.25cm}\includegraphics[width=0.5\columnwidth]{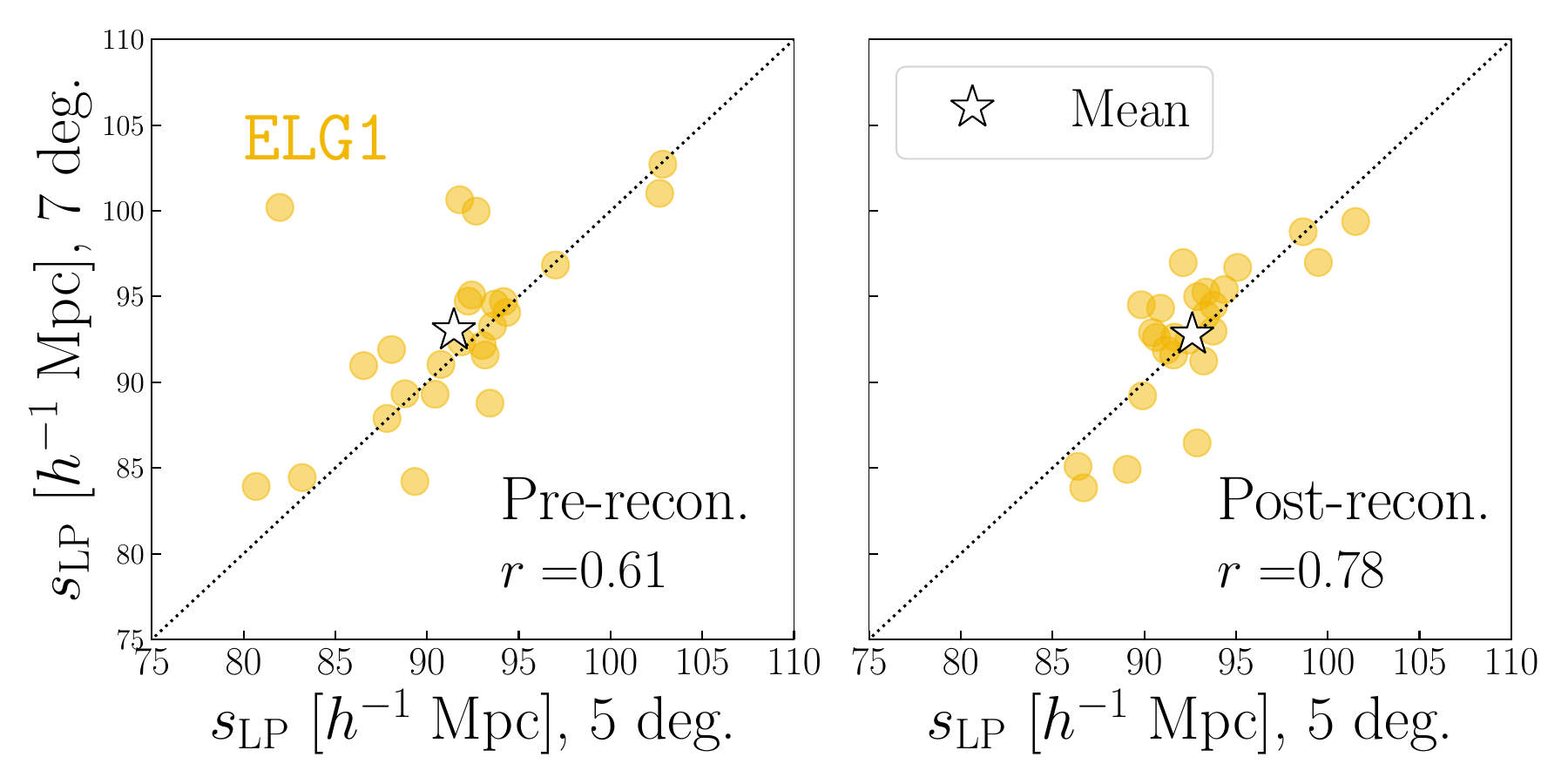} &
      \includegraphics[width=0.5\columnwidth]{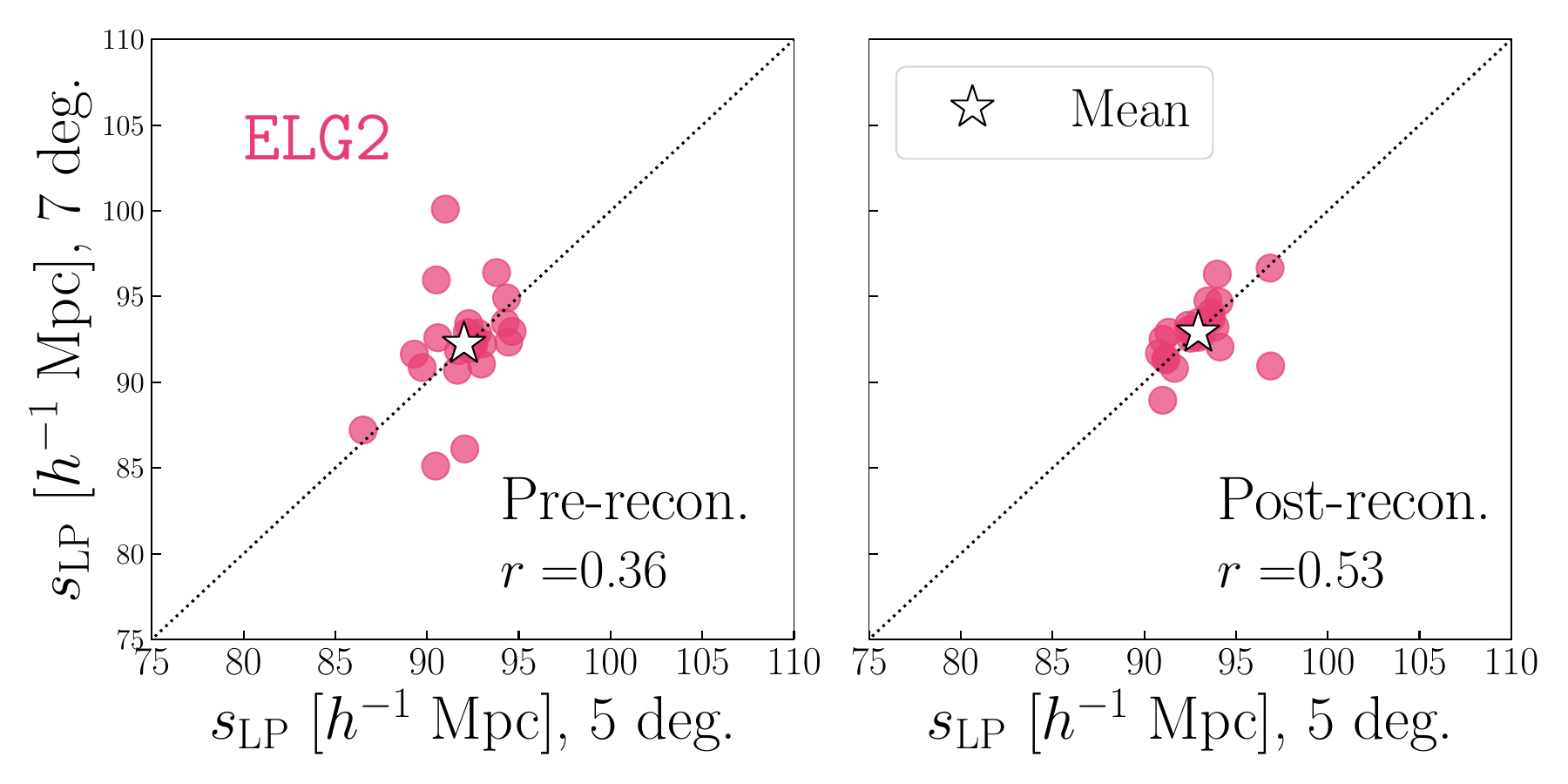}
    \end{tabular}
    \caption{Linear point measurements on correlation functions with bin width 4 \hMpc\ measured on \abacust\ mocks with fifth-degree ($n=5$) and seventh-degree ($n=7$) polynomial interpolations. The left panel for each tracer depicts linear points calculated pre-reconstruction and the right panel, post-reconstruction. We additionally plot the average of the 25 mocks in each panel (white star marker), and annotate the Pearson correlation coefficient $r$ between the two sets of measurements.}
    \label{fig:deg5_vs_deg7}
\end{figure}

All DESI BAO analyses have been performed with $\Delta s=4$\hMpc\, given the availability of analytic covariance matrices in configuration space for this bin width. We compare the linear points measured on 3 \hMpc\ and 4 \hMpc\ binned correlation functions for the 25 \abacust\ mocks for each tracer using fifth degree polynomials in Figure \ref{fig:nbins3_vs_nbins4}.
We compute the Pearson correlation coefficient, $r$, to quantify the scatter between these measurements. The \bgs\ and \lrg\ measurements are highly correlated both pre- and post-reconstruction, whereas the \elg\ measurements contain considerable scatter due to outliers. Upon visual inspection of the linear point fits of the outlier mocks, we find that random noise in the correlation function resulted in vastly different identification of roots. Higher signal-to-noise ratio in DR2 mocks and data is expected to mitigate this issue. Nevertheless, the mean linear point measurements (gray star) in mocks are in agreement regardless of our choice of bin width. We therefore sample the correlation function with 4 \hMpc\ bin widths in our linear point pipeline. We confirm in Figure \ref{fig:deg5_vs_deg7} that fifth degree polynomials are still an appropriate choice despite the change in bin width, given small scatter between the two measurements.  Our findings are consistent with expectations about the order of polynomial \citep{paranjape2022} and dependence on binning \citep{eigencov2024}.

\section{The Linear Point with Poor Signal-to-Noise Ratio}
\label{sec:bgs_bad}

A major drawback of the linear point pipeline, mentioned in Section \ref{sec:calc_lp}, is that the polynomial fit may not be able to identify a peak and dip in the correlation function when the signal-to-noise ratio (SNR) is poor, therefore unable to measure a linear point.
This was explored in \cite{he2023} with MultiDark-Patchy mocks \cite{patchy_mocks}, where the term `reliability' was defined as the percentage of mocks that yielded a measurable linear point. We measure the reliability in \abacust\ mocks and find that post-reconstruction, all mocks for all tracers yield a measurement. However, pre-reconstruction we find 100\% reliability in \lrgt, \lrgth, and both sets of \elg\ mocks, but find that one \bgs\ mock and one \lrgo\ mock fail to yield a measurable linear point. A minimum of 96\% reliability is encouraging; however, we fail to measure a linear point in the pre-reconstruction \bgs\ DR1 correlation function. We explore this below.

In the context of data, we can interpret reliability through the SNR of the data and the resulting polynomial fit. In Section \ref{sec:calc_lp}, we mention that to calculate the error in a single linear point measurement, we generate multiple realizations of polynomial fit to the correlation function using the covariance matrix of the best-fit coefficients. We then calculate the linear point of each realization and report the standard deviation of the samples as the error in the original measurement. Not all realizations will result in a measurement, and so the percentage of failed measurements can serve as a qualitative measure of the SNR. We define the percentage of sampled polynomial fits that did \textit{not} result in a linear point measurement as the failure rate, and present the DR1 failure rates in Table \ref{tab:failure_rates_DR1}.

\begin{figure}
    \centering
    \begin{tabular}{cc}
        \hspace{-0.25cm}\includegraphics[width=0.45\linewidth]{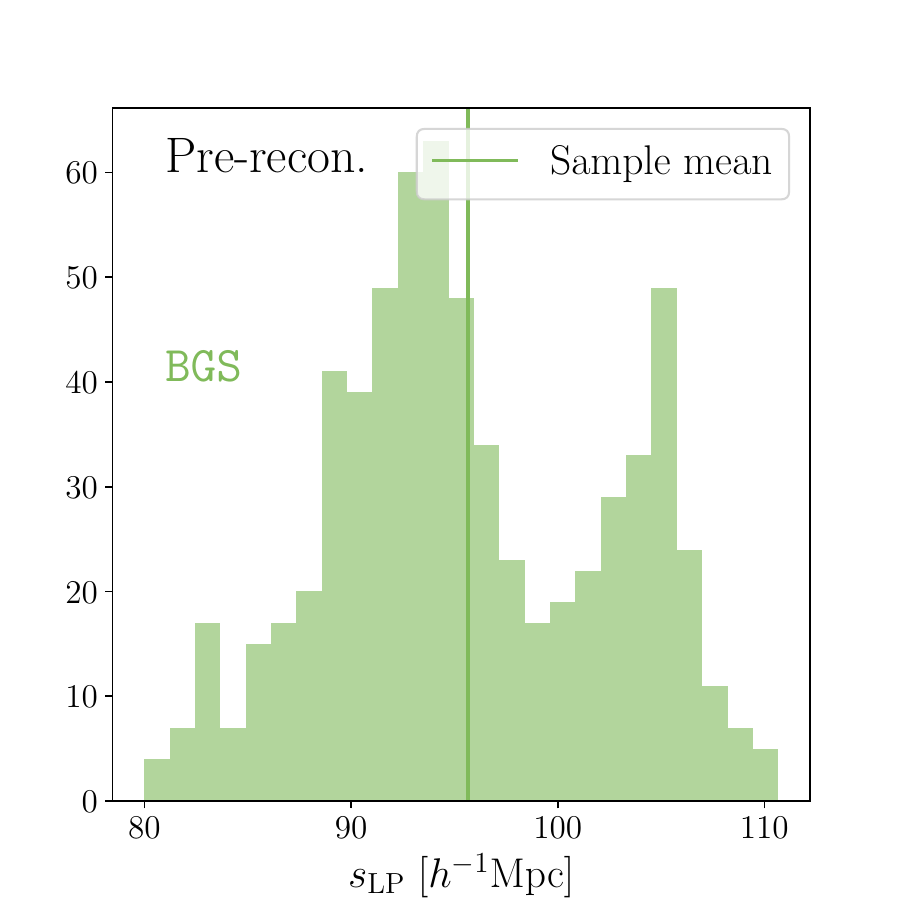} &  \includegraphics[width=0.45\linewidth]{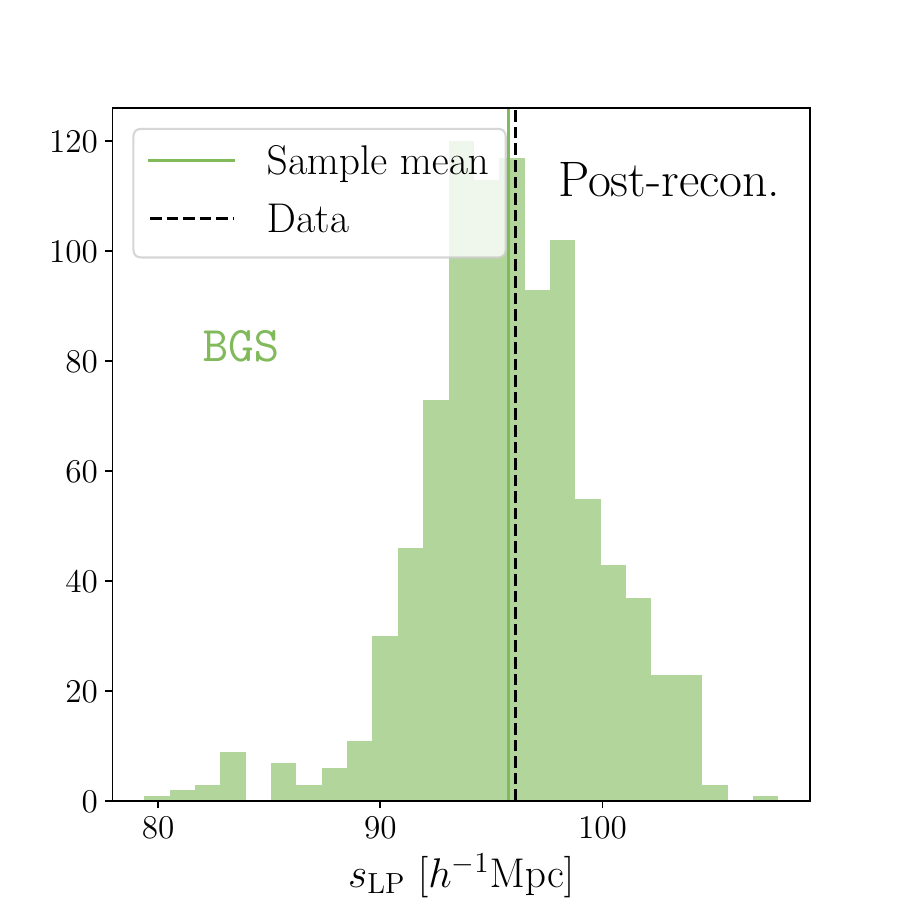}
    \end{tabular}
    \caption{Histograms of the linear points measured on polynomial interpolations of the DR1 \bgs\ correlation functions pre- (left) and post-reconstruction (right), sampled from the covariance matrix of the best-fit polynomial coefficients. The solid vertical line represents the mean of the sample, whereas the black dotted vertical line represents the linear point measured on the original correlation function. The latter is missing from the left panel because we were unable to measure a linear point on the pre-reconstruction \bgs\ correlation function. }
    \label{fig:BGS_hist}
\end{figure}

We see that the \bgs\ correlation function pre-reconstruction has a very high failure rate -- over a third of the polynomial interpolations of the correlation function fail to identify a linear point, which may explain why we did not measure the linear point feature in the DR1 correlation function. In Figure \ref{fig:BGS_hist}, we present a histogram of the linear point values that were measured during this sampling process. For each tracer, we intuitively expect the distribution to be Gaussian centered around the mean of the sample, which should also agree with the linear point originally measured on the DR1 correlation function. While we see this trend in the post-reconstruction \bgs\ samples (despite a relatively higher failure rate), the pre-reconstruction samples are in fact bimodal. This means that due to the poor fit to the correlation function, not only is the linear point not measurable, but the mean of the samples is also an unreliable estimate of the measurement. 

We also note a relatively higher failure rate for pre-reconstruction \elgo\ correlation function; while the linear point is still measured in this case, the relatively high $\chi^2$ value of the fit and the large error bars on the linear point (presented in Table \ref{tab:Y1_results} and Figure \ref{fig:Y1_LP}) hint at poor SNR. 

The linear point pipeline, as it stands, is therefore not optimal for measuring a standard ruler when the correlation function is poorly measured or the errors on those measurements are large. In such cases, template-based fitting approaches may indeed be a necessity. An alternative could be modifying the fitting functions used in the linear point pipeline; in fact, \cite{he2023} explore fitting a fifth-degree polynomial to $s^2\xi(s)$ instead of $\xi(s)$. They find that this improves reliability in Patchy mocks. We explore alternative fitting functions in the following section.

\begin{table}
    \centering
    \begin{tabular}{|l|c|c|c|}
        \hline
        \multicolumn{1}{|c|}{\multirow{2}{*}{Tracer}} &  
        \multirow{2}{*}{Redshift} & 
        \multicolumn{2}{c|}{Failure Rate} \\
        \cline{3-4}
         &  & Pre & Post \\
        \hline
        \bgs & 0.1-0.4 &  34.0\%  & 7.9\% \\
        \lrgo & 0.4-0.6 & 0.6\% & 0.0\% \\
        \lrgt & 0.6-0.8 & 5.9\% & 0.2\% \\
        \lrgth & 0.8-1.1 & 0.0\% &  0.0\% \\ 
        \lrgelg & 0.8-1.1 & 0.0\% &  0.0\% \\ 
        \elgo & 0.8-1.1 & 12.2\% &  0.5\% \\ 
        \elgt & 1.1-1.6 & 0.9\% &  0.0\% \\ 
        \hline
\end{tabular}
\caption{Failure rates for DR1 tracers pre- and post-reconstruction, defined as the percentage of sampled polynomial interpolations of the correlation function that do not yield a linear point measurement.}
\label{tab:failure_rates_DR1}
\end{table}

\section{Alternative Fitting Functions}
\label{sec:alteratives}

As shown above, the polynomial fitting pipeline fails to constrain the linear point under poor SNR conditions. This motivates the search for alternative basis functions for the fitting regime. We require these functions to be purely geometric, and capable of capturing the general shape of the monopole of the two-point correlation function over roughly the 60–120 \hMpc\ range. Two such approaches have been explored in the literature. The first, proposed in \cite{he2023}, fits a fifth-degree polynomial to $s^2\xi$ directly instead of $\xi$, equivalent to fitting a linear combination of terms $a_is^{i-2}$ to the correlation function. The second, developed in \cite{nikakhtar2021, nikakhtar2021_halo, nikakhtar2022_smearing}, fits a linear combination of modified Laguerre functions $\mu_k(x)$, defined as:
\begin{align}
    \mu_{2n}(x)=2n!!L_n^{(1/2)}(-x^2/2) && \mu_{2n-1}(x)=(2n-1)!!\sqrt{\frac{\pi}{2}}L_{n-1/2}^{(1/2)}(-x^2/2) 
    \label{eq:mu_functions}
\end{align}
Here, $L_\beta^{(\alpha)}$ are generalized Laguerre functions, listed in the Appendix of \cite{nikakhtar2021}. We explain why these functions form an acceptable basis in Appendix \ref{sec:laguerre}.
In this section, we compare the linear point obtained from a fifth-degree polynomial fit against those from these two approaches using \abacust\ mocks. For consistency, we use the same fitting range (70-115 \hMpc) and same number of basis functions (six, similar to a fifth degree polynomial).

\begin{figure}[]
    \centering
    \begin{tabular}{ccc}
    \vspace{-.85cm}
       *\hspace{-.5cm}\includegraphics[width=0.33\columnwidth]{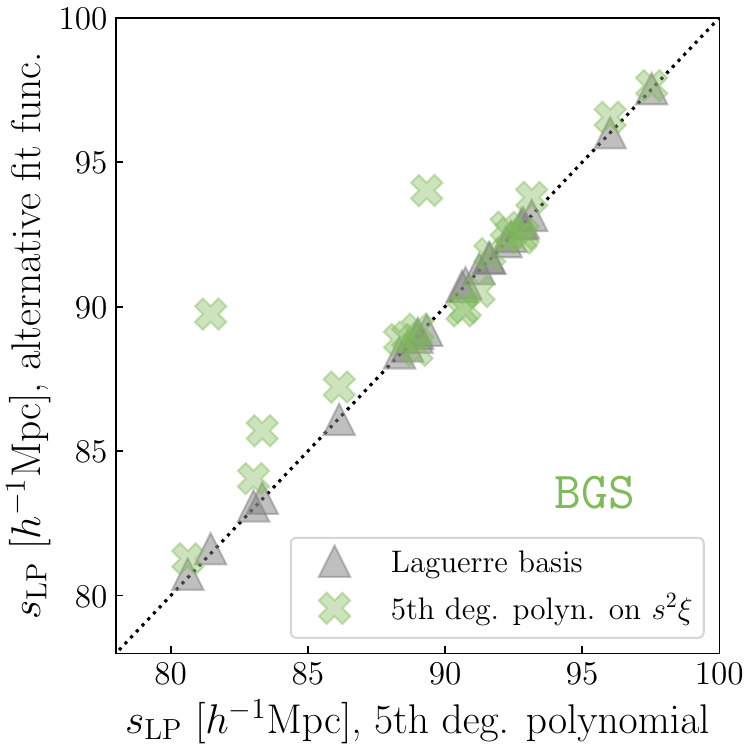} &
       *\hspace{-.5cm}\includegraphics[width=0.32\columnwidth]{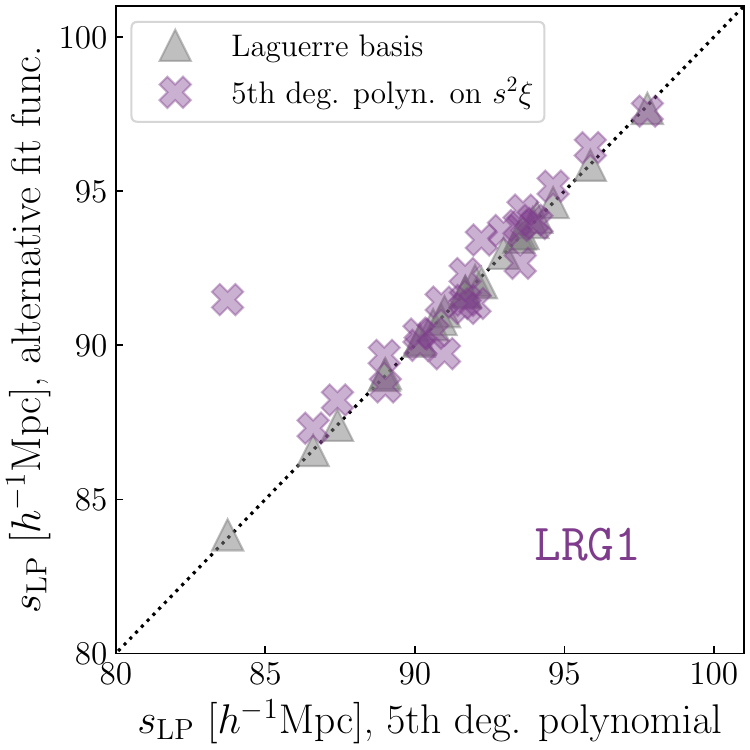}  & 
       *\hspace{-.5cm}\includegraphics[width=0.33\columnwidth]{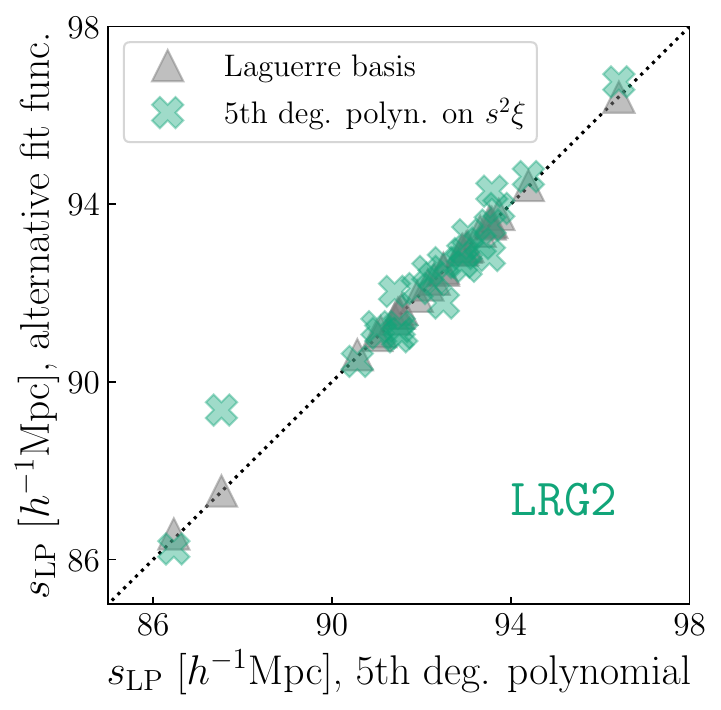} \\ \\ \\
       *\hspace{-.5cm}\includegraphics[width=0.325\columnwidth]{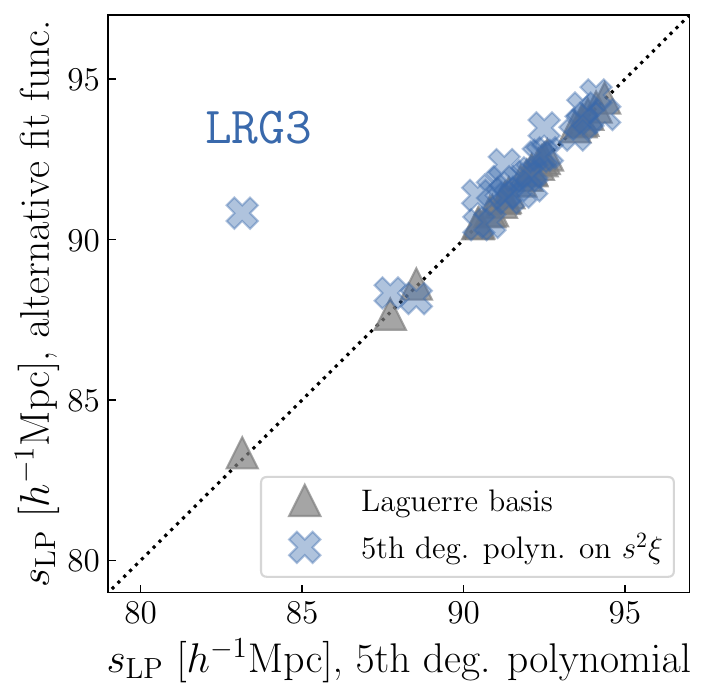} &
       *\hspace{-.5cm}\includegraphics[width=0.338\columnwidth]{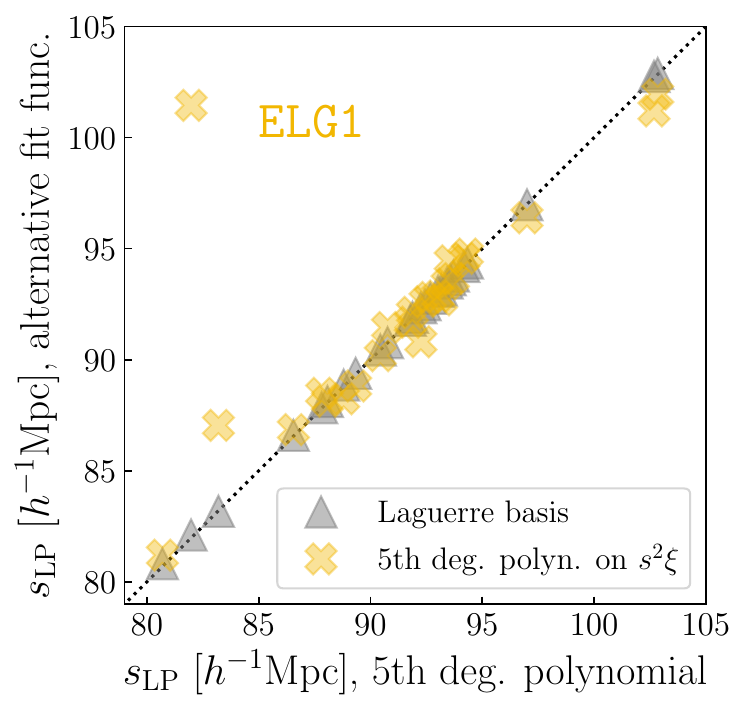} &
       *\hspace{-.5cm}\includegraphics[width=0.33\columnwidth]{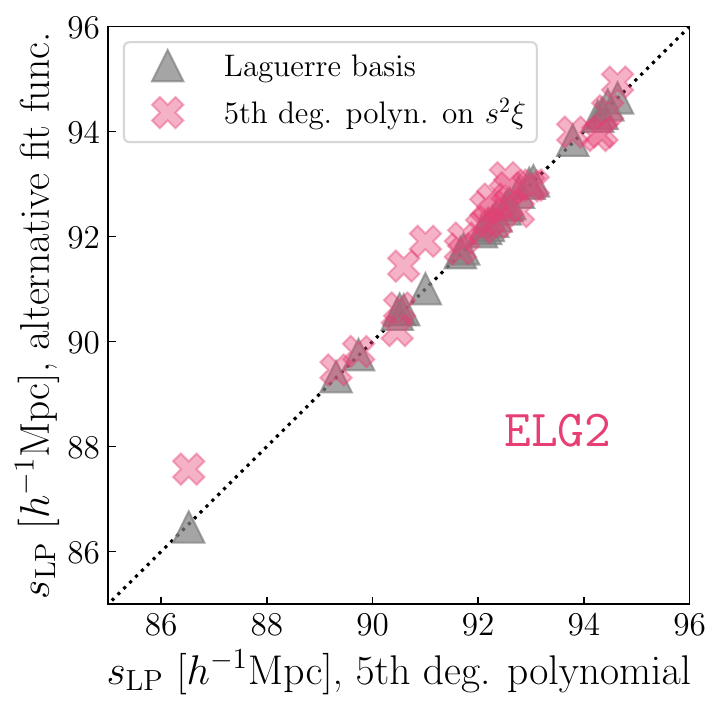} 
    \end{tabular}
    \caption{Linear points calculated on \abacust\ mocks, one panel per tracer, using the standard methodology (fifth-degree polynomial fit to the correlation function $\xi$), compared to those calculated using alternative fitting methods: modified Laguerre functions (gray triangles) and a fifth-degree polynomial fit to $s^2\xi$ (colored crosses). All measurements are pre-reconstruction, and the dotted line marks $x=y$ along which the measurements agree exactly.}
    \label{fig:alt_fits_mocks}
\end{figure}

For each mock, we compute the linear point using the `standard' polynomial method -- fifth-degree polynomial fit to the correlation function $\xi$ -- and the two aforementioned alternative fitting functions. Since the need for alternative fitting functions arises from the poor SNR of the correlation function in the first place (Appendix~\ref{sec:bgs_bad}), we focus only on pre-reconstruction measurements; post-reconstruction correlation functions have consistently stronger BAO signals. The difference between the methods is also more pronounced pre-reconstruction, as seen in previous sections where linear point measurements across the 25 mocks show higher scatter. We depict these measurements for each set of 25 \abacust\ mocks for all tracers in Figure \ref{fig:alt_fits_mocks}.

Across all tracers, the linear points measured using the Laguerre basis (gray triangles in Figure \ref{fig:alt_fits_mocks}) are uniformly consistent with the standard method. This is unsurprising; while odd degree Laguerre polynomials carry error-function terms, the even degree ones reduce to simple polynomials. Measuring the linear point on $s^2\xi$ (colored crosses in Figure \ref{fig:alt_fits_mocks}) rather than on $\xi$ also agrees well with the standard method, albeit with some scatter and outliers. Visual inspection of the outlier fits reveals substantial noise in the correlation functions, so we attribute these larger discrepancies to statistical fluctuations in the mocks themselves. This already hints that with poor SNR, the two methods can diverge. We do however see that all mocks across all tracers yield a linear point measurement when measured on $s^2\xi$, consistent with the results seen in \cite{he2023}.

\begin{figure}[]
    \centering
    \begin{tabular}{cc}
       *\hspace{-.5cm}\includegraphics[width=0.45\columnwidth]{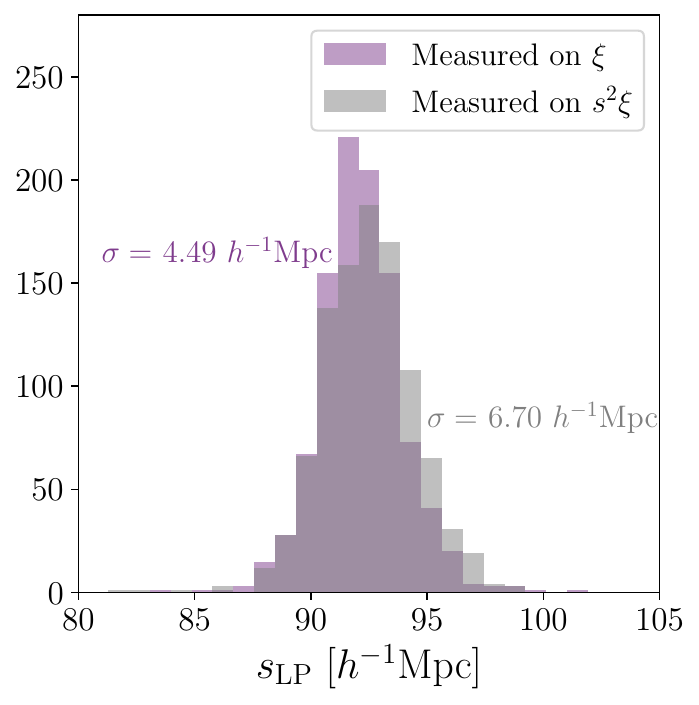}  & 
       *\hspace{-.5cm}\includegraphics[width=0.475\columnwidth]{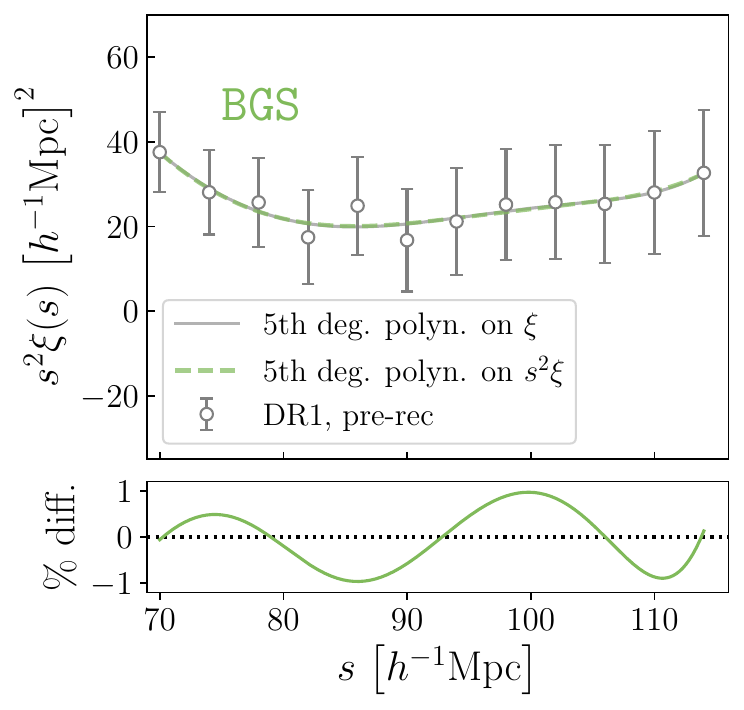} 
    \end{tabular}
    \caption{\emph{Left panel:} The linear points measured using a polynomial fit to the linear theory correlation function $\xi$ (colored) and to $s^2\xi$ (gray) on 1000 samples generated using the DR2 \lrgo\ covariance matrix. Each histogram's standard deviation is annotated in text of the same color. \emph{Right panel:} The pre-reconstruction DR1 \bgs\ correlation function, with a fifth-degree polynomial fit to $\xi$ (gray curve, scaled by $s^2$ for display) compared against the same function fit directly to $s^2\xi$ (dashed colored curve). The bottom panel shows the percent difference between the two curves as a function of $s$.}
\label{fig:DR1_altfits}
\end{figure}

We test further whether measuring the linear point on $\xi$ or $s^2\xi$ is more robust. To do so, we use the linear theory correlation function at $z=0$ and inject noise into it in the form of a covariance matrix (to generate the left side panel of Figure \ref{fig:DR1_altfits}, we use the DR2 \lrgo\ covariance matrix). We generate 1000 samples of the correlation function using the covariance matrix, and for each sample calculate the linear point on $\xi$ and on $s^2\xi$. We find that both measurements yield an average linear point within $1\sigma$ of the fiducial linear point and have 100\% reliability, i.e., each of the 1000 sample yields a measurement. However, the $s^2\xi$ methods yields a larger scatter around the mean, as seen in the left side panel of Figure \ref{fig:DR1_altfits}. We find that this trend persists even when different covariance matrices are used to generate the correlation function samples. We infer that while a fit to $s^2\xi$ is slightly more likely to yield a linear point measurement when SNR is poor, the standard method yields a more precise measurement in cases when the linear point is measurable. 

Finally, we implement both methods on the DR1 \bgs\ sample; we exclude the Laguerre fits, given that all mocks were nearly perfectly consistent with the standard method. We find that both curves visually overlap, with sub-percent residuals over the entire fitting range as shown in right side panel of Figure \ref{fig:DR1_altfits}. More importantly, neither method yields a linear point measurement. 

We have established earlier in this work that linear point measurements made on the more precise DR2 sample, with better SNR than DR1, are more robust; we expect even better SNR with future DESI data releases and upcoming state-of-the-art spectroscopic surveys like Roman. Given that the standard fifth degree polynomial fit to $\xi$ yields a more precise linear point measurement, we need not use the $s^2\xi$ fits to compensate for increasing reliability with poor SNR. We therefore conclude that the standard linear point pipeline is sufficient and appropriate for any future implementations of the linear point pipeline.

\section{Error Analysis in DR1 Mocks and Data}
\label{sec:errors}

We have shown in Sections \ref{sec:mocks} and \ref{sec:results} that the linear point standard ruler, converted to \qisolp, shows excellent agreement with the BAO standard ruler measured using the standard fitting pipeline. However, the scatter in the linear point measurements on \lrg\ and \elg\ mocks is 8-90\% larger post-reconstruction, compared to the errors on \qisolp\ measured on DR1 tracers, which range from 7\% larger to 50\% smaller than those measured on the isotropic BAO scale post-reconstruction.

\begin{figure}[]
    \centering
    \begin{tabular}{ccc}
    \vspace{-.85cm}
       *\hspace{-.5cm}\includegraphics[width=0.33\columnwidth]{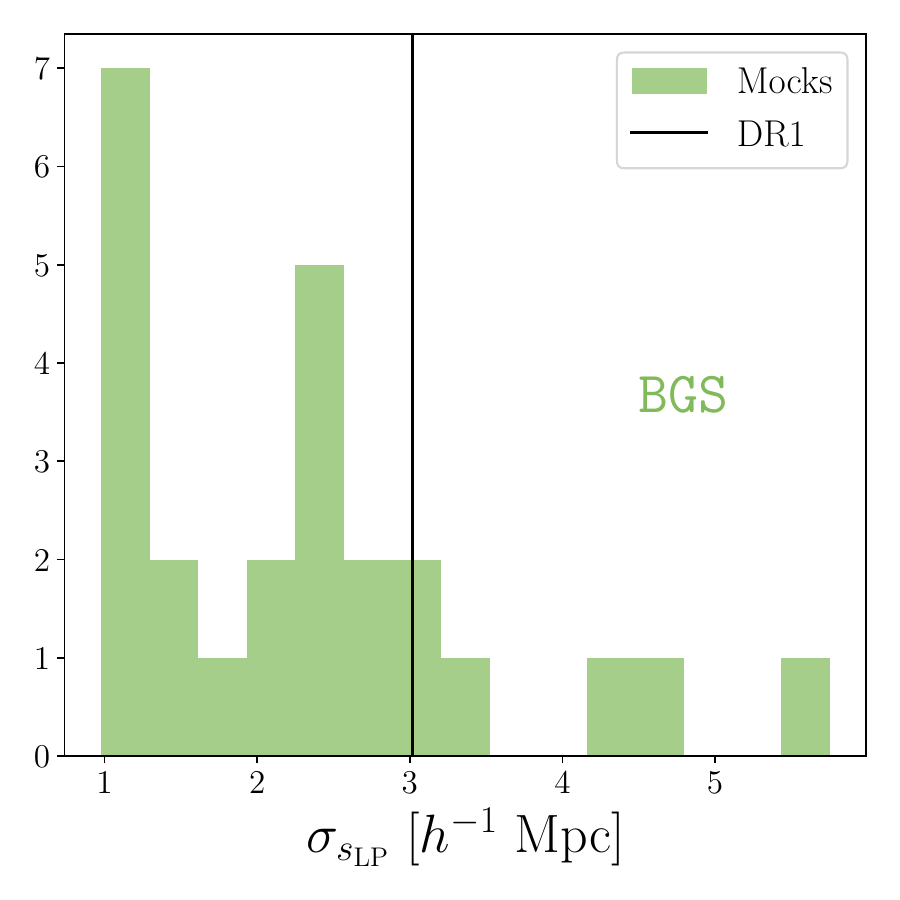} &
       *\hspace{-.5cm}\includegraphics[width=0.33\columnwidth]{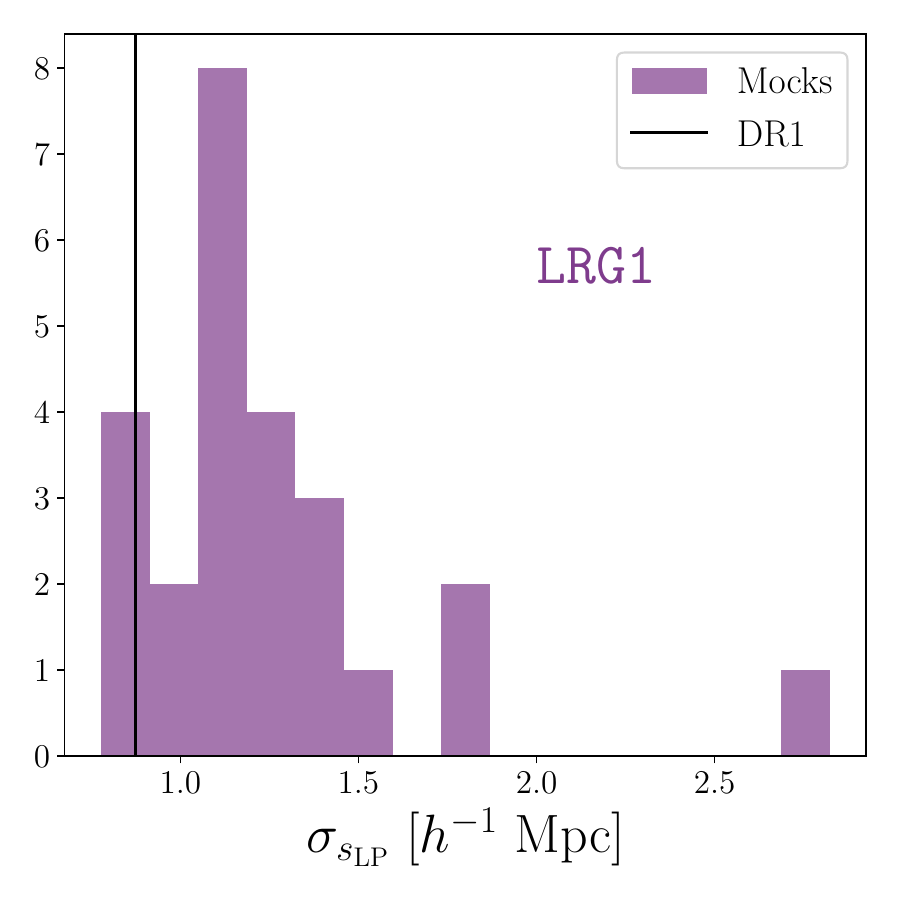}  & 
       *\hspace{-.5cm}\includegraphics[width=0.33\columnwidth]{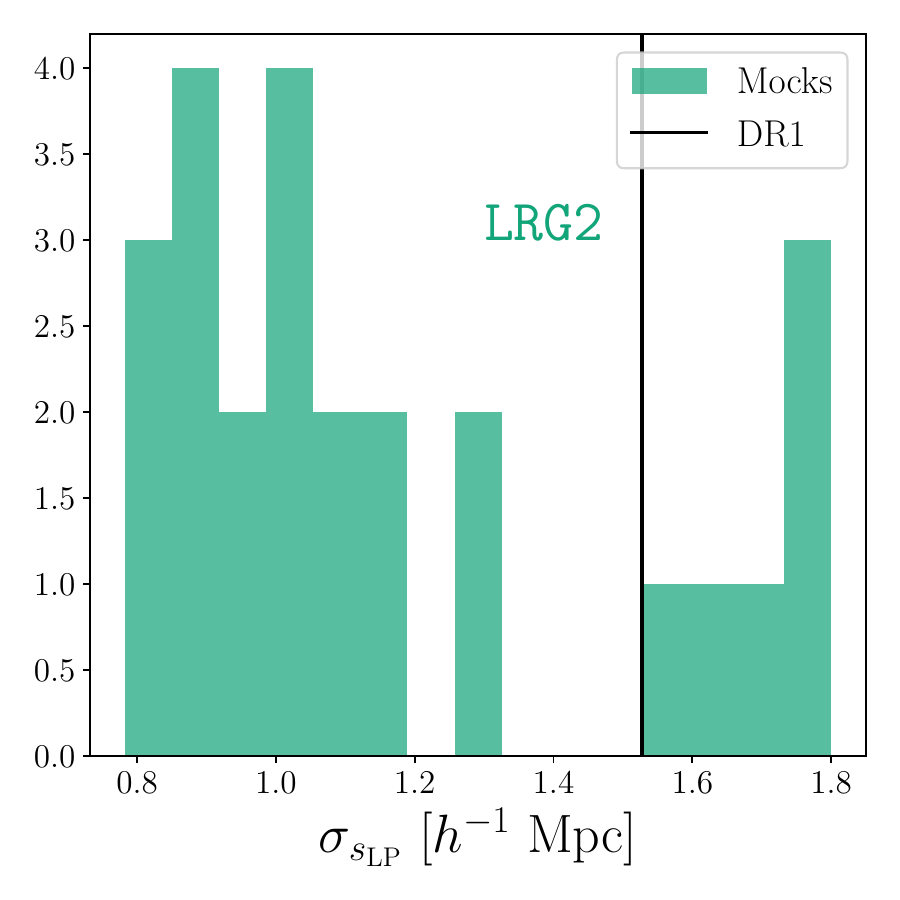} \\
       *\hspace{-.5cm}\includegraphics[width=0.33\columnwidth]{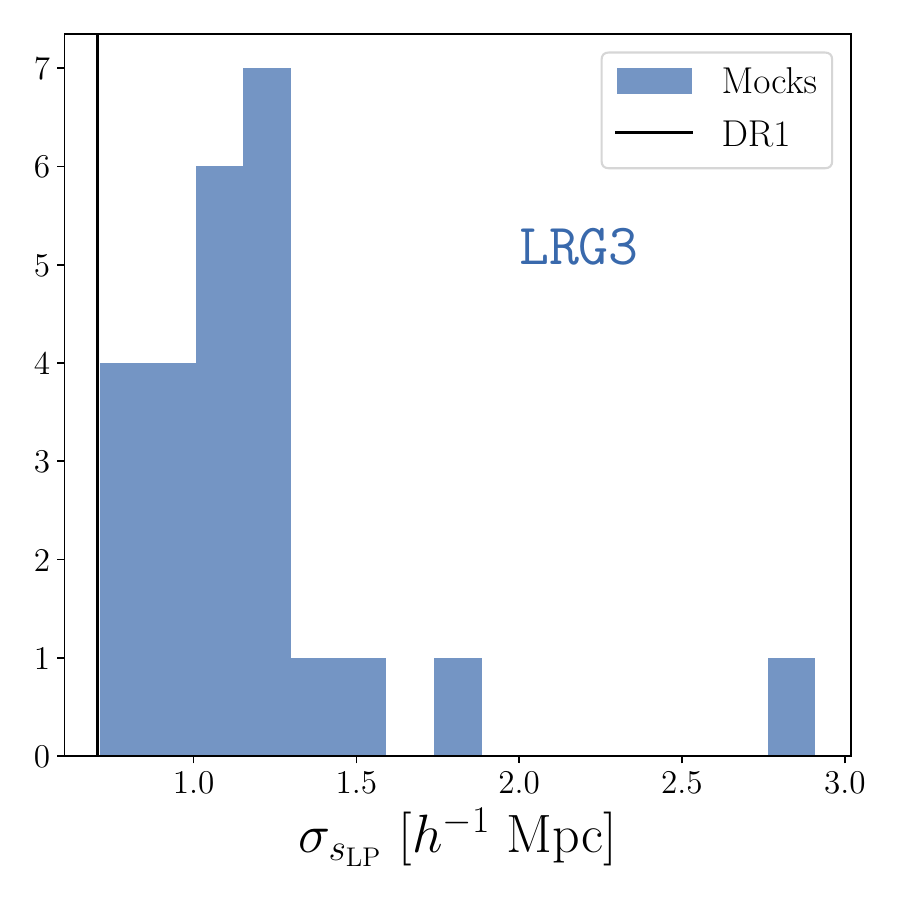} &
       *\hspace{-.5cm}\includegraphics[width=0.33\columnwidth]{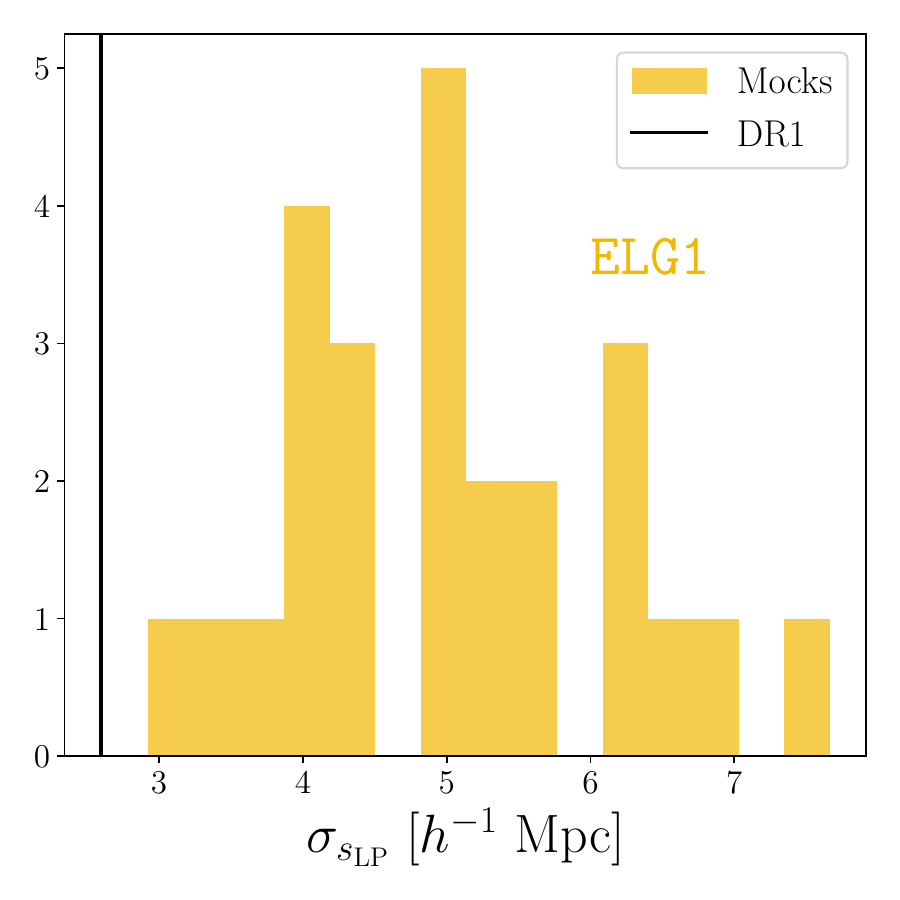} &
       *\hspace{-.5cm}\includegraphics[width=0.33\columnwidth]{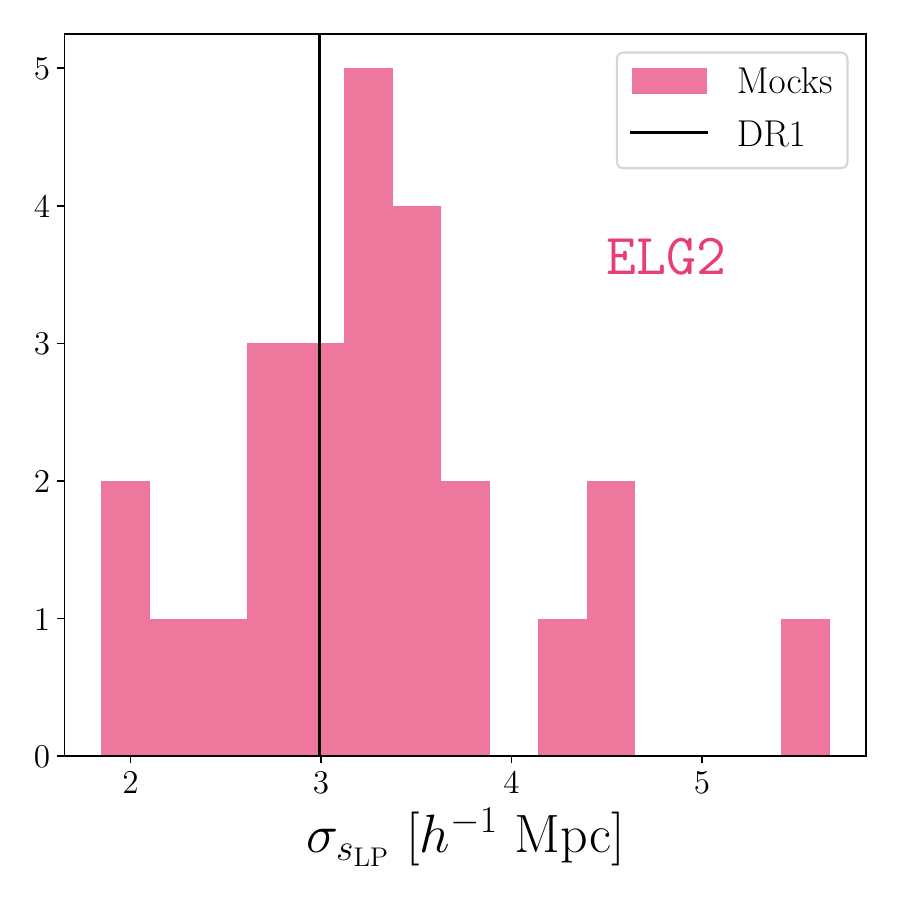} 
    \end{tabular}
    \caption{Histograms of the post-reconstruction errors on the linear point measurements on the 25 \abacust\ mocks for the DESI DR1 \bgs, \lrg, and \elg\ samples, one panel per tracer. The error on the corresponding DR1 linear point measurement is indicated with a solid black vertical line.}
    \label{fig:Y1_vs_mocks_errs_hist}
\end{figure}

We study this discrepancy between mock errors and data errors by calculating the uncertainty $\sigma_{s_\mathrm{LP}}$ on the linear point measured on the 25 mocks and comparing their distribution to the error calculated on the data. We plot the distribution of post-reconstruction errors for the mocks in Figure \ref{fig:Y1_vs_mocks_errs_hist} and indicate the error calculated on the data measurements as a vertical line. There is a large scatter in the error values, and only \lrgo\ and \elgt\ exhibit somewhat Gaussian-like distributions. 
With the exception of \lrgt\ and \elgt, the errors on the data measurements are consistently on the lower end of the distribution of mock errors. 

\begin{figure}[]
    \centering
    \begin{tabular}{ccc}
       *\hspace{-.5cm}\includegraphics[width=0.33\columnwidth]{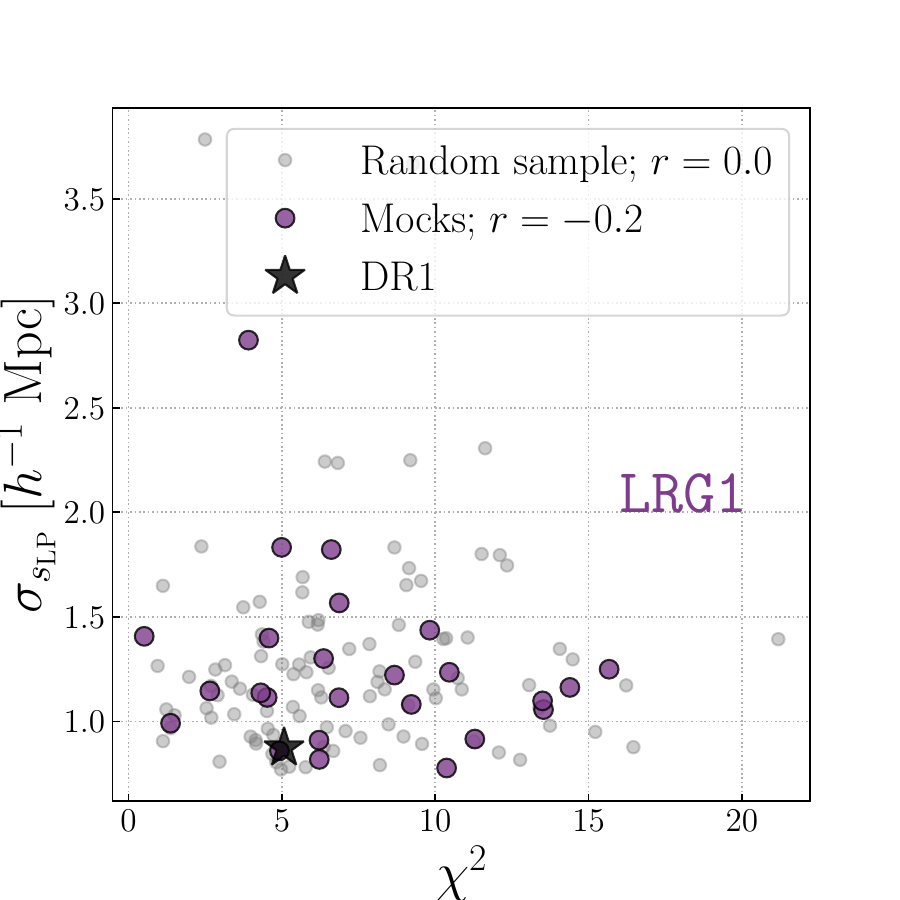}  & 
       *\hspace{-.5cm}\includegraphics[width=0.33\columnwidth]{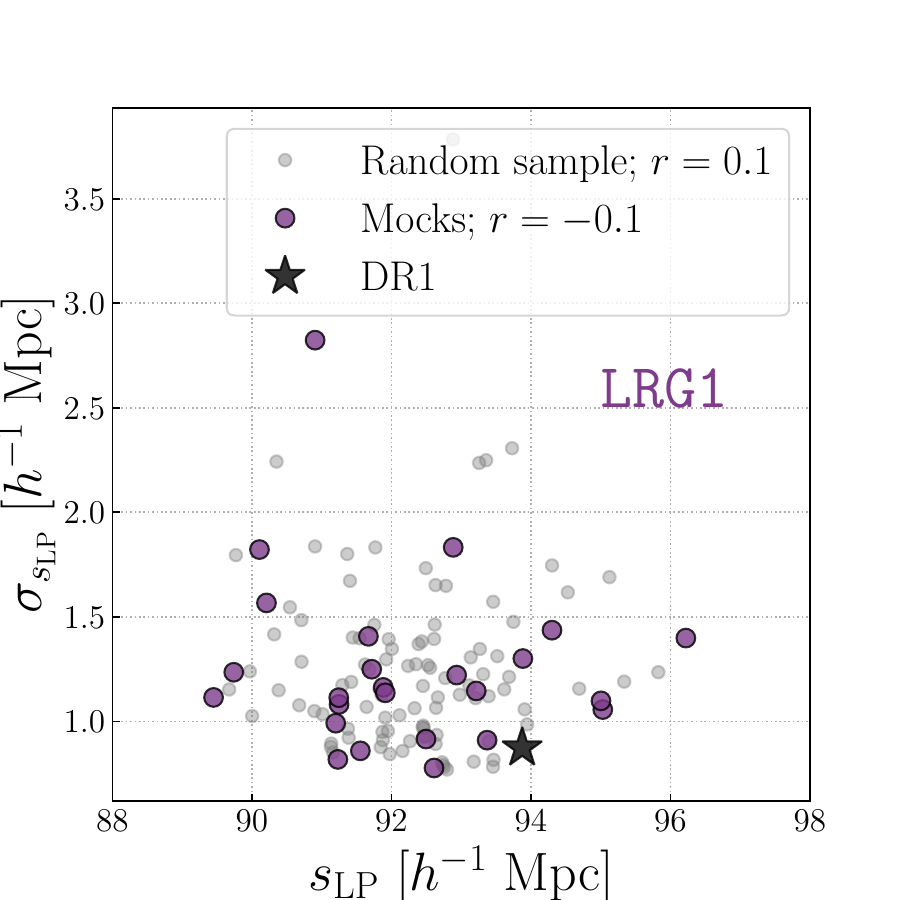} &
       *\hspace{-.5cm}\includegraphics[width=0.33\columnwidth]{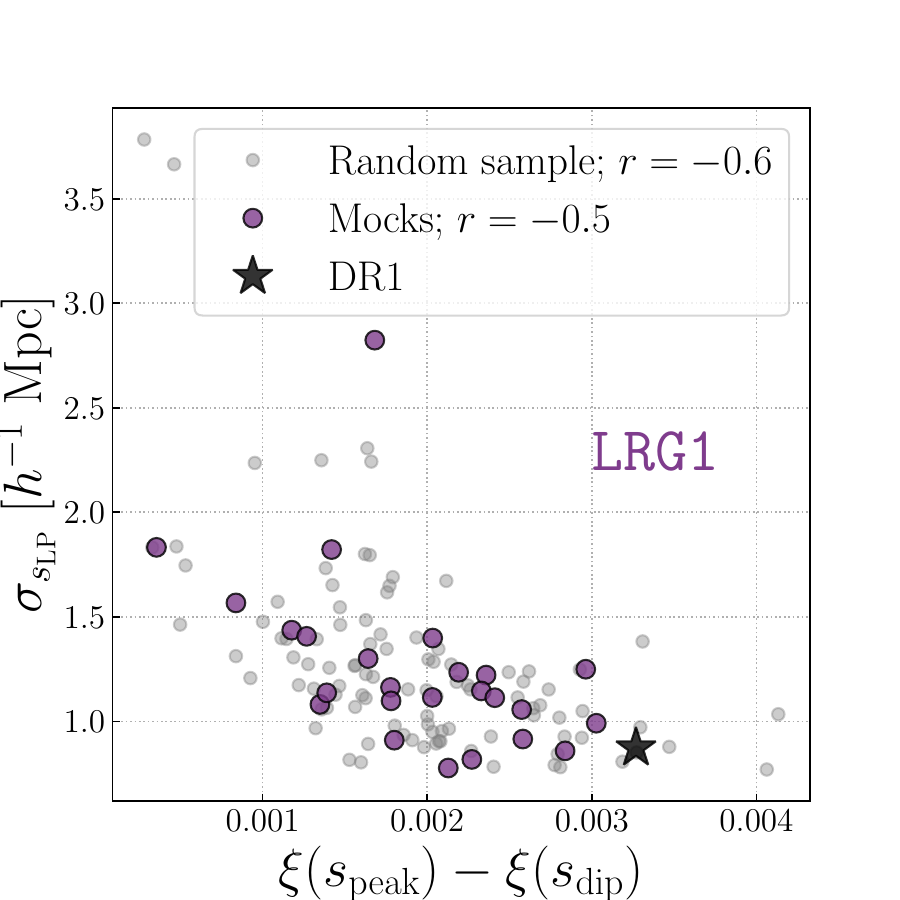} \\
    \end{tabular}
    \caption{The post-reconstruction error on linear point measurements on each of the 25 \lrgo\ \abacust\ mocks plotted as a function of the $\chi^2$ of the polynomial fit (left), the linear point measurement \slp\ (center), and the difference between the amplitudes of the BAO peak $\xi(s_\mathrm{peak})$ and the preceding dip $\xi(s_\mathrm{dip})$ (right). Gray points show the same quantities for a random sample of 100 correlation functions drawn about the mock mean, and the Pearson correlation coefficient $r$ is annotated for both sets. The same quantities calculated on the data from the first data release (DR1) are indicated using a black star marker in each panel. }
\label{fig:LRG1_errs_vs_quantities}
\end{figure}

We further explore any correlations that may arise between $\sigma_{s_\mathrm{LP}}$ and factors that might affect the quality of our measurements, such as the $\chi^2$ of the polynomial fit, the location of the linear point itself, and the amplitude of the BAO peak in the correlation function. We choose to perform these tests using the \lrgo\ mocks, since the distribution of mock errors is well-behaved and the data error is found on the lower end of the mock distribution. We plot the mock errors as a function of each of these quantities in Figure \ref{fig:LRG1_errs_vs_quantities}, and indicate the data error as a black star in each panel. We measure the Pearson correlation coefficient $r$ in each panel to predict the correlation between the two quantities plotted in each panel of the figure.

While we find little to no correlation between the $\chi^2$ and the location of the linear point \slp\ (left-most and central panel of Figure \ref{fig:LRG1_errs_vs_quantities}), we find that the error is negatively correlated with the absolute difference between the size of the BAO peak and the preceding dip, $\xi(s_\mathrm{peak})-\xi(s_\mathrm{dip})$, as shown in the right-side panel of Figure \ref{fig:LRG1_errs_vs_quantities}. 
The scales corresponding to the peak and the dip, $s_\mathrm{peak}$ and $s_\mathrm{dip}$, are computed using the roots of the best-fit polynomial; and the amplitudes, $\xi(s_\mathrm{peak})$ and $\xi(s_\mathrm{dip})$, are computed by evaluating the polynomial interpolation of the correlation function at the peak and dip scales. As indicated in the figure, lower errors on the Y1 linear point measurements are a direct consequence of a more pronounced BAO peak in the data correlation function compared to those in the mocks. This is consistent with previous work \citep{eigencov2024} which argued that the error depends on the sharpness (i.e. curvature) of the peak and dip features:  if the separation between both features is approximately fixed, then larger curvature implies larger  $\xi(s_\mathrm{peak})-\xi(s_\mathrm{dip})$.  
We perform the same tests using linear point errors calculated from a randomly generated sample of 100 correlation functions centered on the mean of the 25 \lrgo\ mock correlation functions using the covariance matrix. These errors are plotted in gray in Figure \ref{fig:LRG1_errs_vs_quantities}, and indicate that the trend seen in the mocks is an accurate representation of a random sample.

\begin{figure}[]
    \centering
    \begin{tabular}{ccc}
    \vspace{-.85cm}
       *\hspace{-.5cm}\includegraphics[width=0.3\textwidth]{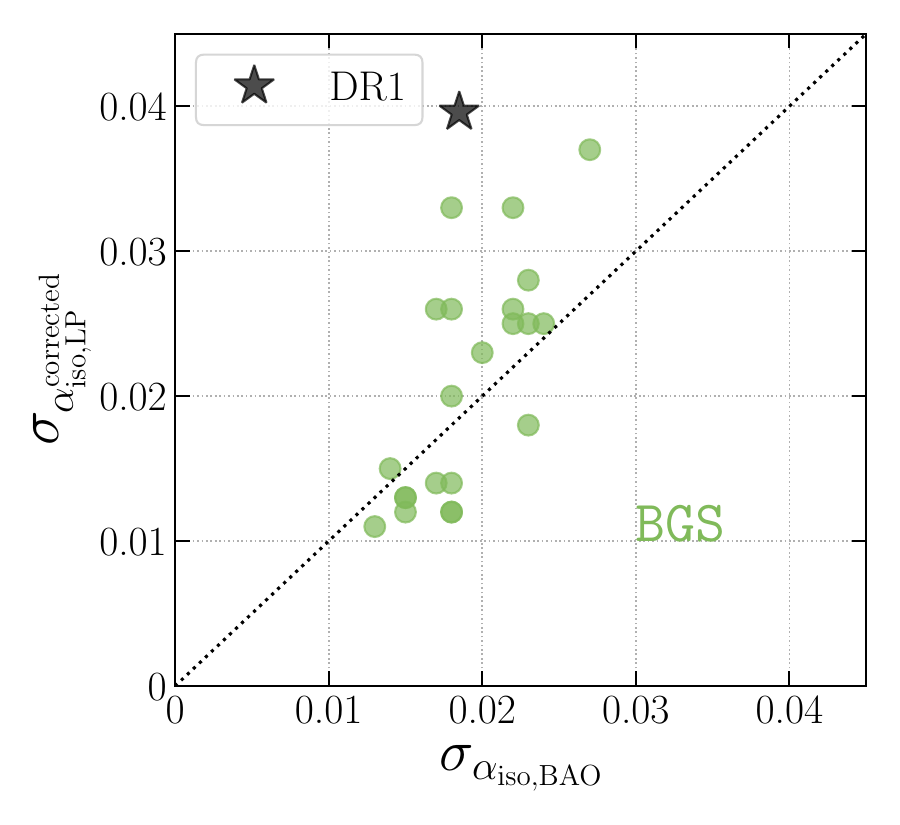} & 
       \includegraphics[width=0.3\textwidth]{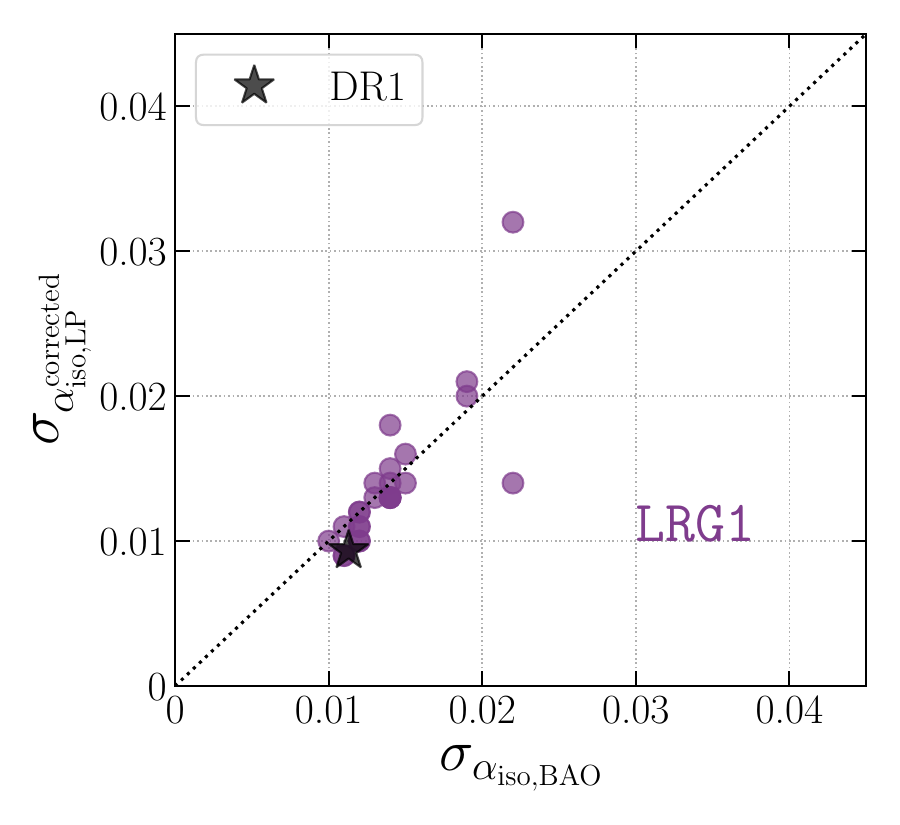}  &
       \includegraphics[width=0.3\textwidth]{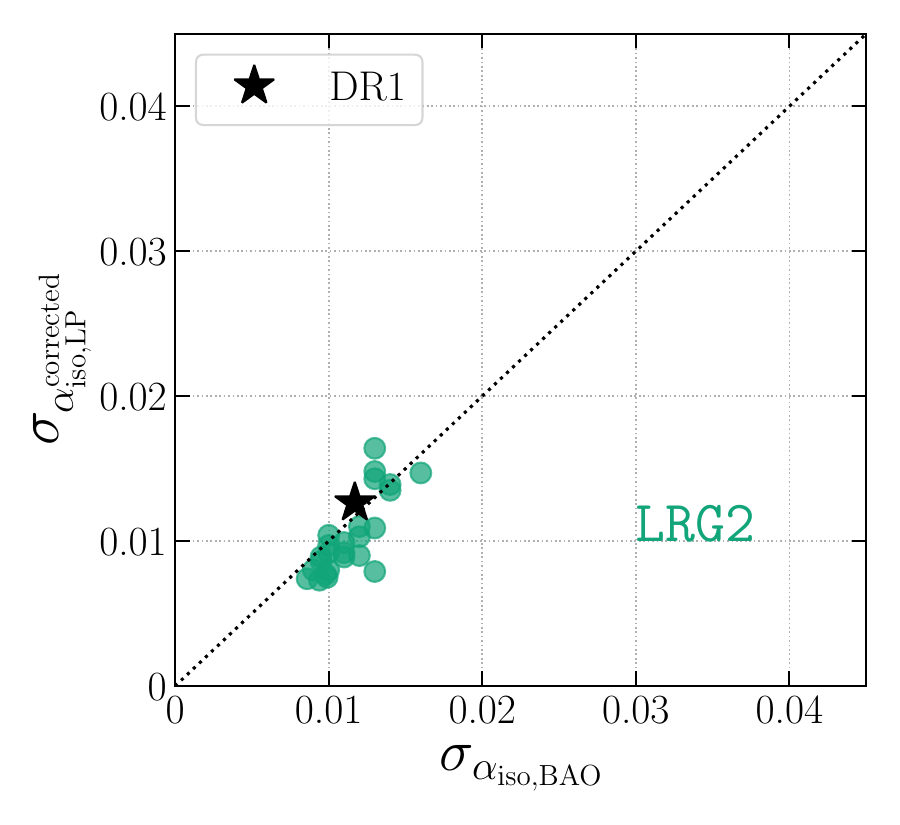} \\
       *\hspace{-.5cm}\includegraphics[width=0.3\textwidth]{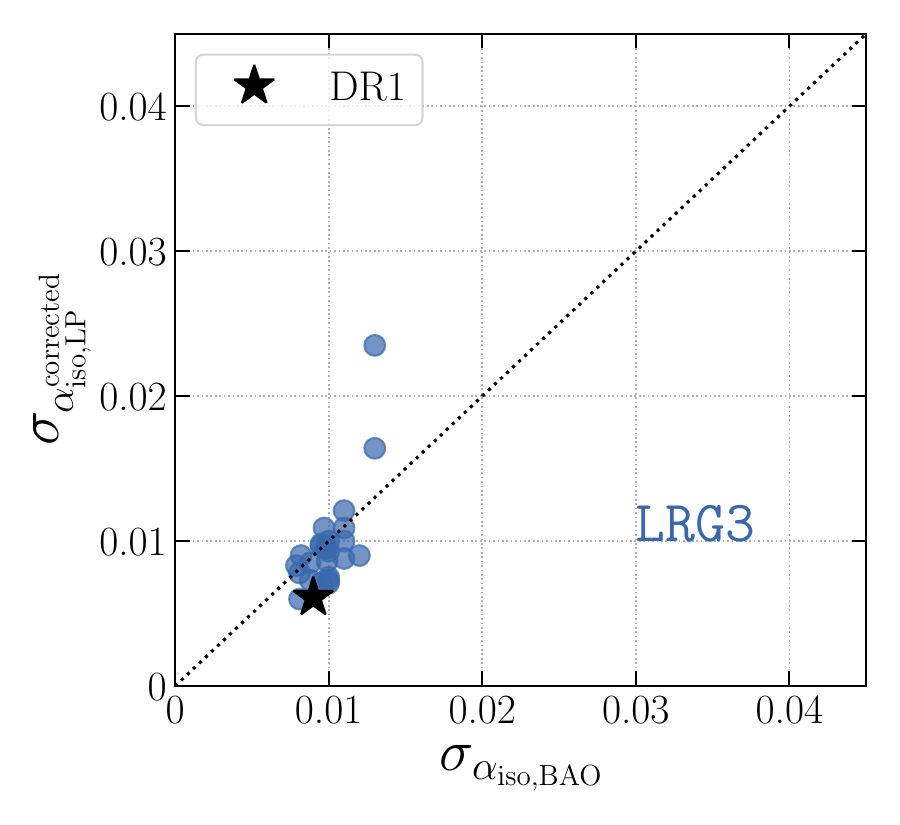} &
        \includegraphics[width=0.3\textwidth]{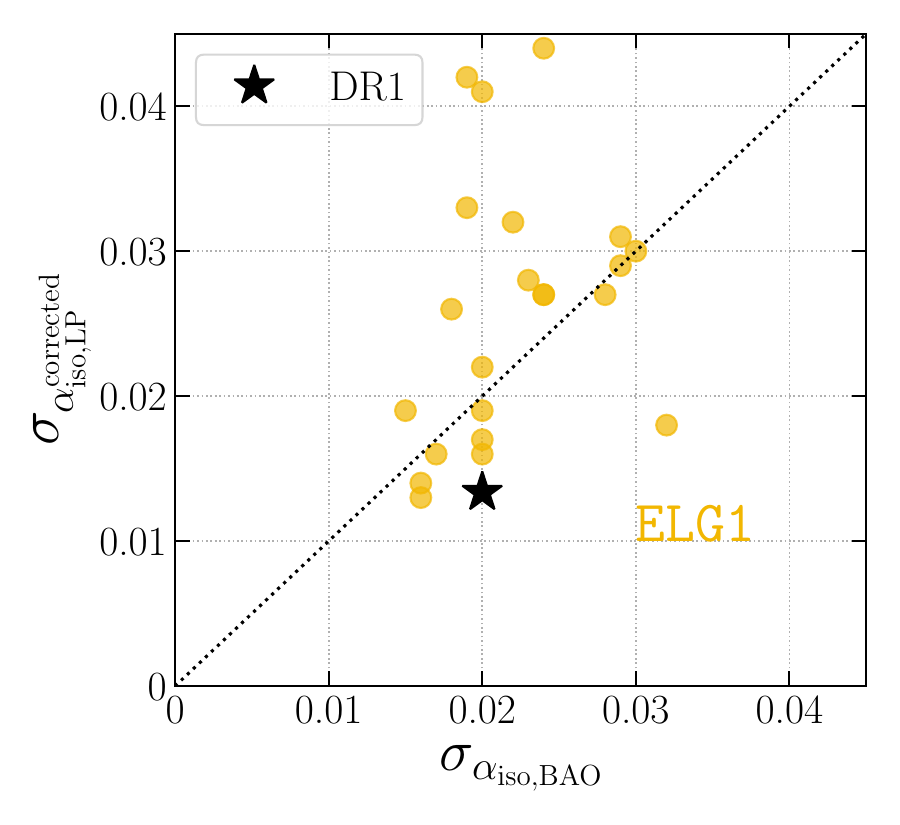} &
        \includegraphics[width=0.3\textwidth]{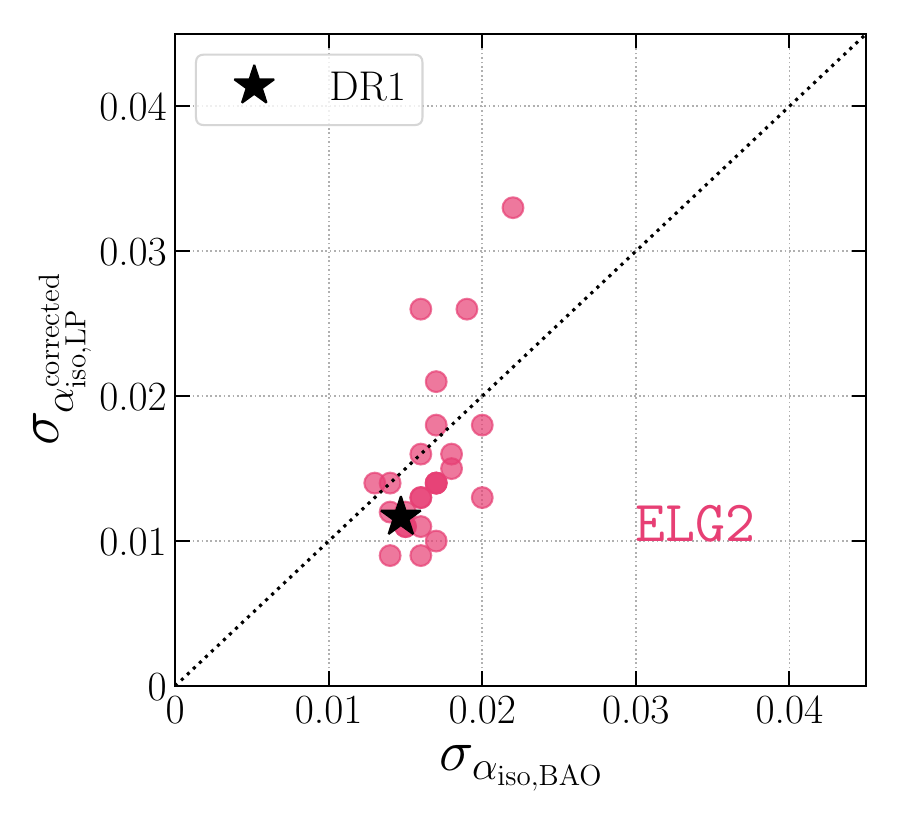} 
        \end{tabular}
    \caption{The errors on the isotropic BAO parameter plotted against the errors on the corresponding linear point parameter for each set of 25 \abacust\ mocks, one panel per tracer. The black star markers indicate the same values calculated on DR1 tracers.}
    \label{fig:Y1_vs_mocks_errs}
\end{figure}

We also test whether smaller errors on the linear point translate to smaller errors in BAO measurements. We convert the post-reconstruction \slp\ errors calculated on each mock to errors on corrected \qisolp\ values ($\sigma_{\alpha_\mathrm{isoLP}^\mathrm{corrected}}$) and plot them against the errors on isotropic BAO measurements ($\sigma_{\alpha_\mathrm{iso,BAO}}$) in Figure \ref{fig:Y1_vs_mocks_errs}. We depict the errors in the Y1 \qisolp\ and \qisobao\ measurements with a black star marker in each panel. We see that the data BAO error is also on the lower end of the mock error distribution, which is consistent with the distribution of \qisolp\ errors.

\section{Laguerre reconstruction}
\label{sec:laguerre}

In large-scale cosmological analysis, the evolved correlation function on BAO scales can be modeled as a Gaussian convolution of the linear theory correlation function (c.f. Eq.~\ref{eq:convolution}). This suggests that an analytical deconvolution of the measured correlation function could serve as an alternative to standard reconstruction algorithms, which typically operate on the matter density field. That is, rather than reconstructing the field and moving particles, this approach directly works with the two-point statistics of the field to recover its linear form, potentially offering a much faster and simpler method. Within the specific range of 60–120 $h^{-1}$ Mpc, the linear theory correlation function can be accurately approximated by a simple odd-degree polynomial. Consequently, the corresponding non-linear correlation function can be effectively fit using a linear combination of generalized Laguerre functions (i.e. Gaussian-convolved polynomials), which are well-suited to capture its behavior in this range, providing a robust mathematical framework for this alternative reconstruction method.

In more detail, if the linear theory correlation function is approximated as a simple $n$th order polynomial
\begin{equation}
    \xi_\mathrm{lin.th.}(r)=\sum_{k=0}^{n}a_k\left(\frac{r}{\sigma}\right)^k,
\label{eq:linear_laguerre}
\end{equation}
where $\sigma$ is set to a fiducial value to keep the coefficients $a_k$ dimensionless, and the non-linear correlation function is just a convolution with a Gaussian of RMS $\Sigma$, then 
\begin{equation}
    \xi_\mathrm{NL}(s) = \sum_{k=0}^{n}c_k\,\mu_k(x),
    \quad {\rm where} \quad x\equiv s/\Sigma,
\label{eq:laguerre}
\end{equation}
$c_k\equiv a_k(\Sigma/\sigma)^k$, and the modified Laguerre functions $\mu_k$ are defined in Eq. \ref{eq:mu_functions}.

The Laguerre reconstruction pipeline leverages this as follows.
When Eq. \ref{eq:laguerre} is fit to the non-linear (observed) correlation function, then the best-fit Laguerre coefficients $c_k$ are used to estimate $a_k=c_k\,(\sigma/\Sigma)^k$.  Substitution of these $a_k$ into Eq. \ref{eq:linear_laguerre} yields the reconstructed linear correlation function \cite{nikakhtar2021, nikakhtar2021_halo}.  

Unlike standard reconstruction algorithms that focus on reconstructing the matter density field, the Laguerre method targets the shape of the monopole of the correlation function \cite[for initial attempts that extend this framework to higher order multipoles, see][]{paranjape2023,paranjape2025}. Additionally, it complements the purely geometric nature of the linear point pipeline, enhancing the toolkit available for BAO analysis with a method that is both efficient and broadly applicable. While model-independent, the Laguerre pipeline contains a non-trivial dependence on the Gaussian smearing kernel $\Sigma$, which is completely degenerate with the fitting coefficients $c_k$ in Eq. \ref{eq:laguerre}. Therefore, a rigorous implementation of Laguerre reconstruction requires the use of informed priors on $\Sigma$, which introduces some cosmology dependence in the pipeline.

\begin{figure}
    \centering
    \begin{tabular}{ccc}
       *\hspace{-.85cm}\includegraphics[width=0.33\textwidth]{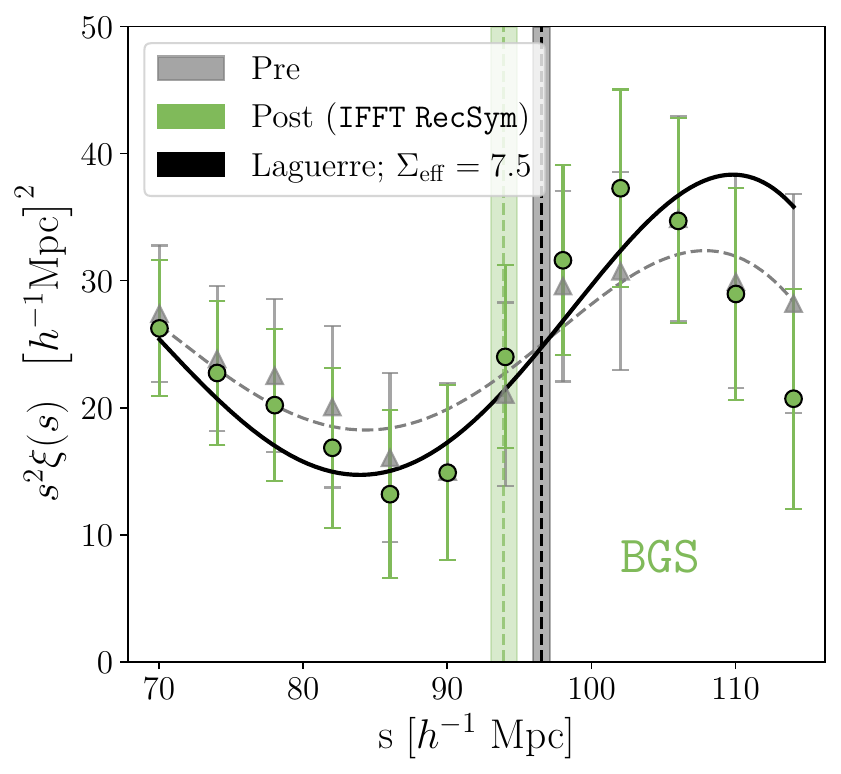} &
       \includegraphics[width=0.33\textwidth]{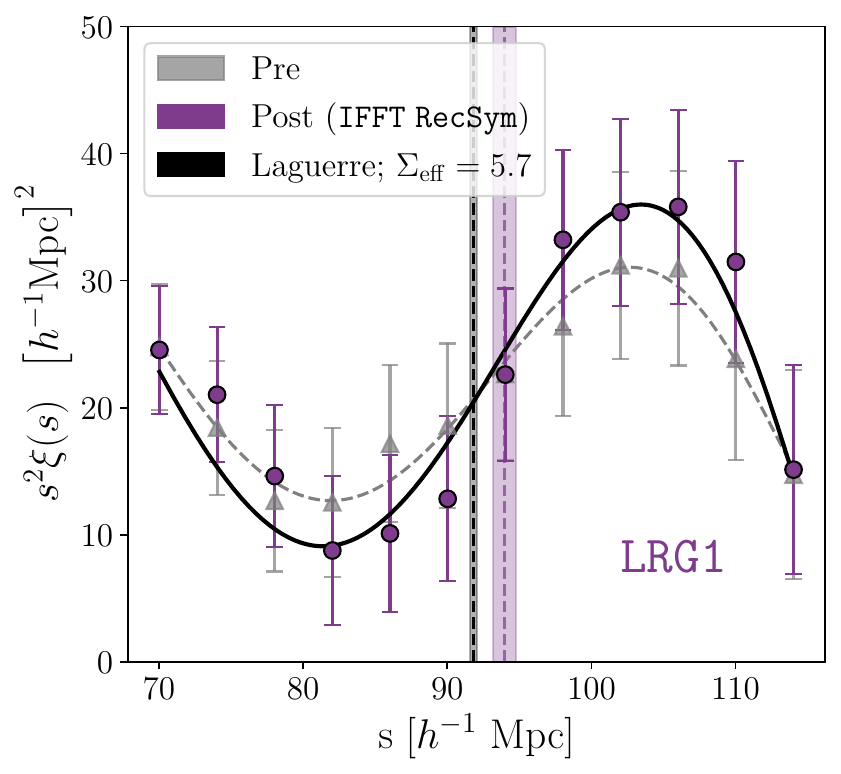}  & \includegraphics[width=0.33\textwidth]{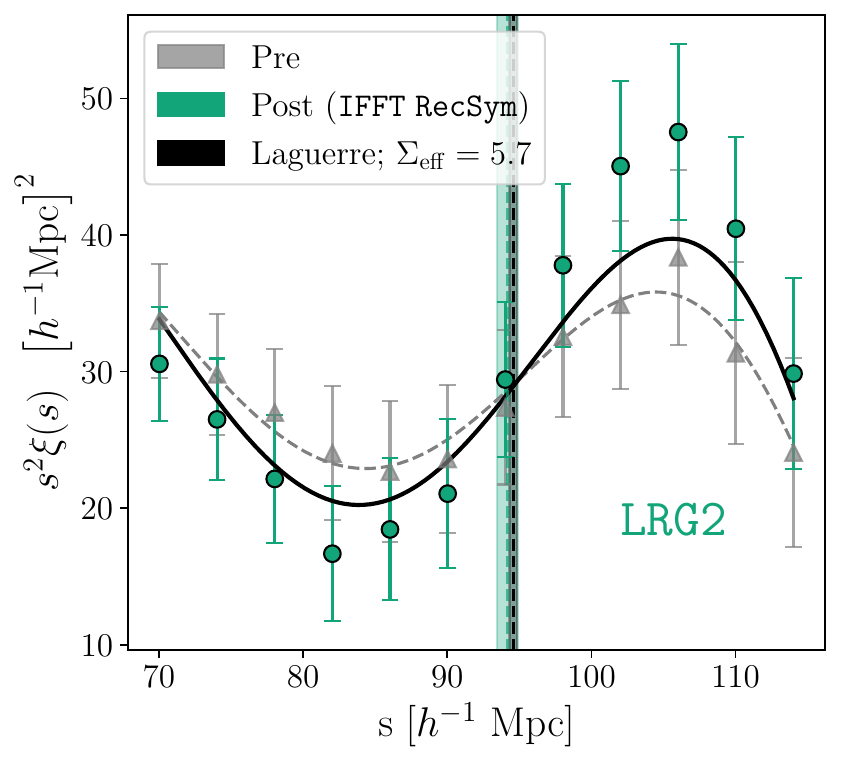} \\
    \end{tabular}
    \begin{tabular}{cc}
    *\hspace{-.7cm}\includegraphics[width=0.33\textwidth]{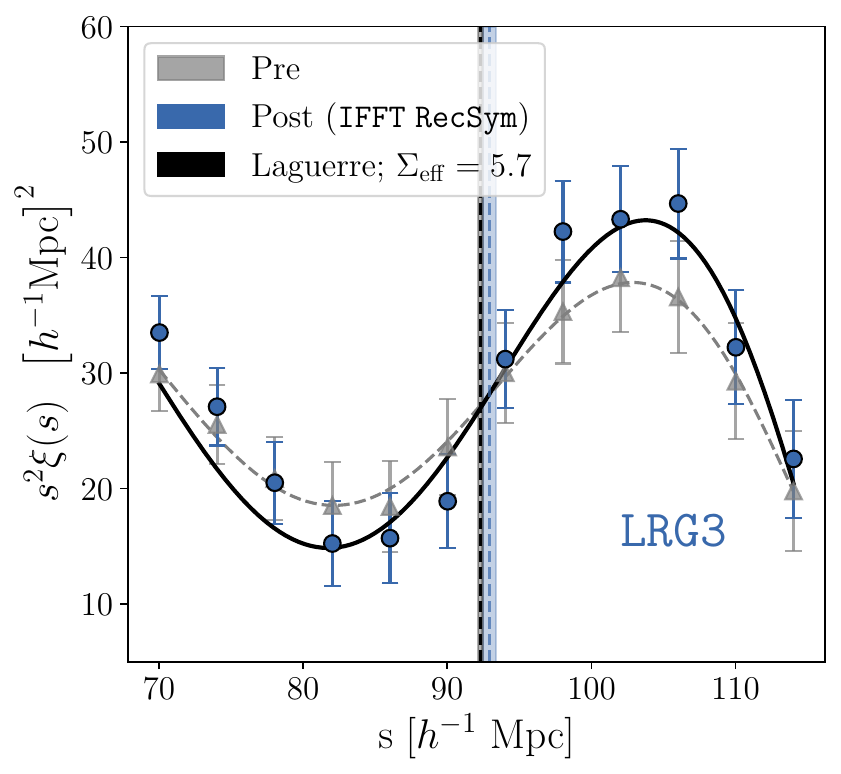} &
    \includegraphics[width=0.33\textwidth]{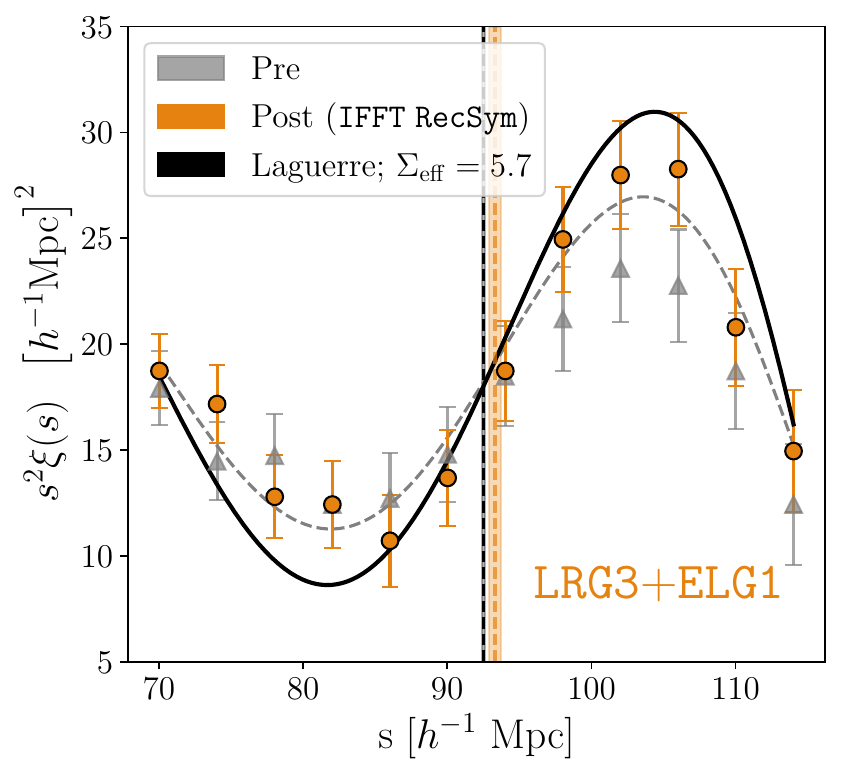} \\
    *\hspace{-.7cm}\includegraphics[width=0.33\textwidth]{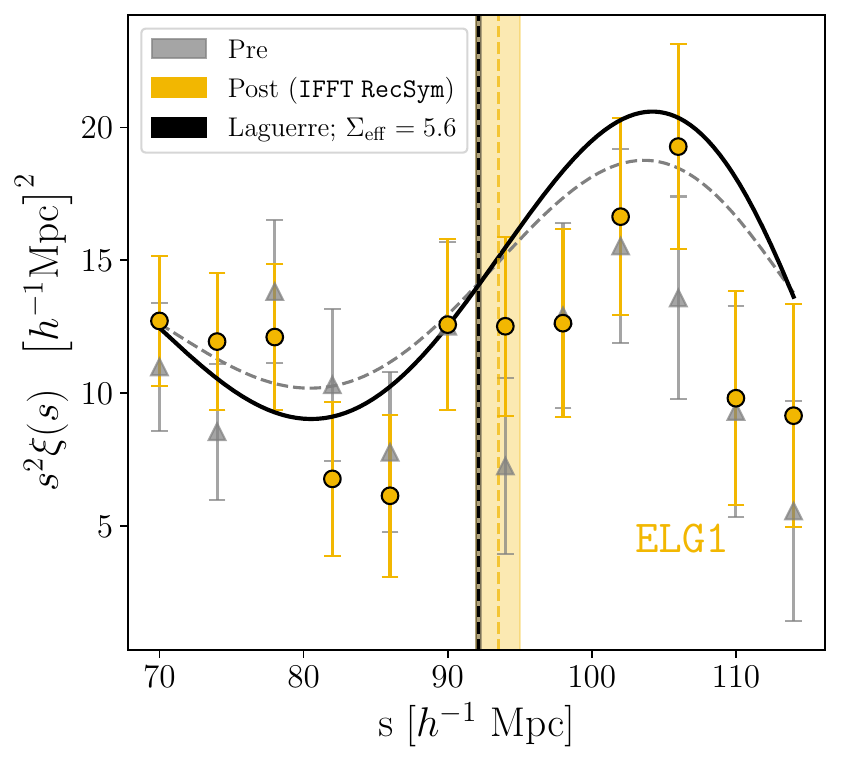} &
    \includegraphics[width=0.33\textwidth]{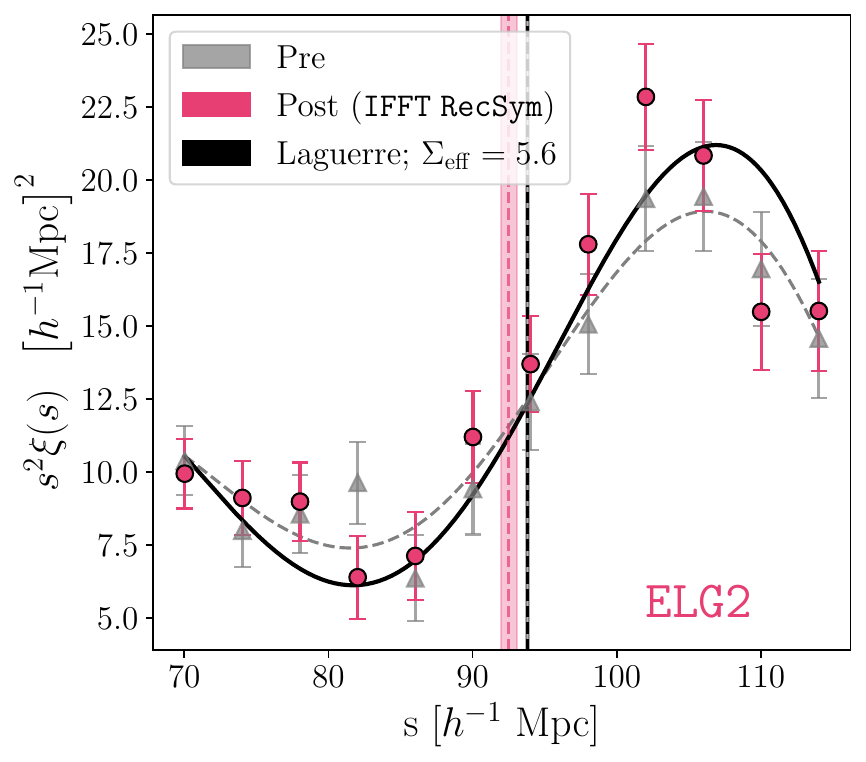}
    \end{tabular}
    \caption{Pre-reconstruction (gray triangles) and post-reconstruction (colored circles) correlation functions for DR2 tracers, one per panel, along with the Laguerre function fit to the pre-reconstruction correlation function (gray dashed curve) and Laguerre reconstructed (i.e. associated simple polynomial) correlation function (solid black curve). The linear points calculated post standard reconstruction (colored dashed vertical line) and post Laguerre reconstruction (black dashed vertical line) are plotted along with their $1\sigma$ uncertainties.}
    \label{fig:Y3_laguerre}
\end{figure}

Here, we present preliminary reconstructions of the observed non-linear correlation function measurements for DR2 targets using the Laguerre method. We use the pre-reconstruction template Gaussian prior on $\Sigma_\mathrm{iso}$ from Table \ref{tab:damping_parameters} to inform our choice of $\Sigma$ in the pipeline. The pre- and post-reconstruction correlation functions, along with the Laguerre reconstructed curve, are shown in Figure \ref{fig:Y3_laguerre}. We observe that Laguerre reconstruction, plotted as a black solid curve, seem to mostly agree with the correlation function calculated post standard reconstruction, shown as colored circles. However, evidently in the case of \lrgt, the deconvolution falls short of the post-reconstruction correlation function, indicating a smaller $\Sigma$ used than what was needed. On the other hand, in the case of \lrgelg, we seem to have overestimated $\Sigma$. This highlights the drawback of the Laguerre reconstruction method -- since the kernel is perfectly degenerate with the fitting coefficients, there is little we can do to mitigate this issue as of now.

\begin{figure}
    \centering
    \begin{tabular}{cc}
        \includegraphics[width=0.45\linewidth]{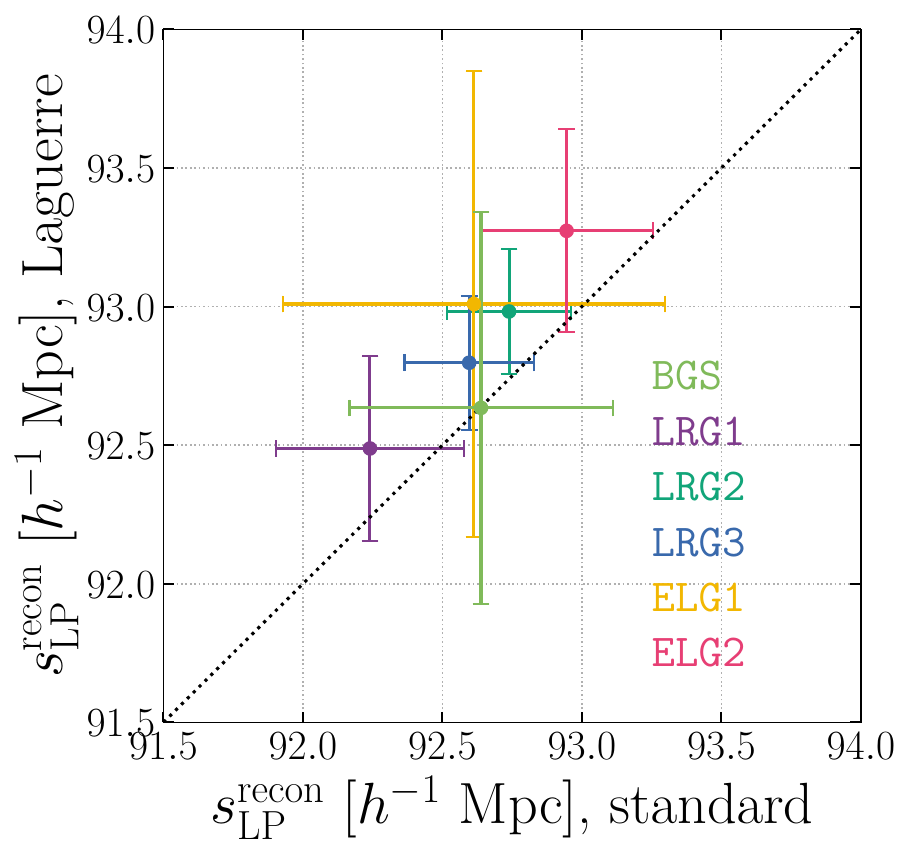} &  \includegraphics[width=0.45\linewidth]{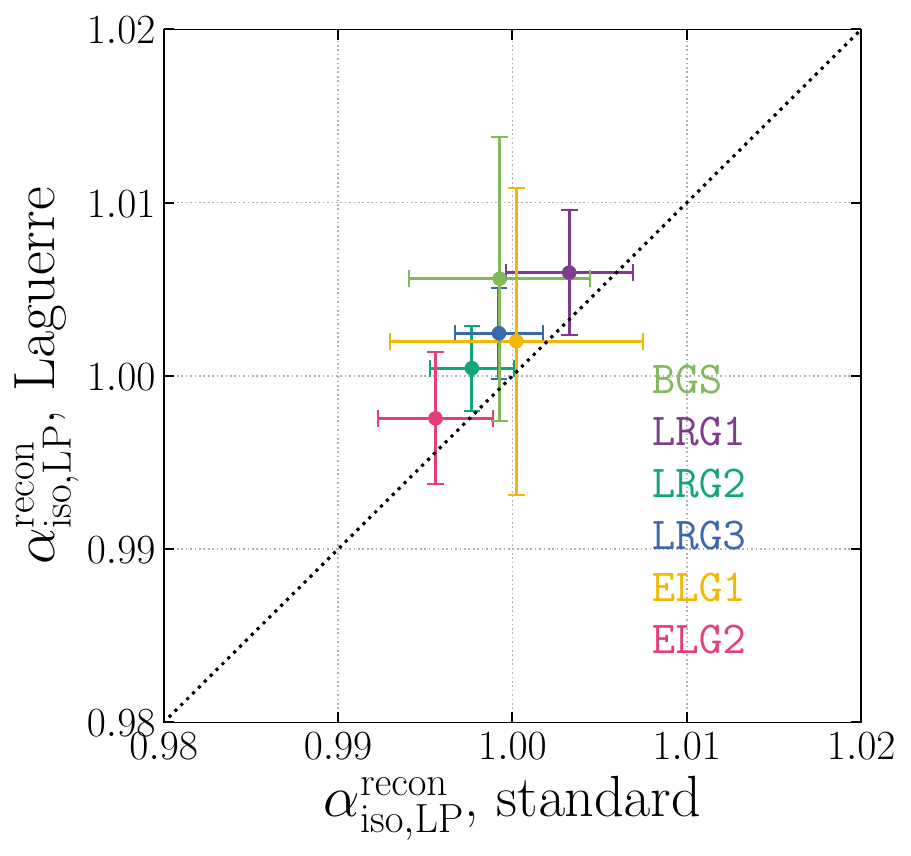}
    \end{tabular}
    \caption{Comparing the linear point measurements made on \lrg\ and \elg\ \abacust\ mocks using standard reconstruction and Laguerre reconstruction. 
    Left panel shows mean linear point measurements using standard post-reconstruction correlation functions ($x$-axis) and Laguerre-reconstructed correlation functions ($y$-axis) with $\Sigma=5.7$ \hMpc. Right panel shows the same measurements converted to \qisolp\ using Eq. \ref{eq:qiso_lp}, with $s_\mathrm{LP}^\mathrm{fid}=92.5$ \hMpc\ to reflect the estimated non-zero smearing the post-reconstruction regime, and retaining $s_\mathrm{LP}^\mathrm{fid}=93.01$ \hMpc\ for Laguerre values. The error bars on each data point are the standard deviations of the 25 measurements, scaled by the square root of the number of mocks. }
    \label{fig:laguerre_mocks}
\end{figure}

We further compute the linear point on the Laguerre-reconstructed correlation functions in \bgs, \lrg, and \elg\ \abacust\ mocks and compare these measurements to those made on standard post-reconstruction correlation functions in the left panel of Figure \ref{fig:laguerre_mocks}. The two measurements show remarkable agreement, albeit with a slight bias of $\sim0.2$\%. This could be attributed to the fact that standard reconstruction retains some non-linear effects as evidenced by non-zero damping in Table \ref{tab:damping_parameters}. We convert the two sets of linear point measurements to \qisolp\ using Eq, \ref{eq:qiso_lp} -- we use $s_\mathrm{LP}^\mathrm{fid}=93.01$ \hMpc\ in the case of Laguerre-reconstructed measurements, and $s_\mathrm{LP}^\mathrm{fid}=92.51$ \hMpc\ for standard reconstruction, since they retain the effects of smearing. We plot these measurements in the right side panel of Figure \ref{fig:laguerre_mocks}, and note excellent correlation between the mean linear point measurements and the errors.

\bibliographystyle{JHEP}
\bibliography{linear_point.bib}

@ARTICLE{crocce_scoccimarro,
       author = {{Crocce}, Mart{\'\i}n and {Scoccimarro}, Rom{\'a}n},
        title = "{Nonlinear evolution of baryon acoustic oscillations}",
      journal = {\prd},
         year = 2008,
        month = jan,
       volume = {77},
       number = {2},
          eid = {023533},
        pages = {023533},
          doi = {10.1103/PhysRevD.77.023533},
archivePrefix = {arXiv},
       eprint = {0704.2783},
 primaryClass = {astro-ph},
       adsurl = {https://ui.adsabs.harvard.edu/abs/2008PhRvD..77b3533C}
}

@ARTICLE{eisenstein2007,
       author = {{Eisenstein}, Daniel J. and {Seo}, Hee-Jong and {White}, Martin},
        title = "{On the Robustness of the Acoustic Scale in the Low-Redshift Clustering of Matter}",
      journal = {\apj},
         year = 2007,
        month = aug,
       volume = {664},
       number = {2},
        pages = {660-674},
          doi = {10.1086/518755},
archivePrefix = {arXiv},
       eprint = {astro-ph/0604361},
 primaryClass = {astro-ph},
       adsurl = {https://ui.adsabs.harvard.edu/abs/2007ApJ...664..660E}
}

@ARTICLE{seo2008,
       author = {{Seo}, Hee-Jong and {Siegel}, Ethan R. and {Eisenstein}, Daniel J. and {White}, Martin},
        title = "{Nonlinear Structure Formation and the Acoustic Scale}",
      journal = {\apj},
         year = 2008,
        month = oct,
       volume = {686},
       number = {1},
        pages = {13-24},
          doi = {10.1086/589921},
archivePrefix = {arXiv},
       eprint = {0805.0117},
 primaryClass = {astro-ph},
       adsurl = {https://ui.adsabs.harvard.edu/abs/2008ApJ...686...13S}
}

@article{FKP,
   title={Power-spectrum analysis of three-dimensional redshift surveys},
   volume={426},
   ISSN={1538-4357},
   url={http://dx.doi.org/10.1086/174036},
   DOI={10.1086/174036},
   journal={\apj},
   publisher={American Astronomical Society},
   author={Feldman, Hume A. and Kaiser, Nick and Peacock, John A.},
   year={1994},
   month=may, pages={23} }

@article{bao_intro_weinberg,
title = "{Observational probes of cosmic acceleration}",
journal = {Physics Reports},
volume = {530},
number = {2},
pages = {87-255},
year = {2013},
note = {Observational Probes of Cosmic Acceleration},
issn = {0370-1573},
doi = {https://doi.org/10.1016/j.physrep.2013.05.001},
url = {https://www.sciencedirect.com/science/article/pii/S0370157313001592},
author = {David H. Weinberg and Michael J. Mortonson and Daniel J. Eisenstein and Christopher Hirata and Adam G. Riess and Eduardo Rozo}
}

@article{bao_intro_eisenstein,
title = {Dark energy and cosmic sound},
journal = {New Astronomy Reviews},
volume = {49},
number = {7},
pages = {360-365},
year = {2005},
note = {Wide-Field Imaging from Space},
issn = {1387-6473},
doi = {https://doi.org/10.1016/j.newar.2005.08.005},
url = {https://www.sciencedirect.com/science/article/pii/S1387647305000850},
author = {D.J. Eisenstein}
}

@article{bao_intro_bassett,
    author = "Bassett, Bruce A. and Hlozek, Renee",
    title = "{Baryon Acoustic Oscillations}",
    eprint = "arXiv:0910.5224",
    archivePrefix = "arXiv",
    primaryClass = "astro-ph.CO",
    month = "10",
    year = "2009"
}

@ARTICLE{peebles,
       author = {{Peebles}, P.~J.~E. and {Yu}, J.~T.},
        title = "{Primeval Adiabatic Perturbation in an Expanding Universe}",
      journal = {\apj},
         year = 1970,
        month = dec,
       volume = {162},
        pages = {815},
          doi = {10.1086/150713},
       adsurl = {https://ui.adsabs.harvard.edu/abs/1970ApJ...162..815P}
}

@article{seo2010,
doi = {10.1088/0004-637X/720/2/1650},
url = {https://dx.doi.org/10.1088/0004-637X/720/2/1650},
year = {2010},
month = {aug},
publisher = {The American Astronomical Society},
volume = {720},
number = {2},
pages = {1650},
author = {Seo, Hee-Jong and Eckel, Jonathan and Eisenstein, Daniel J. and Mehta, Kushal and Metchnik, Marc and Padmanabhan, Nikhil and Pinto, Phillip and Takahashi, Ryuichi and White, Martin and Xu, Xiaoying},
title = {HIGH-PRECISION PREDICTIONS FOR THE ACOUSTIC SCALE IN THE NONLINEAR REGIME},
journal = {\apj}
}

@article{schmittfull2015,
  title = "{Eulerian BAO reconstructions and $N$-point statistics}",
  author = {Schmittfull, Marcel and Feng, Yu and Beutler, Florian and Sherwin, Blake and Chu, Man Yat},
  journal = {Phys. Rev. D},
  volume = {92},
  issue = {12},
  pages = {123522},
  numpages = {24},
  year = {2015},
  month = {Dec},
  publisher = {American Physical Society},
  doi = {10.1103/PhysRevD.92.123522},
  url = {https://link.aps.org/doi/10.1103/PhysRevD.92.123522}
}

@article{ding2018,
    author = {Ding, Zhejie and Seo, Hee-Jong and Vlah, Zvonimir and Feng, Yu and Schmittfull, Marcel and Beutler, Florian},
    title = {Theoretical systematics of Future Baryon Acoustic Oscillation Surveys},
    journal = {\mnras},
    volume = {479},
    number = {1},
    pages = {1021-1054},
    year = {2018},
    month = {05},
    issn = {0035-8711},
    doi = {10.1093/mnras/sty1413},
    url = {https://doi.org/10.1093/mnras/sty1413},
    
}

@article{Sherwin_2019,
doi = {10.1088/1475-7516/2019/02/027},
url = {https://dx.doi.org/10.1088/1475-7516/2019/02/027},
year = {2019},
month = {feb},
publisher = {},
volume = {2019},
number = {02},
pages = {027},
author = {Sherwin, Blake D. and White, Martin},
title = "{The impact of wrong assumptions in BAO reconstruction}",
journal = {\jcap}
}

@article{bernal2020,
  title = {Robustness of baryon acoustic oscillation constraints for early-Universe modifications of $\mathrm{\ensuremath{\Lambda}}\mathrm{CDM}$ cosmology},
  author = {Bernal, Jos\'e Luis and Smith, Tristan L. and Boddy, Kimberly K. and Kamionkowski, Marc},
  journal = {Phys. Rev. D},
  volume = {102},
  issue = {12},
  pages = {123515},
  numpages = {21},
  year = {2020},
  month = {Dec},
  publisher = {American Physical Society},
  doi = {10.1103/PhysRevD.102.123515},
  url = {https://link.aps.org/doi/10.1103/PhysRevD.102.123515}
}

@article{carter2020,
    author = {Carter, Paul and Beutler, Florian and Percival, Will J and DeRose, Joseph and Wechsler, Risa H and Zhao, Cheng},
    title = "{The impact of the fiducial cosmology assumption on BAO distance scale measurements}",
    journal = {\mnras},
    volume = {494},
    number = {2},
    pages = {2076-2089},
    year = {2020},
    month = {03},
    issn = {0035-8711},
    doi = {10.1093/mnras/staa761},
    url = {https://doi.org/10.1093/mnras/staa761},

}

@article{alcock_paczynski,
	title = "{An evolution free test for non-zero cosmological constant}",
	volume = {281},
	copyright = {1979 Springer Nature Limited},
	issn = {1476-4687},
	url = {https://www.nature.com/articles/281358a0},
	doi = {10.1038/281358a0},
	language = {en},
	number = {5730},
	urldate = {2025-02-18},
	journal = {Nature},
	author = {Alcock, Charles and Paczyński, Bohdan},
	month = oct,
	year = {1979},
	note = {Publisher: Nature Publishing Group},
	pages = {358--359},
}

@article{Zeldovich,
    author = {{Zel'dovich}, Ya. B.},
    title = "{Gravitational instability: An approximate theory for large density perturbations.}",
    journal = {\aap},
    year = 1970,
    month = mar,
    volume = {5},
    pages = {84-89},
    adsurl = {https://ui.adsabs.harvard.edu/abs/1970A&A.....5...84Z}
}

@article{LandySzalay,
    author = "Landy, Stephen D. and Szalay, Alexander S.",
    title = "{Bias and variance of angular correlation functions}",
    doi = "10.1086/172900",
    journal = "Astrophys. J.",
    volume = "412",
    pages = "64",
    year = "1993"
}

@article{modifiedLandySzalay,
    author = {Padmanabhan, Nikhil and Xu, Xiaoying and Eisenstein, Daniel J. and Scalzo, Richard and Cuesta, Antonio J. and Mehta, Kushal T. and Kazin, Eyal},
    title = "{A 2 per cent distance to z = 0.35 by reconstructing baryon acoustic oscillations – I. Methods and application to the Sloan Digital Sky Survey}",
    journal = {\mnras},
    volume = {427},
    number = {3},
    pages = {2132-2145},
    year = {2012},
    month = {12},
    issn = {0035-8711},
    doi = {10.1111/j.1365-2966.2012.21888.x},
    url = {https://doi.org/10.1111/j.1365-2966.2012.21888.x},
    
}

@article{planck2018,
   title="{Planck2018 results: VI. Cosmological parameters}",
   volume={641},
   ISSN={1432-0746},
   url={http://dx.doi.org/10.1051/0004-6361/201833910},
   DOI={10.1051/0004-6361/201833910},
   journal={\aap},
   publisher={EDP Sciences},
   author={Aghanim, N. and Akrami, Y. and Ashdown, M. and Aumont, J. and Baccigalupi, C. and Ballardini, M. and Banday, A. J. and Barreiro, R. B. and Bartolo, N. and Basak, S. and Battye, R. and Benabed, K. and Bernard, J.-P. and Bersanelli, M. and Bielewicz, P. and Bock, J. J. and Bond, J. R. and Borrill, J. and Bouchet, F. R. and Boulanger, F. and Bucher, M. and Burigana, C. and Butler, R. C. and Calabrese, E. and Cardoso, J.-F. and Carron, J. and Challinor, A. and Chiang, H. C. and Chluba, J. and Colombo, L. P. L. and Combet, C. and Contreras, D. and Crill, B. P. and Cuttaia, F. and de Bernardis, P. and de Zotti, G. and Delabrouille, J. and Delouis, J.-M. and Di Valentino, E. and Diego, J. M. and Doré, O. and Douspis, M. and Ducout, A. and Dupac, X. and Dusini, S. and Efstathiou, G. and Elsner, F. and Enßlin, T. A. and Eriksen, H. K. and Fantaye, Y. and Farhang, M. and Fergusson, J. and Fernandez-Cobos, R. and Finelli, F. and Forastieri, F. and Frailis, M. and Fraisse, A. A. and Franceschi, E. and Frolov, A. and Galeotta, S. and Galli, S. and Ganga, K. and Génova-Santos, R. T. and Gerbino, M. and Ghosh, T. and González-Nuevo, J. and Górski, K. M. and Gratton, S. and Gruppuso, A. and Gudmundsson, J. E. and Hamann, J. and Handley, W. and Hansen, F. K. and Herranz, D. and Hildebrandt, S. R. and Hivon, E. and Huang, Z. and Jaffe, A. H. and Jones, W. C. and Karakci, A. and Keihänen, E. and Keskitalo, R. and Kiiveri, K. and Kim, J. and Kisner, T. S. and Knox, L. and Krachmalnicoff, N. and Kunz, M. and Kurki-Suonio, H. and Lagache, G. and Lamarre, J.-M. and Lasenby, A. and Lattanzi, M. and Lawrence, C. R. and Le Jeune, M. and Lemos, P. and Lesgourgues, J. and Levrier, F. and Lewis, A. and Liguori, M. and Lilje, P. B. and Lilley, M. and Lindholm, V. and López-Caniego, M. and Lubin, P. M. and Ma, Y.-Z. and Macías-Pérez, J. F. and Maggio, G. and Maino, D. and Mandolesi, N. and Mangilli, A. and Marcos-Caballero, A. and Maris, M. and Martin, P. G. and Martinelli, M. and Martínez-González, E. and Matarrese, S. and Mauri, N. and McEwen, J. D. and Meinhold, P. R. and Melchiorri, A. and Mennella, A. and Migliaccio, M. and Millea, M. and Mitra, S. and Miville-Deschênes, M.-A. and Molinari, D. and Montier, L. and Morgante, G. and Moss, A. and Natoli, P. and Nørgaard-Nielsen, H. U. and Pagano, L. and Paoletti, D. and Partridge, B. and Patanchon, G. and Peiris, H. V. and Perrotta, F. and Pettorino, V. and Piacentini, F. and Polastri, L. and Polenta, G. and Puget, J.-L. and Rachen, J. P. and Reinecke, M. and Remazeilles, M. and Renzi, A. and Rocha, G. and Rosset, C. and Roudier, G. and Rubiño-Martín, J. A. and Ruiz-Granados, B. and Salvati, L. and Sandri, M. and Savelainen, M. and Scott, D. and Shellard, E. P. S. and Sirignano, C. and Sirri, G. and Spencer, L. D. and Sunyaev, R. and Suur-Uski, A.-S. and Tauber, J. A. and Tavagnacco, D. and Tenti, M. and Toffolatti, L. and Tomasi, M. and Trombetti, T. and Valenziano, L. and Valiviita, J. and Van Tent, B. and Vibert, L. and Vielva, P. and Villa, F. and Vittorio, N. and Wandelt, B. D. and Wehus, I. K. and White, M. and White, S. D. M. and Zacchei, A. and Zonca, A.},
   year={2020},
   month=sep, pages={A6} }

@ARTICLE{desi2016_i,
       author = {{DESI Collaboration} and {Aghamousa}, Amir and {Aguilar}, Jessica and {Ahlen}, Steve and {Alam}, Shadab and {Allen}, Lori E. and {Allende Prieto}, Carlos and {Annis}, James and {Bailey}, Stephen and {Balland}, Christophe and {Ballester}, Otger and {Baltay}, Charles and {Beaufore}, Lucas and {Bebek}, Chris and {Beers}, Timothy C. and {Bell}, Eric F. and {Bernal}, Jos{\'e} Luis and {Besuner}, Robert and {Beutler}, Florian and {Blake}, Chris and {Bleuler}, Hannes and {Blomqvist}, Michael and {Blum}, Robert and {Bolton}, Adam S. and {Briceno}, Cesar and {Brooks}, David and {Brownstein}, Joel R. and {Buckley-Geer}, Elizabeth and {Burden}, Angela and {Burtin}, Etienne and {Busca}, Nicolas G. and {Cahn}, Robert N. and {Cai}, Yan-Chuan and {Cardiel-Sas}, Laia and {Carlberg}, Raymond G. and {Carton}, Pierre-Henri and {Casas}, Ricard and {Castander}, Francisco J. and {Cervantes-Cota}, Jorge L. and {Claybaugh}, Todd M. and {Close}, Madeline and {Coker}, Carl T. and {Cole}, Shaun and {Comparat}, Johan and {Cooper}, Andrew P. and {Cousinou}, M. -C. and {Crocce}, Martin and {Cuby}, Jean-Gabriel and {Cunningham}, Daniel P. and {Davis}, Tamara M. and {Dawson}, Kyle S. and {de la Macorra}, Axel and {De Vicente}, Juan and {Delubac}, Timoth{\'e}e and {Derwent}, Mark and {Dey}, Arjun and {Dhungana}, Govinda and {Ding}, Zhejie and {Doel}, Peter and {Duan}, Yutong T. and {Ealet}, Anne and {Edelstein}, Jerry and {Eftekharzadeh}, Sarah and {Eisenstein}, Daniel J. and {Elliott}, Ann and {Escoffier}, St{\'e}phanie and {Evatt}, Matthew and {Fagrelius}, Parker and {Fan}, Xiaohui and {Fanning}, Kevin and {Farahi}, Arya and {Farihi}, Jay and {Favole}, Ginevra and {Feng}, Yu and {Fernandez}, Enrique and {Findlay}, Joseph R. and {Finkbeiner}, Douglas P. and {Fitzpatrick}, Michael J. and {Flaugher}, Brenna and {Flender}, Samuel and {Font-Ribera}, Andreu and {Forero-Romero}, Jaime E. and {Fosalba}, Pablo and {Frenk}, Carlos S. and {Fumagalli}, Michele and {Gaensicke}, Boris T. and {Gallo}, Giuseppe and {Garcia-Bellido}, Juan and {Gaztanaga}, Enrique and {Pietro Gentile Fusillo}, Nicola and {Gerard}, Terry and {Gershkovich}, Irena and {Giannantonio}, Tommaso and {Gillet}, Denis and {Gonzalez-de-Rivera}, Guillermo and {Gonzalez-Perez}, Violeta and {Gott}, Shelby and {Graur}, Or and {Gutierrez}, Gaston and {Guy}, Julien and {Habib}, Salman and {Heetderks}, Henry and {Heetderks}, Ian and {Heitmann}, Katrin and {Hellwing}, Wojciech A. and {Herrera}, David A. and {Ho}, Shirley and {Holland}, Stephen and {Honscheid}, Klaus and {Huff}, Eric and {Hutchinson}, Timothy A. and {Huterer}, Dragan and {Hwang}, Ho Seong and {Illa Laguna}, Joseph Maria and {Ishikawa}, Yuzo and {Jacobs}, Dianna and {Jeffrey}, Niall and {Jelinsky}, Patrick and {Jennings}, Elise and {Jiang}, Linhua and {Jimenez}, Jorge and {Johnson}, Jennifer and {Joyce}, Richard and {Jullo}, Eric and {Juneau}, St{\'e}phanie and {Kama}, Sami and {Karcher}, Armin and {Karkar}, Sonia and {Kehoe}, Robert and {Kennamer}, Noble and {Kent}, Stephen and {Kilbinger}, Martin and {Kim}, Alex G. and {Kirkby}, David and {Kisner}, Theodore and {Kitanidis}, Ellie and {Kneib}, Jean-Paul and {Koposov}, Sergey and {Kovacs}, Eve and {Koyama}, Kazuya and {Kremin}, Anthony and {Kron}, Richard and {Kronig}, Luzius and {Kueter-Young}, Andrea and {Lacey}, Cedric G. and {Lafever}, Robin and {Lahav}, Ofer and {Lambert}, Andrew and {Lampton}, Michael and {Landriau}, Martin and {Lang}, Dustin and {Lauer}, Tod R. and {Le Goff}, Jean-Marc and {Le Guillou}, Laurent and {Le Van Suu}, Auguste and {Lee}, Jae Hyeon and {Lee}, Su-Jeong and {Leitner}, Daniela and {Lesser}, Michael and {Levi}, Michael E. and {L'Huillier}, Benjamin and {Li}, Baojiu and {Liang}, Ming and {Lin}, Huan and {Linder}, Eric and {Loebman}, Sarah R. and {Luki{\'c}}, Zarija and {Ma}, Jun and {MacCrann}, Niall and {Magneville}, Christophe and {Makarem}, Laleh and {Manera}, Marc and {Manser}, Christopher J. and {Marshall}, Robert and {Martini}, Paul and {Massey}, Richard and {Matheson}, Thomas and {McCauley}, Jeremy and {McDonald}, Patrick and {McGreer}, Ian D. and {Meisner}, Aaron and {Metcalfe}, Nigel and {Miller}, Timothy N. and {Miquel}, Ramon and {Moustakas}, John and {Myers}, Adam and {Naik}, Milind and {Newman}, Jeffrey A. and {Nichol}, Robert C. and {Nicola}, Andrina and {Nicolati da Costa}, Luiz and {Nie}, Jundan and {Niz}, Gustavo and {Norberg}, Peder and {Nord}, Brian and {Norman}, Dara and {Nugent}, Peter and {O'Brien}, Thomas and {Oh}, Minji and {Olsen}, Knut A.~G.},
        title = "{The DESI Experiment Part I: Science,Targeting, and Survey Design}",
         year = 2016,
        month = oct,
          eid = {arXiv:1611.00036},
          eprint={arXiv:1611.00036},
        pages = {arXiv:1611.00036},
          doi = {10.48550/arXiv.1611.00036},
archivePrefix = {arXiv},
 primaryClass = {astro-ph.IM},
       adsurl = {https://ui.adsabs.harvard.edu/abs/2016arXiv161100036D}
}

@ARTICLE{desi2016_ii,
      author = "Aghamousa, Amir and others",
    collaboration = "DESI",
    title = "{The DESI Experiment Part II: Instrument Design}",
    eprint = "arXiv:1611.00037",
    archivePrefix = "arXiv",
    primaryClass = "astro-ph.IM",
    reportNumber = "FERMILAB-PUB-16-518-AE",
    month = "10",
    year = "2016"
}

@ARTICLE{Corrector.Miller.2023,
       author = {{Miller}, Timothy N. and {Doel}, Peter and {Gutierrez}, Gaston and {Besuner}, Robert and {Brooks}, David and {Gallo}, Giuseppe and {Heetderks}, Henry and {Jelinsky}, Patrick and {Kent}, Stephen M. and {Lampton}, Michael and {Levi}, Michael E. and {Liang}, Ming and {Meisner}, Aaron and {Sholl}, Michael J. and {Silber}, Joseph Harry and {Sprayberry}, David and {Aguilar}, Jessica Nicole and {de la Macorra}, Axel and {Eisenstein}, Daniel and {Fanning}, Kevin and {Font-Ribera}, Andreu and {Gazta{\~n}aga}, Enrique and {Gontcho A Gontcho}, Satya and {Honscheid}, Klaus and {Jimenez}, Jorge and {Joyce}, Dick and {Kehoe}, Robert and {Kisner}, Theodore and {Kremin}, Anthony and {Landriau}, Martin and {Le Guillou}, Laurent and {Magneville}, Christophe and {Martini}, Paul and {Miquel}, Ramon and {Moustakas}, John and {Nie}, Jundan and {Percival}, Will and {Poppett}, Claire and {Prada}, Francisco and {Rossi}, Graziano and {Schlegel}, David and {Schubnell}, Michael and {Seo}, Hee-Jong and {Sharples}, Ray and {Tarl{\'e}}, Gregory and {Vargas-Maga{\~n}a}, Mariana and {Zhou}, Zhimin and {the DESI Collaboration}},
        title = "{The Optical Corrector for the Dark Energy Spectroscopic Instrument}",
      journal = {\aj},
         year = 2024,
        month = aug,
       volume = {168},
       number = {2},
          eid = {95},
        pages = {95},
          doi = {10.3847/1538-3881/ad45fe},
archivePrefix = {arXiv},
       eprint = {2306.06310},
 primaryClass = {astro-ph.IM},
       adsurl = {https://ui.adsabs.harvard.edu/abs/2024AJ....168...95M}
}

@ARTICLE{FiberSystem.Poppett.2024,
       author = {{Poppett}, Claire and {Tyas}, Luke and {Aguilar}, J. and {Bebek}, Christopher and {Bramall}, D. and {Claybaugh}, T. and {Edelstein}, J. and {Fagrelius}, P. and {Heetderks}, H. and {Jelinsky}, P. and {Jelinsky}, S. and {Lafever}, Robin and {Lambert}, A. and {Lampton}, M. and {Levi}, Michael E. and {Martini}, P. and {Rockosi}, C. and {Schmoll}, J. and {Sharples}, Ray M. and {Sirk}, Martin and {Wishnow}, Edward and {Yu}, Jiaxi and {Ahlen}, S. and {Bault}, A. and {BenZvi}, S. and {Brooks}, D. and {Cole}, S. and {de la Macorra}, A. and {Dey}, Arjun and {Doel}, P. and {Fanning}, K. and {Font-Ribera}, A. and {Forero-Romero}, J.~E. and {Gazta{\~n}aga}, E. and {Gontcho A Gontcho}, S. and {Gonzalez-Morales}, A.~X. and {Hahn}, C. and {Honscheid}, K. and {Jimenez}, J. and {Juneau}, S. and {Kirkby}, D. and {Kremin}, A. and {Landriau}, M. and {Le Guillou}, L. and {Manera}, M. and {Meisner}, A. and {Miquel}, R. and {Moustakas}, J. and {Mueller}, E. and {Mu{\~n}oz-Guti{\'e}rrez}, A. and {Myers}, A.~D. and {Nie}, J. and {Niz}, G. and {Palanque-Delabrouille}, N. and {Percival}, W.~J. and {Prada}, F. and {Rabinowitz}, D. and {Rezaie}, M. and {Rossi}, G. and {Sanchez}, E. and {Schlafly}, Edward F. and {Schlegel}, D. and {Schubnell}, M. and {Seo}, H. and {Sprayberry}, D. and {Tarl{\'e}}, G. and {Vargas-Maga{\~n}a}, M. and {Weaver}, B.~A. and {Zhou}, R.},
        title = "{Overview of the Fiber System for the Dark Energy Spectroscopic Instrument}",
      journal = {\aj},
         year = 2024,
        month = dec,
       volume = {168},
       number = {6},
          eid = {245},
        pages = {245},
          doi = {10.3847/1538-3881/ad76a4},
       adsurl = {https://ui.adsabs.harvard.edu/abs/2024AJ....168..245P}
}

@ARTICLE{desi2022_instrument,
       author = {{DESI Collaboration} and {Abareshi}, B. and {Aguilar}, J. and {Ahlen}, S. and {Alam}, Shadab and {Alexander}, David M. and {Alfarsy}, R. and {Allen}, L. and {Allende Prieto}, C. and {Alves}, O. and {Ameel}, J. and {Armengaud}, E. and {Asorey}, J. and {Aviles}, Alejandro and {Bailey}, S. and {Balaguera-Antol{\'\i}nez}, A. and {Ballester}, O. and {Baltay}, C. and {Bault}, A. and {Beltran}, S.~F. and {Benavides}, B. and {BenZvi}, S. and {Berti}, A. and {Besuner}, R. and {Beutler}, Florian and {Bianchi}, D. and {Blake}, C. and {Blanc}, P. and {Blum}, R. and {Bolton}, A. and {Bose}, S. and {Bramall}, D. and {Brieden}, S. and {Brodzeller}, A. and {Brooks}, D. and {Brownewell}, C. and {Buckley-Geer}, E. and {Cahn}, R.~N. and {Cai}, Z. and {Canning}, R. and {Capasso}, R. and {Carnero Rosell}, A. and {Carton}, P. and {Casas}, R. and {Castander}, F.~J. and {Cervantes-Cota}, J.~L. and {Chabanier}, S. and {Chaussidon}, E. and {Chuang}, C. and {Circosta}, C. and {Cole}, S. and {Cooper}, A.~P. and {da Costa}, L. and {Cousinou}, M. -C. and {Cuceu}, A. and {Davis}, T.~M. and {Dawson}, K. and {de la Cruz-Noriega}, R. and {de la Macorra}, A. and {de Mattia}, A. and {Della Costa}, J. and {Demmer}, P. and {Derwent}, M. and {Dey}, A. and {Dey}, B. and {Dhungana}, G. and {Ding}, Z. and {Dobson}, C. and {Doel}, P. and {Donald-McCann}, J. and {Donaldson}, J. and {Douglass}, K. and {Duan}, Y. and {Dunlop}, P. and {Edelstein}, J. and {Eftekharzadeh}, S. and {Eisenstein}, D.~J. and {Enriquez-Vargas}, M. and {Escoffier}, S. and {Evatt}, M. and {Fagrelius}, P. and {Fan}, X. and {Fanning}, K. and {Fawcett}, V.~A. and {Ferraro}, S. and {Ereza}, J. and {Flaugher}, B. and {Font-Ribera}, A. and {Forero-Romero}, J.~E. and {Frenk}, C.~S. and {Fromenteau}, S. and {G{\"a}nsicke}, B.~T. and {Garcia-Quintero}, C. and {Garrison}, L. and {Gazta{\~n}aga}, E. and {Gerardi}, F. and {Gil-Mar{\'\i}n}, H. and {Gontcho A Gontcho}, S. and {Gonzalez-Morales}, Alma X. and {Gonzalez-de-Rivera}, G. and {Gonzalez-Perez}, V. and {Gordon}, C. and {Graur}, O. and {Green}, D. and {Grove}, C. and {Gruen}, D. and {Gutierrez}, G. and {Guy}, J. and {Hahn}, C. and {Harris}, S. and {Herrera}, D. and {Herrera-Alcantar}, Hiram K. and {Honscheid}, K. and {Howlett}, C. and {Huterer}, D. and {Ir{\v{s}}i{\v{c}}}, V. and {Ishak}, M. and {Jelinsky}, P. and {Jiang}, L. and {Jimenez}, J. and {Jing}, Y.~P. and {Joyce}, R. and {Jullo}, E. and {Juneau}, S. and {Kara{\c{c}}ayl{\i}}, N.~G. and {Karamanis}, M. and {Karcher}, A. and {Karim}, T. and {Kehoe}, R. and {Kent}, S. and {Kirkby}, D. and {Kisner}, T. and {Kitaura}, F. and {Koposov}, S.~E. and {Kov{\'a}cs}, A. and {Kremin}, A. and {Krolewski}, Alex and {L'Huillier}, B. and {Lahav}, O. and {Lambert}, A. and {Lamman}, C. and {Lan}, Ting-Wen and {Landriau}, M. and {Lane}, S. and {Lang}, D. and {Lange}, J.~U. and {Lasker}, J. and {Le Guillou}, L. and {Leauthaud}, A. and {Le Van Suu}, A. and {Levi}, Michael E. and {Li}, T.~S. and {Magneville}, C. and {Manera}, M. and {Manser}, Christopher J. and {Marshall}, B. and {Martini}, Paul and {McCollam}, W. and {McDonald}, P. and {Meisner}, Aaron M. and {Mena-Fern{\'a}ndez}, J. and {Meneses-Rizo}, J. and {Mezcua}, M. and {Miller}, T. and {Miquel}, R. and {Montero-Camacho}, P. and {Moon}, J. and {Moustakas}, J. and {Mueller}, E. and {Mu{\~n}oz-Guti{\'e}rrez}, Andrea and {Myers}, Adam D. and {Nadathur}, S. and {Najita}, J. and {Napolitano}, L. and {Neilsen}, E. and {Newman}, Jeffrey A. and {Nie}, J.~D. and {Ning}, Y. and {Niz}, G. and {Norberg}, P. and {Noriega}, Hern{\'a}n E. and {O'Brien}, T. and {Obuljen}, A. and {Palanque-Delabrouille}, N. and {Palmese}, A. and {Zhiwei}, P. and {Pappalardo}, D. and {PENG}, X. and {Percival}, W.~J. and {Perruchot}, S. and {Pogge}, R. and {Poppett}, C. and {Porredon}, A. and {Prada}, F. and {Prochaska}, J. and {Pucha}, R. and {P{\'e}rez-Fern{\'a}ndez}, A. and {P{\'e}rez-R{\`a}fols}, I. and {Rabinowitz}, D. and {Raichoor}, A.},
        title = "{Overview of the {Instrumentation} for the Dark Energy Spectroscopic Instrument}",
      journal = {\aj},
         year = 2022,
        month = nov,
       volume = {164},
       number = {5},
          eid = {207},
        pages = {207},
          doi = {10.3847/1538-3881/ac882b},
archivePrefix = {arXiv},
       eprint = {2205.10939},
 primaryClass = {astro-ph.IM},
       adsurl = {https://ui.adsabs.harvard.edu/abs/2022AJ....164..207D}
}

@ARTICLE{desi2024_validation,
       author = {{DESI Collaboration} and {Adame}, A.~G. and {Aguilar}, J. and {Ahlen}, S. and {Alam}, S. and {Aldering}, G. and {Alexander}, D.~M. and {Alfarsy}, R. and {Allende Prieto}, C. and {Alvarez}, M. and {Alves}, O. and {Anand}, A. and {Andrade-Oliveira}, F. and {Armengaud}, E. and {Asorey}, J. and {Avila}, S. and {Aviles}, A. and {Bailey}, S. and {Balaguera-Antol{\'\i}nez}, A. and {Ballester}, O. and {Baltay}, C. and {Bault}, A. and {Bautista}, J. and {Behera}, J. and {Beltran}, S.~F. and {BenZvi}, S. and {Beraldo e Silva}, L. and {Bermejo-Climent}, J.~R. and {Berti}, A. and {Besuner}, R. and {Beutler}, F. and {Bianchi}, D. and {Blake}, C. and {Blum}, R. and {Bolton}, A.~S. and {Brieden}, S. and {Brodzeller}, A. and {Brooks}, D. and {Brown}, Z. and {Buckley-Geer}, E. and {Burtin}, E. and {Cabayol-Garcia}, L. and {Cai}, Z. and {Canning}, R. and {Cardiel-Sas}, L. and {Carnero Rosell}, A. and {Castander}, F.~J. and {Cervantes-Cota}, J.~L. and {Chabanier}, S. and {Chaussidon}, E. and {Chaves-Montero}, J. and {Chen}, S. and {Chen}, X. and {Chuang}, C. and {Claybaugh}, T. and {Cole}, S. and {Cooper}, A.~P. and {Cuceu}, A. and {Davis}, T.~M. and {Dawson}, K. and {de Belsunce}, R. and {de la Cruz}, R. and {de la Macorra}, A. and {de Mattia}, A. and {Demina}, R. and {Demirbozan}, U. and {DeRose}, J. and {Dey}, A. and {Dey}, B. and {Dhungana}, G. and {Ding}, J. and {Ding}, Z. and {Doel}, P. and {Doshi}, R. and {Douglass}, K. and {Edge}, A. and {Eftekharzadeh}, S. and {Eisenstein}, D.~J. and {Elliott}, A. and {Escoffier}, S. and {Fagrelius}, P. and {Fan}, X. and {Fanning}, K. and {Fawcett}, V.~A. and {Ferraro}, S. and {Ereza}, J. and {Flaugher}, B. and {Font-Ribera}, A. and {Forero-S{\'a}nchez}, D. and {Forero-Romero}, J.~E. and {Frenk}, C.~S. and {G{\"a}nsicke}, B.~T. and {Garc{\'\i}a}, L. {\'A}. and {Garc{\'\i}a-Bellido}, J. and {Garcia-Quintero}, C. and {Garrison}, L.~H. and {Gil-Mar{\'\i}n}, H. and {Golden-Marx}, J. and {Gontcho A Gontcho}, S. and {Gonzalez-Morales}, A.~X. and {Gonzalez-Perez}, V. and {Gordon}, C. and {Graur}, O. and {Green}, D. and {Gruen}, D. and {Guy}, J. and {Hadzhiyska}, B. and {Hahn}, C. and {Han}, J.~J. and {Hanif}, M.~M.~S. and {Herrera-Alcantar}, H.~K. and {Honscheid}, K. and {Hou}, J. and {Howlett}, C. and {Huterer}, D. and {Ir{\v{s}}i{\v{c}}}, V. and {Ishak}, M. and {Jana}, A. and {Jiang}, L. and {Jimenez}, J. and {Jing}, Y.~P. and {Joudaki}, S. and {Jullo}, E. and {Joyce}, R. and {Juneau}, S. and {Kizhuprakkat}, N. and {Kara{\c{c}}ayl{\i}}, N.~G. and {Karim}, T. and {Kehoe}, R. and {Kent}, S. and {Khederlarian}, A. and {Kim}, S. and {Kirkby}, D. and {Kisner}, T. and {Kitaura}, F. and {Kneib}, J. and {Koposov}, S.~E. and {Kov{\'a}cs}, A. and {Kremin}, A. and {Krolewski}, A. and {L'Huillier}, B. and {Lahav}, O. and {Lambert}, A. and {Lamman}, C. and {Lan}, T. -W. and {Landriau}, M. and {Lang}, D. and {Lange}, J.~U. and {Lasker}, J. and {Le Guillou}, L. and {Leauthaud}, A. and {Levi}, M.~E. and {Li}, T.~S. and {Linder}, E. and {Lyons}, A. and {Magneville}, C. and {Manera}, M. and {Manser}, C.~J. and {Margala}, D. and {Martini}, P. and {McDonald}, P. and {Medina}, G.~E. and {Medina-Varela}, L. and {Meisner}, A. and {Mena-Fern{\'a}ndez}, J. and {Meneses-Rizo}, J. and {Mezcua}, M. and {Miquel}, R. and {Montero-Camacho}, P. and {Moon}, J. and {Moore}, S. and {Moustakas}, J. and {Mueller}, E. and {Mundet}, J. and {Mu{\~n}oz-Guti{\'e}rrez}, A. and {Myers}, A.~D. and {Nadathur}, S. and {Napolitano}, L. and {Neveux}, R. and {Newman}, J.~A. and {Nie}, J. and {Niz}, G. and {Norberg}, P. and {Noriega}, H.~E. and {Paillas}, E. and {Palanque-Delabrouille}, N. and {Palmese}, A. and {Zhiwei}, P. and {Parkinson}, D. and {Penmetsa}, S. and {Percival}, W.~J. and {P{\'e}rez-Fern{\'a}ndez}, A. and {P{\'e}rez-R{\`a}fols}, I. and {Pieri}, M. and {Poppett}, C. and {Porredon}, A. and {Prada}, F. and {Pucha}, R. and {Raichoor}, A. and {Ram{\'\i}rez-P{\'e}rez}, C.},
        title = "{Validation of the Scientific Program for the Dark Energy Spectroscopic Instrument}",
      journal = {\aj},
         year = 2024,
        month = feb,
       volume = {167},
       number = {2},
          eid = {62},
        pages = {62},
          doi = {10.3847/1538-3881/ad0b08},
archivePrefix = {arXiv},
       eprint = {2306.06307},
 primaryClass = {astro-ph.CO},
       adsurl = {https://ui.adsabs.harvard.edu/abs/2024AJ....167...62D}
}

@ARTICLE{desi2023_target_selection,
       author = {{Myers}, Adam D. and {Moustakas}, John and {Bailey}, Stephen and {Weaver}, Benjamin A. and {Cooper}, Andrew P. and {Forero-Romero}, Jaime E. and {Abolfathi}, Bela and {Alexander}, David M. and {Brooks}, David and {Chaussidon}, Edmond and {Chuang}, Chia-Hsun and {Dawson}, Kyle and {Dey}, Arjun and {Dey}, Biprateep and {Dhungana}, Govinda and {Doel}, Peter and {Fanning}, Kevin and {Gazta{\~n}aga}, Enrique and {Gontcho A Gontcho}, Satya and {Gonzalez-Morales}, Alma X. and {Hahn}, ChangHoon and {Herrera-Alcantar}, Hiram K. and {Honscheid}, Klaus and {Ishak}, Mustapha and {Karim}, Tanveer and {Kirkby}, David and {Kisner}, Theodore and {Koposov}, Sergey E. and {Kremin}, Anthony and {Lan}, Ting-Wen and {Landriau}, Martin and {Lang}, Dustin and {Levi}, Michael E. and {Magneville}, Christophe and {Napolitano}, Lucas and {Martini}, Paul and {Meisner}, Aaron and {Newman}, Jeffrey A. and {Palanque-Delabrouille}, Nathalie and {Percival}, Will and {Poppett}, Claire and {Prada}, Francisco and {Raichoor}, Anand and {Ross}, Ashley J. and {Schlafly}, Edward F. and {Schlegel}, David and {Schubnell}, Michael and {Tan}, Ting and {Tarle}, Gregory and {Wilson}, Michael J. and {Y{\`e}che}, Christophe and {Zhou}, Rongpu and {Zhou}, Zhimin and {Zou}, Hu},
        title = "{The Target-selection Pipeline for the Dark Energy Spectroscopic Instrument}",
      journal = {\aj},
         year = 2023,
        month = feb,
       volume = {165},
       number = {2},
          eid = {50},
        pages = {50},
          doi = {10.3847/1538-3881/aca5f9},
archivePrefix = {arXiv},
       eprint = {2208.08518},
 primaryClass = {astro-ph.IM},
       adsurl = {https://ui.adsabs.harvard.edu/abs/2023AJ....165...50M}
}

@article{edr,
   title="{The Early Data Release of the Dark Energy Spectroscopic Instrument}",
   volume={168},
   ISSN={1538-3881},
   url={http://dx.doi.org/10.3847/1538-3881/ad3217},
   DOI={10.3847/1538-3881/ad3217},
   number={2},
   journal={\aj},
   publisher={American Astronomical Society},
   author={{DESI Collaboration} and Adame, A. G. and Aguilar, J. and Ahlen, S. and Alam, S. and Aldering, G. and Alexander, D. M. and Alfarsy, R. and Allende Prieto, C. and Alvarez, M. and Alves, O. and Anand, A. and Andrade-Oliveira, F. and Armengaud, E. and Asorey, J. and Avila, S. and Aviles, A. and Bailey, S. and Balaguera-Antolínez, A. and Ballester, O. and Baltay, C. and Bault, A. and Bautista, J. and Behera, J. and Beltran, S. F. and BenZvi, S. and Beraldo e Silva, L. and Bermejo-Climent, J. R. and Berti, A. and Besuner, R. and Beutler, F. and Bianchi, D. and Blake, C. and Blum, R. and Bolton, A. S. and Brieden, S. and Brodzeller, A. and Brooks, D. and Brown, Z. and Buckley-Geer, E. and Burtin, E. and Cabayol-Garcia, L. and Cai, Z. and Canning, R. and Cardiel-Sas, L. and Carnero Rosell, A. and Castander, F. J. and Cervantes-Cota, J. L. and Chabanier, S. and Chaussidon, E. and Chaves-Montero, J. and Chen, S. and Chen, X. and Chuang, C. and Claybaugh, T. and Cole, S. and Cooper, A. P. and Cuceu, A. and Davis, T. M. and Dawson, K. and de Belsunce, R. and de la Cruz, R. and de la Macorra, A. and Della Costa, J. and de Mattia, A. and Demina, R. and Demirbozan, U. and DeRose, J. and Dey, A. and Dey, B. and Dhungana, G. and Ding, J. and Ding, Z. and Doel, P. and Doshi, R. and Douglass, K. and Edge, A. and Eftekharzadeh, S. and Eisenstein, D. J. and Elliott, A. and Ereza, J. and Escoffier, S. and Fagrelius, P. and Fan, X. and Fanning, K. and Fawcett, V. A. and Ferraro, S. and Flaugher, B. and Font-Ribera, A. and Forero-Romero, J. E. and Forero-Sánchez, D. and Frenk, C. S. and Gänsicke, B. T. and García, L. Á. and García-Bellido, J. and Garcia-Quintero, C. and Garrison, L. H. and Gil-Marín, H. and Golden-Marx, J. and Gontcho A Gontcho, S. and Gonzalez-Morales, A. X. and Gonzalez-Perez, V. and Gordon, C. and Graur, O. and Green, D. and Gruen, D. and Guy, J. and Hadzhiyska, B. and Hahn, C. and Han, J. J. and Hanif, M. M. S and Herrera-Alcantar, H. K. and Honscheid, K. and Hou, J. and Howlett, C. and Huterer, D. and Iršič, V. and Ishak, M. and Jacques, A. and Jana, A. and Jiang, L. and Jimenez, J. and Jing, Y. P. and Joudaki, S. and Joyce, R. and Jullo, E. and Juneau, S. and Karaçaylı, N. G. and Karim, T. and Kehoe, R. and Kent, S. and Khederlarian, A. and Kim, S. and Kirkby, D. and Kisner, T. and Kitaura, F. and Kizhuprakkat, N. and Kneib, J. and Koposov, S. E. and Kovács, A. and Kremin, A. and Krolewski, A. and L’Huillier, B. and Lahav, O. and Lambert, A. and Lamman, C. and Lan, T.-W. and Landriau, M. and Lang, D. and Lange, J. U. and Lasker, J. and Leauthaud, A. and Le Guillou, L. and Levi, M. E. and Li, T. S. and Linder, E. and Lyons, A. and Magneville, C. and Manera, M. and Manser, C. J. and Margala, D. and Martini, P. and McDonald, P. and Medina, G. E. and Medina-Varela, L. and Meisner, A. and Mena-Fernández, J. and Meneses-Rizo, J. and Mezcua, M. and Miquel, R. and Montero-Camacho, P. and Moon, J. and Moore, S. and Moustakas, J. and Mueller, E. and Mundet, J. and Muñoz-Gutiérrez, A. and Myers, A. D. and Nadathur, S. and Napolitano, L. and Neveux, R. and Newman, J. A. and Nie, J. and Nikutta, R. and Niz, G. and Norberg, P. and Noriega, H. E. and Paillas, E. and Palanque-Delabrouille, N. and Palmese, A. and Pan, Z. and Parkinson, D. and Penmetsa, S. and Percival, W. J. and Pérez-Fernández, A. and Pérez-Ràfols, I. and Pieri, M. and Poppett, C. and Porredon, A. and Pothier, S. and Prada, F. and Pucha, R. and Raichoor, A. and Ramírez-Pérez, C. and Ramirez-Solano, S. and Rashkovetskyi, M. and Ravoux, C. and Rocher, A. and Rockosi, C. and Ross, A. J. and Rossi, G. and Ruggeri, R. and Ruhlmann-Kleider, V. and Sabiu, C. G. and Said, K. and Saintonge, A. and Samushia, L. and Sanchez, E. and Saulder, C. and Schaan, E. and Schlafly, E. F. and Schlegel, D. and Scholte, D. and Schubnell, M. and Seo, H. and Shafieloo, A. and Sharples, R. and Sheu, W. and Silber, J. and Sinigaglia, F. and Siudek, M. and Slepian, Z. and Smith, A. and Soumagnac, M. T. and Sprayberry, D. and Stephey, L. and Suárez-Pérez, J. and Sun, Z. and Tan, T. and Tarlé, G. and Tojeiro, R. and Ureña-López, L. A. and Vaisakh, R. and Valcin, D. and Valdes, F. and Valluri, M. and Vargas-Magaña, M. and Variu, A. and Verde, L. and Walther, M. and Wang, B. and Wang, M. S. and Weaver, B. A. and Weaverdyck, N. and Wechsler, R. H. and White, M. and Xie, Y. and Yang, J. and Yèche, C. and Yu, J. and Yuan, S. and Zhang, H. and Zhang, Z. and Zhao, C. and Zheng, Z. and Zhou, R. and Zhou, Z. and Zou, H. and Zou, S. and Zu, Y.},
   year={2024},
   month=jul, pages={58} }

@ARTICLE{desi2023_spectra_processing,
       author = {{Guy}, J. and {Bailey}, S. and {Kremin}, A. and {Alam}, Shadab and {Alexander}, D.~M. and {Allende Prieto}, C. and {BenZvi}, S. and {Bolton}, A.~S. and {Brooks}, D. and {Chaussidon}, E. and {Cooper}, A.~P. and {Dawson}, K. and {de la Macorra}, A. and {Dey}, A. and {Dey}, Biprateep and {Dhungana}, G. and {Eisenstein}, D.~J. and {Font-Ribera}, A. and {Forero-Romero}, J.~E. and {Gazta{\~n}aga}, E. and {Gontcho A Gontcho}, S. and {Green}, D. and {Honscheid}, K. and {Ishak}, M. and {Kehoe}, R. and {Kirkby}, D. and {Kisner}, T. and {Koposov}, Sergey E. and {Lan}, Ting-Wen and {Landriau}, M. and {Le Guillou}, L. and {Levi}, Michael E. and {Magneville}, C. and {Manser}, Christopher J. and {Martini}, P. and {Meisner}, Aaron M. and {Miquel}, R. and {Moustakas}, J. and {Myers}, Adam D. and {Newman}, Jeffrey A. and {Nie}, Jundan and {Palanque-Delabrouille}, N. and {Percival}, W.~J. and {Poppett}, C. and {Prada}, F. and {Raichoor}, A. and {Ravoux}, C. and {Ross}, A.~J. and {Schlafly}, E.~F. and {Schlegel}, D. and {Schubnell}, M. and {Sharples}, Ray M. and {Tarl{\'e}}, Gregory and {Weaver}, B.~A. and {Y{\'e}che}, Christophe and {Zhou}, Rongpu and {Zhou}, Zhimin and {Zou}, H.},
        title = "{The Spectroscopic Data Processing Pipeline for the Dark Energy Spectroscopic Instrument}",
      journal = {\aj},
         year = 2023,
        month = apr,
       volume = {165},
       number = {4},
          eid = {144},
        pages = {144},
          doi = {10.3847/1538-3881/acb212},
archivePrefix = {arXiv},
       eprint = {2209.14482},
 primaryClass = {astro-ph.IM},
       adsurl = {https://ui.adsabs.harvard.edu/abs/2023AJ....165..144G}
}

@ARTICLE{SurveyOps.Schlafly.2023,
       author = {{Schlafly}, Edward F. and {Kirkby}, David and {Schlegel}, David J. and {Myers}, Adam D. and {Raichoor}, Anand and {Dawson}, Kyle and {Aguilar}, Jessica and {Allende Prieto}, Carlos and {Bailey}, Stephen and {BenZvi}, Segev and {Bermejo-Climent}, Jose and {Brooks}, David and {de la Macorra}, Axel and {Dey}, Arjun and {Doel}, Peter and {Fanning}, Kevin and {Font-Ribera}, Andreu and {Forero-Romero}, Jaime E. and {Garc{\'\i}a-Bellido}, Juan and {Gontcho A Gontcho}, Satya and {Guy}, Julien and {Hahn}, ChangHoon and {Honscheid}, Klaus and {Ishak}, Mustapha and {Juneau}, St{\'e}phanie and {Kehoe}, Robert and {Kisner}, Theodore and {Kremin}, Anthony and {Landriau}, Martin and {Lang}, Dustin A. and {Lasker}, James and {Levi}, Michael E. and {Magneville}, Christophe and {Manser}, Christopher J. and {Martini}, Paul and {Meisner}, Aaron M. and {Miquel}, Ramon and {Moustakas}, John and {Newman}, Jeffrey A. and {Nie}, Jundan and {Palanque-Delabrouille}, Nathalie. and {Percival}, Will J. and {Poppett}, Claire and {Rockosi}, Constance and {Ross}, Ashley J. and {Rossi}, Graziano and {Tarl{\'e}}, Gregory and {Weaver}, Benjamin A. and {Y{\`e}che}, Christophe and {Zhou}, Rongpu and {DESI Collaboration}},
        title = "{Survey Operations for the Dark Energy Spectroscopic Instrument}",
      journal = {\aj},
         year = 2023,
        month = dec,
       volume = {166},
       number = {6},
          eid = {259},
        pages = {259},
          doi = {10.3847/1538-3881/ad0832},
archivePrefix = {arXiv},
       eprint = {2306.06309},
 primaryClass = {astro-ph.CO},
       adsurl = {https://ui.adsabs.harvard.edu/abs/2023AJ....166..259S}
}

@article{Hahn_2023_bgs,
   title="{The DESI Bright Galaxy Survey: Final Target Selection, Design, and Validation}",
   volume={165},
   ISSN={1538-3881},
   url={http://dx.doi.org/10.3847/1538-3881/accff8},
   DOI={10.3847/1538-3881/accff8},
   number={6},
   journal={\aj},
   publisher={American Astronomical Society},
   author={Hahn, ChangHoon and Wilson, Michael J. and Ruiz-Macias, Omar and Cole, Shaun and Weinberg, David H. and Moustakas, John and Kremin, Anthony and Tinker, Jeremy L. and Smith, Alex and Wechsler, Risa H. and Ahlen, Steven and Alam, Shadab and Bailey, Stephen and Brooks, David and Cooper, Andrew P. and Davis, Tamara M. and Dawson, Kyle and Dey, Arjun and Dey, Biprateep and Eftekharzadeh, Sarah and Eisenstein, Daniel J. and Fanning, Kevin and Forero-Romero, Jaime E. and Frenk, Carlos S. and Gaztañaga, Enrique and A Gontcho, Satya Gontcho and Guy, Julien and Honscheid, Klaus and Ishak, Mustapha and Juneau, Stéphanie and Kehoe, Robert and Kisner, Theodore and Lan, Ting-Wen and Landriau, Martin and Le Guillou, Laurent and Levi, Michael E. and Magneville, Christophe and Martini, Paul and Meisner, Aaron and Myers, Adam D. and Nie, Jundan and Norberg, Peder and Palanque-Delabrouille, Nathalie and Percival, Will J. and Poppett, Claire and Prada, Francisco and Raichoor, Anand and Ross, Ashley J. and Gaines, Sasha and Saulder, Christoph and Schlafly, Eddie and Schlegel, David and Sierra-Porta, David and Tarle, Gregory and Weaver, Benjamin A. and Yèche, Christophe and Zarrouk, Pauline and Zhou, Rongpu and Zhou, Zhimin and Zou, Hu},
   year={2023},
   month=may, pages={253} }

@ARTICLE{Raichoor_2023_elg,
       author = {{Raichoor}, A. and {Moustakas}, J. and {Newman}, Jeffrey A. and {Karim}, T. and {Ahlen}, S. and {Alam}, Shadab and {Bailey}, S. and {Brooks}, D. and {Dawson}, K. and {de la Macorra}, A. and {de Mattia}, A. and {Dey}, A. and {Dey}, Biprateep and {Dhungana}, G. and {Eftekharzadeh}, S. and {Eisenstein}, D.~J. and {Fanning}, K. and {Font-Ribera}, A. and {Garc{\'\i}a-Bellido}, J. and {Gazta{\~n}aga}, E. and {A Gontcho}, S. Gontcho and {Guy}, J. and {Honscheid}, K. and {Ishak}, M. and {Kehoe}, R. and {Kisner}, T. and {Kremin}, Anthony and {Lan}, Ting-Wen and {Landriau}, M. and {Le Guillou}, L. and {Levi}, Michael E. and {Magneville}, C. and {Manera}, M. and {Martini}, P. and {Meisner}, Aaron M. and {Myers}, Adam D. and {Nie}, Jundan and {Palanque-Delabrouille}, N. and {Percival}, W.~J. and {Poppett}, C. and {Prada}, F. and {Ross}, A.~J. and {Ruhlmann-Kleider}, V. and {Sabiu}, C.~G. and {Schlafly}, E.~F. and {Schlegel}, D. and {Tarl{\'e}}, Gregory and {Weaver}, B.~A. and {Y{\`e}che}, Christophe and {Zhou}, Rongpu and {Zhou}, Zhimin and {Zou}, H.},
        title = "{Target Selection and Validation of DESI Emission Line Galaxies}",
      journal = {\aj},
         year = 2023,
        month = mar,
       volume = {165},
       number = {3},
          eid = {126},
        pages = {126},
          doi = {10.3847/1538-3881/acb213},
archivePrefix = {arXiv},
       eprint = {2208.08513},
 primaryClass = {astro-ph.CO},
       adsurl = {https://ui.adsabs.harvard.edu/abs/2023AJ....165..126R}
}

@article{Chaussidon_2023_qso,
   title="{Target Selection and Validation of DESI Quasars}",
   volume={944},
   ISSN={1538-4357},
   url={http://dx.doi.org/10.3847/1538-4357/acb3c2},
   DOI={10.3847/1538-4357/acb3c2},
   number={1},
   journal={\apj},
   publisher={American Astronomical Society},
   author={Chaussidon, Edmond and Yèche, Christophe and Palanque-Delabrouille, Nathalie and Alexander, David M. and Yang, Jinyi and Ahlen, Steven and Bailey, Stephen and Brooks, David and Cai, Zheng and Chabanier, Solène and Davis, Tamara M. and Dawson, Kyle and de laMacorra, Axel and Dey, Arjun and Dey, Biprateep and Eftekharzadeh, Sarah and Eisenstein, Daniel J. and Fanning, Kevin and Font-Ribera, Andreu and Gaztañaga, Enrique and A Gontcho, Satya Gontcho and Gonzalez-Morales, Alma X. and Guy, Julien and Herrera-Alcantar, Hiram K. and Honscheid, Klaus and Ishak, Mustapha and Jiang, Linhua and Juneau, Stephanie and Kehoe, Robert and Kisner, Theodore and Kovács, Andras and Kremin, Anthony and Lan, Ting-Wen and Landriau, Martin and Le Guillou, Laurent and Levi, Michael E. and Magneville, Christophe and Martini, Paul and Meisner, Aaron M. and Moustakas, John and Muñoz-Gutiérrez, Andrea and Myers, Adam D. and Newman, Jeffrey A. and Nie, Jundan and Percival, Will J. and Poppett, Claire and Prada, Francisco and Raichoor, Anand and Ravoux, Corentin and Ross, Ashley J. and Schlafly, Edward and Schlegel, David and Tan, Ting and Tarlé, Gregory and Zhou, Rongpu and Zhou, Zhimin and Zou, Hu},
   year={2023},
   month=feb, pages={107} }

@article{Zhou_2023_lrg,
   title="{Target Selection and Validation of DESI Luminous Red Galaxies}",
   volume={165},
   ISSN={1538-3881},
   url={http://dx.doi.org/10.3847/1538-3881/aca5fb},
   DOI={10.3847/1538-3881/aca5fb},
   number={2},
   journal={\aj},
   publisher={American Astronomical Society},
   author={Zhou, Rongpu and Dey, Biprateep and Newman, Jeffrey A. and Eisenstein, Daniel J. and Dawson, K. and Bailey, S. and Berti, A. and Guy, J. and Lan, Ting-Wen and Zou, H. and Aguilar, J. and Ahlen, S. and Alam, Shadab and Brooks, D. and de la Macorra, A. and Dey, A. and Dhungana, G. and Fanning, K. and Font-Ribera, A. and Gontcho, S. Gontcho A. and Honscheid, K. and Ishak, Mustapha and Kisner, T. and Kovács, A. and Kremin, A. and Landriau, M. and Levi, Michael E. and Magneville, C. and Manera, Marc and Martini, P. and Meisner, Aaron M. and Miquel, R. and Moustakas, J. and Myers, Adam D. and Nie, Jundan and Palanque-Delabrouille, N. and Percival, W. J. and Poppett, C. and Prada, F. and Raichoor, A. and Ross, A. J. and Schlafly, E. and Schlegel, D. and Schubnell, M. and Tarlé, Gregory and Weaver, B. A. and Wechsler, R. H. and Yéche, Christophe and Zhou, Zhimin},
   year={2023},
   month=jan, pages={58} }

@ARTICLE{DESI_DR1,
       author = {{DESI Collaboration} and {Abdul-Karim}, M. and {Adame}, A.~G. and {Aguado}, D. and {Aguilar}, J. and {Ahlen}, S. and {Alam}, S. and {Aldering}, G. and {Alexander}, D.~M. and {Alfarsy}, R. and {Allen}, L. and {Allende Prieto}, C. and {Alves}, O. and {Anand}, A. and {Andrade}, U. and {Armengaud}, E. and {Avila}, S. and {Aviles}, A. and {Awan}, H. and {Bailey}, S. and {Baleato Lizancos}, A. and {Ballester}, O. and {Bault}, A. and {Bautista}, J. and {BenZvi}, S. and {Beraldo e Silva}, L. and {Bermejo-Climent}, J.~R. and {Beutler}, F. and {Bianchi}, D. and {Blake}, C. and {Blum}, R. and {Bolton}, A.~S. and {Bonici}, M. and {Brieden}, S. and {Brodzeller}, A. and {Brooks}, D. and {Buckley-Geer}, E. and {Burtin}, E. and {Canning}, R. and {Carnero Rosell}, A. and {Carr}, A. and {Carrilho}, P. and {Casas}, L. and {Castander}, F.~J. and {Cereskaite}, R. and {Cervantes-Cota}, J.~L. and {Chaussidon}, E. and {Chaves-Montero}, J. and {Chen}, S. and {Chen}, X. and {Claybaugh}, T. and {Cole}, S. and {Cooper}, A.~P. and {Cousinou}, M. -C. and {Cuceu}, A. and {Davis}, T.~M. and {Dawson}, K.~S. and {de Belsunce}, R. and {de la Cruz}, R. and {de la Macorra}, A. and {de Mattia}, A. and {Deiosso}, N. and {Della Costa}, J. and {Demina}, R. and {Demirbozan}, U. and {DeRose}, J. and {Dey}, A. and {Dey}, B. and {Ding}, J. and {Ding}, Z. and {Doel}, P. and {Douglass}, K. and {Dowicz}, M. and {Ebina}, H. and {Edelstein}, J. and {Eisenstein}, D.~J. and {Elbers}, W. and {Emas}, N. and {Escoffier}, S. and {Fagrelius}, P. and {Fan}, X. and {Fanning}, K. and {Fawcett}, V.~A. and {Fern\textbackslash'andez-Garc\textbackslash'ia}, E. and {Ferraro}, S. and {Findlay}, N. and {Font-Ribera}, A. and {Forero-Romero}, J.~E. and {Forero-S\textbackslash'anchez}, D. and {Frenk}, C.~S. and {G\textbackslash''ansicke}, B.~T. and {Galbany}, L. and {Garc\textbackslash'ia-Bellido}, J. and {Garcia-Quintero}, C. and {Garrison}, L.~H. and {Gazta\textbackslash\raisebox{-0.5ex}\textasciitildenaga}, E. and {Gil-Mar\textbackslash'in}, H. and {Gnedin}, O.~Y. and {Gontcho}, S. Gontcho A and {Gonzalez-Morales}, A.~X. and {Gonzalez-Perez}, V. and {Gordon}, C. and {Graur}, O. and {Green}, D. and {Gruen}, D. and {Gsponer}, R. and {Guandalin}, C. and {Gutierrez}, G. and {Guy}, J. and {Hahn}, C. and {Han}, J.~J. and {Han}, J. and {He}, S. and {Herrera-Alcantar}, H.~K. and {Honscheid}, K. and {Hou}, J. and {Howlett}, C. and {Huterer}, D. and {Ir\textbackslashv\{s\}i\textbackslashv\{c\}}, V. and {Ishak}, M. and {Jacques}, A. and {Jimenez}, J. and {Jing}, Y.~P. and {Joachimi}, B. and {Joudaki}, S. and {Joyce}, R. and {Jullo}, E. and {Juneau}, S. and {Kara\textbackslashc\{c\}ayl\{\textbackslashi\}}, N.~G. and {Karim}, T. and {Kehoe}, R. and {Kent}, S. and {Khederlarian}, A. and {Kirkby}, D. and {Kisner}, T. and {Kitaura}, F. -S. and {Kizhuprakkat}, N. and {Kong}, H. and {Koposov}, S.~E. and {Kremin}, A. and {Krolewski}, A. and {Lahav}, O. and {Lai}, Y. and {Lamman}, C. and {Lan}, T. -W. and {Landriau}, M. and {Lang}, D. and {Lange}, J.~U. and {Lasker}, J. and {Le Goff}, J.~M. and {Le Guillou}, L. and {Leauthaud}, A. and {Levi}, M.~E. and {Li}, S. and {Li}, T.~S. and {Lodha}, K. and {Lokken}, M. and {Luo}, Y. and {Magneville}, C. and {Manera}, M. and {Manser}, C.~J. and {Margala}, D. and {Martini}, P. and {Maus}, M. and {McCullough}, J. and {McDonald}, P. and {Medina}, G.~E. and {Medina-Varela}, L. and {Meisner}, A. and {Mena-Fern\textbackslash'andez}, J. and {Menegas}, A. and {Mezcua}, M. and {Miquel}, R. and {Montero-Camacho}, P. and {Moon}, J. and {Moustakas}, J. and {Mu\textbackslash\raisebox{-0.5ex}\textasciitildenoz-Guti\textbackslash'errez}, A. and {Mu\textbackslash\raisebox{-0.5ex}\textasciitildenoz-Santos}, D. and {Myers}, A.~D. and {Myles}, J. and {Nadathur}, S. and {Najita}, J. and {Napolitano}, L. and {Newman}, J.~A. and {Nikakhtar}, F. and {Nikutta}, R. and {Niz}, G. and {Noriega}, H.~E. and {Padmanabhan}, N. and {Paillas}, E. and {Palanque-Delabrouille}, N. and {Palmese}, A. and {Pan}, J. and {Pan}, Z. and {Parkinson}, D. and {Peacock}, J. and {Percival}, W.~J. and {P\textbackslash'erez-Fern\textbackslash'andez}, A. and {P\textbackslash'erez-R\textbackslash`afols}, I. and {Peterson}, P.},
        title = "{Data Release 1 of the Dark Energy Spectroscopic Instrument}",
        doi = {10.3847/1538-3881/ae4c43},
        url = {https://doi.org/10.3847/1538-3881/ae4c43},
        year = {2026},
        month = {apr},
        publisher = {The American Astronomical Society},
        volume = {171},
        number = {5},
        pages = {285},
        journal={\aj}
}

@article{desi_dr2_validation,
      title="{Validation of the DESI DR2 Measurements of Baryon Acoustic Oscillations from Galaxies and Quasars}", 
      author={U. Andrade and E. Paillas and J. Mena-Fernández and Q. Li and A. J. Ross and S. Nadathur and M. Rashkovetskyi and A. Pérez-Fernández and H. Seo and N. Sanders and O. Alves and X. Chen and N. Deiosso and A. de Mattia and M. White and M. Abdul-Karim and S. Ahlen and E. Armengaud and A. Aviles and D. Bianchi and S. Brieden and A. Brodzeller and D. Brooks and E. Burtin and R. Calderon and R. Canning and A. Carnero Rosell and L. Casas and F. J. Castander and M. Charles and E. Chaussidon and J. Chaves-Montero and T. Claybaugh and S. Cole and A. Cuceu and K. S. Dawson and A. de la Macorra and J. Della Costa and A. Dey and B. Dey and Z. Ding and P. Doel and D. J. Eisenstein and W. Elbers and E. Fernández-García and S. Ferraro and A. Font-Ribera and J. E. Forero-Romero and C. Garcia-Quintero and L. H. Garrison and E. Gaztañaga and H. Gil-Marín and S. Gontcho A Gontcho and A. X. Gonzalez-Morales and C. Gordon and G. Gutierrez and J. Guy and C. Hahn and S. He and H. K. Herrera-Alcantar and K. Honscheid and C. Howlett and D. Huterer and M. Ishak and S. Juneau and R. Kehoe and D. Kirkby and T. Kisner and A. Kremin and O. Lahav and C. Lamman and M. Landriau and L. Le Guillou and A. Leauthaud and M. E. Levi and C. Magneville and M. Manera and P. Martini and W. Matthewson and A. Meisner and R. Miquel and J. Moustakas and A. Muñoz-Gutiérrez and D. Muñoz-Santos and A. D. Myers and L. Napolitano and J. A. Newman and H. E. Noriega and N. Palanque-Delabrouille and J. Pan and W. J. Percival and I. Pérez-Ràfols and C. Poppett and F. Prada and A. Raichoor and C. Ramírez-Pérez and C. Ravoux and G. Rossi and R. Ruggeri and L. Samushia and E. Sanchez and D. Schlegel and M. Schubnell and F. Sinigaglia and D. Sprayberry and T. Tan and G. Tarlé and P. Taylor and W. Turner and R. Vaisakh and M. Vargas-Magaña and M. Walther and B. A. Weaver and M. Wolfson and J. Yu and C. Yèche and P. Zarrouk and R. Zhou and H. Zou},
      year={2025},
      eprint={arXiv:2503.14742},
      archivePrefix={arXiv},
      primaryClass={astro-ph.CO},
      url={https://arxiv.org/abs/2503.14742}, 
}

@article{desi_dr2_extendedDE,
      title="{Extended Dark Energy analysis using DESI DR2 BAO measurements}", 
      author = {Lodha, K. and Calderon, R. and Matthewson, W. L. and Shafieloo, A. and Ishak, M. and Pan, J. and Garcia-Quintero, C. and Huterer, D. and Valogiannis, G. and Ure\~na-L\'opez, L. A. and Kamble, N. V. and Parkinson, D. and Kim, A. G. and Zhao, G. B. and Cervantes-Cota, J. L. and Rohlf, J. and Lozano-Rodr\'{\i}guez, F. and Rom\'an-Herrera, J. O. and Abdul-Karim, M. and Aguilar, J. and Ahlen, S. and Alves, O. and Andrade, U. and Armengaud, E. and Aviles, A. and Behera, J. and BenZvi, S. and Bianchi, D. and Brodzeller, A. and Brooks, D. and Burtin, E. and Canning, R. and Rosell, A. Carnero and Casas, L. and Castander, F. J. and Charles, M. and Chaussidon, E. and Chaves-Montero, J. and Chebat, D. and Claybaugh, T. and Cole, S. and Cuceu, A. and Dawson, K. S. and de la Macorra, A. and de Mattia, A. and Deiosso, N. and Demina, R. and Dey, Arjun and Dey, Biprateep and Ding, Z. and Doel, P. and Eisenstein, D. J. and Elbers, W. and Ferraro, S. and Font-Ribera, A. and Forero-Romero, J. E. and Garrison, Lehman H. and Gazta\~naga, E. and Gil-Mar\'{\i}n, H. and Gontcho, S. Gontcho A. and Gonzalez-Morales, A. X. and Gutierrez, G. and Guy, J. and Hahn, C. and Herbold, M. and Herrera-Alcantar, H. K. and Honscheid, K. and Howlett, C. and Juneau, S. and Kehoe, R. and Kirkby, D. and Kisner, T. and Kremin, A. and Lahav, O. and Lamman, C. and Landriau, M. and Le Guillou, L. and Leauthaud, A. and Levi, M. E. and Li, Q. and Magneville, C. and Manera, M. and Martini, P. and Meisner, A. and Mena-Fern\'andez, J. and Miquel, R. and Moustakas, J. and Santos, D. Mu\~noz and Mu\~noz-Guti\'errez, A. and Myers, A. D. and Nadathur, S. and Niz, G. and Noriega, H. E. and Paillas, E. and Palanque-Delabrouille, N. and Percival, W. J. and Pieri, Matthew M. and Poppett, C. and Prada, F. and P\'erez-Fern\'andez, A. and P\'erez-R\`afols, I. and Ram\'{\i}rez-P\'erez, C. and Rashkovetskyi, M. and Ravoux, C. and Ross, A. J. and Rossi, G. and Ruhlmann-Kleider, V. and Samushia, L. and Sanchez, E. and Schlegel, D. and Schubnell, M. and Seo, H. and Sinigaglia, F. and Sprayberry, D. and Tan, T. and Tarl\'e, G. and Taylor, P. and Turner, W. and Vargas-Maga\~na, M. and Walther, M. and Weaver, B. A. and Wolfson, M. and Y\`eche, C. and Zarrouk, P. and Zhou, R. and Zou, H.},
  journal = {Phys. Rev. D},
  volume = {112},
  issue = {8},
  pages = {083511},
  numpages = {27},
  year = {2025},
  month = {Oct},
  publisher = {American Physical Society},
  doi = {10.1103/w4c6-1r5j},
  url = {https://link.aps.org/doi/10.1103/w4c6-1r5j}
}

@article{Anand_2024,
   title="{Archetype-based Redshift Estimation for the Dark Energy Spectroscopic Instrument Survey}",
   volume={168},
   ISSN={1538-3881},
   url={http://dx.doi.org/10.3847/1538-3881/ad60c2},
   DOI={10.3847/1538-3881/ad60c2},
   number={3},
   journal={\aj},
   publisher={American Astronomical Society},
   author={Anand, Abhijeet and Guy, Julien and Bailey, Stephen and Moustakas, John and Aguilar, J. and Ahlen, S. and Bolton, A. S. and Brodzeller, A. and Brooks, D. and Claybaugh, T. and Cole, S. and de la Macorra, A. and Dey, Biprateep and Fanning, K. and Forero-Romero, J. E. and Gaztañaga, E. and Gontcho A Gontcho, S. and Gutierrez, G. and Honscheid, K. and Howlett, C. and Juneau, S. and Kirkby, D. and Kisner, T. and Kremin, A. and Lambert, A. and Landriau, M. and Le Guillou, L. and Manera, M. and Meisner, A. and Miquel, R. and Mueller, E. and Niz, G. and Palanque-Delabrouille, N. and Percival, W. J. and Poppett, C. and Prada, F. and Raichoor, A. and Rezaie, M. and Rossi, G. and Sanchez, E. and Schlafly, E. F. and Schlegel, D. and Schubnell, M. and Sprayberry, D. and Tarlé, G. and Warner, C. and Weaver, B. A. and Zhou, R. and Zou, H.},
   year={2024},
   month=aug, pages={124} }

@article{EZmock,
   title="{EZmocks: extending the Zel’dovich approximation to generate mock galaxy catalogues with accurate clustering statistics}",
   volume={446},
   ISSN={0035-8711},
   url={http://dx.doi.org/10.1093/mnras/stu2301},
   DOI={10.1093/mnras/stu2301},
   number={3},
   journal={\mnras},
   publisher={Oxford University Press (OUP)},
   author={Chuang, Chia-Hsun and Kitaura, Francisco-Shu and Prada, Francisco and Zhao, Cheng and Yepes, Gustavo},
   year={2014},
   month=nov, pages={2621–2628} }

@article{abacus,
    author = {Garrison, Lehman H and Eisenstein, Daniel J and Ferrer, Douglas and Maksimova, Nina A and Pinto, Philip A},
    title = "{The ABACUS cosmological N-body code}",
    journal = {\mnras},
    volume = {508},
    number = {1},
    pages = {575-596},
    year = {2021},
    month = {09},
    issn = {0035-8711},
    doi = {10.1093/mnras/stab2482},
    url = {https://doi.org/10.1093/mnras/stab2482},
    
}

@article{abacussummit ,
    author = {Maksimova, Nina A and Garrison, Lehman H and Eisenstein, Daniel J and Hadzhiyska, Boryana and Bose, Sownak and Satterthwaite, Thomas P},
    title = "{AbacusSummit: a massive set of high-accuracy, high-resolution N-body simulations}",
    journal = {\mnras},
    volume = {508},
    number = {3},
    pages = {4017-4037},
    year = {2021},
    month = {09},
    issn = {0035-8711},
    doi = {10.1093/mnras/stab2484},
    url = {https://doi.org/10.1093/mnras/stab2484},

}

@ARTICLE{desi2024_ii,
       doi = {10.1088/1475-7516/2025/07/017},
url = {https://doi.org/10.1088/1475-7516/2025/07/017},
year = {2025},
month = {jul},
publisher = {IOP Publishing},
volume = {2025},
number = {07},
pages = {017},
author = {{DESI Collaboration} and Adame, A.G. and Aguilar, J. and Ahlen, S. and Alam, S. and Alexander, D.M. and Alvarez, M. and Alves, O. and Anand, A. and Andrade, U. and Armengaud, E. and Avila, S. and Aviles, A. and Awan, H. and Bailey, S. and Baltay, C. and Bault, A. and Behera, J. and BenZvi, S. and Beutler, F. and Bianchi, D. and Blake, C. and Blum, R. and Brieden, S. and Brodzeller, A. and Brooks, D. and Brown, Z. and Buckley-Geer, E. and Burtin, E. and Calderon, R. and Canning, R. and Carnero Rosell, A. and Cereskaite, R. and Cervantes-Cota, J.L. and Chabanier, S. and Chaussidon, E. and Chaves-Montero, J. and Chen, S. and Chen, X. and Claybaugh, T. and Cole, S. and Cuceu, A. and Davis, T.M. and Dawson, K. and de la Macorra, A. and de Mattia, A. and Deiosso, N. and Demina, R. and Dey, A. and Dey, B. and Ding, Z. and Doel, P. and Edelstein, J. and Eftekharzadeh, S. and Eisenstein, D.J. and Elliott, A. and Fagrelius, P. and Fanning, K. and Ferraro, S. and Ereza, J. and Findlay, N. and Flaugher, B. and Font-Ribera, A. and Forero-Sánchez, D. and Forero-Romero, J.E. and Frenk, C.S. and Garcia-Quintero, C. and Gaztañaga, E. and Gil-Marín, H. and Gontcho, S.Gontcho A. and Gonzalez-Morales, A.X. and Gonzalez-Perez, V. and Gordon, C. and Green, D. and Gruen, D. and Gsponer, R. and Gutierrez, G. and Guy, J. and Hadzhiyska, B. and Hahn, C. and Hanif, M.M.S. and Herrera-Alcantar, H.K. and Honscheid, K. and Hou, J. and Howlett, C. and Huterer, D. and Iršič, V. and Ishak, M. and Juneau, S. and Karaçaylı, N.G. and Kehoe, R. and Kent, S. and Kirkby, D. and Kitaura, F.-S. and Kong, H. and Kremin, A. and Krolewski, A. and Lai, Y. and Lan, T.-W. and Landriau, M. and Lang, D. and Lasker, J. and Le Goff, J.M. and Le Guillou, L. and Leauthaud, A. and Levi, M.E. and Li, T.S. and Lodha, K. and Magneville, C. and Manera, M. and Margala, D. and Martini, P. and Maus, M. and McDonald, P. and Medina-Varela, L. and Meisner, A. and Mena-Fernández, J. and Miquel, R. and Moon, J. and Moore, S. and Moustakas, J. and Mudur, N. and Mueller, E. and Muñoz-Gutiérrez, A. and Myers, A.D. and Nadathur, S. and Napolitano, L. and Neveux, R. and Newman, J.A. and Nguyen, N.M. and Nie, J. and Niz, G. and Noriega, H.E. and Padmanabhan, N. and Paillas, E. and Palanque-Delabrouille, N. and Pan, J. and Penmetsa, S. and Percival, W.J. and Pieri, M.M. and Pinon, M. and Poppett, C. and Porredon, A. and Prada, F. and Pérez-Fernández, A. and Pérez-Ràfols, I. and Rabinowitz, D. and Raichoor, A. and Ramírez-Pérez, C. and Ramirez-Solano, S. and Rashkovetskyi, M. and Ravoux, C. and Rezaie, M. and Rich, J. and Rocher, A. and Rockosi, C. and Roe, N.A. and Rosado-Marin, A. and Ross, A.J. and Rossi, G. and Ruggeri, R. and Ruhlmann-Kleider, V. and Samushia, L. and Sanchez, E. and Saulder, C. and Schlafly, E.F. and Schlegel, D. and Scholte, D. and Schubnell, M. and Seo, H. and Sharples, R. and Silber, J. and Slosar, A. and Smith, A. and Sprayberry, D. and Tan, T. and Tarlé, G. and Trusov, S. and Vaisakh, R. and Valcin, D. and Valdes, F. and Vargas-Magaña, M. and Verde, L. and Walther, M. and Wang, B. and Wang, M.S. and Weaver, B.A. and Weaverdyck, N. and Wechsler, R.H. and Weinberg, D.H. and White, M. and Wilson, M.J. and Yu, J. and Yu, Y. and Yuan, S. and Yèche, C. and Zaborowski, E.A. and Zarrouk, P. and Zhang, H. and Zhao, C. and Zhao, R. and Zhou, R. and Zou, H. and The DESI collaboration},
title = "{DESI 2024 II: sample definitions, characteristics, and two-point clustering statistics}",
journal = {\jcap},
}

@article{desi2024_iii,
doi = {10.1088/1475-7516/2025/04/012},
url = {https://doi.org/10.1088/1475-7516/2025/04/012},
year = {2025},
month = {apr},
publisher = {IOP Publishing},
volume = {2025},
number = {04},
pages = {012},
author = {{DESI Collaboration} and Adame, A.G. and Aguilar, J. and Ahlen, S. and Alam, S. and Alexander, D.M. and Alvarez, M. and Alves, O. and Anand, A. and Andrade, U. and Armengaud, E. and Avila, S. and Aviles, A. and Awan, H. and Bailey, S. and Baltay, C. and Bault, A. and Behera, J. and BenZvi, S. and Beutler, F. and Bianchi, D. and Blake, C. and Blum, R. and Brieden, S. and Brodzeller, A. and Brooks, D. and Buckley-Geer, E. and Burtin, E. and Calderon, R. and Canning, R. and Carnero Rosell, A. and Cereskaite, R. and Cervantes-Cota, J.L. and Chabanier, S. and Chaussidon, E. and Chaves-Montero, J. and Chen, S. and Chen, X. and Claybaugh, T. and Cole, S. and Cuceu, A. and Davis, T.M. and Dawson, K. and de la Macorra, A. and de Mattia, A. and Deiosso, N. and Dey, A. and Dey, B. and Ding, Z. and Doel, P. and Edelstein, J. and Eftekharzadeh, S. and Eisenstein, D.J. and Elliott, A. and Fagrelius, P. and Fanning, K. and Ferraro, S. and Ereza, J. and Findlay, N. and Flaugher, B. and Font-Ribera, A. and Forero-Sánchez, D. and Forero-Romero, J.E. and Garcia-Quintero, C. and Gaztañaga, E. and Gil-Marín, H. and Gontcho, S.Gontcho A. and Gonzalez-Morales, A.X. and Gonzalez-Perez, V. and Gordon, C. and Green, D. and Gruen, D. and Gsponer, R. and Gutierrez, G. and Guy, J. and Hadzhiyska, B. and Hahn, C. and Hanif, M.M.S. and Herrera-Alcantar, H.K. and Honscheid, K. and Howlett, C. and Huterer, D. and Iršič, V. and Ishak, M. and Juneau, S. and Karaçaylı, N.G. and Kehoe, R. and Kent, S. and Kirkby, D. and Kong, H. and Kremin, A. and Krolewski, A. and Lai, Y. and Lan, T.-W. and Landriau, M. and Lang, D. and Lasker, J. and Le Goff, J.M. and Le Guillou, L. and Leauthaud, A. and Levi, M.E. and Li, T.S. and Linder, E. and Lodha, K. and Magneville, C. and Manera, M. and Margala, D. and Martini, P. and Maus, M. and McDonald, P. and Medina-Varela, L. and Meisner, A. and Mena-Fernández, J. and Miquel, R. and Moon, J. and Moore, S. and Moustakas, J. and Mueller, E. and Muñoz-Gutiérrez, A. and Myers, A.D. and Nadathur, S. and Napolitano, L. and Neveux, R. and Newman, J.A. and Nguyen, N.M. and Nie, J. and Niz, G. and Noriega, H.E. and Padmanabhan, N. and Paillas, E. and Palanque-Delabrouille, N. and Pan, J. and Penmetsa, S. and Percival, W.J. and Pieri, M.M. and Pinon, M. and Poppett, C. and Porredon, A. and Prada, F. and Pérez-Fernández, A. and Pérez-Ràfols, I. and Rabinowitz, D. and Raichoor, A. and Ramírez-Pérez, C. and Ramirez-Solano, S. and Rashkovetskyi, M. and Ravoux, C. and Rezaie, M. and Rich, J. and Rocher, A. and Rockosi, C. and Roe, N.A. and Rosado-Marin, A. and Ross, A.J. and Rossi, G. and Ruggeri, R. and Ruhlmann-Kleider, V. and Samushia, L. and Sanchez, E. and Saulder, C. and Schlafly, E.F. and Schlegel, D. and Schubnell, M. and Seo, H. and Sharples, R. and Silber, J. and Slosar, A. and Smith, A. and Sprayberry, D. and Swanson, J. and Tan, T. and Tarlé, G. and Trusov, S. and Vaisakh, R. and Valcin, D. and Valdes, F. and Vargas-Magaña, M. and Verde, L. and Walther, M. and Wang, B. and Wang, M.S. and Weaver, B.A. and Weaverdyck, N. and Wechsler, R.H. and Weinberg, D.H. and White, M. and Wilson, M.J. and Yu, J. and Yu, Y. and Yuan, S. and Yèche, C. and Zaborowski, E.A. and Zarrouk, P. and Zhang, H. and Zhao, C. and Zhao, R. and Zhou, R. and Zou, H. and The DESI collaboration},
title = "{DESI 2024 III: baryon acoustic oscillations from galaxies and quasars}",
journal = {\jcap},
}

@ARTICLE{desi2024_v,
       doi = {10.1088/1475-7516/2025/09/008},
url = {https://doi.org/10.1088/1475-7516/2025/09/008},
year = {2025},
month = {sep},
publisher = {IOP Publishing},
volume = {2025},
number = {09},
pages = {008},
author = {{DESI Collaboration} and Adame, A.G. and Aguilar, J. and Ahlen, S. and Alam, S. and Alexander, D.M. and Alvarez, M. and Alves, O. and Anand, A. and Andrade, U. and Armengaud, E. and Avila, S. and Aviles, A. and Awan, H. and Bailey, S. and Baltay, C. and Bault, A. and Behera, J. and BenZvi, S. and Beutler, F. and Bianchi, D. and Blake, C. and Blum, R. and Brieden, S. and Brodzeller, A. and Brooks, D. and Buckley-Geer, E. and Burtin, E. and Calderon, R. and Canning, R. and Carnero Rosell, A. and Cereskaite, R. and Cervantes-Cota, J.L. and Chabanier, S. and Chaussidon, E. and Chaves-Montero, J. and Chen, S. and Chen, X. and Claybaugh, T. and Cole, S. and Cuceu, A. and Davis, T.M. and Dawson, K. and de la Macorra, A. and de Mattia, A. and Deiosso, N. and Dey, A. and Dey, B. and Ding, Z. and Doel, P. and Edelstein, J. and Eftekharzadeh, S. and Eisenstein, D.J. and Elliott, A. and Fagrelius, P. and Fanning, K. and Ferraro, S. and Ereza, J. and Findlay, N. and Flaugher, B. and Font-Ribera, A. and Forero-Sánchez, D. and Forero-Romero, J.E. and Garcia-Quintero, C. and Garrison, L.H. and Gaztañaga, E. and Gil-Marín, H. and Gontcho, S.Gontcho A. and Gonzalez-Morales, A.X. and Gonzalez-Perez, V. and Gordon, C. and Green, D. and Gruen, D. and Gsponer, R. and Gutierrez, G. and Guy, J. and Hadzhiyska, B. and Hahn, C. and Hanif, M.M.S. and Herrera-Alcantar, H.K. and Honscheid, K. and Howlett, C. and Huterer, D. and Iršič, V. and Ishak, M. and Juneau, S. and Karaçaylı, N.G. and Kehoe, R. and Kent, S. and Kirkby, D. and Kong, H. and Koposov, S.E. and Kremin, A. and Krolewski, A. and Lai, Y. and Lan, T.-W. and Landriau, M. and Lang, D. and Lasker, J. and Le Goff, J.M. and Le Guillou, L. and Leauthaud, A. and Levi, M.E. and Li, T.S. and Lodha, K. and Magneville, C. and Manera, M. and Margala, D. and Martini, P. and Maus, M. and McDonald, P. and Medina-Varela, L. and Meisner, A. and Mena-Fernández, J. and Miquel, R. and Moon, J. and Moore, S. and Moustakas, J. and Mueller, E. and Muñoz-Gutiérrez, A. and Myers, A.D. and Nadathur, S. and Napolitano, L. and Neveux, R. and Newman, J.A. and Nguyen, N.M. and Nie, J. and Niz, G. and Noriega, H.E. and Padmanabhan, N. and Paillas, E. and Palanque-Delabrouille, N. and Pan, J. and Penmetsa, S. and Percival, W.J. and Pieri, M.M. and Pinon, M. and Poppett, C. and Porredon, A. and Prada, F. and Pérez-Fernández, A. and Pérez-Ràfols, I. and Rabinowitz, D. and Raichoor, A. and Ramírez-Pérez, C. and Ramirez-Solano, S. and Rashkovetskyi, M. and Ravoux, C. and Rezaie, M. and Rich, J. and Rocher, A. and Rockosi, C. and Rodríguez-Martínez, F. and Roe, N.A. and Rosado-Marin, A. and Ross, A.J. and Rossi, G. and Ruggeri, R. and Ruhlmann-Kleider, V. and Samushia, L. and Sanchez, E. and Saulder, C. and Schlafly, E.F. and Schlegel, D. and Schubnell, M. and Seo, H. and Sharples, R. and Silber, J. and Slosar, A. and Smith, A. and Sprayberry, D. and Tan, T. and Tarlé, G. and Trusov, S. and Vaisakh, R. and Valcin, D. and Valdes, F. and Vargas-Magaña, M. and Verde, L. and Walther, M. and Wang, B. and Wang, M.S. and Weaver, B.A. and Weaverdyck, N. and Wechsler, R.H. and Weinberg, D.H. and White, M. and Wilson, M.J. and Yu, J. and Yu, Y. and Yuan, S. and Yèche, C. and Zaborowski, E.A. and Zarrouk, P. and Zhang, H. and Zhao, C. and Zhao, R. and Zhou, R. and Zou, H. and The DESI collaboration},
title = "{DESI 2024 V: Full-Shape galaxy clustering from galaxies and quasars}",
journal = {\jcap},
}

@article{desi2024_vi,
   title="{DESI 2024 VI:  cosmological constraints from the measurements of baryon acoustic oscillations}",
   volume={2025},
   ISSN={1475-7516},
   url={http://dx.doi.org/10.1088/1475-7516/2025/02/021},
   DOI={10.1088/1475-7516/2025/02/021},
   number={02},
   journal={\jcap},
   publisher={IOP Publishing},
   author={{DESI Collaboration} and Adame, A.G. and Aguilar, J. and Ahlen, S. and Alam, S. and Alexander, D.M. and Alvarez, M. and Alves, O. and Anand, A. and Andrade, U. and Armengaud, E. and Avila, S. and Aviles, A. and Awan, H. and Bahr-Kalus, B. and Bailey, S. and Baltay, C. and Bault, A. and Behera, J. and BenZvi, S. and Bera, A. and Beutler, F. and Bianchi, D. and Blake, C. and Blum, R. and Brieden, S. and Brodzeller, A. and Brooks, D. and Buckley-Geer, E. and Burtin, E. and Calderon, R. and Canning, R. and Carnero Rosell, A. and Cereskaite, R. and Cervantes-Cota, J.L. and Chabanier, S. and Chaussidon, E. and Chaves-Montero, J. and Chen, S. and Chen, X. and Claybaugh, T. and Cole, S. and Cuceu, A. and Davis, T.M. and Dawson, K. and de la Macorra, A. and de Mattia, A. and Deiosso, N. and Dey, A. and Dey, B. and Ding, Z. and Doel, P. and Edelstein, J. and Eftekharzadeh, S. and Eisenstein, D.J. and Elliott, A. and Fagrelius, P. and Fanning, K. and Ferraro, S. and Ereza, J. and Findlay, N. and Flaugher, B. and Font-Ribera, A. and Forero-Sánchez, D. and Forero-Romero, J.E. and Frenk, C.S. and Garcia-Quintero, C. and Gaztañaga, E. and Gil-Marín, H. and Gontcho, S.Gontcho A. and Gonzalez-Morales, A.X. and Gonzalez-Perez, V. and Gordon, C. and Green, D. and Gruen, D. and Gsponer, R. and Gutierrez, G. and Guy, J. and Hadzhiyska, B. and Hahn, C. and Hanif, M.M.S. and Herrera-Alcantar, H.K. and Honscheid, K. and Howlett, C. and Huterer, D. and Iršič, V. and Ishak, M. and Juneau, S. and Karaçaylı, N.G. and Kehoe, R. and Kent, S. and Kirkby, D. and Kremin, A. and Krolewski, A. and Lai, Y. and Lan, T.-W. and Landriau, M. and Lang, D. and Lasker, J. and Le Goff, J.M. and Le Guillou, L. and Leauthaud, A. and Levi, M.E. and Li, T.S. and Linder, E. and Lodha, K. and Magneville, C. and Manera, M. and Margala, D. and Martini, P. and Maus, M. and McDonald, P. and Medina-Varela, L. and Meisner, A. and Mena-Fernández, J. and Miquel, R. and Moon, J. and Moore, S. and Moustakas, J. and Mueller, E. and Muñoz-Gutiérrez, A. and Myers, A.D. and Nadathur, S. and Napolitano, L. and Neveux, R. and Newman, J.A. and Nguyen, N.M. and Nie, J. and Niz, G. and Noriega, H.E. and Padmanabhan, N. and Paillas, E. and Palanque-Delabrouille, N. and Pan, J. and Penmetsa, S. and Percival, W.J. and Pieri, M.M. and Pinon, M. and Poppett, C. and Porredon, A. and Prada, F. and Pérez-Fernández, A. and Pérez-Ràfols, I. and Rabinowitz, D. and Raichoor, A. and Ramírez-Pérez, C. and Ramirez-Solano, S. and Rashkovetskyi, M. and Ravoux, C. and Rezaie, M. and Rich, J. and Rocher, A. and Rockosi, C. and Roe, N.A. and Rosado-Marin, A. and Ross, A.J. and Rossi, G. and Ruggeri, R. and Ruhlmann-Kleider, V. and Samushia, L. and Sanchez, E. and Saulder, C. and Schlafly, E.F. and Schlegel, D. and Schubnell, M. and Seo, H. and Shafieloo, A. and Sharples, R. and Silber, J. and Slosar, A. and Smith, A. and Sprayberry, D. and Tan, T. and Tarlé, G. and Taylor, P. and Trusov, S. and Ureña-López, L.A. and Vaisakh, R. and Valcin, D. and Valdes, F. and Vargas-Magaña, M. and Verde, L. and Walther, M. and Wang, B. and Wang, M.S. and Weaver, B.A. and Weaverdyck, N. and Wechsler, R.H. and Weinberg, D.H. and White, M. and Yu, J. and Yu, Y. and Yuan, S. and Yèche, C. and Zaborowski, E.A. and Zarrouk, P. and Zhang, H. and Zhao, C. and Zhao, R. and Zhou, R. and Zhuang, T. and Zou, H.},
   year={2025},
   month=feb, pages={021} }

@ARTICLE{desi2024_vii,
       doi = {10.1088/1475-7516/2025/07/028},
url = {https://doi.org/10.1088/1475-7516/2025/07/028},
year = {2025},
month = {jul},
publisher = {IOP Publishing},
volume = {2025},
number = {07},
pages = {028},
author = {{DESI Collaboration} and Adame, A.G. and Aguilar, J. and Ahlen, S. and Alam, S. and Alexander, D.M. and Allende Prieto, C. and Alvarez, M. and Alves, O. and Anand, A. and Andrade, U. and Armengaud, E. and Avila, S. and Aviles, A. and Awan, H. and Bahr-Kalus, B. and Bailey, S. and Baltay, C. and Bault, A. and Behera, J. and BenZvi, S. and Beutler, F. and Bianchi, D. and Blake, C. and Blum, R. and Bonici, M. and Brieden, S. and Brodzeller, A. and Brooks, D. and Buckley-Geer, E. and Burtin, E. and Calderon, R. and Canning, R. and Carnero Rosell, A. and Cereskaite, R. and Cervantes-Cota, J.L. and Chabanier, S. and Chaussidon, E. and Chaves-Montero, J. and Chebat, D. and Chen, S. and Chen, X. and Claybaugh, T. and Cole, S. and Cuceu, A. and Davis, T.M. and Dawson, K. and de la Macorra, A. and de Mattia, A. and Deiosso, N. and Dey, A. and Dey, B. and Ding, Z. and Doel, P. and Edelstein, J. and Eftekharzadeh, S. and Eisenstein, D.J. and Elbers, W. and Elliott, A. and Fagrelius, P. and Fanning, K. and Ferraro, S. and Ereza, J. and Findlay, N. and Flaugher, B. and Font-Ribera, A. and Forero-Sánchez, D. and Forero-Romero, J.E. and Frenk, C.S. and Garcia-Quintero, C. and Garrison, L.H. and Gaztañaga, E. and Gil-Marín, H. and Gontcho, S.Gontcho A. and Gonzalez-Morales, A.X. and Gonzalez-Perez, V. and Gordon, C. and Green, D. and Gruen, D. and Gsponer, R. and Gutierrez, G. and Guy, J. and Hadzhiyska, B. and Hahn, C. and Hanif, M.M.S. and Herrera-Alcantar, H.K. and Honscheid, K. and Howlett, C. and Huterer, D. and Iršič, V. and Ishak, M. and Joyce, R. and Juneau, S. and Karaçaylı, N.G. and Kehoe, R. and Kent, S. and Kirkby, D. and Kong, H. and Koposov, S.E. and Kremin, A. and Krolewski, A. and Lahav, O. and Lai, Y. and Lan, T.-W. and Landriau, M. and Lang, D. and Lasker, J. and Le Goff, J.M. and Le Guillou, L. and Leauthaud, A. and Levi, M.E. and Li, T.S. and Lodha, K. and Magneville, C. and Manera, M. and Margala, D. and Martini, P. and Matthewson, W. and Maus, M. and McDonald, P. and Medina-Varela, L. and Meisner, A. and Mena-Fernández, J. and Miquel, R. and Moon, J. and Moore, S. and Moustakas, J. and Mudur, N. and Mueller, E. and Muñoz-Gutiérrez, A. and Myers, A.D. and Nadathur, S. and Napolitano, L. and Neveux, R. and Newman, J.A. and Nguyen, N.M. and Nie, J. and Niz, G. and Noriega, H.E. and Padmanabhan, N. and Paillas, E. and Palanque-Delabrouille, N. and Pan, J. and Penmetsa, S. and Percival, W.J. and Pieri, M.M. and Pinon, M. and Poppett, C. and Porredon, A. and Prada, F. and Pérez-Fernández, A. and Pérez-Ràfols, I. and Rabinowitz, D. and Raichoor, A. and Ramírez-Pérez, C. and Ramirez-Solano, S. and Rashkovetskyi, M. and Ravoux, C. and Rezaie, M. and Rich, J. and Rocher, A. and Rockosi, C. and Roe, N.A. and Rosado-Marin, A. and Ross, A.J. and Rossi, G. and Ruggeri, R. and Ruhlmann-Kleider, V. and Samushia, L. and Sanchez, E. and Saulder, C. and Schlafly, E.F. and Schlegel, D. and Schubnell, M. and Seo, H. and Shafieloo, A. and Sharples, R. and Silber, J. and Slosar, A. and Smith, A. and Sprayberry, D. and Tan, T. and Tarlé, G. and Taylor, P. and Trusov, S. and Vaisakh, R. and Valcin, D. and Valdes, F. and Valogiannis, G. and Vargas-Magaña, M. and Verde, L. and Walther, M. and Wang, B. and Wang, M.S. and Weaver, B.A. and Weaverdyck, N. and Wechsler, R.H. and Weinberg, D.H. and White, M. and Wilson, M.J. and Yi, L. and Yu, J. and Yu, Y. and Yuan, S. and Yèche, C. and Zaborowski, E.A. and Zarrouk, P. and Zhang, H. and Zhao, C. and Zhao, R. and Zhou, R. and Zhuang, T. and Zou, H. and The DESI collaboration},
title = "{DESI 2024 VII: cosmological constraints from the full-shape modeling of clustering measurements}",
journal = {\jcap},
}

@article{Ross_2025_LSS,
doi = {10.1088/1475-7516/2025/01/125},
url = {https://dx.doi.org/10.1088/1475-7516/2025/01/125},
year = {2025},
month = {jan},
publisher = {IOP Publishing},
volume = {2025},
number = {01},
pages = {125},
author = {Ross, A.J. and Aguilar, J. and Ahlen, S. and Alam, S. and Anand, A. and Bailey, S. and Bianchi, D. and Brieden, S. and Brooks, D. and Burtin, E. and Carnero Rosell, A. and Chaussidon, E. and Claybaugh, T. and Cole, S. and Dawson, K. and de la Macorra, A. and de Mattia, A. and Dey, A. and Dey, B. and Doel, P. and Fanning, K. and Ferraro, S. and Ereza, J. and Font-Ribera, A. and Forero-Romero, J.E. and Gaztañaga, E. and Gil-Marín, H. and Gontcho, S.Gontcho A. and Gonzalez-Morales, A.X. and Guy, J. and Hahn, C. and Heydenreich, S. and Honscheid, K. and Howlett, C. and Ishak, M. and Karim, T. and Kirkby, D. and Kisner, T. and Kong, H. and Kremin, A. and Krolewski, A. and Lambert, A. and Landriau, M. and Lasker, J. and Guillou, L.L. and Levi, M.E. and Manera, M. and Martini, P. and McDonald, P. and Meisner, A. and Miquel, R. and Moon, J. and Moustakas, J. and Muñoz-Gutiérrez, A. and Myers, A.D. and Nadathur, S. and Napolitano, L. and Newman, J.A. and Nie, J. and Niz, G. and Palanque-Delabrouille, N. and Percival, W.J. and Poppett, C. and Prada, F. and Raichoor, A. and Ravoux, C. and Rezaie, M. and Rosado-Marin, A. and Rossi, G. and Samushia, L. and Sanchez, E. and Schlafly, E.F. and Schlegel, D. and Seo, H. and Smith, A. and Sprayberry, D. and Tarlé, G. and Valcin, D. and Vargas-Magaña, M. and Weaver, B.A. and Wilson, M.J. and Yu, J. and Zarrouk, P. and Zhao, C. and Zhou, R. and Zou, H.},
title = "{The construction of large-scale structure catalogs for the Dark Energy Spectroscopic Instrument}",
journal = {\jcap}
}

@article{altmtl,
    author = {Lasker, J. and others},
    title = "{Production of alternate realizations of DESI fiber assignment for unbiased clustering measurement in data and simulations}",
    eprint = "2404.03006",
    archivePrefix = "arXiv",
    primaryClass = "astro-ph.CO",
    reportNumber = "FERMILAB-PUB-24-0167-PPD",
    doi = "10.1088/1475-7516/2025/01/127",
    journal = "\jcap",
    volume = "01",
    pages = "127",
    year = "2025"
}

@article{ffa,
      doi = {10.1088/1475-7516/2025/04/074},
url = {https://doi.org/10.1088/1475-7516/2025/04/074},
year = {2025},
month = {apr},
publisher = {IOP Publishing},
volume = {2025},
number = {04},
pages = {074},
author = {Bianchi, D. and Hanif, M.M.S. and Carnero Rosell, A. and Lasker, J. and Ross, A.J. and Pinon, M. and de Mattia, A. and White, M. and Ahlen, S. and Bailey, S. and Brooks, D. and Burtin, E. and Chaussidon, E. and Claybaugh, T. and Cole, S. and de la Macorra, A. and Ferraro, S. and Font-Ribera, A. and Forero-Romero, J.E. and Gaztañaga, E. and Gontcho, S.Gontcho A. and Gutierrez, G. and Guy, J. and Hahn, C. and Honscheid, K. and Howlett, C. and Juneau, S. and Kirkby, D. and Kisner, T. and Kremin, A. and Landriau, M. and Le Guillou, L. and Levi, M.E. and McDonald, P. and Meisner, A. and Miquel, R. and Moustakas, J. and Palanque-Delabrouille, N. and Percival, W.J. and Prada, F. and Pérez-Ràfols, I. and Raichoor, A. and Rossi, G. and Sanchez, E. and Schlegel, D. and Schubnell, M. and Sharples, R. and Silber, J. and Sprayberry, D. and Tarlé, G. and Vargas-Magaña, M. and Weaver, B.A. and Zarrouk, P. and Zhou, R. and Zou, H.},
title = "{Characterization of DESI fiber assignment incompleteness effect on 2-point clustering and mitigation methods for DR1 analysis}",
journal = {\jcap},
}

@article{desi_dr2_bao,
      title="{DESI DR2 Results II: Measurements of Baryon Acoustic Oscillations and Cosmological Constraints}", 
      author = {Abdul Karim, M. and Aguilar, J. and Ahlen, S. and Alam, S. and Allen, L. and Prieto, C. Allende and Alves, O. and Anand, A. and Andrade, U. and Armengaud, E. and Aviles, A. and Bailey, S. and Baltay, C. and Bansal, P. and Bault, A. and Behera, J. and BenZvi, S. and Bianchi, D. and Blake, C. and Brieden, S. and Brodzeller, A. and Brooks, D. and Buckley-Geer, E. and Burtin, E. and Calderon, R. and Canning, R. and Rosell, A. Carnero and Carrilho, P. and Casas, L. and Castander, F. J. and Charles, M. and Chaussidon, E. and Chaves-Montero, J. and Chebat, D. and Chen, X. and Claybaugh, T. and Cole, S. and Cooper, A. P. and Cuceu, A. and Dawson, K. S. and de la Macorra, A. and de Mattia, A. and Deiosso, N. and Della Costa, J. and Demina, R. and Dey, A. and Dey, B. and Ding, Z. and Doel, P. and Edelstein, J. and Eisenstein, D. J. and Elbers, W. and Fagrelius, P. and Fanning, K. and Fern\'andez-Garc\'{\i}a, E. and Ferraro, S. and Font-Ribera, A. and Forero-Romero, J. E. and Frenk, C. S. and Garcia-Quintero, C. and Garrison, L. H. and Gazta\~naga, E. and Gil-Mar\'{\i}n, H. and Gontcho, S. Gontcho A. and Gonzalez, D. and Gonzalez-Morales, A. X. and Gordon, C. and Green, D. and Gutierrez, G. and Guy, J. and Hadzhiyska, B. and Hahn, C. and He, S. and Herbold, M. and Herrera-Alcantar, H. K. and Ho, M.-F. and Honscheid, K. and Howlett, C. and Huterer, D. and Ishak, M. and Juneau, S. and Kamble, N. V. and Kara\ifmmode \mbox{\c{c}}\else \c{c}\fi{}ayl��, N. G. and Kehoe, R. and Kent, S. and Kim, A. G. and Kirkby, D. and Kisner, T. and Koposov, S. E. and Kremin, A. and Krolewski, A. and Lahav, O. and Lamman, C. and Landriau, M. and Lang, D. and Lasker, J. and Le Goff, J. M. and Le Guillou, L. and Leauthaud, A. and Levi, M. E. and Li, Q. and Li, T. S. and Lodha, K. and Lokken, M. and Lozano-Rodr\'{\i}guez, F. and Magneville, C. and Manera, M. and Martini, P. and Matthewson, W. L. and Meisner, A. and Mena-Fern\'andez, J. and Menegas, A. and Mergulh\~ao, T. and Miquel, R. and Moustakas, J. and Mu\~noz-Guti\'errez, A. and Mu\~noz-Santos, D. and Myers, A. D. and Nadathur, S. and Naidoo, K. and Napolitano, L. and Newman, J. A. and Niz, G. and Noriega, H. E. and Paillas, E. and Palanque-Delabrouille, N. and Pan, J. and Peacock, J. A. and Ibanez, M. P. and Percival, W. J. and P\'erez-Fern\'andez, A. and P\'erez-R\`afols, I. and Pieri, M. M. and Poppett, C. and Prada, F. and Rabinowitz, D. and Raichoor, A. and Ram\'{\i}rez-P\'erez, C. and Rashkovetskyi, M. and Ravoux, C. and Rich, J. and Rocher, A. and Rockosi, C. and Rohlf, J. and Rom\'an-Herrera, J. O. and Ross, A. J. and Rossi, G. and Ruggeri, R. and Ruhlmann-Kleider, V. and Samushia, L. and Sanchez, E. and Sanders, N. and Schlegel, D. and Schubnell, M. and Seo, H. and Shafieloo, A. and Sharples, R. and Silber, J. and Sinigaglia, F. and Sprayberry, D. and Tan, T. and Tarl\'e, G. and Taylor, P. and Turner, W. and Ure\~na-L\'opez, L. A. and Vaisakh, R. and Valdes, F. and Valogiannis, G. and Vargas-Maga\~na, M. and Verde, L. and Walther, M. and Weaver, B. A. and Weinberg, D. H. and White, M. and Wolfson, M. and Y\`eche, C. and Yu, J. and Zaborowski, E. A. and Zarrouk, P. and Zhai, Z. and Zhang, H. and Zhao, C. and Zhao, G. B. and Zhou, R. and Zou, H.},
  collaboration = {DESI},
  journal = {Phys. Rev. D},
  volume = {112},
  issue = {8},
  pages = {083515},
  numpages = {40},
  year = {2025},
  month = {Oct},
  publisher = {American Physical Society},
  doi = {10.1103/tr6y-kpc6},
  url = {https://link.aps.org/doi/10.1103/tr6y-kpc6}
}

@ARTICLE{padmanabhan2008,
       author = {{Padmanabhan}, Nikhil and {White}, Martin},
        title = "{Constraining anisotropic baryon oscillations}",
      journal = {\prd},
         year = 2008,
        month = jun,
       volume = {77},
       number = {12},
          eid = {123540},
        pages = {123540},
          doi = {10.1103/PhysRevD.77.123540},
archivePrefix = {arXiv},
       eprint = {0804.0799},
 primaryClass = {astro-ph},
       adsurl = {https://ui.adsabs.harvard.edu/abs/2008PhRvD..77l3540P}
}

@ARTICLE{eisenstein2007_recon,
       author = {{Eisenstein}, Daniel J. and {Seo}, Hee-Jong and {Sirko}, Edwin and {Spergel}, David N.},
        title = "{Improving Cosmological Distance Measurements by Reconstruction of the Baryon Acoustic Peak}",
      journal = {\apj},
         year = 2007,
        month = aug,
       volume = {664},
       number = {2},
        pages = {675-679},
          doi = {10.1086/518712},
archivePrefix = {arXiv},
       eprint = {astro-ph/0604362},
 primaryClass = {astro-ph},
       adsurl = {https://ui.adsabs.harvard.edu/abs/2007ApJ...664..675E}
}

@ARTICLE{padmanabhan2009,
       author = {{Padmanabhan}, Nikhil and {White}, Martin and {Cohn}, J.~D.},
        title = "{Reconstructing baryon oscillations: A Lagrangian theory perspective}",
      journal = {\prd},
         year = 2009,
        month = mar,
       volume = {79},
       number = {6},
          eid = {063523},
        pages = {063523},
          doi = {10.1103/PhysRevD.79.063523},
archivePrefix = {arXiv},
       eprint = {0812.2905},
 primaryClass = {astro-ph},
       adsurl = {https://ui.adsabs.harvard.edu/abs/2009PhRvD..79f3523P}
}

@ARTICLE{white2015_recsym,
       author = {{White}, Martin},
        title = "{Reconstruction within the Zeldovich approximation}",
      journal = {\mnras},
         year = 2015,
        month = jul,
       volume = {450},
       number = {4},
        pages = {3822-3828},
          doi = {10.1093/mnras/stv842},
archivePrefix = {arXiv},
       eprint = {1504.03677},
 primaryClass = {astro-ph.CO},
       adsurl = {https://ui.adsabs.harvard.edu/abs/2015MNRAS.450.3822W}
}

@ARTICLE{burden2015_ifft,
       author = {{Burden}, A. and {Percival}, W.~J. and {Howlett}, C.},
        title = "{Reconstruction in Fourier space}",
      journal = {\mnras},
         year = 2015,
        month = oct,
       volume = {453},
       number = {1},
        pages = {456-468},
          doi = {10.1093/mnras/stv1581},
archivePrefix = {arXiv},
       eprint = {1504.02591},
 primaryClass = {astro-ph.CO},
       adsurl = {https://ui.adsabs.harvard.edu/abs/2015MNRAS.453..456B}
}

@article{chen2024_bao,
	title = "{Baryon acoustic oscillation theory and modelling systematics for the DESI 2024 results}",
	volume = {534},
	copyright = {https://creativecommons.org/licenses/by/4.0/},
	issn = {0035-8711, 1365-2966},
	url = {https://academic.oup.com/mnras/article/534/1/544/7750629},
	doi = {10.1093/mnras/stae2090},
	language = {en},
	number = {1},
	urldate = {2025-02-18},
	journal = {\mnras},
	author = {Chen, S -F and Howlett, C and White, M and McDonald, P and Ross, A J and Seo, H -J and Padmanabhan, N and Aguilar, J and Ahlen, S and Alam, S and Alves, O and Andrade, U and Blum, R and Brooks, D and Chen, X and Cole, S and Dawson, K and de la Macorra, A and Dey, A and Ding, Z and Doel, P and Ferraro, S and Font-Ribera, A and Forero-Sánchez, D and Forero-Romero, J E and Garcia-Quintero, C and Gaztañaga, E and Gontcho, S G A and Hanif, M M S and Honscheid, K and Kisner, T and Kremin, A and Lambert, A and Landriau, M and Levi, M E and Manera, M and Meisner, A and Mena-Fernández, J and Miquel, R and Munoz-Gutierrez, A and Paillas, E and Palanque-Delabrouille, N and Percival, W J and Pérez-Fernández, A and Prada, F and Rashkovetskyi, M and Rezaie, M and Rosado-Marin, A and Rossi, G and Ruggeri, R and Sanchez, E and Schlegel, D and Silber, J and Tarlé, G and Vargas-Magaña, M and Weaver, B A and Yu, J and Yuan, S and Zhou, R and Zhou, Z},
	month = sep,
	year = {2024},
	pages = {544--574},
}

@article{optimal_recon,
      doi = {10.1088/1475-7516/2025/01/142},
url = {https://doi.org/10.1088/1475-7516/2025/01/142},
year = {2025},
month = {jan},
publisher = {IOP Publishing},
volume = {2025},
number = {01},
pages = {142},
author = {Paillas, E. and Ding, Z. and Chen, X. and Seo, H. and Padmanabhan, N. and de Mattia, A. and Ross, A.J. and Nadathur, S. and Howlett, C. and Aguilar, J. and Ahlen, S. and Alves, O. and Andrade, U. and Brooks, D. and Buckley-Geer, E. and Burtin, E. and Chen, S. and Claybaugh, T. and Cole, S. and Dawson, K. and de la Macorra, A. and Dey, Arjun and Doel, P. and Fanning, K. and Ferraro, S. and Forero-Romero, J.E. and Garcia-Quintero, C. and Gaztañaga, E. and Gil-Marín, H. and Gontcho, S.Gontcho A. and Gutierrez, G. and Hahn, C. and Hanif, M.M.S. and Honscheid, K. and Ishak, M. and Kehoe, R. and Kremin, A. and Landriau, M. and Le Guillou, L. and Levi, M.E. and Manera, M. and Martini, P. and Medina-Varela, L. and Meisner, A. and Mena-Fernández, J. and Miquel, R. and Moustakas, J. and Mueller, E. and Muñoz-Gutiérrez, A. and Myers, A.D. and Newman, J.A. and Nie, J. and Niz, G. and Palanque-Delabrouille, N. and Percival, W.J. and Poppett, C. and Prada, F. and Pérez-Fernández, A. and Rashkovetskyi, M. and Rezaie, M. and Rosado-Marin, A. and Rossi, G. and Ruggeri, R. and Sanchez, E. and Saulder, C. and Schlafly, E.F. and Schlegel, D. and Schubnell, M. and Sprayberry, D. and Tarlé, G. and Valcin, D. and Vargas-Magaña, M. and Yu, J. and Yuan, S. and Zhou, R. and Zou, H.},
title = "{Optimal reconstruction of baryon acoustic oscillations for DESI 2024}",
journal = {\jcap} 
}

@ARTICLE{chen2023_cnn,
       author = {{Chen}, Xinyi and {Zhu}, Fangzhou and {Gaines}, Sasha and {Padmanabhan}, Nikhil},
        title = "{Effective cosmic density field reconstruction with convolutional neural network}",
      journal = {\mnras},
         year = 2023,
        month = aug,
       volume = {523},
       number = {4},
        pages = {6272-6281},
          doi = {10.1093/mnras/stad1868},
archivePrefix = {arXiv},
       eprint = {2306.10538},
 primaryClass = {astro-ph.CO},
       adsurl = {https://ui.adsabs.harvard.edu/abs/2023MNRAS.523.6272C}
}

@ARTICLE{shallue2023_cnn,
       author = {{Shallue}, Christopher J. and {Eisenstein}, Daniel J.},
        title = "{Reconstructing cosmological initial conditions from late-time structure with convolutional neural networks}",
      journal = {\mnras},
         year = 2023,
        month = apr,
       volume = {520},
       number = {4},
        pages = {6256-6267},
          doi = {10.1093/mnras/stad528},
archivePrefix = {arXiv},
       eprint = {2207.12511},
 primaryClass = {astro-ph.CO},
       adsurl = {https://ui.adsabs.harvard.edu/abs/2023MNRAS.520.6256S}
}

@ARTICLE{parker2025_ml,
       author = {{Parker}, Liam and {Bayer}, Adrian E. and {Seljak}, Uro{\v{s}}},
        title = "{Initial conditions from galaxies: machine-learning subgrid correction to standard reconstruction}",
      journal = {\jcap},
         year = 2025,
        month = sep,
       volume = {2025},
       number = {9},
          eid = {039},
        pages = {039},
          doi = {10.1088/1475-7516/2025/09/039},
archivePrefix = {arXiv},
       eprint = {2504.01092},
 primaryClass = {astro-ph.CO},
       adsurl = {https://ui.adsabs.harvard.edu/abs/2025JCAP...09..039P}
}

@ARTICLE{chen2024_bao_algos,
       doi = {10.1088/1475-7516/2026/05/001},
url = {https://doi.org/10.1088/1475-7516/2026/05/001},
year = {2026},
month = {may},
publisher = {IOP Publishing},
volume = {2026},
number = {05},
pages = {001},
author = {Chen, X. and Ding, Z. and Paillas, E. and Nadathur, S. and Seo, H. and Chen, S. and Padmanabhan, N. and White, M. and de Mattia, A. and McDonald, P. and Ross, A.J. and Variu, A. and Carnero Rosell, A. and Hadzhiyska, B. and Hanif, M.M.S. and Forero-Sánchez, D. and Ahlen, S. and Alves, O. and Andrade, U. and BenZvi, S. and Bianchi, D. and Brooks, D. and Chaussidon, E. and Claybaugh, T. and de la Macorra, A. and Dey, Biprateep and Fanning, K. and Ferraro, S. and Font-Ribera, A. and Forero-Romero, J.E. and Garcia-Quintero, C. and Gaztañaga, E. and Gontcho A Gontcho, S. and Gutierrez, G. and Hahn, C. and Honscheid, K. and Juneau, S. and Kehoe, R. and Kirkby, D. and Kisner, T. and Kremin, A. and Levi, M.E. and Meisner, A. and Mena-Fernández, J. and Miquel, R. and Moustakas, J. and Muñoz-Gutiérrez, A. and Nikakhtar, F. and Palanque-Delabrouille, N. and Percival, W.J. and Prada, F. and Pérez-Ràfols, I. and Rashkovetskyi, M. and Rossi, G. and Ruggeri, R. and Sanchez, E. and Saulder, C. and Schlegel, D. and Schubnell, M. and Smith, A. and Sprayberry, D. and Tarlé, G. and Valcin, D. and Vargas-Magaña, M. and Weaver, B.A. and Yuan, S. and Zhou, R.},
title = {Extensive analysis of reconstruction algorithms for DESI 2024 baryon acoustic oscillations},
journal = {\jcap},
}

@article{desi_recon_fiducial_cosmo,
      title="{Fiducial-Cosmology-dependent systematics for the DESI 2024 BAO Analysis}", 
      author={A. Pérez-Fernández and L. Medina-Varela and R. Ruggeri and M. Vargas-Magaña and H. Seo and N. Padmanabhan and M. Ishak and J. Aguilar and S. Ahlen and S. Alam and O. Alves and S. Brieden and D. Brooks and A. Carnero Rosell and X. Chen and T. Claybaugh and S. Cole and K. Dawson and A. de la Macorra and A. de Mattia and Arjun Dey and Z. Ding and P. Doel and K. Fanning and C. Garcia-Quintero and E. Gaztañaga and S. Gontcho A Gontcho and G. Gutierrez and K. Honscheid and S. Juneau and D. Kirkby and T. Kisner and A. Lambert and M. Landriau and J. Lasker and L. Le Guillou and M. Manera and P. Martini and A. Meisner and J. Mena-Fernández and R. Miquel and J. Moustakas and A. D. Myers and S. Nadathur and J. A. Newman and G. Niz and E. Paillas and N. Palanque-Delabrouille and W. J. Percival and C. Poppett and F. Prada and M. Rashkovetskyi and A. Rocher and G. Rossi and A. Sanchez and E. Sanchez and M. Schubnell and D. Sprayberry and G. Tarlé and D. Valcin and B. A. Weaver and J. Yu and H. Zou},
      year = "2025",
    month = jan,
    day = "30",
    doi = "10.1088/1475-7516/2025/01/144",
    language = "English",
    volume = "2025",
    pages = "1--36",
    journal = {\jcap},
    issn = "1475-7516",
    publisher = "IOP Publishing",
    number = "1",
}

@ARTICLE{hada_einsenstein2018_iterative,
       author = {{Hada}, Ryuichiro and {Eisenstein}, Daniel J.},
        title = "{An iterative reconstruction of cosmological initial density fields}",
      journal = {\mnras},
         year = 2018,
        month = aug,
       volume = {478},
       number = {2},
        pages = {1866-1874},
          doi = {10.1093/mnras/sty1203},
archivePrefix = {arXiv},
       eprint = {1804.04738},
 primaryClass = {astro-ph.CO},
       adsurl = {https://ui.adsabs.harvard.edu/abs/2018MNRAS.478.1866H}
}

@ARTICLE{chen2024_iterative,
       author = {{Chen}, Xinyi and {Padmanabhan}, Nikhil},
        title = "{Analysis of an iterative reconstruction method in comparison of the standard reconstruction method}",
      journal = {\mnras},
         year = 2024,
        month = oct,
       volume = {534},
       number = {2},
        pages = {1490-1503},
          doi = {10.1093/mnras/stae2180},
archivePrefix = {arXiv},
       eprint = {2311.09531},
 primaryClass = {astro-ph.CO},
       adsurl = {https://ui.adsabs.harvard.edu/abs/2024MNRAS.534.1490C}
}

@ARTICLE{schmittful2017_iterative,
       author = {{Schmittfull}, Marcel and {Baldauf}, Tobias and {Zaldarriaga}, Matias},
        title = "{Iterative initial condition reconstruction}",
      journal = {\prd},
         year = 2017,
        month = jul,
       volume = {96},
       number = {2},
          eid = {023505},
        pages = {023505},
          doi = {10.1103/PhysRevD.96.023505},
archivePrefix = {arXiv},
       eprint = {1704.06634},
 primaryClass = {astro-ph.CO},
       adsurl = {https://ui.adsabs.harvard.edu/abs/2017PhRvD..96b3505S}
}

@ARTICLE{seo2022_iterative,
       author = {{Seo}, Hee-Jong and {Ota}, Atsuhisa and {Schmittfull}, Marcel and {Saito}, Shun and {Beutler}, Florian},
        title = "{Iterative reconstruction excursions for Baryon Acoustic Oscillations and beyond}",
      journal = {\mnras},
         year = 2022,
        month = apr,
       volume = {511},
       number = {2},
        pages = {1557-1573},
          doi = {10.1093/mnras/stac082},
archivePrefix = {arXiv},
       eprint = {2106.00530},
 primaryClass = {astro-ph.CO},
       adsurl = {https://ui.adsabs.harvard.edu/abs/2022MNRAS.511.1557S}
}

@misc{Alves2024prep,
  author = "Alves, Otavio and {DESI Collaboration}",
  title  = "{Analytical covariance matrices of DESI galaxy power spectrum multipoles}",
  year   = "{(in prep.)}"
}

@article{Wadekar:2019rdu,
    author = "Wadekar, Digvijay and Scoccimarro, Roman",
    title = "{Galaxy power spectrum multipoles covariance in perturbation theory}",
    eprint = "1910.02914",
    archivePrefix = "arXiv",
    primaryClass = "astro-ph.CO",
    doi = "10.1103/PhysRevD.102.123517",
    journal = "Phys. Rev. D",
    volume = "102",
    number = "12",
    pages = "123517",
    year = "2020"
}

@article{Kobayashi:2023vpu,
    author = "Kobayashi, Yosuke",
    title = "{Fast computation of the non-Gaussian covariance of redshift-space galaxy power spectrum multipoles}",
    eprint = "2308.08593",
    archivePrefix = "arXiv",
    primaryClass = "astro-ph.CO",
    doi = "10.1103/PhysRevD.108.103512",
    journal = "Phys. Rev. D",
    volume = "108",
    number = "10",
    pages = "103512",
    year = "2023"
}

@ARTICLE{rascalc2024,
       author = {{Rashkovetskyi}, M. and {Forero-S{\'a}nchez}, D. and {de Mattia}, A. and {Eisenstein}, D.~J. and {Padmanabhan}, N. and {Seo}, H. and {Ross}, A.~J. and {Aguilar}, J. and {Ahlen}, S. and {Alves}, O. and {Andrade}, U. and {Brooks}, D. and {Burtin}, E. and {Chen}, X. and {Claybaugh}, T. and {Cole}, S. and {de la Macorra}, A. and {Ding}, Z. and {Doel}, P. and {Fanning}, K. and {Ferraro}, S. and {Font-Ribera}, A. and {Forero-Romero}, J.~E. and {Garcia-Quintero}, C. and {Gil-Mar{\'\i}n}, H. and {Gontcho A Gontcho}, S. and {Gonzalez-Morales}, A.~X. and {Gutierrez}, G. and {Honscheid}, K. and {Howlett}, C. and {Juneau}, S. and {Kremin}, A. and {Le Guillou}, L. and {Manera}, M. and {Medina-Varela}, L. and {Mena-Fern{\'a}ndez}, J. and {Miquel}, R. and {Mueller}, E. and {Mu{\~n}oz-Guti{\'e}rrez}, A. and {Myers}, A.~D. and {Nie}, J. and {Niz}, G. and {Paillas}, E. and {Percival}, W.~J. and {Poppett}, C. and {P{\'e}rez-Fern{\'a}ndez}, A. and {Rezaie}, M. and {Rosado-Marin}, A. and {Rossi}, G. and {Ruggeri}, R. and {Sanchez}, E. and {Saulder}, C. and {Schlegel}, D. and {Schubnell}, M. and {Sprayberry}, D. and {Tarl{\'e}}, G. and {Weaver}, B.~A. and {Yu}, J. and {Zhao}, C. and {Zou}, H.},
        title = "{Semi-analytical covariance matrices for two-point correlation function for DESI 2024 data}",
      journal = {\jcap},
         year = 2025,
        month = jan,
       volume = {2025},
       number = {1},
          eid = {145},
        pages = {145},
          doi = {10.1088/1475-7516/2025/01/145},
archivePrefix = {arXiv},
       eprint = {2404.03007},
 primaryClass = {astro-ph.CO},
       adsurl = {https://ui.adsabs.harvard.edu/abs/2025JCAP...01..145R}
}

@ARTICLE{rascalc_validation,
       author = {{Rashkovetskyi}, Michael and {Eisenstein}, Daniel J. and {Aguilar}, Jessica Nicole and {Brooks}, David and {Claybaugh}, Todd and {Cole}, Shaun and {Dawson}, Kyle and {de la Macorra}, Axel and {Doel}, Peter and {Fanning}, Kevin and {Font-Ribera}, Andreu and {Forero-Romero}, Jaime E. and {Gontcho A Gontcho}, Satya and {Hahn}, ChangHoon and {Honscheid}, Klaus and {Kehoe}, Robert and {Kisner}, Theodore and {Landriau}, Martin and {Levi}, Michael and {Manera}, Marc and {Miquel}, Ramon and {Moon}, Jeongin and {Nadathur}, Seshadri and {Nie}, Jundan and {Poppett}, Claire and {Ross}, Ashley J. and {Rossi}, Graziano and {Sanchez}, Eusebio and {Saulder}, Christoph and {Schubnell}, Michael and {Seo}, Hee-Jong and {Tarle}, Gregory and {Valcin}, David and {Weaver}, Benjamin Alan and {Zhao}, Cheng and {Zhou}, Zhimin and {Zou}, Hu},
        title = "{Validation of semi-analytical, semi-empirical covariance matrices for two-point correlation function for early DESI data}",
      journal = {\mnras},
         year = 2023,
        month = sep,
       volume = {524},
       number = {3},
        pages = {3894-3911},
          doi = {10.1093/mnras/stad2078},
archivePrefix = {arXiv},
       eprint = {2306.06320},
 primaryClass = {astro-ph.CO},
       adsurl = {https://ui.adsabs.harvard.edu/abs/2023MNRAS.524.3894R}
}

@ARTICLE{analytic_numeric_cov_comparison,
       doi = {10.1088/1475-7516/2025/04/055},
url = {https://doi.org/10.1088/1475-7516/2025/04/055},
year = {2025},
month = {apr},
publisher = {IOP Publishing},
volume = {2025},
number = {04},
pages = {055},
author = {Forero-Sánchez, D. and Rashkovetskyi, M. and Alves, O. and de Mattia, A. and Padmanabhan, N. and Seo, H. and Nadathur, S. and Ross, A.J. and Gil-Marín, H. and Zarrouk, P. and Yu, J. and Ding, Z. and Andrade, U. and Chen, X. and Garcia-Quintero, C. and Mena-Fernández, J. and Ahlen, S. and Bianchi, D. and Brooks, D. and Burtin, E. and Chaussidon, E. and Claybaugh, T. and Cole, S. and de la Macorra, A. and Enriquez-Vargas, M. and Gaztañaga, E. and Gutierrez, G. and Honscheid, K. and Howlett, C. and Kisner, T. and Landriau, M. and Le Guillou, L. and Levi, M.E. and Miquel, R. and Moustakas, J. and Palanque-Delabrouille, N. and Percival, W.J. and Pérez-Ràfols, I. and Rossi, G. and Sanchez, E. and Schlegel, D. and Schubnell, M. and Sprayberry, D. and Tarlé, G. and Vargas-Magaña, M. and Weaver, B.A. and Zou, H.},
title = "{Analytical and EZmock covariance validation for the DESI 2024 results}",
journal = {\jcap},

}

@ARTICLE{analytic_cov_oconnell2016,
       author = {{O'Connell}, Ross and {Eisenstein}, Daniel and {Vargas}, Mariana and {Ho}, Shirley and {Padmanabhan}, Nikhil},
        title = "{Large covariance matrices: smooth models from the two-point correlation function}",
      journal = {\mnras},
         year = 2016,
        month = nov,
       volume = {462},
       number = {3},
        pages = {2681-2694},
          doi = {10.1093/mnras/stw1821},
archivePrefix = {arXiv},
       eprint = {1510.01740},
 primaryClass = {astro-ph.CO},
       adsurl = {https://ui.adsabs.harvard.edu/abs/2016MNRAS.462.2681O}
}

@ARTICLE{analytic_cov_oconnell2019,
       author = {{O'Connell}, Ross and {Eisenstein}, Daniel J.},
        title = "{Large covariance matrices: accurate models without mocks}",
      journal = {\mnras},
         year = 2019,
        month = aug,
       volume = {487},
       number = {2},
        pages = {2701-2717},
          doi = {10.1093/mnras/stz1359},
archivePrefix = {arXiv},
       eprint = {1808.05978},
 primaryClass = {astro-ph.CO},
       adsurl = {https://ui.adsabs.harvard.edu/abs/2019MNRAS.487.2701O}
}

@ARTICLE{analytic_cov_philcox2019,
       author = {{Philcox}, Oliver H.~E. and {Eisenstein}, Daniel J.},
        title = "{Estimating covariance matrices for two- and three-point correlation function moments in Arbitrary Survey Geometries}",
      journal = {\mnras},
         year = 2019,
        month = dec,
       volume = {490},
       number = {4},
        pages = {5931-5951},
          doi = {10.1093/mnras/stz2896},
archivePrefix = {arXiv},
       eprint = {1910.04764},
 primaryClass = {astro-ph.CO},
       adsurl = {https://ui.adsabs.harvard.edu/abs/2019MNRAS.490.5931P}
}

@ARTICLE{rascalc2020,
       author = {{Philcox}, Oliver H.~E. and {Eisenstein}, Daniel J. and {O'Connell}, Ross and {Wiegand}, Alexander},
        title = "{RASCALC: a jackknife approach to estimating single- and multitracer galaxy covariance matrices}",
      journal = {\mnras},
         year = 2020,
        month = jan,
       volume = {491},
       number = {3},
        pages = {3290-3317},
          doi = {10.1093/mnras/stz3218},
archivePrefix = {arXiv},
       eprint = {1904.11070},
 primaryClass = {astro-ph.CO},
       adsurl = {https://ui.adsabs.harvard.edu/abs/2020MNRAS.491.3290P}
}

@ARTICLE{eppur2008,
       author = {{Smith}, Robert E. and {Scoccimarro}, Rom{\'a}n and {Sheth}, Ravi K.},
        title = "{Motion of the acoustic peak in the correlation function}",
      journal = {\prd},
         year = 2008,
        month = feb,
       volume = {77},
       number = {4},
          eid = {043525},
        pages = {043525},
          doi = {10.1103/PhysRevD.77.043525},
archivePrefix = {arXiv},
       eprint = {astro-ph/0703620},
 primaryClass = {astro-ph},
       adsurl = {https://ui.adsabs.harvard.edu/abs/2008PhRvD..77d3525S}
}

@ARTICLE{gaussPoisson2016,
       author = {{Grieb}, Jan Niklas and {S{\'a}nchez}, Ariel G. and {Salazar-Albornoz}, Salvador and {Dalla Vecchia}, Claudio},
        title = "{Gaussian covariance matrices for anisotropic galaxy clustering measurements}",
      journal = {\mnras},
         year = 2016,
        month = apr,
       volume = {457},
       number = {2},
        pages = {1577-1592},
          doi = {10.1093/mnras/stw065},
archivePrefix = {arXiv},
       eprint = {1509.04293},
 primaryClass = {astro-ph.CO},
       adsurl = {https://ui.adsabs.harvard.edu/abs/2016MNRAS.457.1577G}
}

@ARTICLE{eigencov2024,
       author = {{Lee}, Jaemyoung (Jason) and {Nikakhtar}, Farnik and {Paranjape}, Aseem and {Sheth}, Ravi K.},
        title = "{Eigen-decomposition of covariance matrices: An application to the BAO linear point}",
      journal = {\prd},
         year = 2024,
        month = nov,
       volume = {110},
       number = {10},
          eid = {103515},
        pages = {103515},
          doi = {10.1103/PhysRevD.110.103515},
archivePrefix = {arXiv},
       eprint = {2407.04692},
 primaryClass = {astro-ph.CO},
       adsurl = {https://ui.adsabs.harvard.edu/abs/2024PhRvD.110j3515L}
}

@article{class,
   title="{The Cosmic Linear Anisotropy Solving System (CLASS).
 Part II: Approximation schemes}",
   volume={2011},
   ISSN={1475-7516},
   url={http://dx.doi.org/10.1088/1475-7516/2011/07/034},
   DOI={10.1088/1475-7516/2011/07/034},
   number={07},
   journal={\jcap},
   publisher={IOP Publishing},
   author={Diego Blas and Julien Lesgourgues and Thomas Tram},
   year={2011},
   month=jul, pages={034–034} }

@misc{jax,
  author = {James Bradbury and Roy Frostig and Peter Hawkins and Matthew James Johnson and Chris Leary and Dougal Maclaurin and George Necula and Adam Paszke and Jake Vander{P}las and Skye Wanderman-{M}ilne and Qiao Zhang},
  title = {{JAX}: composable transformations of {P}ython+{N}um{P}y programs},
  addurl = {http://github.com/jax-ml/jax},
  version = {0.3.13},
  year = {2018},
}

@article{emcee,
   title="{emcee: The MCMC Hammer}",
   volume={125},
   ISSN={1538-3873},
   url={http://dx.doi.org/10.1086/670067},
   DOI={10.1086/670067},
   number={925},
   journal={Publications of the Astronomical Society of the Pacific},
   publisher={IOP Publishing},
   author={Foreman-Mackey, Daniel and Hogg, David W. and Lang, Dustin and Goodman, Jonathan},
   year={2013},
   month=mar, pages={306–312} }

@ARTICLE{corrfunc1,
    author = {{Sinha}, Manodeep and {Garrison}, Lehman H.},
    title = "{CORRFUNC - a suite of blazing fast correlation functions on
    the CPU}",
    journal = {\mnras},
    year = "2020",
    month = "Jan",
    volume = {491},
    number = {2},
    pages = {3022-3041},
    doi = {10.1093/mnras/stz3157},
    adsurl =
    {https://ui.adsabs.harvard.edu/abs/2020MNRAS.491.3022S}
}

@InProceedings{corrfunc2,
    author="Sinha, Manodeep and Garrison, Lehman",
    editor="Majumdar, Amit and Arora, Ritu",
    title="CORRFUNC: Blazing Fast Correlation Functions with AVX512F SIMD Intrinsics",
    booktitle="Software Challenges to Exascale Computing",
    year="2019",
    publisher="Springer Singapore",
    address="Singapore",
    pages="3-20",
    isbn="978-981-13-7729-7"
}

@article{anselmi2016_lp,
    author = {Anselmi, Stefano and Starkman, Glenn D. and Sheth, Ravi K.},
    title = "{Beating non-linearities: improving the baryon acoustic oscillations with the linear point}",
    journal = {\mnras},
    volume = {455},
    number = {3},
    pages = {2474-2483},
    year = {2015},
    month = {11},
    issn = {0035-8711},
    doi = {10.1093/mnras/stv2436},
    url = {https://doi.org/10.1093/mnras/stv2436},
    
}

@article{anselmi2017_sdss,
  title="{Galaxy Correlation Functions Provide a More Robust Cosmological Standard Ruler.}",
  author={Stefano Anselmi and Glenn D. Starkman and Pier Stefano Corasaniti and Ravi K. Sheth and Idit Zehavi},
  journal={\prl},
  year={2017},
  volume={121 2},
  doi = {10.1103/PhysRevLett.121.021302},
  url={https://api.semanticscholar.org/CorpusID:51941773}
}

@article{anselmi2018_validation,
  title = "{Linear point standard ruler for galaxy survey data: Validation with mock catalogs}",
  author = {Anselmi, Stefano and Corasaniti, Pier-Stefano and Starkman, Glenn D. and Sheth, Ravi K. and Zehavi, Idit},
  journal = {Phys. Rev. D},
  volume = {98},
  issue = {2},
  pages = {023527},
  numpages = {9},
  year = {2018},
  month = {Jul},
  publisher = {American Physical Society},
  doi = {10.1103/PhysRevD.98.023527},
  url = {https://link.aps.org/doi/10.1103/PhysRevD.98.023527}
}

@article{odwyer2020,
  title = {Linear point and sound horizon as purely geometric standard rulers},
  author = {O'Dwyer, M\'arcio and Anselmi, Stefano and Starkman, Glenn D. and Corasaniti, Pier-Stefano and Sheth, Ravi K. and Zehavi, Idit},
  journal = {Phys. Rev. D},
  volume = {101},
  issue = {8},
  pages = {083517},
  numpages = {13},
  year = {2020},
  month = {Apr},
  publisher = {American Physical Society},
  doi = {10.1103/PhysRevD.101.083517},
  url = {https://link.aps.org/doi/10.1103/PhysRevD.101.083517}
}

@article{anselmi2019_distance_inference,
	title = "{Cosmic distance inference from purely geometric BAO methods: Linear point standard ruler and correlation function model fitting}",
	volume = {99},
	issn = {2470-0010, 2470-0029},
	shorttitle = {Cosmic distance inference from purely geometric {BAO} methods},
	url = {https://link.aps.org/doi/10.1103/PhysRevD.99.123515},
	doi = {10.1103/PhysRevD.99.123515},
	language = {en},
	number = {12},
	urldate = {2024-10-09},
	journal = {Phys. Rev. D},
	author = {Anselmi, Stefano and Corasaniti, Pier-Stefano and Sanchez, Ariel G. and Starkman, Glenn D. and Sheth, Ravi K. and Zehavi, Idit},
	month = jun,
	year = {2019},
	pages = {123515},
}

@article{anselmi2023_cosmological_forecasts,
	title = "{Cosmological forecasts for future galaxy surveys with the linear point standard ruler: {Toward} consistent {BAO} analyses far from a fiducial cosmology}",
	volume = {107},
	issn = {2470-0010, 2470-0029},
	shorttitle = {Cosmological forecasts for future galaxy surveys with the linear point standard ruler},
	url = {http://arxiv.org/abs/2205.09098},
	doi = {10.1103/PhysRevD.107.123506},
	number = {12},
	urldate = {2025-02-18},
	journal = {Phys. Rev. D},
	author = {Anselmi, Stefano and Starkman, Glenn D. and Renzi, Alessandro},
	month = jun,
	year = {2023},
	note = {arXiv:2205.09098 [astro-ph]},
	pages = {123506},
}

@ARTICLE{he2023,
       author = {{He}, Mengfan and {Zhao}, Cheng and {Shan}, Huanyuan},
        title = "{Cosmological constraints with the linear point from the BOSS survey}",
      journal = {\mnras},
         year = 2023,
        month = oct,
       volume = {525},
       number = {2},
        pages = {1746-1757},
          doi = {10.1093/mnras/stad2207},
archivePrefix = {arXiv},
       eprint = {2303.10661},
 primaryClass = {astro-ph.CO},
       adsurl = {https://ui.adsabs.harvard.edu/abs/2023MNRAS.525.1746H}
}

@article{nikakhtar2021,
  title = {Laguerre reconstruction of the correlation function on baryon acoustic oscillation scales},
  author = {Nikakhtar, Farnik and Sheth, Ravi K. and Zehavi, Idit},
  journal = {Phys. Rev. D},
  volume = {104},
  issue = {4},
  pages = {043530},
  numpages = {16},
  year = {2021},
  month = {Aug},
  publisher = {American Physical Society},
  doi = {10.1103/PhysRevD.104.043530},
  url = {https://link.aps.org/doi/10.1103/PhysRevD.104.043530}
}

@article{nikakhtar2021_halo,
    author = "Nikakhtar, Farnik and Sheth, Ravi K. and Zehavi, Idit",
    title = "{Laguerre reconstruction of the BAO feature in halo-based mock galaxy catalogues}",
    eprint = "2107.12537",
    archivePrefix = "arXiv",
    primaryClass = "astro-ph.CO",
    doi = "10.1103/PhysRevD.104.063504",
    journal = "Phys. Rev. D",
    volume = "104",
    number = "6",
    pages = "063504",
    year = "2021"
}

@article{nikakhtar2022_smearing,
  title = "{Smearing scale in Laguerre reconstructions of the correlation function}",
  author = {Nikakhtar, Farnik and Sheth, Ravi K. and Zehavi, Idit},
  journal = {Phys. Rev. D},
  volume = {105},
  issue = {4},
  pages = {043536},
  numpages = {12},
  year = {2022},
  month = {Feb},
  publisher = {American Physical Society},
  doi = {10.1103/PhysRevD.105.043536},
  url = {https://link.aps.org/doi/10.1103/PhysRevD.105.043536}
}

@article{paranjape2022,
       author = {{Paranjape}, Aseem and {Sheth}, Ravi K.},
        title = "{Bayesian evidence comparison for distance scale estimates}",
      journal = {\mnras},
         year = 2022,
        month = dec,
       volume = {517},
       number = {4},
        pages = {4696-4704},
          doi = {10.1093/mnras/stac2984},
archivePrefix = {arXiv},
       eprint = {2209.00668},
 primaryClass = {astro-ph.CO},
       adsurl = {https://ui.adsabs.harvard.edu/abs/2022MNRAS.517.4696P}
}

@ARTICLE{paranjape2023,
       author = {{Paranjape}, Aseem and {Sheth}, Ravi K.},
        title = "{Model-agnostic cosmological constraints from the baryon acoustic oscillation feature in redshift space}",
      journal = {\mnras},
         year = 2023,
        month = nov,
       volume = {526},
       number = {1},
        pages = {700-716},
          doi = {10.1093/mnras/stad2741},
archivePrefix = {arXiv},
       eprint = {2304.09198},
 primaryClass = {astro-ph.CO},
       adsurl = {https://ui.adsabs.harvard.edu/abs/2023MNRAS.526..700P}
}

@ARTICLE{paranjape2025,
       doi = {10.1088/1475-7516/2025/10/031},
url = {https://doi.org/10.1088/1475-7516/2025/10/031},
year = {2025},
month = {oct},
publisher = {IOP Publishing},
volume = {2025},
number = {10},
pages = {031},
author = {Paranjape, Aseem and Sheth, Ravi K.},
title = "{Scale-dependent bias and mode coupling in redshift-space clustering near the BAO scale}",
journal = {\jcap},
}

@article{ot_nikakhtar2022,
   title="{Optimal Transport Reconstruction of Baryon Acoustic Oscillations}",
   volume={129},
   ISSN={1079-7114},
   url={http://dx.doi.org/10.1103/PhysRevLett.129.251101},
   DOI={10.1103/physrevlett.129.251101},
   number={25},
   journal={\prl},
   publisher={American Physical Society (APS)},
   author={Nikakhtar, Farnik and Sheth, Ravi K. and Lévy, Bruno and Mohayaee, Roya},
   year={2022},
   month=dec }

@article{ot_von_Hausegger2022,
   title="{Accurate Baryon Acoustic Oscillations Reconstruction via Semidiscrete Optimal Transport}",
   volume={128},
   ISSN={1079-7114},
   url={http://dx.doi.org/10.1103/PhysRevLett.128.201302},
   DOI={10.1103/physrevlett.128.201302},
   number={20},
   journal={\prl},
   publisher={American Physical Society (APS)},
   author={von Hausegger, Sebastian and Lévy, Bruno and Mohayaee, Roya},
   year={2022},
   month=may }

@article{ot_nikakhtar2023,
  title = "{Optimal transport reconstruction of biased tracers in redshift space}",
  author = {Nikakhtar, Farnik and Padmanabhan, Nikhil and L\'evy, Bruno and Sheth, Ravi K. and Mohayaee, Roya},
  journal = {Phys. Rev. D},
  volume = {108},
  issue = {8},
  pages = {083534},
  numpages = {15},
  year = {2023},
  month = {Oct},
  publisher = {American Physical Society},
  doi = {10.1103/PhysRevD.108.083534},
  url = {https://link.aps.org/doi/10.1103/PhysRevD.108.083534}
}

@article{ot_nikakhtar2024,
  title = "{Displacement field analysis via optimal transport: Multitracer approach to cosmological reconstruction}",
  author = {Nikakhtar, Farnik and Sheth, Ravi K. and Padmanabhan, Nikhil and L\'evy, Bruno and Mohayaee, Roya},
  journal = {Phys. Rev. D},
  volume = {109},
  issue = {12},
  pages = {123512},
  numpages = {12},
  year = {2024},
  month = {Jun},
  publisher = {American Physical Society},
  doi = {10.1103/PhysRevD.109.123512},
  url = {https://link.aps.org/doi/10.1103/PhysRevD.109.123512}
}

@article{ot_levy2021,
    author = {Levy, Bruno and Mohayaee, Roya and von Hausegger, Sebastian},
    title = {A fast semidiscrete optimal transport algorithm for a unique reconstruction of the early Universe},
    journal = {\mnras},
    volume = {506},
    number = {1},
    pages = {1165-1185},
    year = {2021},
    month = {06},
    issn = {0035-8711},
    doi = {10.1093/mnras/stab1676},
    url = {https://doi.org/10.1093/mnras/stab1676},
    
}

@article{eBOSS,
  title = "{Completed SDSS-IV extended Baryon Oscillation Spectroscopic Survey: Cosmological implications from two decades of spectroscopic surveys at the Apache Point Observatory}",
  author = {Alam, Shadab and Aubert, Marie and Avila, Santiago and Balland, Christophe and Bautista, Julian E. and Bershady, Matthew A. and Bizyaev, Dmitry and Blanton, Michael R. and Bolton, Adam S. and Bovy, Jo and Brinkmann, Jonathan and Brownstein, Joel R. and Burtin, Etienne and Chabanier, Sol\`ene and Chapman, Michael J. and Choi, Peter Doohyun and Chuang, Chia-Hsun and Comparat, Johan and Cousinou, Marie-Claude and Cuceu, Andrei and Dawson, Kyle S. and de la Torre, Sylvain and de Mattia, Arnaud and Agathe, Victoria de Sainte and des Bourboux, H\'elion du Mas and Escoffier, Stephanie and Etourneau, Thomas and Farr, James and Font-Ribera, Andreu and Frinchaboy, Peter M. and Fromenteau, Sebastien and Gil-Mar\'{\i}n, H\'ector and Le Goff, Jean-Marc and Gonzalez-Morales, Alma X. and Gonzalez-Perez, Violeta and Grabowski, Kathleen and Guy, Julien and Hawken, Adam J. and Hou, Jiamin and Kong, Hui and Parker, James and Klaene, Mark and Kneib, Jean-Paul and Lin, Sicheng and Long, Daniel and Lyke, Brad W. and de la Macorra, Axel and Martini, Paul and Masters, Karen and Mohammad, Faizan G. and Moon, Jeongin and Mueller, Eva-Maria and Mu\~noz-Guti\'errez, Andrea and Myers, Adam D. and Nadathur, Seshadri and Neveux, Richard and Newman, Jeffrey A. and Noterdaeme, Pasquier and Oravetz, Audrey and Oravetz, Daniel and Palanque-Delabrouille, Nathalie and Pan, Kaike and Paviot, Romain and Percival, Will J. and P\'erez-R\`afols, Ignasi and Petitjean, Patrick and Pieri, Matthew M. and Prakash, Abhishek and Raichoor, Anand and Ravoux, Corentin and Rezaie, Mehdi and Rich, James and Ross, Ashley J. and Rossi, Graziano and Ruggeri, Rossana and Ruhlmann-Kleider, Vanina and S\'anchez, Ariel G. and S\'anchez, F. Javier and S\'anchez-Gallego, Jos\'e R. and Sayres, Conor and Schneider, Donald P. and Seo, Hee-Jong and Shafieloo, Arman and Slosar, An\ifmmode \check{z}\else \v{z}\fi{}e and Smith, Alex and Stermer, Julianna and Tamone, Amelie and Tinker, Jeremy L. and Tojeiro, Rita and Vargas-Maga\~na, Mariana and Variu, Andrei and Wang, Yuting and Weaver, Benjamin A. and Weijmans, Anne-Marie and Y\`eche, Christophe and Zarrouk, Pauline and Zhao, Cheng and Zhao, Gong-Bo and Zheng, Zheng},
  journal = {Phys. Rev. D},
  volume = {103},
  issue = {8},
  pages = {083533},
  numpages = {43},
  year = {2021},
  month = {Apr},
  publisher = {American Physical Society},
  doi = {10.1103/PhysRevD.103.083533},
  url = {https://link.aps.org/doi/10.1103/PhysRevD.103.083533}
}

@article{sdss,
doi = {10.1086/466512},
url = {https://dx.doi.org/10.1086/466512},
year = {2005},
month = {nov},
publisher = {},
volume = {633},
number = {2},
pages = {560},
author = {Eisenstein, Daniel J. and Zehavi, Idit and Hogg, David W. and Scoccimarro, Roman and Blanton, Michael R. and Nichol, Robert C. and Scranton, Ryan and Seo, Hee-Jong and Tegmark, Max and Zheng, Zheng and Anderson, Scott F. and Annis, Jim and Bahcall, Neta and Brinkmann, Jon and Burles, Scott and Castander, Francisco J. and Connolly, Andrew and Csabai, Istvan and Doi, Mamoru and Fukugita, Masataka and Frieman, Joshua A. and Glazebrook, Karl and Gunn, James E. and Hendry, John S. and Hennessy, Gregory and Ivezić, Zeljko and Kent, Stephen and Knapp, Gillian R. and Lin, Huan and Loh, Yeong-Shang and Lupton, Robert H. and Margon, Bruce and McKay, Timothy A. and Meiksin, Avery and Munn, Jeffery A. and Pope, Adrian and Richmond, Michael W. and Schlegel, David and Schneider, Donald P. and Shimasaku, Kazuhiro and Stoughton, Christopher and Strauss, Michael A. and SubbaRao, Mark and Szalay, Alexander S. and Szapudi, István and Tucker, Douglas L. and Yanny, Brian and York, Donald G.},
title = "{Detection of the Baryon Acoustic Peak in the Large-Scale Correlation Function of SDSS Luminous Red Galaxies}",
journal = {\apj}
}

@article{2dFGRS,
    author = {Cole, Shaun and Percival, Will J. and Peacock, John A. and Norberg, Peder and Baugh, Carlton M. and Frenk, Carlos S. and Baldry, Ivan and Bland-Hawthorn, Joss and Bridges, Terry and Cannon, Russell and Colless, Matthew and Collins, Chris and Couch, Warrick and Cross, Nicholas J. G. and Dalton, Gavin and Eke, Vincent R. and de Propris, Roberto and Driver, Simon P. and Efstathiou, George and Ellis, Richard S. and Glazebrook, Karl and Jackson, Carole and Jenkins, Adrian and Lahav, Ofer and Lewis, Ian and Lumsden, Stuart and Maddox, Steve and Madgwick, Darren and Peterson, Bruce A. and Sutherland, Will and Taylor, Keith and The 2dFGRS Team},
    title = "{The 2dF Galaxy Redshift Survey: power-spectrum analysis of the final data set and cosmological implications}",
    journal = {\mnras},
    volume = {362},
    number = {2},
    pages = {505-534},
    year = {2005},
    month = {09},
    issn = {0035-8711},
    doi = {10.1111/j.1365-2966.2005.09318.x},
    url = {https://doi.org/10.1111/j.1365-2966.2005.09318.x}
}

@article{6dFGRS,
   title="{The 6dF Galaxy Survey: baryon acoustic oscillations and the local Hubble constant: 6dFGS: BAOs and the local Hubble constant}",
   volume={416},
   ISSN={0035-8711},
   url={http://dx.doi.org/10.1111/j.1365-2966.2011.19250.x},
   DOI={10.1111/j.1365-2966.2011.19250.x},
   number={4},
   journal={\mnras},
   publisher={Oxford University Press (OUP)},
   author={Beutler, Florian and Blake, Chris and Colless, Matthew and Jones, D. Heath and Staveley-Smith, Lister and Campbell, Lachlan and Parker, Quentin and Saunders, Will and Watson, Fred},
   year={2011},
   month=jul, pages={3017–3032} }

@article{boss,
    author = {Alam, Shadab and Ata, Metin and Bailey, Stephen and Beutler, Florian and Bizyaev, Dmitry and Blazek, Jonathan A. and Bolton, Adam S. and Brownstein, Joel R. and Burden, Angela and Chuang, Chia-Hsun and Comparat, Johan and Cuesta, Antonio J. and Dawson, Kyle S. and Eisenstein, Daniel J. and Escoffier, Stephanie and Gil-Marín, Héctor and Grieb, Jan Niklas and Hand, Nick and Ho, Shirley and Kinemuchi, Karen and Kirkby, David and Kitaura, Francisco and Malanushenko, Elena and Malanushenko, Viktor and Maraston, Claudia and McBride, Cameron K. and Nichol, Robert C. and Olmstead, Matthew D. and Oravetz, Daniel and Padmanabhan, Nikhil and Palanque-Delabrouille, Nathalie and Pan, Kaike and Pellejero-Ibanez, Marcos and Percival, Will J. and Petitjean, Patrick and Prada, Francisco and Price-Whelan, Adrian M. and Reid, Beth A. and Rodríguez-Torres, Sergio A. and Roe, Natalie A. and Ross, Ashley J. and Ross, Nicholas P. and Rossi, Graziano and Rubiño-Martín, Jose Alberto and Saito, Shun and Salazar-Albornoz, Salvador and Samushia, Lado and Sánchez, Ariel G. and Satpathy, Siddharth and Schlegel, David J. and Schneider, Donald P. and Scóccola, Claudia G. and Seo, Hee-Jong and Sheldon, Erin S. and Simmons, Audrey and Slosar, Anže and Strauss, Michael A. and Swanson, Molly E. C. and Thomas, Daniel and Tinker, Jeremy L. and Tojeiro, Rita and Magaña, Mariana Vargas and Vazquez, Jose Alberto and Verde, Licia and Wake, David A. and Wang, Yuting and Weinberg, David H. and White, Martin and Wood-Vasey, W. Michael and Yèche, Christophe and Zehavi, Idit and Zhai, Zhongxu and Zhao, Gong-Bo},
    title = "{The clustering of galaxies in the completed SDSS-III Baryon Oscillation Spectroscopic Survey: cosmological analysis of the DR12 galaxy sample}",
    journal = {\mnras},
    volume = {470},
    number = {3},
    pages = {2617-2652},
    year = {2017},
    month = {03},
    issn = {0035-8711},
    doi = {10.1093/mnras/stx721},
    url = {https://doi.org/10.1093/mnras/stx721}
}

@article{wigglez,
    author = {Blake, Chris and Brough, Sarah and Colless, Matthew and Contreras, Carlos and Couch, Warrick and Croom, Scott and Croton, Darren and Davis, Tamara M. and Drinkwater, Michael J. and Forster, Karl and Gilbank, David and Gladders, Mike and Glazebrook, Karl and Jelliffe, Ben and Jurek, Russell J. and Li, I-hui and Madore, Barry and Martin, D. Christopher and Pimbblet, Kevin and Poole, Gregory B. and Pracy, Michael and Sharp, Rob and Wisnioski, Emily and Woods, David and Wyder, Ted K. and Yee, H. K. C.},
    title = "{The WiggleZ Dark Energy Survey: joint measurements of the expansion and growth history at z {$<$} 1}",
    journal = {\mnras},
    volume = {425},
    number = {1},
    pages = {405-414},
    year = {2012},
    month = {09},
    issn = {0035-8711},
    doi = {10.1111/j.1365-2966.2012.21473.x},
    url = {https://doi.org/10.1111/j.1365-2966.2012.21473.x},
}

@ARTICLE{boss_dr12,
       author = {{Reid}, Beth and {Ho}, Shirley and {Padmanabhan}, Nikhil and {Percival}, Will J. and {Tinker}, Jeremy and {Tojeiro}, Rita and {White}, Martin and {Eisenstein}, Daniel J. and {Maraston}, Claudia and {Ross}, Ashley J. and {S{\'a}nchez}, Ariel G. and {Schlegel}, David and {Sheldon}, Erin and {Strauss}, Michael A. and {Thomas}, Daniel and {Wake}, David and {Beutler}, Florian and {Bizyaev}, Dmitry and {Bolton}, Adam S. and {Brownstein}, Joel R. and {Chuang}, Chia-Hsun and {Dawson}, Kyle and {Harding}, Paul and {Kitaura}, Francisco-Shu and {Leauthaud}, Alexie and {Masters}, Karen and {McBride}, Cameron K. and {More}, Surhud and {Olmstead}, Matthew D. and {Oravetz}, Daniel and {Nuza}, Sebasti{\'a}n E. and {Pan}, Kaike and {Parejko}, John and {Pforr}, Janine and {Prada}, Francisco and {Rodr{\'\i}guez-Torres}, Sergio and {Salazar-Albornoz}, Salvador and {Samushia}, Lado and {Schneider}, Donald P. and {Sc{\'o}ccola}, Claudia G. and {Simmons}, Audrey and {Vargas-Magana}, Mariana},
        title = "{SDSS-III Baryon Oscillation Spectroscopic Survey Data Release 12: galaxy target selection and large-scale structure catalogues}",
      journal = {\mnras},
         year = 2016,
        month = jan,
       volume = {455},
       number = {2},
        pages = {1553-1573},
          doi = {10.1093/mnras/stv2382},
archivePrefix = {arXiv},
       eprint = {1509.06529},
 primaryClass = {astro-ph.CO},
       adsurl = {https://ui.adsabs.harvard.edu/abs/2016MNRAS.455.1553R}
}

@article{des_y6,
  title = "{Dark Energy Survey: A 2.1\% measurement of the angular baryonic acoustic oscillation scale at redshift ${z}_{\mathrm{eff}}=0.85$ from the final dataset}",
  author = {Abbott, T. M. C. and Adamow, M. and Aguena, M. and Allam, S. and Alves, O. and Amon, A. and Andrade-Oliveira, F. and Asorey, J. and Avila, S. and Bacon, D. and Bechtol, K. and Bernstein, G. M. and Bertin, E. and Blazek, J. and Bocquet, S. and Brooks, D. and Burke, D. L. and Camacho, H. and Carnero Rosell, A. and Carollo, D. and Carr, A. and Carretero, J. and Castander, F. J. and Cawthon, R. and Chan, K. C. and Chang, C. and Conselice, C. and Costanzi, M. and Crocce, M. and da Costa, L. N. and Pereira, M. E. S. and Davis, T. M. and De Vicente, J. and Deiosso, N. and Desai, S. and Diehl, H. T. and Dodelson, S. and Doux, C. and Drlica-Wagner, A. and Elvin-Poole, J. and Everett, S. and Ferrero, I. and Fert\'e, A. and Flaugher, B. and Fosalba, P. and Frieman, J. and Garc\'{\i}a-Bellido, J. and Gaztanaga, E. and Giannini, G. and Glazebrook, K. and Gruendl, R. A. and Gutierrez, G. and Hartley, W. G. and Hinton, S. R. and Hollowood, D. L. and Honscheid, K. and Huterer, D. and James, D. J. and Kent, S. and Kuehn, K. and Lahav, O. and Lee, S. and Lewis, G. F. and Lidman, C. and Lima, M. and Lin, H. and Malik, U. and Maraston, C. and Marshall, J. L. and Martini, P. and Mena-Fern\'andez, J. and Menanteau, F. and Miquel, R. and Mohr, J. J. and Myles, J. and M\"oller, A. and Nichol, R. C. and Ogando, R. L. C. and Palmese, A. and Percival, W. J. and Pieres, A. and Plazas Malag\'on, A. A. and Porredon, A. and Prat, J. and Rodr\'{\i}guez-Monroy, M. and Romer, A. K. and Roodman, A. and Rosenfeld, R. and Ross, A. J. and Rykoff, E. S. and Sako, M. and Samuroff, S. and S\'anchez, C. and Sanchez, E. and Sanchez Cid, D. and Santiago, B. and Schubnell, M. and Sevilla-Noarbe, I. and Sheldon, E. and Smith, M. and Suchyta, E. and Swanson, M. E. C. and Tarle, G. and Thomas, D. and To, C. and Toribio San Cipriano, L. and Troxel, M. A. and Tucker, B. E. and Tucker, D. L. and Walker, A. R. and Weaverdyck, N. and Weller, J. and Wiseman, P. and Yanny, B.},
  collaboration = {DES},
  journal = {Phys. Rev. D},
  volume = {110},
  issue = {6},
  pages = {063515},
  numpages = {39},
  year = {2024},
  month = {Sep},
  publisher = {American Physical Society},
  doi = {10.1103/PhysRevD.110.063515},
  url = {https://link.aps.org/doi/10.1103/PhysRevD.110.063515}
}

@article{euclid_bao,
    author = "Duret, V. and others",
    collaboration = "Euclid",
    title = "{Euclid preparation. BAO analysis of photometric galaxy clustering in configuration space}",
    eprint = "arXiv:2503.11621",
    archivePrefix = "arXiv",
    primaryClass = "astro-ph.CO",
    month = "3",
    year = "2025"
}

@article{qpm,
   title={Mock galaxy catalogues using the quick particle mesh method},
   volume={437},
   ISSN={1365-2966},
   url={http://dx.doi.org/10.1093/mnras/stt2071},
   DOI={10.1093/mnras/stt2071},
   number={3},
   journal={\mnras},
   publisher={Oxford University Press (OUP)},
   author={White, Martin and Tinker, Jeremy L. and McBride, Cameron K.},
   year={2013},
   month=nov, pages={2594–2606} }

@article{sdss_ezmock,
   title="{The completed SDSS-IV extended Baryon Oscillation Spectroscopic Survey: 1000 multi-tracer mock catalogues with redshift evolution and systematics for galaxies and quasars of the final data release}",
   volume={503},
   ISSN={1365-2966},
   url={http://dx.doi.org/10.1093/mnras/stab510},
   DOI={10.1093/mnras/stab510},
   number={1},
   journal={\mnras},
   publisher={Oxford University Press (OUP)},
   author={Zhao, Cheng and Chuang, Chia-Hsun and Bautista, Julian and de Mattia, Arnaud and Raichoor, Anand and Ross, Ashley J and Hou, Jiamin and Neveux, Richard and Tao, Charling and Burtin, Etienne and Dawson, Kyle S and de la Torre, Sylvain and Gil-Marín, Héctor and Kneib, Jean-Paul and Percival, Will J and Rossi, Graziano and Tamone, Amélie and Tinker, Jeremy L and Zhao, Gong-Bo and Alam, Shadab and Mueller, Eva-Maria},
   year={2021},
   month=feb, pages={1149–1173} }

@article{sdss_recon,
    author = {Vargas-Magaña, Mariana and Ho, Shirley and Fromenteau, Sebastien. and Cuesta, Antonio. J.},
    title = "{The clustering of galaxies in the SDSS-III Baryon Oscillation Spectroscopic Survey: effect of smoothing of density field on reconstruction and anisotropic BAO analysis}",
    journal = {\mnras},
    volume = {467},
    number = {2},
    pages = {2331-2348},
    year = {2017},
    month = {01},
    issn = {0035-8711},
    doi = {10.1093/mnras/stx048},
    url = {https://doi.org/10.1093/mnras/stx048},

}

\end{document}